\pdfoutput=1
\documentclass[a4paper,11pt]{article}

\usepackage{jcappub}

\usepackage[utf8]{inputenc}
\usepackage{lmodern}
\usepackage{amsfonts}
\usepackage{dsfont}
\usepackage{siunitx}
\usepackage[version=4]{mhchem}
\usepackage{placeins}
\usepackage{framed}
\usepackage{lipsum}
\usepackage{xspace}
\usepackage[dvipsnames]{xcolor}
\usepackage{hyperref}
\usepackage{graphicx}
\usepackage{float}
\usepackage{dcolumn}
\usepackage{bm}
\usepackage{lmodern}
\usepackage{comment} 
\usepackage{soul}
\usepackage{hyphenat}
\usepackage{subcaption}
\usepackage{multirow}

\usepackage[showframe=false]{geometry}
\usepackage{changepage}

\usepackage{booktabs,threeparttable,makecell}
\usepackage{bbold}
\usepackage{pifont}

\definecolor{myorange}{RGB}{199,146,32}

\newcommand{\be}{\begin{equation}}
\newcommand{\ee}{\end{equation}}
\newcommand{\bea}{\begin{eqnarray}}
\newcommand{\eea}{\end{eqnarray}}
\newcommand{\BiFeSe}{(Bi$_{1-x}$Fe$_x$)$_2$Se$_3$\xspace}
\newcommand{\MnBiTe}[3]{Mn$_{#1}$Bi$_{#2}$Te$_{#3}$\xspace}

\newcommand{\safemaths}[1]{\ensuremath{#1}\xspace}

\newcommand{\order}{\mathcal{O}}
\newcommand{\eul}{\safemaths{\mathrm{e}}}
\newcommand{\dd}{\mathrm{d}}

\newcommand{\Mpl}{M_\text{Pl}}
\newcommand{\WpsqHz}{\W/\sqrt{\Hz}}
\newcommand{\alphaEM}{\alpha}

\newcommand{\Vuc}{V_\text{u.c.}}
\newcommand{\dyAQ}{\safemaths{\delta\Theta}}
\newcommand{\stAQ}{\safemaths{\Theta^0}}
\newcommand{\makeAQ}[1]{\safemaths{#1_\Theta}}
\newcommand{\mAQ}{\makeAQ{m}}
\newcommand{\kAQ}{\makeAQ{k}}
\newcommand{\nAQ}{\makeAQ{n}}
\newcommand{\fAQ}{\makeAQ{f}}

\newcommand{\vdm}{v_\text{\tiny DM}}
\newcommand{\mDA}{\safemaths{m_a}}
\newcommand{\fDA}{\safemaths{f_a}}
\newcommand{\gDA}{\safemaths{g_{a\gamma}}}

\newcommand{\kp}{\safemaths{k_p}}

\newcommand{\GammaM}{\safemaths{\Gamma_m}}
\newcommand{\GammaRho}{\safemaths{\Gamma_\rho}}

\newcommand{\vc}[1]{\safemaths{\bm{#1}}}
\newcommand{\mat}[1]{\safemaths{\mathbf{#1}}}

\newcommand{\Bext}{\safemaths{B_\text{e}}}

\newcommand{\omLO}{\omega_\text{\tiny LO}}
\newcommand{\defaultSetupTMI}{We use typical material values for a Mn$_2$Bi$_2$Te$_5$ TMI, cf.\ table~\ref{tab:derived_params} and eq.~\eqref{eqn:b_meV}, and $n = 5$ and $\mu = 1$\xspace}
\newcommand{\defaultSetup}{\defaultSetupTMI. The external $B$-field is $B_e = \SI{2}{\tesla}$\xspace}

\title{Axion Quasiparticles for Axion Dark Matter Detection}

\author[a,b,c]{Jan Sch\"{u}tte-Engel,}
\author[d]{David J. E. Marsh,}
\author[e,f]{Alexander J. Millar,}
\author[g]{Akihiko Sekine,}
\author[h]{Francesca Chadha-Day,}
\author[d]{Sebastian Hoof,}
\author[i,j]{Mazhar N. Ali,}
\author[k]{Kin Chung Fong,}
\author[l]{Edward Hardy,}
\author[m,n]{and Libor \v{S}mejkal}

\affiliation[a]{Department of Physics, University of Illinois at Urbana-Champaign,\\ Urbana, IL 61801, U.S.A.}
\affiliation[b]{Illinois Center for Advanced Studies of the Universe,University of Illinois at Urbana-Champaign, Urbana, IL 61801, U.S.A.}
\affiliation[c]{Hamburg University, 22761 Hamburg, Germany}
\affiliation[d]{Institut f\"{u}r Astrophysik, Georg-August-Universit\"{a}t G\"{o}ttingen, Friedrich-Hund-Platz~1, 37077 G\"{o}ttingen, Germany}
\affiliation[e]{The Oskar Klein Centre for Cosmoparticle Physics, Department of Physics, Stockholm University, AlbaNova, 10691 Stockholm, Sweden}
\affiliation[f]{Nordita, KTH Royal Institute of Technology and Stockholm University, Roslagstullsbacken 23, 10691 Stockholm, Sweden}
\affiliation[g]{RIKEN Center for Emergent Matter Science, Wako, Saitama 351-0198, Japan}
\affiliation[h]{Department of Physics, University of Durham, South Rd, \\ Durham DH1 3LE, United Kingdom}
\affiliation[i]{Max Planck Institute of Microstructure Physics, Weinberg 2, 06120 Halle | Germany}
\affiliation[j]{Kavli Institute of Nanoscience, Delft University of Technology, Delft, Netherlands}
\affiliation[k]{Raytheon BBN Technologies, Quantum Engineering and Computing, \\ Cambridge, Massachusetts 02138, USA}
\affiliation[l]{Mathematical Sciences, The University of Liverpool, \\ Liverpool, L69 7ZL, United Kingdom}
\affiliation[m]{Institut f\"{u}r Physik, Johannes Gutenberg Universit\"{a}t Mainz, 55128 Mainz, Germany}
\affiliation[n]{Institute of Physics, Czech Academy of Sciences, Cukrovarnick\'{a} 10, 162 00,\\ Praha 6, Czech Republic}

\emailAdd{jans@illinois.edu}
\emailAdd{david.j.marsh@kcl.ac.uk}
\emailAdd{alexander.millar@fysik.su.se}
\emailAdd{akihiko.sekine1@gmail.com}
\emailAdd{francesca.chadha-day@durham.ac.uk}
\emailAdd{hoof@uni-goettingen.de}
\emailAdd{maz@berkeley.edu}
\emailAdd{fongkc@gmail.com}
\emailAdd{edward.hardy@liverpool.ac.uk}
\emailAdd{lsmejkal@uni-mainz.de}

\abstract{It has been suggested that certain antiferromagnetic topological insulators contain axion quasiparticles (AQs), and that such materials could be used to detect axion dark matter (DM). The AQ is a longitudinal antiferromagnetic spin fluctuation coupled to the electromagnetic  Chern-Simons term, which, in the presence of an applied magnetic field, leads to mass mixing between the AQ and the electric field. The electromagnetic boundary conditions and transmission and reflection coefficients are computed. A model for including losses into this system is presented, and the resulting linewidth is computed. It is shown how transmission spectroscopy can be used to measure the resonant frequencies and damping coefficients of the material, and demonstrate conclusively the existence of the AQ. The dispersion relation and boundary conditions permit resonant conversion of axion DM into THz photons in a material volume that is independent of the resonant frequency, which is tuneable via an applied magnetic field.  A parameter study for axion DM detection is performed, computing boost amplitudes and bandwidths using realistic material properties including loss. The proposal could allow for detection of axion DM in the mass range between 1 and 10 meV using current and near future technology. \\

Preprints: IPPP/20/78, NORDITA-2021-007}

\begin{document}
\maketitle
\flushbottom

\section{Introduction}\label{sec:intro}

The quantum chromodynamics (QCD) axion~\cite{pecceiquinn1977,weinberg1978,wilczek1978} solves the charge-parity ($\mathcal{C}\mathcal{P}$) problem of the strong nuclear force~\cite{Vafa:1984xg,Afach:2015sja,Hook:2018dlk}, and is a plausible candidate~\cite{1983PhLB..120..133A,1983PhLB..120..137D,1983PhLB..120..127P} to compose the dark matter~(DM) in the cosmos~\cite{2016A&A...594A..13P}. The axion mass is bounded from above~\cite{2008LNP...741...51R,Chang:2018rso,1906.11844} and below~\cite{2015PhRvD..91h4011A,Stott:2018opm} by astrophysical constraints~(for reviews, see Refs.~\cite{my_review,2017JCAP...10..010G,1801.08127,DiLuzio:2020wdo}, and Appendix~\ref{sec:axion_dm}), placing it in the range
\begin{equation}
	\SI{1}{peV} \lesssim \mDA \lesssim \SI{20}{\meV} \, .
	\label{eqn:allowed_axion_mass}
\end{equation}

The local DM density is known from stellar motions in the Milky~Way~\cite{PhysRevD.98.030001}. Assuming axions comprise all the (local) DM, the axion number density is given by $n_a = \rho_\text{loc}/\mDA$. Due to the very small axion mass, the number density is very large and axions can be modelled as a coherent classical field, $\phi$. The field value is:
\be
	\phi = \Phi \cos(\mDA t) \, ,
	\label{eqn:dark_axion_field}
\ee
where $\Phi$ is Rayleigh-distributed~\cite{Foster:2017hbq,1905.13650} with mean $\sqrt{2\rho_\text{loc}}/\mDA$ and linewidth $\Delta \omega/\omega\sim 10^{-6}$ given by the Maxwell-Boltzmann distribution of axion velocities around the local galactic circular speed, $v_\text{loc}\approx \SI{200}{\km/\s}$ (see e.g.\ refs.~\cite{OHare:2017yze,Foster:2017hbq}).

Axions couple to electromagnetism via the interaction $\mathcal{L}=\gDA\phi \vc{E}\cdot\vc{B}$. Thus, in the presence of an applied magnetic field, $\vc{B}_0$, the DM axion field in Eq.~\eqref{eqn:dark_axion_field} acts as a source for the electric field, $\vc{E}$. This is the inverse Primakoff process for axions, and leads to axion-photon conversion in a magnetic field. The rate of axion-photon conversion depends on the unknown value of the coupling $\gDA$ and happens at an unknown frequency $\omega = \mDA\pm\Delta\omega$. For the QCD axion (as opposed to a generic ``axion like particle''~\cite{my_review}) the mass and coupling are linearly related, $\gDA\propto\mDA$, although different models for the Peccei-Quinn~\cite{pecceiquinn1977} charges of fundamental fermions predict different values for the constant of proportionality. The two historical reference models of Kim-Shifman-Vainshtein-Zhakarov (KSVZ)~\cite{1979PhRvL..43..103K,1980NuPhB.166..493S} and Dine-Fischler-Srednicki-Zhitnitsky (DFSZ)~\cite{Zhitnitsky:1980tq,1981PhLB..104..199D} span a narrow range, while more recent generalisations with non-minimal particle content allow for more variation~\cite{DiLuzio:2016sbl,1705.05370,DiLuzio:2020wdo}.

The axion-photon coupling $\gDA$ is constrained by a large number of null-results from experimental searches and astrophysical considerations~\cite{PhysRevD.98.030001}. For experimentally allowed values of $(\mDA,\gDA)$, and accessible magnetic field strengths, the photon production rate in vacuum is unobservably small. The power can be increased in two basic ways. If the conversion happens along the surface of a magnetized mirror, then the produced photons can be focused onto a detector~\cite{2013JCAP...04..016H}. This approach is broadband, and does not depend on the axion mass. Reaching sensitivity to the QCD~axion requires very large mirrors, very sensitive detectors, and control over environmental noise. Alternatively, the signal can be resonantly or coherently enhanced (e.g.\ Refs.~\cite{Braine:2019fqb,1983PhRvL..51.1415S,Braine:2019fqb,TheMADMAXWorkingGroup:2016hpc,2014PhRvX...4b1030B,Melcon:2020xvj,Semertzidis:2019gkj,Backes:2020ajv,Ouellet:2018beu,Michimura:2019qxr,Crescini:2020cvl}). These approaches are narrow band, and require tuning to the unknown DM axion frequency. 

Depending on the model of early Universe cosmology, and the evolution of the axion field at high temperatures $T\gg \SI{1}{\MeV}$, the entire allowed mass range Eq.~\eqref{eqn:allowed_axion_mass} can plausibly explain the observed DM abundance. The mass range near \SI{1}{\meV} (corresponding to frequencies in the low \si{\THz}) is favoured in some models of axion cosmology (see Appendix~\ref{sec:axion_dm}), but is challenging experimentally due to the lack of large volume, tuneable THz resonators, and efficient, low-noise, large bandwidth \si{\THz} detectors. 

In Ref.~\cite{2019PhRvL.123l1601M} (Paper~I) we proposed an experimental scheme to detect axion DM using axion-quasiparticle (AQ) materials based on topological magnetic insulators (TMIs)~\cite{2010NatPh...6..284L}, a proposal we called ``TOORAD'' for ``TOpolOgical Resonant Axion Detection''. Since Li et al.~\cite{2010NatPh...6..284L} first proposed to realise axion quasiparticles in the antiferromagnetic topological insulator (AF-TI) \ce{Fe}-doped Bismuth Selenide, \BiFeSe, the quest to realise related materials in the lab has picked up incredible pace. A currently favoured candidate \MnBiTe{2}{2}{5}~\cite{Zhang_2020}, is, however, yet to be fabricated successfully. AQ materials allow the possibility to explore aspects of axion physics in the laboratory~\cite{PhysRevLett.58.1799}. The AQ resonance hyrbidises with the electric field forming an axion-polariton~\cite{2010NatPh...6..284L}. The polariton frequency is of order the AF anisotropy field, with typical values $\mathcal{O}(\SI{1}{meV})$, and is tuneable with applied static field $B$~\cite{2019PhRvL.123l1601M}. This proposal opens the possibility for large volume THz resonance, easily tuneable with an applied magnetic field, thus overcoming the first hurdle to detection of meV axions. The proposal makes use of the current interest in manufacture of low noise, high efficiency single photon detectors (SPDs) in THz~\cite{2019_Lee}. The development of such detectors has benefits for sub mm astronomy and cosmology, as well as application to other DM direct detection experiments~\cite{2013JCAP...04..016H}.

The present paper expands on the ideas outlined in Paper~I with more in depth modelling and calculations. A guide to the results is given below.

\subsection*{Axion Quasiparticle Materials}

We begin with a detailed treatment of the materials science, and outline a scheme to prove the existence of AQs in TMIs, and measure their parameters.

\begin{itemize}

\item We introduce the basic model for the equations coupling the electric field and the AQ. There are two parameters that determine the model: the AQ mass, $\mAQ$, and the decay constant, $\fAQ$, as summarised in Section~\ref{sec:general_remarks}. 

\item Next  in Section~\ref{sec:af_ti}, we clarify the microscopic model for AQs in TMIs. We begin with the symmetry criteria, followed by a microscopic model based on the Dirac Hamiltonian. The AQ is the longitudinal fluctuataion of the antiferromagnetic order parameter in the Hubbard model. The Appendix summarises the related phenomenon of antiferromagnetic resonance and transverse magnons in the effective field theory of the Heisenberg model.

\item Both $\mAQ$ and $\fAQ$ can be estimated from known material properties. We consider \BiFeSe, the candidate material from Paper~I and Ref.~\cite{2010NatPh...6..284L}, and also the more recent candidate material \MnBiTe{2}{2}{5}~\cite{Zhang_2020}. The results of this study are given in Tables~\ref{tab:material_params} and \ref{tab:derived_params}. 

\item We next consider sources of loss. The largest sources of loss are identified to be conductive losses to the electric field, and crystal and magnetic domain induced line broadening for the AQ. The loss model is summarised in Table~\ref{tab:losses}.

\item Using the model thus developed, we present a computation of the transmission spectrum of an AQ material. The spectrum shows two peaks due to the mixing of the electric field and the AQ, the locations of which can be used to measure the parameters $\mAQ$ and $\fAQ$. The width of the resonances provides a measurement of the loss parameters on resonance, which cannot otherwise be identified from existing measurements. Such a measurement can be performed using THz time domain spectroscopy~\cite{2017PhRvL.119v7201L}. The procedure is shown schematically in Fig.~\ref{fig:ATR}

\end{itemize}

\subsection*{Axion Dark Matter Detection}

\begin{itemize}
\item Axion DM acts as a source to the AQ model developed in the previous sections. Axion-photon conversion in a magnetic field sources photons, which hybridize with the AQ forming polaritons, and thus acquire an effective mass. It is shown that this model can be treated in the same way as a dielectric haloscope~\cite{Millar:2016cjp}. The resonance in the polariton spectrum leads to an effective refractive index $n<1$, and an enhancement of the axion-induced electric field, see Fig.~\ref{fig:DA_AQ_P_interpretation}. 
\item We compute the power boost amplitude, $\beta(\omega)$, for a range of plausible values for the model parameters, losses, and material thickness. See, for example, Fig.~\ref{fig:DA_AQ_P_mixing_Signal_Losses_main}. 
\item The power enhancement is driven by the material thickness, $d$, which should \emph{exceed} the wavelength of emitted photons. When losses are included, we identify a maximum thickness above which the power enhancement decreases due to the finite skin-depth. See Fig.~\ref{fig:betaloss}.
\item We perform forecasts for the limits on axion DM parameter space, $(\mDA,\gDA)$, that can be obtained for a range of plausible material and THz detector parameters. We identify pessimistic and optimistic possibilities for the discovery reach, summarised in Fig.~\ref{fig:summary_plot}.
\end{itemize}

We use units $\hbar=c=k_B=1$ throughout most of the text, in combination with SI where appropriate.

\section{Axion Quasiparticle Materials}
\label{sec:AQ-Materials}

\subsection{General Remarks}\label{sec:general_remarks}
Axion quasiparticles (AQs) are defined, for our purposes, as a degree of freedom, denoted by \dyAQ, coupled to the electromagnetic Chern-Simons term:
\be
	S_{\text{topo}} = \frac{\alphaEM}{\pi}\int \dd^4x \; (\dyAQ + \stAQ) \, \vc{E} \cdot \vc{B} \, ,
\label{eqn:chern_simons}
\ee
where \stAQ is the constant electromagnetic Chern-Simons term, equal to zero in ordinary insulators and $\pi$ in topological insulators~(TIs). In these materials, surface currents are accounted for by inclusion of a non-zero value for \stAQ (the topological magneto-electric effect due to the Hall conductivity~\cite{PhysRevLett.102.146805,TongNotes}). In the presence of a dynamical AQ field, $\dyAQ$, the static vacuum value $\stAQ$ is allowed to take on a continuum of values between $0$ and $\pi$. The total axion field is denoted by $\Theta=\dyAQ + \stAQ$. We review these concepts further below, for a detailed presentation see Ref.~\cite{Sekine2020}.


The dynamics of the AQs are described by~\cite{2010NatPh...6..284L,Sekine:2014xva}
\be
S_\Theta = \frac{\fAQ^2}{2} \int d^4x\left[\left(\partial_t\dyAQ\right)^2-\left(v_i\partial_i\dyAQ\right)^2-\mAQ^2\dyAQ^2\right],
\label{eqn:ActionKleinGordon}
\ee
where $\fAQ$, $v_i$ and $\mAQ$ are the stiffness, velocity and mass of the AQ. The velocities $v_i$ are of the order of the spin wave speed in typical antiferromagnets, $v_s\sim 10^{-4}c$, see for example Ref.~\cite{Pickart}. In the coupled equations of motion for the electric field and the AQ (see Section~\ref{sec:transmission}), $\fAQ$ enters in the combination\footnote{Note that we use the Lorentz-Heaviside convention, where $\SI{1}{\tesla} \approx \SI{195}{\eV^2}$.}
\be
b = \frac{\alphaEM}{\pi\sqrt{2}}\frac{\Bext}{\sqrt{\epsilon} \fAQ} = \SI{1.6}{\meV} \, \left(\frac{25}{\epsilon_1} \right)^{1/2} \left(\frac{\Bext}{\SI{2}{\tesla}}\right) \left(\frac{\SI{70}{\eV}}{\fAQ}\right)\, .
\label{eqn:b_meV}
\ee
 
In addition to the action for the AQ we consider electromagnetic fields governed by Maxwell's equations in media, which depend on the complex valued dielectric function, $\tilde{\epsilon} = \epsilon_1+i\epsilon_2= \epsilon_1+i\sigma/\omega$ (where $\sigma$ is the conductivity), and magnetic susceptibility, $\chi_m$. Where there is no room for confusion we use $\epsilon_1=\epsilon$ in some of the following. The phenomenological model also requires the specification of a loss matrix, $\mat{\Gamma}$.

\subsection{Realisation in Dirac Quasiparticle Antiferromagnets}\label{sec:af_ti}
The idea to realise axion electrodynamics in solids was originally developed by Wilczek~\cite{PhysRevLett.58.1799} who, however, could not identify a magnetic solid that breaks parity and time-reversal while preserving its combination: as we will see, necessary conditions for AQs. Recent developments in nonmagnetic and magnetic electronic topological phases of matter, and study of the topological magnetoelectric effect associated with the Chern-Simons term in magnetoelectrics~\cite{Hasan2010,Qi2011} have led to the identification of several routes to realise axion electrodynamics in energy bands of magnetic topological insulators and Dirac quasiparticle antiferromagnets. The electronic, magnetic, topological energy bands can couple to spin fluctuations, and thus generate a dynamical axion phase on the electromagnetic Chern-Simons term. 

In this section we discuss the Dirac quasiparticle model of AQs in electronic energy bands. We compare the symmetry criteria for static and dynamical axion topological antiferromagnets, and discuss the most prominent material candidates. 

\subsubsection{Symmetry criteria for static and dynamical magnetic axion insulators}
The topological $\Theta$ term is called also an \textit{axion angle} as it can take any value between 0 and $2\pi$. The operations of charge conjugation $\mathcal{C}$, parity $\mathcal{P}$ (known as inversion symmetry in condensed matter, a terminology we adopt throughout this section to distinguish it from other types of parity operation in solids), and time-reversal $\mathcal{T}$ are the discrete symmetries constraining the values of $\Theta$, and which define the properties of fundamental forces in nature via the $\mathcal{CPT}$ theorem. 
$\mathcal{CP}$ breaking means that the physical laws are not invariant under combination of interchanging particle with its antiparticle with inverting the spatial coordinates. If $\Theta \neq 0,\pi$, then $\mathcal{CP}$ is violated. The combined $\mathcal{CPT}$ symmetry is believed to be preserved (i.e. the so called $\mathcal{CPT}$ theorem) and thus the violation of $\mathcal{CP}$ implies the violation of $\mathcal{T}$ symmetry, i.e. the reversal of the time coordinate, and thus particle motion. Realisation of $\mathcal{CP}$-broken theory and axion electrodynamics with non-quantized axion angle can be achieved in materials with broken $\mathcal{T}$ symmetry~\cite{PhysRevLett.58.1799,Wan2012,Varnava2018}. In materials, magnetic ordering can break the $\mathcal{T}$ symmetry. In this section we will discuss the symmetries of magnetic axion insulators which exhibit nonzero pseudoscalar axion quasiparticle $\Theta$ (we use capital letter to label the solid state quasiparticle axion to distinguish it from the DM axion).  

The nonzero axion response can be find in subgroup of conventional and topological magnetoelectric materials. The conventional magneto-electric polarizability tensor is defined as\cite{PhysRevLett.102.146805,Nenno2020}: 
\begin{equation}
\alpha_{i j}=\left(\partial P_{i} / \partial B_{j}\right)_{\mathbf{E}}=\left(\partial M_{j} / \partial E_{i}\right)_{\mathbf{B}}.
\end{equation}
Here $P_{i}, B_{j}, M_{j} $, and $E_{i}$ are electric polarisation, magnetic field, magnetization, and electric field. The magnetoelectric polarizability tensor can be decomposed as~\cite{PhysRevLett.102.146805}:
\begin{equation}
\alpha_{i j}=\tilde{\alpha}_{i j}+\frac{\Theta e^{2}}{2 \pi h} \delta_{i j},
\end{equation}
where the first term is the non-diagonal part of the tensor arising from spin, orbital and ionic contribution~\cite{Fiebig2009}. The second term is the diagonal pseudoscalar part of the coupling related to the axion angle $\Theta$. 

We will now review symmetry criteria for nonzero axion quasiparticle $\Theta$. 
In solid state potentials, discrete symmetries impose severe constraint on the existence and form of the topological axion angle~\cite{Varnava2020}, and provide robust insight into the topological characterisation of the energy bands~\cite{Turner2012,Fang2012b,PhysRevX.7.041069,Watanabe2018c,Elcoro2020,PhysRevB.103.245127}.  
The topological classification assigns two insulators into the same category as long as it is possible to connect the two corresponding Hamiltonians by a continuous deformation without closing an energy gap and while preserving all symmetries~\cite{Hasan2010,Qi2011}. 

Three symmetry based strategies attracted great interest in recent decades. First, solid state quantum field theory considers parity, chiral, and particle-hole symmetries, which are relevant for rather strongly correlated states of matter such as superconductors and lead to abstract multidimensional classification~\cite{Ryu2010}. Second, more numerically feasible symmetry analysis of Wannnier band structure. The Wannier band~structure refers to mixes of real and momentum space band structure, with hybrid Wannier charge centres, which encodes the topological character of given states~\cite{Gresch2016,Varnava2018,Varnava2020}. The formulation is particularly useful for first-principle calculations of the axion angle. 
Third, we can use space group or magnetic space group symmetries to derive symmetry indicators of single particle energy bands~\cite{Smejkal2016,Tang2016,Watanabe2018c,Varnava2020,Elcoro2020}.

The $\boldsymbol{E}.\boldsymbol{B}$ term is odd under time-reversal symmetry, inversion symmetry and any improper rotations, e.g. mirror symmetries~\cite{Fang2012b,Varnava2020}. If the crystal has such a symmetry: 
\begin{equation}
\Theta=-\Theta.
\end{equation}
The symmetry constraint would force any periodic function to vanish. However, $\Theta$ is periodic angle defined only modulo $2\pi$ and thus these symmetries enforce only 
\begin{equation}
\Theta=0\text{, or }\pi.
\end{equation}
When none of these symmetries is present $\Theta$ can be still non-quantized.

\begin{figure}[t!]
\hspace*{0cm}
    \includegraphics[width=1\textwidth]{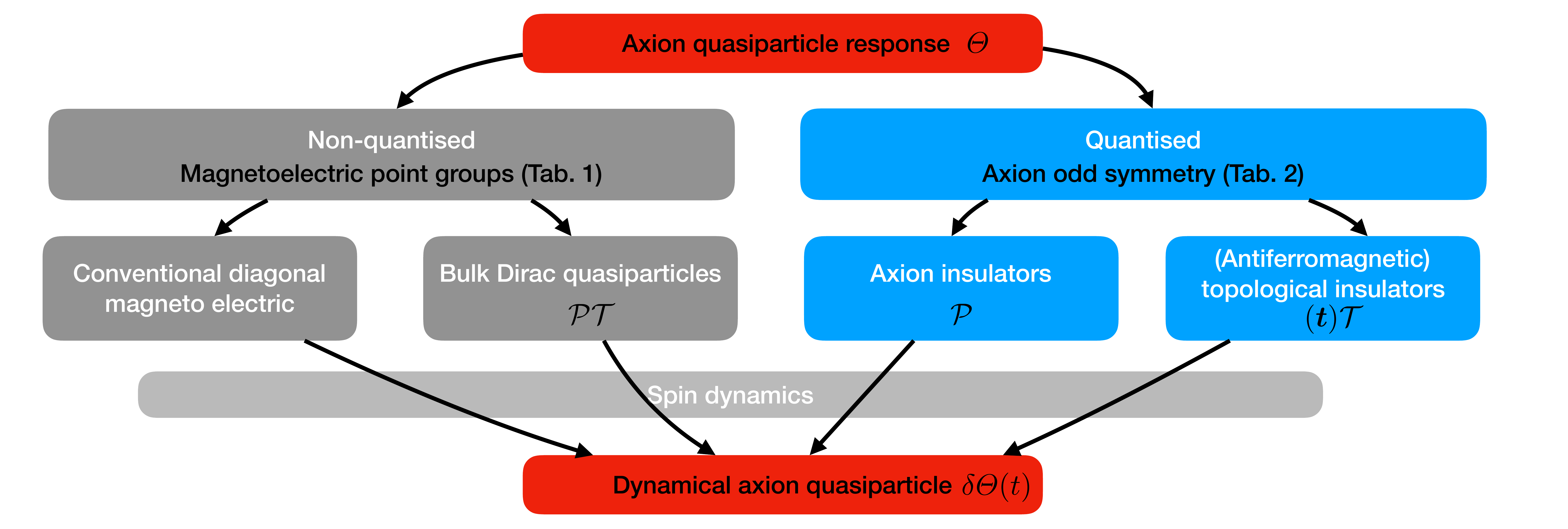}\\   
\caption{Flowchart of generating dynamical axion quasiparticles with four building bock systems.}
\label{fig:af-ti:f0}
\end{figure}

Based on the \textit{magnetic} symmetry classification, we can distinguish four classes of pseudoscalar magneto-electric axion response materials shown in Fig.~\ref{fig:af-ti:f0}. First two classes are conventional magnetoelectrics~\cite{DZYALOSHINSKII1959} and dynamical axion insulators~\cite{2010NatPh...6..284L} with nonzero pseudoscalar part of the magnetoelectric polarisability tensor (and combined $\mathcal{PT}$ symmetry~\cite{Tang2016,Smejkal2016} such as Mn$_{\text{2}}$Bi$_{\text{2}}$Te$_{\text{5}}$~\cite{Zhang_2020}). Second two classes are the topological insulators and axion insulators with quantized magnetoelectric response such as doped Bi$_{\text{2}}$Se$_{\text{3}}$ or MnBi$_{\text{2}}$Te$_{\text{4}}$~\cite{2010NatPh...6..284L,Hasan2010,Qi2011,Nenno2020,Sekine2020}. 

The nonquantized value of $\Theta$ can be find in subset of 58 magnetic point groups allowing for general magnetoelectric response. We summarise in Tab.~\ref{Tab1}~ only the 40 mangetic point groups which allow for the nonzero diagonal magnetoelectric response elements \cite{Siratori1992,Rivera2009}. We also list whether the material has allowed ferromagnetism (FM, 12 magnetic point groups) or is enforced by the point group symmetry to be antiferromagnetic (AF, 28 point groups)~\cite{Bradley} together with several material examples. We see that the magnetoelectric response can be anisotropic what was confirmed experimentally~\cite{Folen1961}. Note that the third row of the Tab.~\ref{Tab1} gives zero trace. This analysis excludes from pseudoscalar magnetoelectric coupling materials which do exhibit only traceless magnetoelectric coupling. When the system breaks $\mathcal{P}$ and $\mathcal{T}$ but preserves its combination, it can host also bulk Dirac quasiparticles~\cite{Smejkal2016}. We mark the  $\mathcal{PT}$ symmetric magnetoelectric pseudoscalar point groups by brackets in Tab.~\ref{Tab1}.

\begin{table}[]
\begin{tabular}{cccc}
Components & FM MPG  & AF MPG & Material \\ \hline \hline
$\left(\begin{array}{lll}\alpha_{xx} & \alpha_{yy}& \alpha_{zz} \end{array}\right)$ & $1$ $2$ $m^{\prime}$ $m^{\prime}m^{\prime}2$  & ($\overline{1}^{\prime}$) ($\boldsymbol{2}/\boldsymbol{m}^{\prime}$)  $222$ ($m^{\prime}m^{\prime}m^{\prime}$) & (Fe,Bi)$_{\text{2}}$Se$_{\text{3}}$ \\  \hline
 & $3$  $4$ $6$ & ($\overline{3}^{\prime}$) $\overline{4}^{\prime}$ $4/m^{\prime}$ $\overline{6}^{\prime}$  $6/m^{\prime}$ &  \\ 
$\left(\begin{array}{lll}\alpha_{xx} & \alpha_{xx} & \alpha_{zz}\end{array}\right)$ &  $3m^{\prime}$  $4m^{\prime}m^{\prime}$   & $32$ ($\overline{\boldsymbol{3}}^{\prime}\boldsymbol{m}^{\prime}$) $422$ $\overline{4}^{\prime}2m^{\prime}$ ($4/m^{\prime}m^{\prime}m^{\prime}$)  & Cr${_\text{2}}$O$_{\text{3}}$\cite{Siratori1992}  \\
&    $6m^{\prime}m^{\prime}$ &  $622$ $\overline{6}^{\prime}m^{\prime}2$  ($6/m^{\prime}m^{\prime}m^{\prime}$)  &  \\ \hline
$\left(\begin{array}{lll}\alpha_{xx} & -\alpha_{xx} & 0 \end{array}\right)$ & $\overline{4}$, $\overline{4}2^{\prime}m^{\prime}$ & $4^{\prime}$ ($4^{\prime}/m^{\prime}$),  $4^{\prime}22^{\prime}$ $4^{\prime}mm^{\prime}$ $\overline{4}m2$ ($4^{\prime}/m^{\prime}mm^{\prime}$)  &   \\  \hline
$\left(\begin{array}{lll}\alpha_{xx} & \alpha_{xx} & \alpha_{xx} \end{array}\right)$ &  & $23$, ($m^{\prime}\overline{3}^{\prime}$), $432$, $\overline{4}^{\prime}3m^{\prime}$ ($m^{\prime}\overline{3}^{\prime}m^{\prime}$) &   \\ \hline \hline
\end{tabular}
\caption{Table of antiferromagnetic and ferromagnetic nonquantized axion magnetoelectric symmetry groups and candidate material. In the first column we show only the diagonal part of magnetoelectric polarizability tensor $\alpha_{ij}$. The symbols $\overline{1}$ and $1^{\prime}$ mark spatial inversion and time-reversal symmetry, respectively. FM and AF MPG refers to ferromagnetic and antiferromagnetic magnetic point group~\cite{Bradley}.}
\label{Tab1}
\end{table}
\begin{figure}[t!]
\hspace*{0cm}
    \includegraphics[width=1\textwidth]{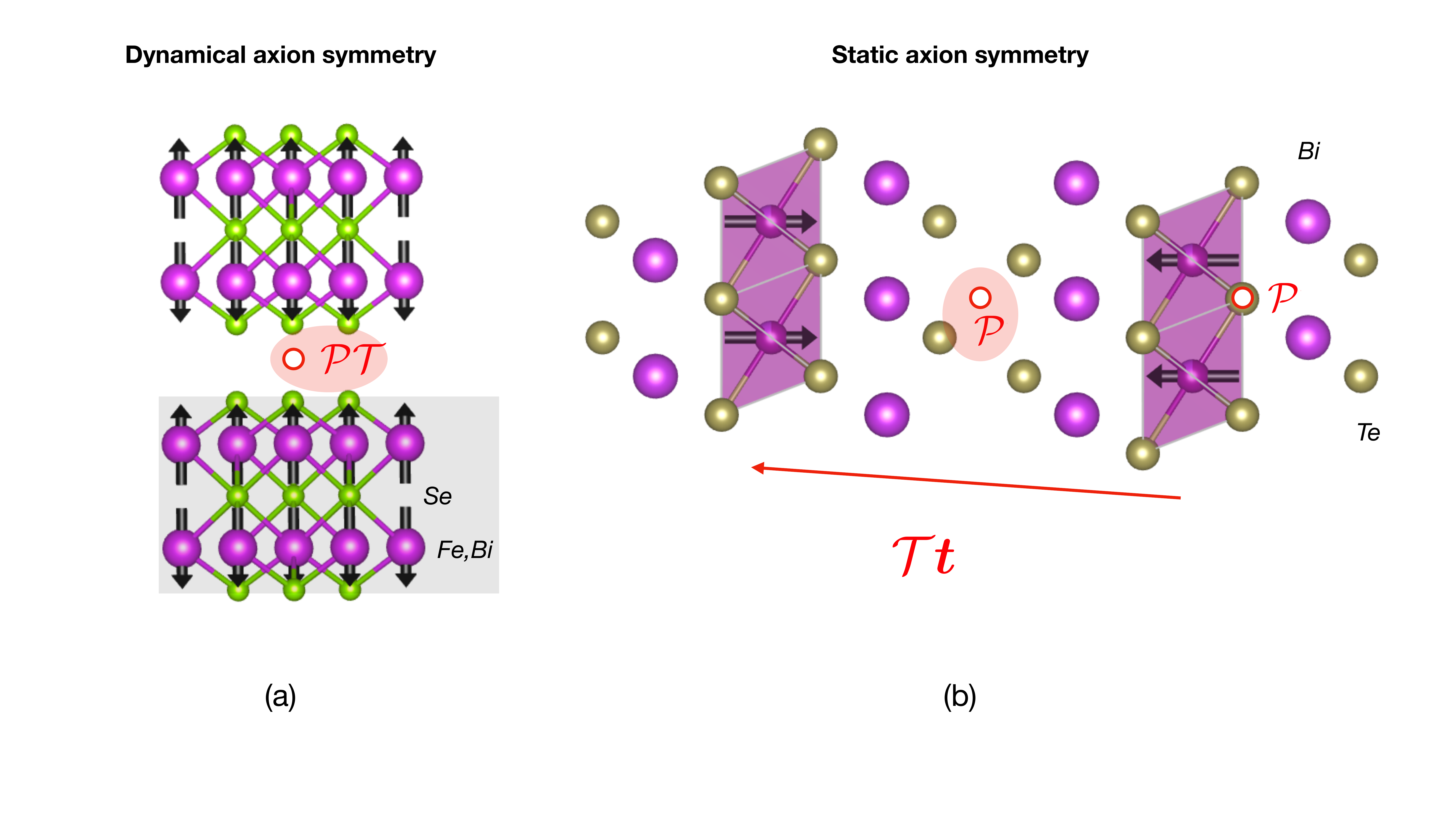}\\   
\caption{(a) Crystal structure of topological insulator Bi$_2$Se$_3$ consist from quintuple layers forming rhombohedral unit cell. Antiferromagnetism of the magnetically doped (Fe,Bi)$_2$Se$_3$ breaks spatial inversion $\mathcal{P}$ and time-reversal $\mathcal{T}$ symmetry, but preserve combined $\mathcal{PT}$ symmetry. (b) Crystal structure of intrinsic antiferromagnetic axion insulator MnBi$_2$Te$_4$. The quantized value of axion angle is protected by the inversion symmetry $\mathcal{P}$ (we maek two inversion symmetry points in the lattice). The system exhibits also partial unit cell translation $\boldsymbol{t}$ combined with time-reversal symmetry $\mathcal{T}$.}
\label{fig:af-ti:f1}
\end{figure}

In topological insulators, such as Bi$_2$Se$_3$ (the nonmagnetic phase of crystal shown in Fig.~\ref{fig:af-ti:f1}(a)) and Bi$_2$Te$_3$, the presence of $\mathcal{T}$ symmetry in combination with nontriivial band inversion ensures the axion angle $\Theta$ to be $\pi$~\cite{Wu2016a},  requires zero surface Hall conductivity, and the topological magneto-electric effect~\cite{Armitage2019}. 
The topological magneto-electric effect in topological insulators refers to a quantized magneto-electric response, and has been observed also by magneto-optical measurements~\cite{Wu2016a}. In fact, the quantization of $\Theta$ in non-magnetic topological insulators can be taken as defining property of topological insulators~\cite{Armitage2019}. Recently, also antiferromagnetic topological insulator~\cite{Mong2010} was found in \MnBiTe{}{2}{4}~\cite{Otrokov2019}. Antiferromagnetic topological insulator state is protected by time-reversal symmetry coupled with partial unit cell translation $\boldsymbol{t}$ as we show in Fig.~\ref{fig:af-ti:f1}(b).

The \textit{static} axion insulators are magnetic topological insulators, such as \MnBiTe{}{2}{4}~\cite{Li2019a,Liu2020b}, which break $\mathcal{T}$ symmetry via the presence of a magnetic ion (in this case, Mn). However, they exhibit axion response with $\Theta={\rm \pi}$, protected by the presence of axion odd symmetries such as inversions, see inversion centre in Fig.~\ref{fig:af-ti:f1}(b), or crystalline symmetries. The axion-odd symmetries are the symmetries which reverse the sign of $\Theta$ and support the so called $\mathbb{Z}_2$ classification~\cite{Kane2005b,Turner2012,Fang2012b}. Among the additional axion-odd symmetries are improper rotations, and antiunitary proper rotations (for instance rotation combined with time-reversal). In the Table~\ref{tab:symmetry_class_axion}, we list axion angle quantizing symmetry operations, $g$. We decompose the symmetry operation $g=g_{\parallel}\circ g_{\perp}$ into the parts $g_{\parallel}$ and $g_{\perp}$ which are parallel and perpendicular (in the surface plane) to the given surface normal $\hat{z}$\cite{Varnava2020}. We remark that we list the point group operation, but in general we need to pay attention to the nonsymmoprhic partial translations of the group, for details see~\cite{Varnava2020}.

\begin{table}
\centering
\begin{tabular}{ccc}
\hline\hline
$g$ & $g_{\|}$ & $g_{\perp}$ \\ \hline \text { Operations reversing } $\hat{\mathbf{z}}$ & & \\ $\mathcal{M}_{z}$ & \multirow{4}{*}{ $\mathcal{M}_{z}$} & $\mathcal{E}$ \\ $\mathcal{P}$ & & $\mathcal{C}_{2}$ \\ $\mathcal{S}_{3,4,6}$ &  & $\mathcal{C}_{3,4,6}$ \\ $\bar{\mathcal{C}}_{2}\mathcal{T}$ &  & $\mathcal{M}_{d}\mathcal{T}$ \\ \text { Operations preserving } $\hat{\mathbf{z}}$ & & \\ $\mathcal{E}\mathcal{T}$ & \multirow{4}{*}{ $\mathcal{E}$}  & $\mathcal{E}\mathcal{T}$ \\ $\mathcal{C}_{2,3,4,6}\mathcal{T}$ &  & $\mathcal{C}_{2,3,4}\mathcal{T}$ \\ $\mathcal{M}_{d}$ &  & $\mathcal{M}_{d}$ \\ \hline\hline
\end{tabular}
\caption{Axion angle quantizing symmetries. $\mathcal{E}$, $\mathcal{P}$,  $\mathcal{M}_{z}$, $\mathcal{M}_{d}$,  $\mathcal{C}_{2,3,4,6}$, $\mathcal{S}_{3,4,6}$, and $\mathcal{T}$ mark unitary symmmetry operations of identity, inversion, mirror parallel and perpendicular to surface normal $\hat{\boldsymbol{z}}$, rotational axis,  improper rotations, and time-reversal, respectively. Overbar marks inversion. Adapted after~\cite{Varnava2020}.}
\label{tab:symmetry_class_axion}
\end{table}

Finally, the \textit{dynamical} axion insulator allows for nonquantized dynamical axion angle.
The dynamics of the axion angle was suggested to be induced by chiral magnetic effect, antiferromangetic resonance~\cite{Zhang_2020}, longitudinal spin fluctuations~\cite{2010NatPh...6..284L} in an antiferromagnet or spin fluctuations in paramagnetic state~\cite{Wang2016Z}. 
In Fig.~\ref{fig:af-ti:f1}(a), we show an example of lattice with dynamical axion insulator state - Fe-doped \BiFeSe with $\mathcal{PT}$ symmetric crystal. Here, the antiferromagnetism breaks the inversion and time-reversal symmetries of the Bi$_2$Se$_3$ crystal. The symmetry breaking is desribed by mass term $M_{5}$ which corresponds to the band-gap in surface state. The combined $\mathcal{PT}$ symmetry is in the \BiFeSe crystal preseved and enforces Kramers degenerate bands. This can be seen by acting $\mathcal{PT}$ symmetry on the Bloch state to show that these two states have the same energy and are orthoghonal~\cite{Herring1966,Ramazashvili2008,Fang2013,Tang2016,Smejkal2016}. The presence of $\mathcal{PT}$ allows for antiferromagnetic Dirac quasipaticles~\cite{Tang2016,Smejkal2016} with plethora of unconventional and practically useful response such as large anisotropic magnetoresistance~\cite{Smejkal2016,Smejkal2017b}.
We discuss the material physics requirements for dynamical axion insulator in the section followed by section on minimal effective model of a dynamical axion insulator.

\subsubsection{Material candidates}
In this section we list requirements for a dynamical axion insulator which is also suitable for dark axion detection~\cite{Marsh:2018zyw}. In addition to constraints comming from requiring dynamical axion quasiparticles, we need to ensure  strong coupling of the $\Theta$ magneto-electric response to the fluctuations of the magnetic order parameter. The concept was originally developed for the longitudinal fluctuations in the N\'{e}el order parameter in magnetically doped  topological insulators (\BiFeSe in Ref.~\cite{2010NatPh...6..284L}) and recently extended into the instrinsic antiferromagnet \MnBiTe{2}{2}{5}~\cite{Zhang_2020}. 
This dynamical axion field is quite weak due to the low magnetoelectric coupling and trivial electronic structure in conventional materials such as Cr$_{\text{2}}$O$_{\text{3}}$\cite{Malashevich2012} and BiFeO$_{\text{3}}$~\cite{Coh2011} with $\Theta$ = $10^{-3}$ and $10^{-4}$, respectively, see Tab.~\ref{Axion_materials}. The dynamical axion effect (i.e. a large $\Theta$ response to external perturbations) can be enhanced in the proximity of the topological phase transitions\cite{Zhang_2020}. We now summarise the material criteria for a dynamical axion quasiparticles for detecting dark matter axion:
\begin{itemize}
\item Nonzero dynamical axion angle. The material symmetry allows for dynamical axion insulator state and axion spin density wave~\cite{2010NatPh...6..284L,Wang2016Z,Zhang_2020} with mass in the range of meV. This is one of the main advantage of using axion quasiparticles in antiferromagnets for detecting light and weakly interacting DM axions~\cite{2019PhRvL.123l1601M}.
\item Large bulk band-gap~\cite{Wan2012}. The material is in bulk semicondcuting or insulating with a large bulk band-gap, without disturbing bulk metallic states. In turn its low energy physics is governed solely by the axion coupling.
\item Topological mass term $M_{5}$  should be of order of dynamical axion fluctuation mass $m_{a}$ to ensure resonance with DM axion~\cite{2010NatPh...6..284L,2019PhRvL.123l1601M}. 
\item Large fluctiation in axion angle. This can be achieved close to the magnetic and topological phase transition as $\delta \Theta(\boldsymbol{r},t)=\delta M_{5}/g$, and $1/g\sim 1/\mathcal{M}(0)$~\cite{Zhang_2020}. The topological phase transition should be approached from the topological side.  Practically, one can tune this mass term by alloying. The alloying can effectively tune the strength of the spin orbit interaction.  However, the proximity close to the magnetic transition can compromise narrow linewidth, see next point. 
\item Trade-off among narrow linewidth and sensitivity. Narrow linewidth of the axion respones where the thermal fluction and scattering are supressed. This imposis tmeperature constraints (i.e. $T\ll T_N$, $T\ll m_5$). In contrast, enhanced response close the mangetic phase transition could enhance sensitivity. 
\item Robust magnetic ordering with elevated critical (N\'eel) temperature. 
\item As we will see in chapter about power output we need large spin-flop fields ($> 1$\,T) again favouring antiferromagnetic ordering.
\item Linear coupling of magnetic fluctuations to generate measurable axion polariton which is used for the detection of the dark matter axion. Li et al.~\cite{2010NatPh...6..284L} has used linear coupling of the longitudinal spin wave mode. From this perspective, the relatively streightforward generations  of dynamical axion by chiral magnetic effect or (anti)ferromagnetic resonance are not suitable. For the conventional transeversal spin waves would produce rahter quadratic coupling. This point is an open problem, however, the antiferromangetic spin density wave states~\cite{PhysRevB.84.195138} with longitudinal component are also possible candidates at the moment.
\item Magnetic and relativistic chemistry of low energy state manifold~\cite{Wan2012}: 3d states ensuring magnetism and time-reversal symmetry breaking and heavy elements with strong atomic spin-orbit interaction. Low energy states of common topological insulators are often heavy p-states which have low correlations and do not support magnetism. 
\end{itemize}
The last point can be justified by considering limitation of existing axion insulators proposals. 3d and 4d elements do have large electronic correlations but rather small spin-orbit interaction and thus it is difficult to tune the system into/close to the topological state. 4f and 5f elements pose heavy masses and narrow bands with excpetion of rate Kondo topological insulators~\cite{Dzero2010}. 5d pyrochlore~\cite{Wan2011} and spinel~\cite{Wan2012} elements are computationally predicted to host axion states within relatively small window of correlations strength complicating manufacturing the material. 

To summarize, the most promissing material systems are intrinsic antriferomagnetic axion insulators~\cite{Li2019a}, magnetically doped topological insulators~\cite{2010NatPh...6..284L}, certain conventional magnetoelectrics~\cite{Wang2011Z} and heterostructures of topological insulators~\cite{PhysRevB.102.121107}. The $\mathcal{PT}$ symmetric antiferromagnetism seems to be favourable over ferromagnetism as it naturaly provides for Dirac quasiparticles with tunable axion quasiparticles masses, longitudinal spin waves, larger spin-flop fields, elevated N\'eel temperatures, possibility to combine chemistry required for magnetism and spin-orbit coupling in single material platform. 
We list some of the promissing building block materials and systems for dynamical axion quasiparticles in the Tab.~\ref{Axion_materials}. Besides listing materials which are directly dynamical axion insulators we added also materials which can be used as starting configurations to build the dynamical axion insulator, for instance, by alloying of the static axion insulators. 

\begin{table}
\caption{Table of magneto-electric insulating material classes and candidates. FM (AF) marks (anti)ferromagnetism, $T_C$ the critical temperature, and $\Delta$ the bulk band gap.}
\label{Axion_materials}
\renewcommand{\arraystretch}{1.15}
\begin{tabular}{ccccc}
\toprule
\textbf{Phase} & \textbf{Material class} & \textbf{$\boldsymbol{T_C}$~(K)} & \textbf{$\boldsymbol\Delta$~[meV]}  & $\boldsymbol{\Theta$} \\
\midrule
\multirow{2}{*}{Magnetoelectric} & BiFeO$_{\text{3}}$ & 643 & 950 & \num{0.9e-4} \cite{Coh2011} \\
 & Cr${_\text{2}}$O$_{\text{3}}$ & 343 & 1300 & \num{1.3e-3} \cite{Malashevich2012} \\
\midrule
Magnet/TIs & CrI$_{\text{3}}$/Bi$_{\text{2}}$Se$_{\text{3}}$/MnBi$_{\text{2}}$Se$_{\text{4}}$ & \textgreater{}10 & 5.6 & $\pi$ \cite{PhysRevB.102.121107} \\
\multirow{2}{*}{Intrinsic $\mathcal{P}$ AFs} & MnBi$_{\text{2}}$Te$_{\text{4}}$ & \textless{}25 & \textless 220 & $\pi$ \\
 & EuIn(Sn)$_{\text{2}}$As(P)$_{\text{2}}$ & 16 & \textless{}100 & $\pi$ \\ \hline
Doped TIs & Cr(Fe)-Bi$_{\text{2}}$Se$_{\text{3}}$ & $\sim$10~\cite{PhysRevB.92.104405} & $\sim$30 & nonquantized \\
Intrinsic $\mathcal{PT}$ AFs & Mn(Eu)$_{\text{2}}$Bi$_{\text{2}}$Te$_{\text{5}}$ & 6 & $\sim$50 & $0.83\,\pi$~\cite{PhysRevB.102.121107}\\ \bottomrule
\end{tabular}
\end{table}

We emphasize that the bulk energy bands encode the information about the dynamical axion insulator response, and its surface states~\cite{2010NatPh...6..284L}. 
We can see this on expression for the intrinsic magnetoelectric susceptibility, axion coupling, can be calculated in the Bloch representation as~\cite{2010NatPh...6..284L}:
\begin{equation}
\Theta=-\frac{1}{4 \pi} \int_{\mathrm{BZ}} d^{3} k \epsilon^{\alpha \beta \gamma} \operatorname{Tr}\left[\mathcal{A}_{\alpha} \partial_{\beta} \mathcal{A}_{\gamma}-i \frac{2}{3} \mathcal{A}_{\alpha} \mathcal{A}_{\beta} \mathcal{A}_{\gamma}\right].
\end{equation}
%
Here we explictily see the axion angle relation to the non-abelian Berry connection $\mathcal{A}_{\alpha, n m}(\mathbf{k})=\left\langle u_{n \mathbf{k}}\left|i \partial_{k_{\alpha}}\right| u_{m}\right\rangle$ constructed from the Bloch functions $\left|u_{n \mathbf{k}}\right\rangle$. The trace is over occupied valence bands. 

The first-principle calculations of the axion angle is reserch topic on its own~\cite{Varnava2018,Varnava2020}. For the sake of brevity we will adopt here simpler approach. We can use first-prinicple calculations and symmetry analysis to identify and parametrize low energy effective Hamiltonian for which the calculation of axion angle and its dynamical response is numerically less demanding. We will now describe dynamical axion quasiparticle model which is applicable to Fe-doped Bi$_{\text{2}}$Se$_{\text{3}}$~\cite{2010NatPh...6..284L} and intrinsic antiferromagnet Mn$_{\text{2}}$Bi$_{\text{2}}$Te$_{\text{5}}$~\cite{Zhang_2020} and also heterostructures~\cite{PhysRevB.102.121107}.

\subsubsection{Dirac model of axion quasiparticles}
We can derive the minimal model of dynamical axion insulator starting from the Dirac quasiparticle model for the bulk states of topological insulator Bi$_{\text{2}}$Se$_{\text{3}}$~\cite{Zhang2009a}. The low energy physics can be captured by four-band Hamiltonian in the basis of bonding and antibonding Bi $p_{z}$ states $\left|P 2_{z}^{-}, \uparrow(\downarrow)\right\rangle \text { and }\left|P 1_{z}^{+}, \uparrow(\downarrow)\right\rangle$~\cite{Zhang2009a,2010NatPh...6..284L,Zhang_2020}:
\begin{equation}
\mathcal{H}_{\mathrm{Dirac}}=\epsilon_{0}(\vc{k})+\sum_{a=1}^{5} d_{a}(\vc{k}) \Gamma_{a}.
\end{equation}
Here $\Gamma_{a}$ refer to the Dirac matrices representation: 
\begin{equation}
\Gamma_{(1,2,3,4,5)}=\left(\sigma_{x} \otimes s_{x} \, , \sigma_{x}, \otimes s_{y}, \sigma_{y} \otimes I_{2 \times 2}, \sigma_{z} \otimes I_{2 \times 2} \, , \sigma_{x} \otimes s_{z}\right)
\label{alpha5-Dirac0}
\end{equation}
in the basis $(|P 1_{z}^{+}, \uparrow\rangle,|P 1_{z}^{+}, \downarrow\rangle, |P 2_{z}^{-}, \uparrow\rangle,|P 2_{z}^{-}, \downarrow\rangle)$. $\sigma$ and $s$ are orbital and spin Pauli matrices.
The $4\times 4$ matrices $\Gamma_a$ satisfy the Clifford algebra $\{\Gamma_a,\Gamma_b\}=2\delta_{ab}$ with $\Gamma_5=\Gamma_1\Gamma_2\Gamma_3\Gamma_4$.
This model can be tuned to the trivial ($\Theta=0$) or topological insulator state ($\Theta=\pi$). To induce nonzero $\delta\Theta$ and dynamical axion state we need to add $\mathcal{P}$ and $\mathcal{T}$ symmetry breaking terms due to the antiferromagnetism. 

The crystal momentum dependent coefficients take the form: 
\begin{eqnarray}
d_{1,2,3}(\boldsymbol{k}) &=& A_{1,2,3}(\boldsymbol{k})+m_{x,y,z}, \\
\epsilon_{0}(\boldsymbol{k})&=&C+2 D_{1}+4 D_{2}-2 D_{1} \cos k_{z}-2 D_{2}\left(\cos k_{x}+\cos k_{y}\right) \\
d_{4}(\boldsymbol{k})=\mathcal{M}(\mathbf{k})&=&M_{0}-2 B_{1}-4 B_{2}+2 B_{1} \cos k_{z}+2 B_{2}\left(\cos k_{x}+\cos k_{y}\right) \\
d_{5}(\boldsymbol{k})=M_{5}.
\label{perham}
\end{eqnarray}
Here the fourth term $\mathcal{M}(\boldsymbol{k})$ controls the topological phase transition from the trivial to topological insulator, is invariant under $\mathcal{T}$, and we denote $\mathcal{M}(\boldsymbol{k}=0)=M_{0}$.
The topological insulating phase is achieved when $M,B_{1},B_{2}>0$~\cite{Zhang2009a}.
 The symmetry breaking terms are the masses $m_{x,y,z}$ and $M_{5}$ (a $\mathcal{C}\mathcal{P}$-odd chiral mass term).
We see that the spatial inversion $\mathcal{P}=\sigma_{z}\otimes I_{2x2} $ and time-reversal operators $\mathcal{T}=iI_{2x2}\otimes s_{y}\mathcal{K}$ do not commute with the Hamiltonian, while their combination does. Here $\mathcal{K}$ is complex conjugation.  

\begin{figure}[t!]
\hspace*{0cm}
    \includegraphics[width=1\textwidth]{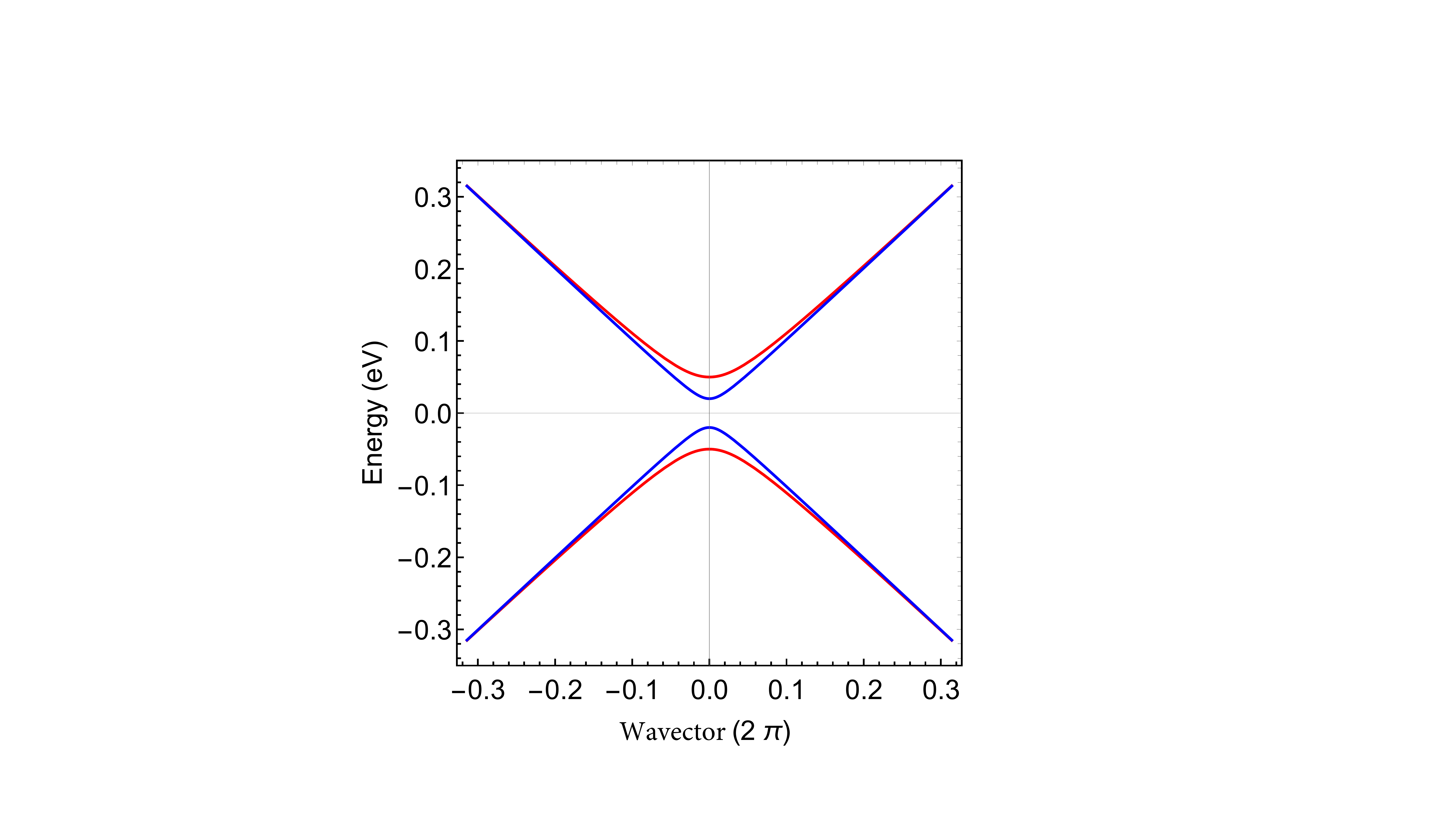}\\   
\caption{(a) Minimal model energy bands of edge states (red) of dynamical axion insulator with parity and time-reversal breaking antiferromangetic term $M_{5}=0.5M_0$, where $M_0$ is bulk (blue) Dirac band-gap. The wavector corresponds to given valley index $f$ in the text.}
\label{fig:af-ti:f2}
\end{figure}

Only the last mass term $M_{5}$ induces linear perturbations to $\Theta$ as we will show further, and without loss of generality one can set $m_{x,y,z}=0$.
In turn, the $M_{5}$ term opens a surface band gap in the surface states Dirac Hamiltonian as we show in Fig.~\ref{fig:af-ti:f2}. The $A, B, C$, $D$, and masses $M, M_{5}$ constants are material dependent and can be determined by fitting the electronic structure calculated from the first-principles~\cite{Zhang2009a,2010NatPh...6..284L,PhysRevB.102.121107,Liu2010}. We also remark, that for calculating the complete response of the material we need to know the full periodic Hamiltonian Eq.~\eqref{perham}. 

When its sufficient to study small wavector excitations we can use continuum variant, $\boldsymbol{k},\boldsymbol{p}$-expansion, around momentum points $X_f$, where $\bm{q}=\bm{k}-X_f$:
\begin{align}
\mathcal{H}_f(\bm{q})=q_x\alpha_1+q_y\alpha_2+q_z\alpha_3+M_0\alpha_4+M_{5f}\alpha_5.
\label{alpha5-Dirac}
\end{align}
Here we use the standard Dirac equation basis:
\begin{equation}
\beta=\alpha_{4}=\left(\begin{array}{cc}
I & 0 \\
0 & -I
\end{array}\right), \alpha^{i=1,2,3}=\left(\begin{array}{cc}
0 & \sigma_{i} \\
-\sigma_{i} & 0
\end{array}\right) \Rightarrow \alpha_{5}=\left(\begin{array}{cc}
0 & I \\
I & 0
\end{array}\right).
\end{equation}
Furthermore, the subscript $f$ denotes the valley degree of freedom in the low-energy electronic band of the system, and can be understood as the  Dirac quasiparticle \textit{flavour}. In the AFI phase of the Bi$_2$Se$_3$ family doped with magnetic impurities such as Fe~\cite{2010NatPh...6..284L}, there is a single Dirac fermion and $M_{5,1}=M_{5,2}=0$, $M_{5,3}=-(2/3)Un_z$.
(In the AFI phase of the Fu-Kane-Mele-Hubbard (FKMH) model~\cite{Sekine:2014xva}, there are three Dirac fermions and $M_{5,a}=Un_a$ $(a=x,y,z)$.) Here, $\bm{n}=(\langle \vc{S}_A-\vc{S}_B)/2=n_x\bm{e}_x+n_y\bm{e}_y+n_z\bm{e}_z$ denotes the mean-field AF order parameter (i.e., the N\'{e}el field, with $\vc{S}_{A,B}$ the spin of ions on $A$ and $B$-type lattice sites) and $U$ is the on-site electron-electron interaction strength (i.e. the Hubbard term, see below). The kinetic term $\sum_{\mu=1}^3q_\mu\alpha_\mu$ is spin-dependent as a consequence of spin-orbit coupling.

We derive the effective action consisting of the N\'{e}el field $\bm{n}=\bm{n}_0+\delta\bm{n}$ and an external electromagnetic potential $A_\mu$, where $\bm{n}_0$ denotes the ground state of the N\'{e}el field and $\delta\bm{n}$ denotes the fluctuation due to excitations. 
For this purpose, it is convenient to adopt a perturbative method.
We start with the total action of an AF insulator described by Eq. (\ref{alpha5-Dirac}) with an external electromagnetic potential $A_\mu$:
\begin{align}
S_{\rm eff}[\psi,\bar{\psi},\bm{n},A_\mu]=\int dtd^3 r\sum_f\bar{\psi}_f(\bm{r},t)\left[i\gamma^\mu D_\mu-M_0+i\gamma^5M_{5f}\right]\psi_f(\bm{r},t),\label{Eq-S1}
\end{align}
where $t$ is real time, $\psi_f(\bm{r},t)$ is a four-component spinor, $\bar{\psi}_f=\psi_f^\dagger\gamma^0$, $D_\mu=\partial_\mu+ieA_\mu$, and we have used the fact that $\alpha_4=\gamma^0$, $\alpha_5=-i\gamma^0\gamma^5$ and $\alpha_j=\gamma^0\gamma^j$ ($j=1,2,3$).
Here, the gamma matrices satisfy the identities $\{\gamma^\mu,\gamma^5\}=0$ and $\{\gamma^\mu,\gamma^\nu\}=2g^{\mu\nu}$ with $g^{\mu\nu}=\mathrm{diag}(1,-1,-1,-1)$.
By integrating out the fermionic field $\psi_f$, we obtain the effective action $W_{\rm eff}$ for $\bm{n}$ and $A_\mu$ as
\begin{align}
Z[\bm{n},A_\mu]&=\int\mathcal{D}[\psi,\bar{\psi}]e^{iS_{\rm eff}}=\exp\left\{\sum_f\mathrm{Tr}\ln\left[G_{0f}^{-1}(1+G_{0f}V_f)\right]\right\}\nonumber\\
&=\exp\left[\sum_f\mathrm{Tr}\left(\ln G_{0f}^{-1}\right)-\sum_f\sum_{n=1}^\infty\frac{1}{n}\mathrm{Tr}\left(-G_{0f}V_f\right)^n\right]\nonumber\\
&\equiv e^{iW_{\rm eff}[\bm{n},A_\mu]}.
\end{align}
In order to obtain the action of the low-energy spin-wave excitation, i.e., the AF magnon, we set the Green's function of the unperturbed part as $G_{0f}=(i\gamma^\mu\partial_\mu-M_0+i\gamma^5M_{5f})^{-1}$, and the perturbation term as $V_f=-e\gamma^\mu A_\mu+i\gamma^5\delta M_{5f}$.
Note that we have used $i\gamma^\mu D_\mu-M_0+i\gamma^5(M_{5f}+\delta M_{5f})=G_{0f}^{-1}+V_f$.
In the random phase approximation, the leading-order terms read
\begin{align}
iW_{\rm eff}[\bm{n},A_\mu]=-\frac{1}{2}\sum_f\mathrm{Tr}\left(G_{0f}i\gamma^5\delta M_{5f}\right)^2+\sum_f\mathrm{Tr}\left[\left(G_{0f}e\gamma^\mu A_\mu\right)^2\left(G_{0f}i\gamma^5\delta M_{5f}\right)\right],\label{Eq-S4}
\end{align}
where the first and second terms on the right-hand side correspond to a bubble-type diagram and a triangle-type digram, respectively.

To compute the traces of the gamma matrices we use the following identities:
\begin{align}
\mathrm{tr}(\gamma^\mu)=\mathrm{tr}(\gamma^5)=0,\ \ \ \mathrm{tr}(\gamma^\mu\gamma^\nu)=4g^{\mu\nu},\ \ \ \mathrm{tr}(\gamma^\mu\gamma^\nu\gamma^5)=0,\ \ \ \mathrm{tr}(\gamma^\mu\gamma^\nu\gamma^\rho\gamma^\sigma\gamma^5)=-4i\epsilon^{\mu\nu\rho\sigma}.
\end{align}
The first term in Eq. (\ref{Eq-S4}) is given explicitly by
\begin{align}
\mathrm{Tr}\left(G_{0f}i\gamma^5\delta M_{5f}\right)^2
&=\int\frac{d^4q}{(2\pi)^4}\int\frac{d^4k}{(2\pi)^4}\mathrm{tr}\left[G_{0f}(k)i\gamma^5\delta M_{5f}(q)G_{0f}(k+q)i\gamma^5\delta M_{5f}(-q)\right]\nonumber\\
&=4\int\frac{d^4q}{(2\pi)^4}\int\frac{d^4k}{(2\pi)^4}\frac{[k_\mu(k+q)^\mu-M_0^2+M_{5f}^2]\delta M_{5f}(q)\delta M_{5f}(-q)}{(k^2-M_0^2-M_{5f}^2)[(k+q)^2-M_0^2-M_{5f}^2]}\nonumber\\
&\equiv \int\frac{d^4q}{(2\pi)^4}\Pi_f(q)\delta M_{5f}(q)\delta M_{5f}(-q),\label{Eq-S5}
\end{align}
where $k^2=g^{\mu\nu}k_\mu k_\nu=k_\mu k^\mu=k_0^2-\bm{k}^2$.
We have used $G_{0f}(k)=(\gamma^\mu k_\mu+M_0+i\gamma^5M_{5f})/(k^2-M_0^2-M_{5f}^2)$, $\{\gamma^\mu,\gamma^5\}=0$, and $\{\gamma^\mu,\gamma^\nu\}=2g^{\mu\nu}$.
After performing a contour integration, we arrive at the action of the form
\begin{align}
\sum_f\mathrm{Tr}\left(G_{0f}i\gamma^5\delta M_{5f}\right)^2=i\sum_f J_f\int dtd^3r \left[(\partial_\mu \delta M_{5f})(\partial^\mu \delta M_{5f})-m_f^2(\delta M_{5f})^2\right],
\label{action-Neel-field}
\end{align}
where $J_f$ and $m_f$ are the stiffness and mass of the spin-wave excitation mode, which are given respectively by~\cite{2010NatPh...6..284L}
\begin{align}
J_f=\left.\frac{\partial^2\Pi_f(q)}{\partial q_0^2}\right|_{q\to 0}=\int_{\mathrm{BZ}}\frac{d^3k}{(2\pi)^3}\frac{\sum_{i=1}^4 d_i^2}{16|d|^5},
\label{eqn:Jf_microscopic}
\end{align}
\begin{align}
J_fm_f^2=\left.\Pi_f(q)\right|_{q\to 0}=M_{5f}^2\int_{\mathrm{BZ}}\frac{d^3k}{(2\pi)^3}\frac{1}{4|d|^3},
\label{eqn:Jf_mf_microscopic}
\end{align}
where $|d|=\sqrt{\sum_{a=1}^5 d_a^2}$, $q\to 0$ indicates the limit of both $q_0\to 0$ and $\bm{q}\to\bm{0}$, and here $f$ denotes the flavour.
Equation~(\ref{action-Neel-field}) is nothing but the action of the N\'{e}el field described by the non-linear sigma model~\cite{Haldane1983}.
In the present low-energy effective model [Eq. (\ref{Eq-S1})], the information on the anisotropy of the N\'{e}el field is not included.
On the other hand, many (actual) AF insulators have the easy-axis anisotropy.
Hence the term $\sum_f m_f^2\delta n_f^2$ will be replaced by a term like $m^2(\delta\bm{n}\cdot\bm{e}_A)^2$ with $\bm{e}_A$ denoting the easy axis.
The second term in Eq. (\ref{Eq-S4}) is the triangle anomaly, which gives the Chern-Simons term.
The final result is~\cite{Nagaosa1996,Hosur2010}
\begin{align}
\sum_f\mathrm{Tr}\left[\left(G_{0f}e\gamma^\mu A_\mu\right)^2\left(G_{0f}i\gamma^5\delta M_{5f}\right)\right]&=i\frac{\alphaEM}{2\pi}\int dtd^3r\dyAQ(\bm{r},t)\epsilon^{\mu\nu\rho\lambda}\partial_\mu A_\nu \partial_\rho A_\lambda\nonumber\\
&=i\frac{\alphaEM}{\pi}\int dtd^3 r \dyAQ(\bm{r},t) \bm{E}\cdot\bm{B},
\label{action-theta-term}
\end{align}
where~\cite{Sekine:2014xva}
\be
\dyAQ(\bm{r},t)=\sum_f\tan^{-1}\left[\frac{M_f+\delta M_{5f}(\bm{r},t)}{M_0}\right]-\sum_f\tan^{-1}\left[\frac{M_f}{M_0}\right]\approx-\sum_f \frac{\delta M_{5f}(\bm{r},t)}{M_0}\, .
\label{eqn:AQ_from_n}
\ee
From Eq.~(\ref{action-theta-term}) we find that the fluctuation of the $\gamma^5$ mass $M_{5f}$ behaves just as a dynamical axion field. From Eqs.~(\ref{action-Neel-field}) and (\ref{action-theta-term}), we finally arrive at the action of the AQ~\cite{2010NatPh...6..284L,PhysRevLett.116.096401}: 
\begin{align}
S_{\rm AQ}=M_0^2J\int dtd^3r \left[(\partial_\mu \dyAQ)(\partial^\mu \dyAQ)-\mAQ^2\dyAQ^2\right]+\frac{\alphaEM}{\pi}\int dtd^3 r \dyAQ(\bm{r},t) \bm{E}\cdot\bm{B}\, , 
\label{eqn:AQ_effective_action}
\end{align}
where have used that for systems described by the Dirac Hamiltonian (Eq.~\eqref{alpha5-Dirac}) the quantity labelled $g$ in the action given in Ref.~\cite{2010NatPh...6..284L} can be set equal to the bulk band gap $M_0$. We identify the decay constant as $M_0^2 J = {\fAQ}^2/2$, and note that the spin wave speed appears in the spatial derivatives by choice of units.

\subsection{AQ as Longitudinal Magnon}

For concreteness, let us consider the AF insulator phase of \BiFeSe\, and \MnBiTe{2}{2}{5} such that there is a single degree of freedom with $M_{5,1}=M_{5,2}=0$, and $M_{5,3}=-(2/3)Un_\parallel$, where $n_\parallel$ is parallel to the easy-axis anisotropy. In terms of the AF order parameters the AQ is given by expanding eq.~\eqref{eqn:AQ_from_n}, leading to
\be
\dyAQ \approx \frac{2U}{3M_0}\delta n_\parallel\, .
\ee
Thus, we see that the AQ $\dyAQ$ is the longitudinal fluctuation in the AF order. 

The EFT of \emph{transverse} magnons is presented in  Appendix~\ref{appendix:eft_afmr}, and is based on the Heisenberg model. The Heisenberg model is the strong coupling limit of the Hubbard model used to describe the AQ, but nonetheless it provides some insight into the physics, which we discuss briefly. The EFT describes the AF order parameter, $\vc{n}$. Let us denote the components of $\vc{n}$ as $n_\parallel$ along the easy-axis and $n_{\bot,1,2}$ orthogonal to it. In the EFT we have that: 
\be
\delta n_{\parallel} \approx -\frac{\delta n_{\bot,1}^2}{2}-\frac{\delta n_{\bot,2}^2}{2}\, .
\ee
Thus the AQ is related \emph{non-linearly} to the transverse magnons of the Heisenberg EFT. 

In the Dirac model for the AQ, the interaction between $\dyAQ$ and electromagnetism is given entirely by the chiral anomaly, i.e. the interaction $\dyAQ\vc{E}\cdot\vc{B}$. On the other hand the Heisenberg EFT contains the spin interaction $\mathcal{L}_{\rm em} = \mu_B \vc{s}\cdot\vc{H}$, with $\vc{s}=\dot{\vc{n}}\times\vc{n}$ at leading order. As we have just established, however, the Heisenberg model fields are not linearly related to the AQ in the Hubbard model with $t/U\ll 1$. We therefore neglect the interaction $\mathcal{L}_{\rm em}$ in our subsequent calculations based on the effective action Eq.~\eqref{eqn:AQ_effective_action}. If only the axion, $\dyAQ\propto \delta n_\parallel$, is present, then indeed $\dot{\vc{n}}\times\vc{n}=0$. 

However, if the AFMR fields $\delta n_\bot$ are also excited, then $\mathcal{L}_{\rm em}$ mixes the fields and leads to the Kittel shift in the frequencies of these fields $\omega = \mu_B H_0+\sqrt{m_s^2+v^2k^2}$ (see Appendix~\ref{appendix:eft_afmr}). The Kittel shift would also mix the AFMR fields with the axion. It is not clear to us how to model these two effects, the AQ and AFMR with an applied field, at the same time because the two descriptions are valid in opposite regimes of the Hubbard model parameters. The splitting $\mu_B H_0\ll m_s$ for fields $H_0\sim 1 \text{ T}$, and so our subsequent results would not be changed drastically by such an effect. Nevertheless, the splitting may be possible to observe experimentally if it is present. This remains an open question.

We have not been able to derive an EFT for the AQ longitudinal magnon along the same lines as the EFT of AFMR given in the Appendix. One possibility for such a theory generalises the AF-ordering to a general spin density wave ordering vector $\vc{Q}$. In this case, one arrives at a quadratic Lagrangian for the transverse and longitudinal magnons\footnote{Other approaches to the longitudinal mode include Refs.~\cite{2014JPhCS.529a2020X, PhysRevB.100.024428}.} with coupling to external sources~\cite{PhysRevB.84.195138}. However, in addition to these desired ingredients there are also spinor degrees of freedom, the ``holons'' describing the spin-charge separation. Another possibility, which we suggest, is to generalise the N\'{e}el order parameter to an $SU(2)$ doublet with the AQ a Goldstone boson associated to a Chiral $U(1)$ subgroup under which the Dirac quasiparticles are charged.

\subsection{Parameter Estimation}\label{sec:params}

Three unknown quantities determine the AQ model: the mass~$\mAQ$, decay constant~$\fAQ$, and speed $v_s$~(from the spatial derivatives, giving the wave speed). We generally work in the limit $v_s\ll c$ and ignore the magnon dispersion relative to the $E$-field. This leaves two parameters, $\mAQ$ and $\fAQ$. We show in detail in section~\ref{sec:boundary_conditions} how both $\mAQ$ and $\fAQ$ can be determined experimentally from the polariton resonances and gap via transmission spectroscopy (related to the total reflectance measurement proposed by Ref.~\cite{2010NatPh...6..284L}). In this section, however, we wish to estimate these parameters from known material properties. 

We consider two candidate materials, firstly the magnetically doped TI \BiFeSe\, of Ref.~\cite{2010NatPh...6..284L}. Reference~\cite{PhysRevB.88.235131} considered a number of different TIs doped with different magnetic ions, and found that only \BiFeSe is both antiferromagnetic and insulating. \BiFeSe has been successfully fabricated. However, the magentism is fragile due to the doping (required around 3.5\%), and the region of the phase diagram exhibiting the AQ is small. Therefore, we also consider the new class of intrinsically magnetic TIs, \MnBiTe{x}{y}{z}, of which only \MnBiTe{2}{2}{5} is thought to contain an AQ, but has yet to be fabricated. Material properties for both cases are listed in Table~\ref{tab:material_params}, while the derived parameters are given in Table~\ref{tab:derived_params}. Our estimates for the derived parameters are discussed in the following. 

\begin{table}
\caption{Material parameters for AQ materials. \MnBiTe{2}{2}{5} has not been experimentally realised, and parameters in the references are calculated \emph{ab initio}, rather than measured. Values in parentheses were assumed in Ref.~\cite{2010NatPh...6..284L}. We assume the dielectric constants for both materials are equal to the undoped Bi$_2$Se$_3$ extrapolated to low energy~\cite{doi:10.1063/1.3466552,CaoWang}.\label{tab:material_params}}
\renewcommand{\arraystretch}{1.15}
\centering
\begin{threeparttable}
\begin{tabular}{l l l l l l}
\toprule
Symbol & Name & \multicolumn{2}{l}{\BiFeSe} & \multicolumn{2}{l}{\MnBiTe{2}{2}{5}} \\ 
\midrule
 $\mu_B H_E$ & Exchange & \SI{1}{\meV} & \cite{2012PhRvL.109z6405Z}  & \SI{0.8}{\meV} & \cite{PhysRevB.102.121107} \\  
 $\mu_B H_A$ & Anisotropy & \SI{16}{\meV} & \cite{PhysRevB.88.235131} & \SI{0.1}{\meV} & \cite{PhysRevB.102.121107} \\ \\ 
 $\Vuc$ & Unit cell volume & 440  \AA$^3$ & & 270 \AA$^3$ & \\
 $U$ & Hubbard term & \SI{3}{\eV} & \cite{PhysRevB.88.235131} & \SI{3}{\eV} & \cite{PhysRevB.102.121107} \\
 $M_0$ & Bulk band gap & \SI{0.03}{\eV}~(\SI{0.2}{\eV}) & \cite{PhysRevB.88.235131} & 0.05 eV& \cite{PhysRevB.102.121107} \\
 $t$ & Nearest neighbour hopping\tnote{a} & \SI{0.04}{\eV}  & & \SI{0.04}{\eV} & \\
 $S$ & Magnetic moment & 4.99 & \cite{PhysRevB.88.235131} & 4.59 & \cite{PhysRevB.102.121107}\\
  $T_N$ & N\'{e}el temperature & \SI{10}{\K} & \cite{PhysRevB.92.104405} & \SI{6}{\K}\tnote{b}\\
   $\epsilon_1$ & Dielectric constant & 25~(100) & & 25 & \\
\bottomrule
\end{tabular}
 \begin{tablenotes}\footnotesize
 	\item[a] The hopping parameters $t$ are derived from $H_E$ assuming half-filling.
 	\item[b] Estimated from the Liechtenstein magnetic force theorem, $T_N=3\mu_B H_E/2 k_B$~\cite{1987JMMM...67...65L}.
		\end{tablenotes}
\end{threeparttable}
\end{table}

\begin{table}
	 \caption{Derived AQ parameters. ``Material 1'' is our best approximation to \BiFeSe. We report the results of Ref.~\cite{2010NatPh...6..284L}, who assumed a cubic lattice to evaluate the band integrals, but rescaled by our values of $M_0$. We use a combination of normalisation to the cubic lattice result, and the material properties in Table~\ref{tab:material_params} to estimate the parameters for ``Material 2'', our best approximation to \MnBiTe{2}{2}{5}.\label{tab:derived_params}}
	\center
\begin{tabular}{ l l l l l}
\toprule
Symbol & Name & Equations & ``Material 1'' & ``Material 2'' \\ 
\midrule
  $\mAQ$ & AQ mass & \eqref{eqn:fQ_Hubbard}, \eqref{eqn:mAQ_est} & 2 meV & \SI{1.8}{\meV}\\
 $\fAQ$ & AQ decay constant & \eqref{eqn:mAQ_microscopic},~\eqref{eqn:fAQ_est} & \SI{30}{\eV} & \SI{70}{\eV}  \\
\bottomrule 
\end{tabular}
\end{table}

The microscopic model for the AQ is derived from the Hubbard model in the weak coupling limit. In the Hubbard model, one allows hopping of spins between lattice sites. The Hubbard Hamiltonian is:
\begin{equation}
H = -t \sum_{\langle i j \rangle, \sigma} a^{\dagger}_{i \sigma} a_{j \sigma} + U \sum_{i} n_{i \uparrow} n_{i \downarrow},
\end{equation}
where $a^{\dagger}_{i \sigma}$ and $a_{i \sigma}$ are the creation and annihilation operators for a spin $\sigma$ at lattice site $i$ and the first sum is over nearest neighbour sites. $n_{i \uparrow}$ and $n_{i \downarrow}$ are the spin up and spin down density operators for the $i$th lattice site. The first term describes the kinetic energy of the system, whose scale is given by the hopping parameter $t$.  The second term describes the interaction between spins on the same site, with scale given by the Hubbard term $U$. In the limit of half filling and $U \gg t$, the Hubbard model is equivalent to a Heisenberg model with $J_H \sim t^2/U$ \cite{Fazekas}. The exchange field is related to the Heisenberg Hamiltonian via Eq.~\eqref{eqn:heisenberg} as 
\be
H_E = \frac{2 S J_H}{g \mu_B} \, ,
\label{eqn:exchangeField}
\ee
where $S$ is the ion spin and $g$ is the spectroscopic splitting factor \cite{doi:10.1063/1.5109132}~(see Appendix~\ref{appendix:AFMR} for more details). This relation was used in Table~\ref{tab:material_params} to set the hopping parameter $t$ given $U$ , $S$ and $\mu_B H_E$ and taking $g=2$.

The electron band energies~$d_i$, Eqs.~\eqref{eqn:Jf_microscopic}, \eqref{eqn:Jf_mf_microscopic}, appearing in the microscopic model are normalized with respect to $t$. The Brillouin zone (BZ) momentum, $k$, on the other hand, is normalised with respect to the unit cell. This suggests normalizing the integrals Eqs.~(\ref{eqn:Jf_microscopic},\ref{eqn:Jf_mf_microscopic}) as (we consider only the case with a single Dirac fermion from now on and drop the subscript $f$):
\be
J=\int_{\rm BZ} \frac{{\rm d}^3k}{(2\pi)^3}\frac{\sum_i d_i^2}{16|d|^3} =  \frac{\mathcal{I}_1}{\Vuc t^3}\, , 
\label{eqn:BZ_integral_1}
\ee
(note that this $J$ is not Heisenberg $J_H$, in fact $J_H\propto 1/J$) and
\be
\mAQ^2J=M_{5}^2\int_{\rm BZ} \frac{{\rm d}^3k}{(2\pi)^3}\frac{1}{4|d|^3} =  M_{5}^2\frac{\mathcal{I}_2}{\Vuc t^3}\, , 
\label{eqn:BZ_integral_2}
\ee
where $\Vuc$ is the volume of the unit cell. It then follows that the AQ mass is: 
\be
\mAQ= M_{5}\sqrt{\frac{\mathcal{I}_2}{\mathcal{I}_1}}=\frac{2SU}{3}\sqrt{\frac{\mathcal{I}_2}{\mathcal{I}_1}}\, .
\label{eqn:mAQ_microscopic}
\ee
Notice that for an exact Dirac dispersion for $d$, the integrals over the BZ vanish if the Dirac mass, $M_0$, vanishes, as we expect from the Gell--Mann-Oakes-Renner relation~\cite{PhysRev.175.2195}. However, these integrals should be evaluated for $d$'s computed in the full theory, i.e. \emph{ab initio} density functional theory for the Hubbard model.

In the full theory, the normalized integrals $\mathcal{I}$ depend on the ratio $t/U$. In terms of the Hubbard model parameters we have $M_{5}=(2/3)U n_z$, where $n_z=S$ is the normalised AF order. The decay constant is: 
\be
{\fAQ}^2 = 2 M_0^2 \frac{\mathcal{I}_1}{\Vuc t^3}\, , \label{eqn:fQ_Hubbard}
\ee
Using a cubic lattice model, Ref.~\cite{2010NatPh...6..284L} computed the BZ integrals for \BiFeSe. The integrals depend on the ratio $t/U$, so we can also use this result for \MnBiTe{2}{2}{5} if we extract the values of the normalised integrals. 

Reference~\cite{2010NatPh...6..284L} report $b=\SI{0.2}{\meV}$ at 2 T and $\mAQ=\SI{2}{\meV}$. Ref.~\cite{2010NatPh...6..284L} assumed values for the dielectric constant (taken at the gap instead of near the spin wave resonance) and bulk band gap (taken from the model without doping~\cite{Zhang2009a}) of \BiFeSe, which we wish to update (in Table~\ref{tab:material_params}, the values assumed by Ref.~\cite{2010NatPh...6..284L} are given in parentheses). Fortunately, both of these quantities can be factored out of the relevant expressions to arrive simply with the normalised integrals. We find:
\be
\mathcal{I}_1= 4\times 10^{-7} \, , \quad \mathcal{I}_2 = 4\mathcal{I}_1\times 10^{-8}\, .
\ee
Leading to the derived model parameters:
\begin{align}
\fAQ &= \SI{30}{\eV}\left(\frac{M_0}{\SI{0.03}{\eV}}\right)^{0.5}\left(\frac{\Vuc}{440\text{\AA}^3}\right)^{-0.5}\left(\frac{t}{\SI{0.04}{\eV}}\right)^{-1.5}\left(\frac{\mathcal{I}_1}{4\times 10^{-7}}\right)^{0.5} \label{eqn:fAQ_est}\\
\mAQ &= \SI{2}{\meV}\left(\frac{S}{4.99}\right)\left(\frac{U}{\SI{3}{\eV}}\right)\left(\frac{\mathcal{I}_2/\mathcal{I}_1}{4\times10^{-8}}\right)^{0.5}\, .\label{eqn:mAQ_est}
\end{align}
The derived parameters are presented in Table~\ref{tab:derived_params}, where we adopt the less committal names ``Material 1'' and ``Material 2'' for \BiFeSe\, and \MnBiTe{2}{2}{5} respectively, to acknowledge the limitations of our estimates.

Note that in Table~\ref{tab:material_params} we quote the anisotropy field $\mu H_A$, but that this plays no role in our estimation of the AQ parameters. The anisotropy field in fact determines the \emph{transverse} magnon masses (see Appendix~\ref{appendix:eft_afmr}), and not the mass of the longituninal AQ. In Paper~I we mistakenly assumed to use the transverse magnon mass for the AQ (along with a doping fudge factor). The transverse and longitudinal modes turn out to have similar masses. While we do not know of a fundamental reason for this coincidence, they are both clearly governed by the same $\mathcal{O}(\text{meV})$ magnetic energy scales. 

Finally, we mention the important \emph{spin flop transition} (for a detailed description and bibliography, see ref.~\cite{PhysRevB.66.214410}). Large magnetic fields cause spins to align and induce net magnetization. The magnetization increases linearly for fields larger than the spin-flop field, $H_{\rm SF}$, eventually destroying the AF order. The spin flop field for \MnBiTe{}{2}{4} is 3.5 T~\cite{Otrokov2019}. In easy axis systems, the AF order is destroyed completely when the magnetization saturates. This occurs at the \emph{spin flip transition} for fields larger than the exchange field, $H_E$. Large applied fields that destroy AF order will also destroy the AQ. For the exchange fields given in table~\ref{tab:material_params} we expect these transitions to happen in the many Tesla regime. In the following we consider fields up to 10 T for illustration.

\subsection{Damping and Losses}\label{sec:damping}
As discussed below, the magnon and photon losses are crucial in determining how effective an AQ material is for detection of DM.  In order to detect the AQ and measure its properties, it is essential that any experiment is carried out at temperatures below the N\'{e}el temperature. Fortunately both candidate materials have $T_N>4$\,K, and so initial measurements can be made at more accessible liquid Helium temperatures. As we discuss below, there are at least two sources of loss (conductance, and magnon scattering) that become less important at low temperatures. When using AQ materials to search for DM, it could therefore be advantageous to operate at $T\ll \omega_a$ dilution refrigerator temperatures. 

\begin{table}
\caption{Summary of the loss model. Elements of $\mat{\Gamma}$ are specified before diagonalising the kinetic term (see Eq.~\ref{eq:eom_with_losses}). The only $E$-field loss is the conductance. The total AQ loss is given by the sum of the remaining terms. Loss channels deemed negligible include AQ decay to photons, AQ-photon scattering, and off-diagonal losses.\label{tab:losses}}
\renewcommand{\arraystretch}{1.15}
\centering
\begin{threeparttable}
{\small
\begin{tabular}{ l l l l l}
\toprule
Type of losses & Symbol & Parameterisation & Reference values & Comments  \\ 
\midrule
Conductance & $\Gamma_\rho$ & $\epsilon_2 \,\omega$ & $10^{-4}\,\omega$ & \makecell[l]{Extrapolated from\\ optical wavelengths} \\
Gilbert damping & $\Gamma_{\rm lin}$ & $\alpha_\text{G} \, (1+\chi_m^{-1}) \, \omega$ & $10^{-9}\,\omega$ & $\chi_m$ given in Ref.~\cite{PhysRevX.9.041038}\tnote{\textdagger}\\
Magnon scattering & $\Gamma_{4m}$ & n/a &  & \makecell[l]{Boltzmann-suppressed\\ for $T<m$} \\
\makecell[l]{Impurities\\ \hspace{0.5em}\& domains} & $\Gamma_{\rm cryst.}$ & $(\delta L/L) \, \omega$ & $[10^{-4},10^{-3}]\,\omega$ & \makecell[l]{Typical impurity scale\\ $L\sim \SI{1}{\um}$~\cite{PhysRevX.9.041038}} \\
\bottomrule
 \end{tabular}
}
\begin{tablenotes}\footnotesize
	\item[\textdagger] We thank Chang Liu for providing this result.
\end{tablenotes}
\end{threeparttable}
\end{table}

\subsubsection{Resistivity and the Dielectric Function}

\begin{figure}
\centering
\includegraphics[width=0.7\textwidth]{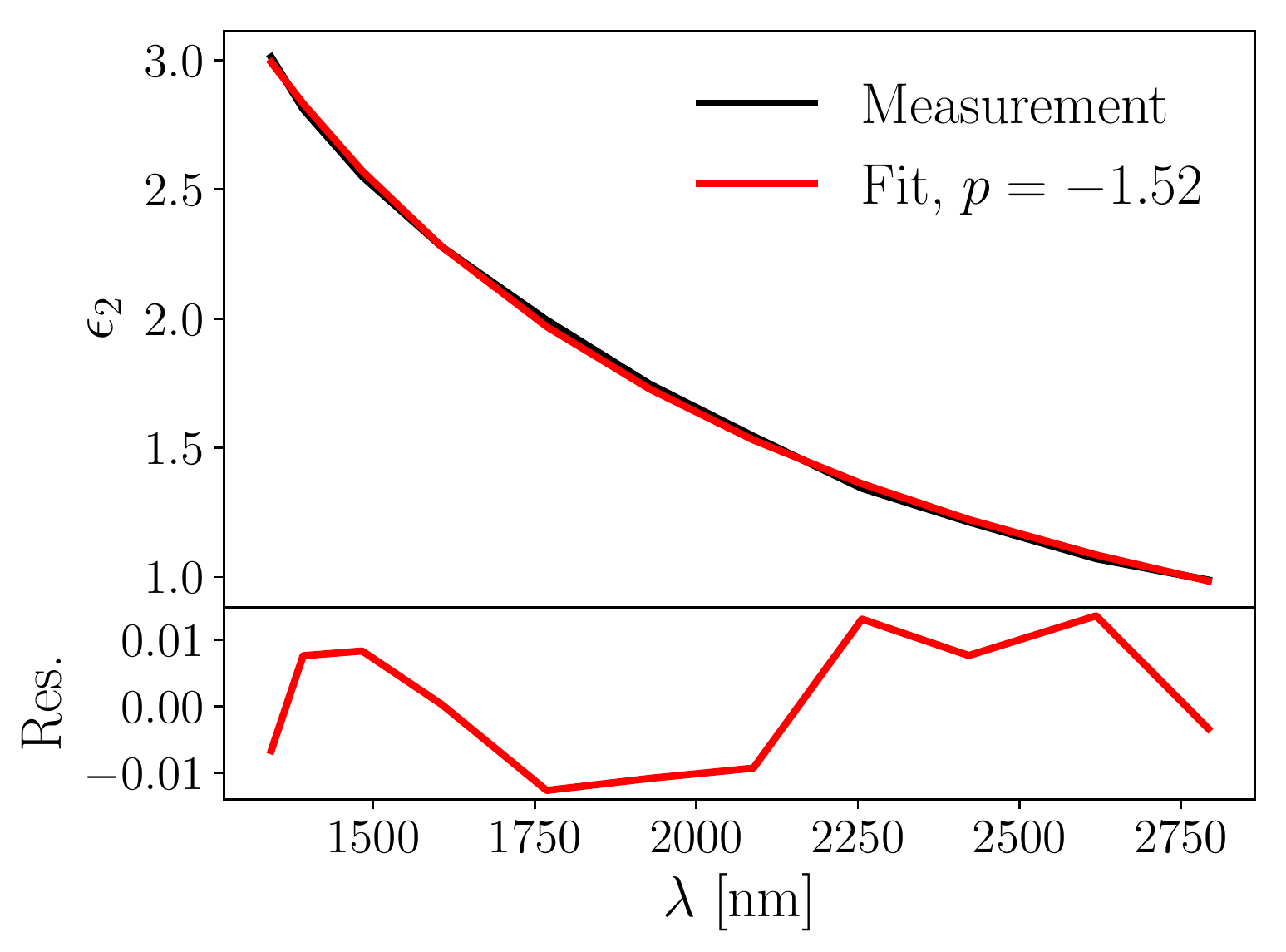}
\caption{Bi$_2$Se$_3$ dielectric function, $\epsilon_2$, as a function of wavelength $\lambda$ in the optical regime ($\omega\approx 1$ eV). Measurements are from Ref.~\cite{doi:10.1063/1.3466552,CaoWang} for the more favourable trigonal case. The results are well fit by a powerlaw, $\epsilon_2\propto \lambda^p$ with $p=-1.52$.}
\label{fig:epsilon_fit}
\end{figure}

Material conductance (inverse resistivity) appears in the $E$-field equations of motion as a damping term $\Gamma_\rho=1/\rho = \SI{0.6}{\meV}\,[\rho/\si{\ohm \cm})]^{-1}$, from which we see that a resonance near $\SI{1}{\meV}$ requires $\rho\gg \SI{1}{\ohm\cm})$ for $Q=\omega/\Gamma\gg 1$. For a resonance involving the electric field, one requires large resistance, i.e. low conductance.

Ref.~\cite{2011PhRvB..84g5316R} measure $\rho$ in the optical ($\omega\sim \SI{1}{\eV}$) at $T\approx \SI{1}{\K}$ of $\rho = \SI{2e-3}{\ohm\cm}$ for undoped Bi$_2$Se$_3$, lowering to $\rho = \SI{5e-4}{\ohm\cm}$ with doping. However, it is shown that annealing the TI at high $T$ can increase $\rho$ to be as large as \SI{1}{\ohm\cm}. For \MnBiTe{}{2}{4} the situation is similar, with two different measurements giving a longitudinal $\rho\approx \SI{e-3}{\ohm\cm}$ at $T\sim \mathcal{O}(\text{few\,\si{\K}})$~\cite{PhysRevMaterials.3.064202}. In the case of \MnBiTe{}{2}{4}, resistivity can be raised by doping with antimony~(\ce{Sb})~\cite{2020arXiv200409123M}. Even so, topological insulators are actually very poor insulators at typical electronic frequencies.  

The measurements of bulk $\rho$ for both Bi$_2$Se$_3$ and \MnBiTe{}{2}{4} are taken at high energy near the band gap around \SI{1}{\eV}, and far from the spin wave resonance frequency at low energies. References~\cite{doi:10.1063/1.3466552,CaoWang} studied the dielectric function of Bi$_2$Se$_3$ as a function of probe wavelength for the trigonal and orthorhombic phases. The complex dielectric function is $\tilde{\epsilon}(\omega) = \epsilon_1-i\epsilon_2$. For energies below the gap, $E\lesssim \SI{1}{\eV}$, $\epsilon_1$ has value around 25 at the longest wavelenths measured and is only slowly decreasing, while $\epsilon_2$ tends to zero rapidly at large wavelengths in the trigonal case (which is thus more favourable for our purposes). The value of $\epsilon_1$ is considerably smaller than the $\epsilon_1=100$ estimate used in Paper~I and assumed in Ref.~\cite{2010NatPh...6..284L}. As we show below, smaller values of $\epsilon_1$ are highly desirable for DM detection. 

The resistivity is given by $\rho(\omega)=1/[\omega\epsilon_2(\omega)]$. A narrow linewidth on resonance requires to $\epsilon_2(\omega_+)\ll 1$. Measurements in Ref.~\cite{CaoWang} extend to a maximum wavelength 2800 nm where $\epsilon_2\sim 1$. A simple power law extrapolation to THz wavelengths gives $\epsilon_2(\SI{1}{\meV})=9.5\times 10^{-5}$ (see fig.~\ref{fig:epsilon_fit}). Thus, the resistivity on the polariton resonance at wavelengths of order \SI{1}{\mm} is significantly higher than the bulk measurements in the optical. The value of $\epsilon_2$ is different for different crystal structures of Bi$_2$Se$_3$, and we consider only the most favourable case with the highest resistivity. We take the value $\epsilon_2=10^{-4}$ as a reference scale, however, we do not include any further frequency dependence, which would certainly be different for different materials, such as \MnBiTe{2}{2}{5}. The resistivity on resonance can be determined from the linewidth as measured by THz transmission spectroscopy, as we demonstrate in Section~\ref{sec:transmission}.

\subsubsection{Magnon Losses}

As we have discussed, the AQ is not described by the same EFT as ordinary AF-magnons. However, due to the relation between the AQ and the magnon fluctuation, we use the well-studied magnon case as a means to assess the possible magnitude of the axion linewidth, and the qualitative possibilities. Furthermore, as we will see, the dominant contribution is estimated to be due to material impurities, which do not depend on the microscopic model for the AQ. We split the magnon losses into different contributions:
\be
\GammaM = \sum_i\Gamma_i ,
\ee
where the index $i$ sums over terms defined in the following subsections. 

Ref.~\cite{LvovBook} gives a comprehensive account of non-linear wave dynamics relevant to the magnon linewidth. Early works on magnon scattering and linewidth include Ref.~\cite{PhysRevB.3.961}. The recent pioneering work of Refs.~\cite{Bayrakci1926,PhysRevLett.111.017204} showed how neutron diffraction with energy resolution down to 1~$\mu$eV can be used to confirm the theoretical predictions for the AF-magnon linewidth, and the dependence on temperature and momentum across the whole Brillouin zone, including many of the contributions discussed in the following. We focus on a few channels for losses, by means of example, closely following Ref.~\cite{PhysRevLett.111.017204}. Scattering channels that we have not considered include AF-magnon-ferromagnetic magnon scattering, and magnon-phonon scattering: these are discussed in e.g. Ref.~\cite{LvovBook}. 

In the present work, we are only concerned with the $q\approx 0$ mode at $T\ll T_N$, where many contributions can be neglected. In this regime, as we show in Section~\ref{sec:transmission}, the total AQ contribution to the linewidth can be measured using THz transmission spectroscopy.

\subsubsection*{``Linear'' Losses and Gilbert Damping, $\Gamma_{\rm lin}$}

Losses are historically incorporated for spin waves by the introduction of the phenomenological Gilbert damping term into the Landau-Lifshitz equation, making the Landau-Lifshitz-Gilbert (LLG) equation. Gilbert damping is a linear loss, since it simply represents decay of spin waves due to torque. There is not a universally accepted first principles model of Gilbert damping. One possible model is presented in Ref.~\cite{PhysRevLett.102.137601}, where Gilbert damping is shown to arise due to spin orbit coupling in the Dirac equation (other models include Refs.~\cite{PhysRevB.75.214420,PhysRevB.79.064404}). In this case, the damping term is written as:
\be
\Gamma_{\rm lin} = \alpha_{\rm G}(1+\chi_m^{-1})\omega \, . \quad \alpha_{\rm G} = \frac{e \mu \Sigma_s}{8 m_e^2}\, ,
\label{eqn:gilbert_def}
\ee 
where $m_e$ is the electron mass and $\chi_m$ is the dimensionless magnetic susceptibility (volume susceptibility in SI units), and $\Sigma_s=S/\Vuc$. The dimensionless prefactor $\alpha_{\rm G}$ is of order $10^{-12}$ for \BiFeSe\, and \MnBiTe{2}{2}{5}. The value of $\chi_m$ was measured for \MnBiTe{}{2}{4} in Ref.~\cite{PhysRevX.9.041038} and found to be of order $\chi_m\approx 10^{-3}$ for $T<T_N$ (see Table~\ref{tab:losses}). Thus the relative width, $\Gamma/\omega$, is of order $10^{-9}$, which is negligible compared to the other sources of loss in the following. Furthermore, $\chi_m$ is small enough to be neglected in the magnetic permeability (with $c=1$), $\mu_m=1+\chi_m$, which we fix to unity.

\subsubsection*{Magnon-Magnon Scattering, $\Gamma_{4m}$}
\begin{figure}
\centering
    \includegraphics[width=0.44\textwidth]{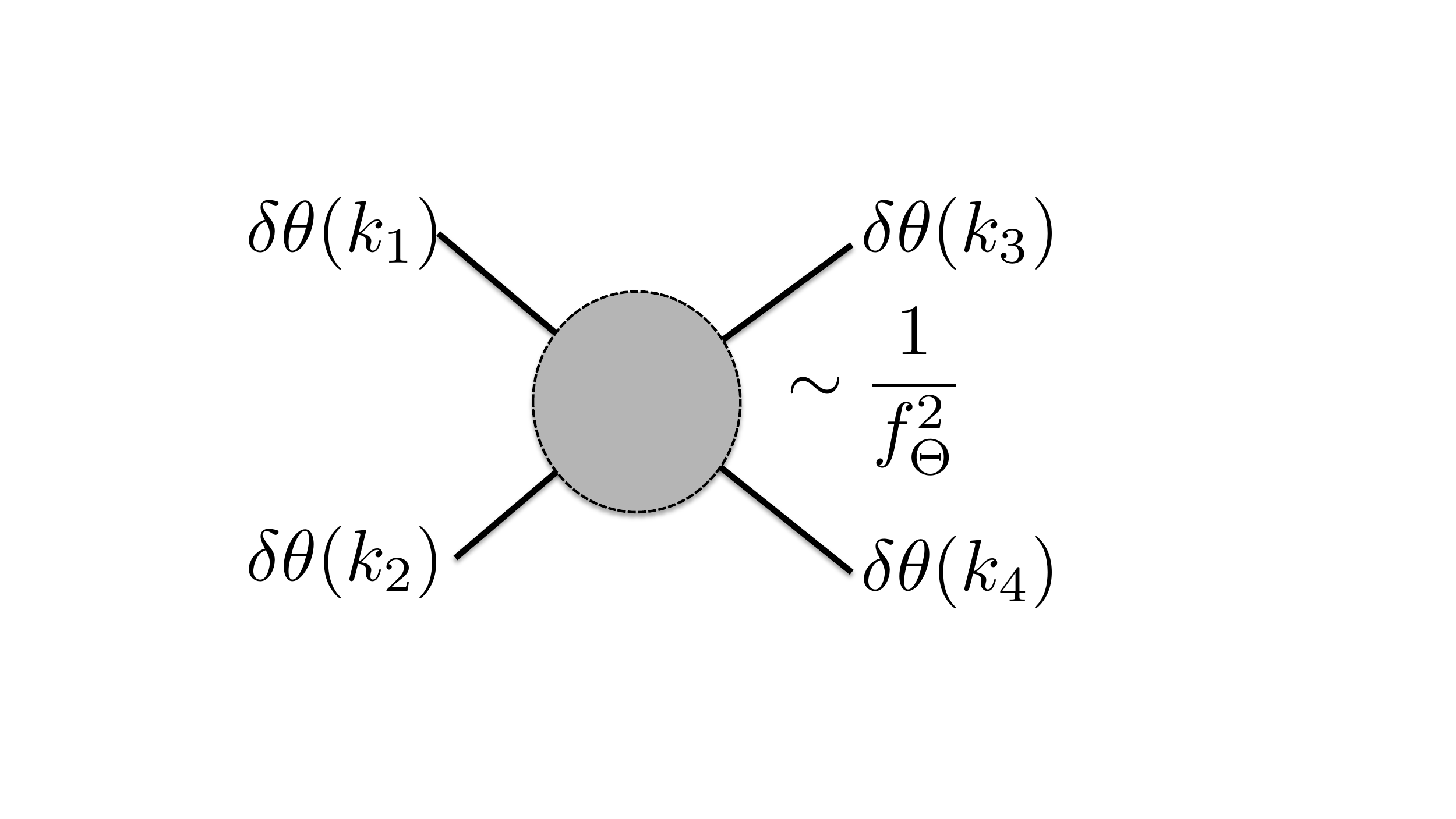}
    \includegraphics[width=0.54\textwidth]{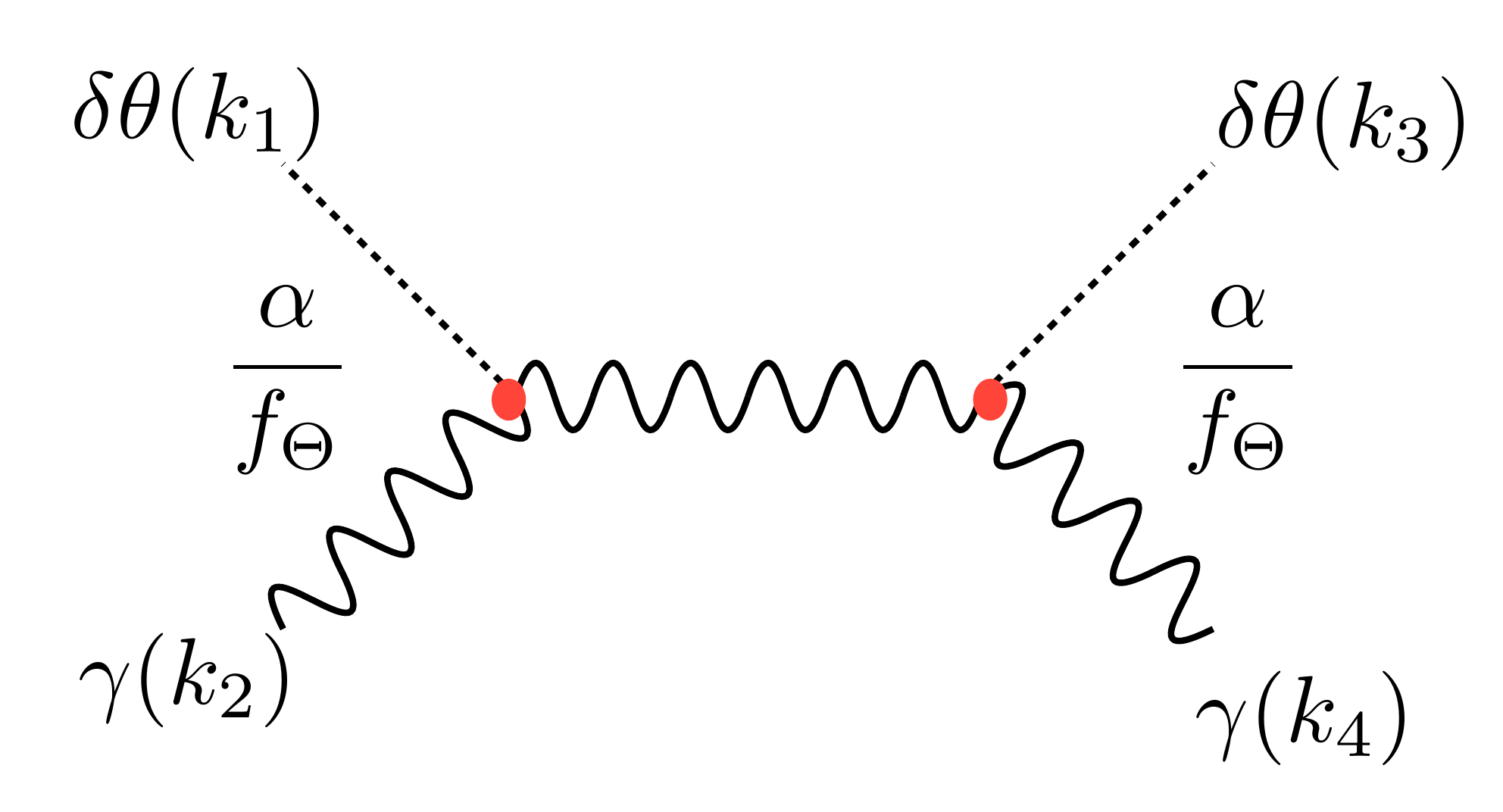}
\caption{\emph{Left}: Four magnon scattering. In EFT, the amplitude can be calculated as shown in Ref.~\cite{Hofmann:1998pp}. As shown in Ref.~\cite{PhysRevLett.111.017204}, it is the leading contribution to the magnon linewidth for $T\sim T_N$. For temperatures far below the spin wave mass, this term is Boltzmann suppressed. \emph{Right}: Feynman diagram for the $s$-channel of the process Eq.~\eqref{eqn:axion_photon_scattering} mediated by the axion term in the Lagrangian. The parametric dependence of the vertex factors is shown in red. This process is suppressed by two powers of $\alphaEM$ compared to the four magnon amplitude, Eq.~\eqref{eqn:magnon_amplitude}.}
    \label{fig:feynman}
\end{figure}

Reference~\cite{PhysRevLett.111.017204} showed that two-to-two magnon scattering is the dominant contribution to the linewidth above $\sim 10$ K in the antiferromagnets Rb$_2$MnF$_4$ and MnF$_2$ as measured by neutron scattering. The linewidth at 10 K due to this process is $\Gamma_m\approx \SI{10}{\mu eV}$, falling rapidly at lower temperatures. We will show how this behaviour arises below. Indeed, as noted in \cite{Bayrakci1926}, for $q \rightarrow 0$ and $T \rightarrow 0$, all scattering contributions to the magnon linewidth vanish. Ref \cite{Bayrakci1926} also find that this is true for scattering between the magnon and longitudinal spin fluctuations such as the axion. We find it useful to derive in some detail the scattering contribution to the linewidth, and demonstrate why it vanishes at low temperature, since this is the most well understood part of our loss model.

The magnon modes obey a Boltzmann equation. Mode coupling via non-linearities induces an effective lifetime for any initial configuration. Mode coupling arises from the four-magnon amplitude:
\be
\delta\theta (k_1)+\delta\theta(k_2)\longleftrightarrow \delta\theta (k_3)+\delta\theta(k_4)\, ,
\label{eqn:four_magnon_1}
\ee
which has matrix element $\mathcal{M}(k_1,k_2,k_3,k_4)$, and is shown in Fig.~\ref{fig:feynman}.  The state with momentum $k_1$ is the mode in the condensate of interest, $k_2$ is a thermal magnon. Magnons $k_3$ and $k_4$ are modes scattered out of the condensate, and thus losses. This matrix element appears in the collisional Boltzmann equation for the magnon distribution function $f_1\equiv f(k_1)$ as (see e.g. Ref.~\cite{2003moco.book.....D}):
\begin{eqnarray}
\frac{{\rm d}f_1}{{\rm d}t} &=& \int \prod_{i=2}^4{\rm d}\Phi_i (2\pi)^4\delta^{(3)}(k_1+k_2-k_3-k_4)\delta (\omega_1+\omega_2-\omega_3-\omega_4)\nonumber \\ 
&\,&|\mathcal{M}|^2[f_3f_4(1+f_1)(1+f_2)-f_1f_2(1+f_3)(1+f_4)]\, , \\
&=& -\int \frac{{\rm d}^3k_2}{(2\pi)^3}f_1f_2|v_1-v_2| \sigma\, ,
\label{eqn:boltzmann}
\end{eqnarray}
where ${\rm d}\Phi_i = \frac{d^3 k_i}{(2 \pi)^3}$ is the \emph {non-relativistic} phase space element for state with momentum $k_i$, the Dirac delta's enforce energy-momentum conservation, $k_i$ represents the 3-momentum of the ith particle, and the $f_i$ factors assume the particles are bosons. Ref.~\cite{PhysRevLett.111.017204} caution that when such integrals are evaluated numerically, one should be careful to include the Umklapp processes, related to conservation of crystal momentum.v

We formulate the integral non-relativistically as the material picks out a preferred frame for the magnons. In the second line, the first term in the square brackets represents production of states $k_1,k_2$ (the inverse process in Eq.~\ref{eqn:four_magnon_1}), while the second term represents losses. In the last line we have assumed $f_3=f_4=0$ for unoccupied final states, and used the definition of the differential cross section (this structure is familiar from particle physics scattering theory~\cite{1995iqft.book.....P}). Ref.~\cite{LvovBook} derives an equivalent equation beginning from the LLG equation, which also shows this non-linear loss term explicitly in terms of the four-magnon amplitude.

Eq.~\eqref{eqn:boltzmann} is the collisional Boltzmann equation, $\partial_t f_1 = C[f_1]$, where $C[f_1]$ is the scattering integral. Factoring out $f_1$ for the condensate, the scattering integral takes the form $C[f_1]\sim 1/\tau$ and we identify the relaxation time $\tau$ for the distribution function to change significantly from its initial state. This gives the result that:
\be
\Gamma_{4m} =1/\tau\sim \langle\sigma v\rangle\, ,
\ee
where the angle brackets denote the thermal average, i.e. phase space integral with the thermal distribution $f_2$. 

Magnons can be described by EFT, as discussed in Appendix~\ref{appendix:eft_afmr}. The four magnon amplitude is given by the equivalent of the QCD pion amplitude evaluated around non-zero quark masses~\cite{Hofmann:1998pp}.
\begin{multline}
\mathcal{M} = \frac{1}{4 \sqrt{\omega_1 \omega_2 \omega_3 \omega_4}} \frac{v_m^4}{F_2^2} \lbrace \delta^{ab} \delta^{cd} \left( \frac{2}{v_m^2} \omega_1 \omega_2 -2 k_1 \cdot k_2 + m_m^2 \right) \\
+ \delta^{ac} \delta^{bd} \left( - \frac{2}{v_m^2} \omega_1 \omega_3 +2 k_1 \cdot k_3 + m_m^2 \right) + \delta^{ad} \delta^{bc} \left( - \frac{2}{v_m^2} \omega_1 \omega_4 +2 k_1 \cdot k_4 + m_m^2 \right) \rbrace,
\label{eqn:magnon_amplitude}
\end{multline}
where $a, b, c, d = 1, 2$ denote the magnon polarizations, $v_m$ is the magnon velocity and $m_m$ is the magnon mass.

This is the amplitude appropriate to a non-relativistic normalization, with 1 particle per unit volume rather than the usual $2 \omega$ particles per unit volume in relativistic quantum mechanics.  

In this case, the cross section is related to the T-matrix above as~\cite{Hofmann:1998pp}:
\be 
d \sigma  = \frac{|\mathcal{M}|^2}{v} (2 \pi)^4 \delta^4(k_1 + k_2 - k_3 - k_4) \frac{d^3 k_3}{(2 \pi)^3}  \frac{d^3 k_4}{(2 \pi)^3},
\ee
where $v = |v_1 - v_2|$ is the relative velocity of the incoming particles.

We integrate over $k_3$ and $k_4$ to obtain the total cross section for given incoming momenta $k_1$ and $k_2$:
\be 
\sigma(k_1,k_2) = \int \frac{d^3 k_3}{(2 \pi)^3} \frac{d^3 k_4}{(2 \pi)^3} d \sigma = \int \frac{d k_4}{(2 \pi)^3} d \Omega k_4^2\frac{| \mathcal{M}|^2}{v} (2 \pi) \delta(\omega_1 + \omega_2 - \omega_3 - \omega_4)) \bigg \rvert_{(k_1+k_2 - k_3 - k_4 = 0)},
\ee
where $\omega_4 = \sqrt{k_4^2 v_m^2 + m_m^2}$ and $\omega_4 = \sqrt{(k_1 + k_2 - k_4)^2 v_m^2 + m_m^2}$. At this point in the calculation, we might be tempted to move to the centre of mass frame. However, this would change the magnon velocity $v_m$, with the new magnon velocity depending on $\Omega$, leading to a magnon dispersion relation that depends on $\Omega$. Therefore, it is in our best interests to remain in the rest frame of the material. We thus obtain the differential cross section:
\be 
\frac{d \sigma}{d \Omega} = \frac{1}{(2 \pi)^2} \frac{1}{v} \frac{|\mathcal{M}|^2 k_4^2}{\frac{k_3}{\omega_3} + \frac{k_4}{\omega_4}},
\ee 
where $k_3,\omega_3, k_4, \omega_4$ are defined by conservation of energy and momentum for a given $\Omega$. \\

Now let us consider the scaling of $\Gamma_{4m}$ with temperature $T$.  We note first that the factor of $v$ in $\Gamma_{4m}$ is cancelled by the factor of $\frac{1}{v}$ in $\frac{d \sigma}{d \Omega}$. We will focus first on the scaling of the line widths measured in \cite{PhysRevLett.111.017204} at temperatures from $3$\,K to $0.8 T_N$ for magnons with momentum $k_1 = 0$ to $k_1 = q_ZB$ at the edge of the zone boundary. The contributions of $T$ to $\Gamma_{4m}$ are as follows:
\begin{itemize}
\item Thermal magnons have an energy set by $T$. We assume that $T \gtrsim m_m$, such that thermal magnons can be excited. We therefore take $\omega_2 \sim k_2 \sim T$.

\item The scaling of the outgoing momenta with $T$ depends on the relative sizes of $T$ and $\omega_1$. The energy at the zone boundary in \cite{PhysRevLett.111.017204} is \SI{6.6}{\meV} for Rb\textsubscript{2}MnF\textsubscript{4} and \SI{6.3}{\meV} for MnF\textsubscript{2}, while the temperature ranges from $3\,\text{K} = \SI{0.26}{\meV}$ to $0.8\, T_N$, corresponding to \SI{2.6}{\meV} and \SI{4.6}{\meV} respectively. Therefore both cases where $T > \omega_1$ and cases where $\omega_1 > T$ are measured. When $T \gg \omega_1$, the temperature provides most of the energy in the scattering process and we have $k_3,\omega_3,k_4,\omega_4 \sim T$. When $T \ll \omega_1$, the energy of the damped magnon provides most of the energy in the scattering process and we have $k_3,\omega_3,k_4,\omega_4 \sim \omega_1$.

\item The \emph{number} of thermal magnons also scales with $T$. Assuming that there is no significant mass gap at $k_2=0$ for the magnons considered in \cite{PhysRevLett.111.017204}, we have $\int d^3 k_2 f_2 \sim T^3$, as for a black body. 

\item We have also $\int d^3 k_3 d^3 k_4 \sim T^2$ when $T \gg \omega_1$ from the factor of $k_4^2$ in the phase space integral. 

\item As $T \gtrsim m_m$, the $\omega_1$ and $k_1$ terms in $\mathcal{M}$ dominate. Using the scalings above, this gives $\mathcal{M} \sim T^{-1/2}$.

\end{itemize}

Putting these elements together, we find $\Gamma_{4m} \sim T^3 T^2 T^{-1} = T^4$ for $\omega_1 \ll T$ and $\Gamma_{4m} \sim T^3 T^{-1} = T^2$ for $\omega_1 \gg T$. We can compare this prediction with the measured result in Figure 4 in \cite{PhysRevLett.111.017204}. For low magnon wavenumber $q$ (corresponding to low $\omega_1$, we have $\Gamma_{4m} \sim T^4$ as expected. As $q$ is increased, the scaling with $T$ decreases towards $\Gamma_{4m} \sim T^2$ as predicted. However, the measured $\Gamma_{4m} \sim T$ when $q=0$ case is not explained by this analysis.

We would also expect that for temperatures much lower than the magnon mass, very few thermal magnons would be excited, and $\Gamma_{4m}$ would be exponentially suppressed. For a magnon mass $m_m \sim 1$ meV, this corresponds to $T < 10$ K. 

Starting from the Boltzman equation, we have argued that magnon-magnon scattering decays with $T$, reproducing the experimentally observed trends in \cite{PhysRevLett.111.017204}, and is then exponentially suppressed at temperatures below the magnon mass. The scattering contribution to the antiferromagnetic magnon linewidth is calculated analytically for several low $T$ regimes in \cite{PhysRevB.3.961}. This yields a power law fall off with $T$ in each case. 

We therefore conclude that, at low $T$, and particularly for temperatures below the magnon mass, the magnon scattering contribution to the linewidth is negligible.

\subsubsection*{Axion-Photon Scattering, $\Gamma_{\gamma m}$}

Scattering of magnons from thermal photons contributes to the magnon line-width $\Gamma_m$. This process is induced by the four particle amplitude\:
\be
\delta\theta (k_1)+\gamma(k_2)\longleftrightarrow \delta\theta (k_3)+\gamma(k_4)\, ,
\label{eqn:axion_photon_scattering}
\ee
i.e. magnon/AQ-photon scattering mediated by the Chern-Simons interaction, Eq.~\eqref{eqn:chern_simons}. Inspecting the Feynman diagram, Fig.~\ref{fig:feynman} (right panel), this amplitude is suppressed by two powers of the fine structure constant $\alphaEM$ with respect to the four magnon amplitude, and so we do not expect magnon-photon scattering to be significant compared with magnon-magnon scattering. The inverse process, scattering thermal magnons from the electric field, is similarly suppressed, and thus likely to be subdominant to conductive losses to $\vc{E}$.

\subsubsection*{Axion Lifetime, $\Gamma_{m\gamma\gamma}$}

The Chern-Simons interaction leads to direct decay of an AQ into two photons. The contribution to the width is:
\be
\Gamma_{m\gamma\gamma} = \frac{\alpha^2}{256\pi^3}\frac{m_s^3}{\fAQ^2} = \SI{6.7e-22}{\eV}\left(\frac{m_s}{\text{meV}}\right)^3\left(\frac{\SI{100}{eV}}{\fAQ}\right)^2\, ,
\ee
corresponding to a lifetime on the order of months. This process can be safely neglected compared to all other scales in the problem.

\subsubsection*{Off-Diagonal Losses}
The off-diagonal terms in the loss  correspond to loss terms of the form $\frac{d \phi_k}{d t} \sim \langle A_k \rangle$ and $\frac{d A_k}{d t} \sim \langle \phi_k \rangle$. As we generically expect $\frac{d \phi_k}{d t} \sim \langle \phi_k \rangle$ and $\frac{d A_k}{d t} \sim \langle A_k \rangle$, these will all be of the form:

\begin{equation}
\frac{d \phi_k}{d t} \sim \langle A_k \rangle \langle \phi_k \rangle,
\end{equation}

\begin{equation}
\frac{d A_k}{d t} \sim \langle A_k \rangle \langle \phi_k \rangle.
\end{equation}

These off-diagonal loss terms are therefore not present at linear order in the perturbations $A_k$ and $\phi_k$.

\subsubsection*{Impurities and Domains, $\Gamma_{\rm cryst.}$}

At low temperatures, the dominant contribution to the magnon linewidth in Ref.~\cite{PhysRevLett.111.017204} is attributed to scattering of magnons off magnetic domains and crystal impurities, which is $T$-independent . 

In the simplest picture, scattering from magnetic domains leads to a lifetime:
\be
\tau \sim \frac{L_{\rm mag.}}{2v(q)}\, ,
\label{eqn:naive_domain_scattering}
\ee
where $v(q)$ is the velocity of the mode with momentum $q$, and $L_{\rm mag.}$ is the size of the domain. The Ref.~\cite{PhysRevX.9.041038} crystals of \MnBiTe{}{2}{4}\, have estimated magnetic domain size $L_{\rm mag.}\sim 1 \, \mu{\rm m}$. We require the axion-polariton to propagate at least through the thickness, $d$, of the sample, and thus magnetic domains appear to strongly affect the skin depth and resonance width of axion-quasiparticle dominated polaritons in the limit $d\gg L_{\rm mag.}$.

However, in the  $q\to0$ limit the magnon wavelength exceeds the size of a domain and  Eq.~\eqref{eqn:naive_domain_scattering} ceases to apply. Furthermore we consider the limit $v_s=0$ and ignore the magnon propagation compared to the electric field. It is currently unknown how scattering from domains will affect such long wavelength mixed modes. On one hand, it may be that the domain walls appear as small scale fluctuations that decouple from large wavelength modes. Conversely, given that the domain walls disrupt the short range interactions that support the small $q$ magnons it is possible that they have non-trivial effects despite the scale separation. 

A second $T$-independent contribution to the linewidth, which is expected to remain in the $q\to0$ limit, is due to scattering from impurities. This was accounted for in Ref.~\cite{PhysRevLett.111.017204} with the simple phenomenological model for the impurity density:
\be
\Gamma_{\rm cryst.} = \left(\frac{\delta L}{L}\right)\omega(k)\,
\label{eqn:impurity_lineiwdth}
\ee
where $\delta L$ is the lattice constant, and $L$ is the spacing between impurities, thus $\delta L/L$ is the average number of lattice sites between impurities. The model Eq.~\eqref{eqn:impurity_lineiwdth} accounts in the same manner for magnetic and crystal impurities. In Ref.~\cite{PhysRevX.9.041038} the crystal impurities occur on the same scale as the magnetic domains, $L_{\rm cryst.}\sim 1 \, \mu{\rm m}$, while $\delta L\sim (V_{\rm u.c.})^{1/3}\sim 6$~\AA\, leading to:
\be
\Gamma_{\rm cryst.} = 7\times 10^{-4}\omega\left(\frac{\delta L}{6\text{ \AA}}\right)\left(\frac{1 \,\mu{\rm m}}{L_{\rm cryst.}}\right)\, .
\ee

We estimate that crystal impurity scattering is the dominant contribution to the AQ linewidth in the regime of interest. Given the lack of conclusive calculations or measurements in the literature (or even, as far as we can tell, a detailed model), we regard this as a question best resolved by experimental studies. Indeed, an understanding of the dynamics of small $q$ magnons and axion-polaritons is an interesting off-shoot of the studies proposed in Section~\ref{sec:AQ_calculations}. However, given the importance of this linewidth contribution to our proposed dark matter search, we must adopt a reference value. We adopt the range given in Table~\ref{tab:losses}, $\Gamma_{\rm cryst.}\in [10^{-4},10^{-3}]\,$meV, corresponding to impurity separations of order $1\,\mu$m.

\section{Discovering the Axion Quasiparticle}\label{sec:AQ_calculations}

\begin{figure}
   \hspace{-0.5in}

	\center
   \includegraphics[width=1.0\textwidth]{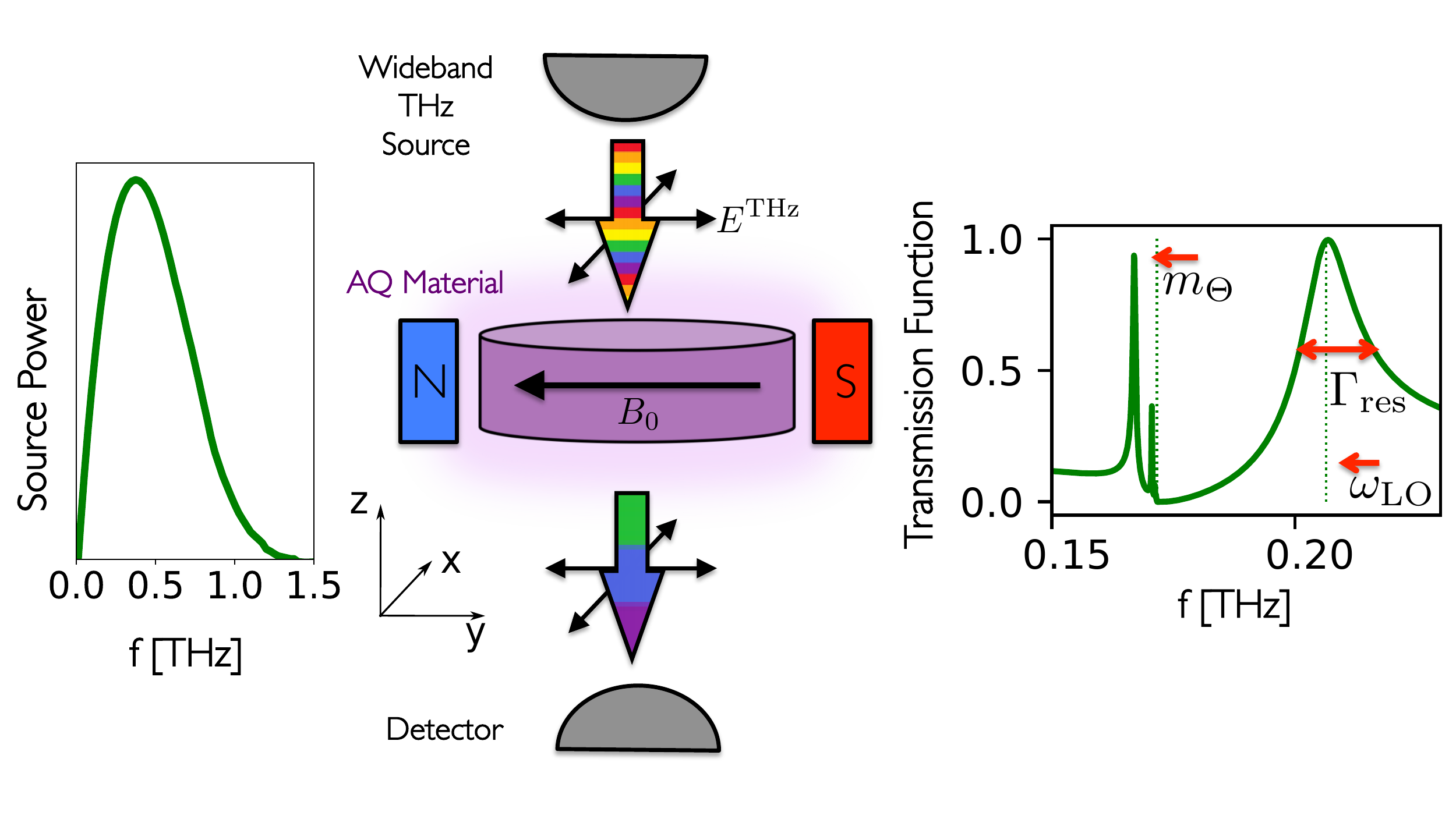}
  	\caption{Proposed transmission experiment to detect the axion-polariton. \emph{Left:} THz source power spectrum. \emph{Centre:} Transmission experiment concept. A source field, which propagates along the negative $z$-direction is incident on a TMI. An external $B$-field $B_e$ is applied parallel to the TMI surface. If AQs exist in the material, the dispersion relation has a gap where no propagating modes exist, thus altering the spectrum of the transmitted radiation. \emph{Right:} Theoretical transmission spectrum. The green line corresponds to the case where a dynamical AQ is present. The gap is indicated by the vertical green dotted lines. The width on resonance, $\Gamma_{\rm res}$, serves to measure the polariton losses.}
  	\label{fig:ATR}
\end{figure}

One of the methods proposed by Ref.~\cite{2010NatPh...6..284L} to detect the presence of AQs in TMIs was total-reflectance measurement, and idea we explore further here. In the following we show to compute the transmission function of TMIs using axion electrodynamics. The transmission function is shown to display a gap, leading to total reflectance. Furthermore, by using a wideband THz source, such a measurement can also determine the axion-polariton resonant frequencies, and loss parameters. The concept of this \emph{THz transmission spectroscopy} measurement is shown in Fig.~\ref{fig:ATR}. Similar measurements have been performed on antiferromagents (e.g. Ref.~\cite{2017PhRvL.119v7201L}), which demonstrate AFMR and determine the magnon linewidth (losses on resonance) for an electromagnetic source.\,\footnote{Crucially, for our purposes, such a measurement uses precisely the same physics (oscillating $E$-field source) as occurs for dark axion detection. This is in contrast to neutron scattering of antiferromagnets (e.g. Ref.~\cite{PhysRevLett.111.017204}), which determines the linewidth for a different excitation mechanism.} Such a measurement has not to date been performed on any AQ candidate material.

\subsection{Axion Electrodynamics and Boundary Conditions}\label{sec:boundary_conditions}
In this section, we review the axion-Maxwell equations for TMIs. We then derive a one-dimensional model as well as the correct interface conditions for all fields involved. Based on the one-dimensional model, we compute the reflection and transmission coefficients for incoming THz~radiation.

\subsubsection{General formulation}\label{sec:Quasiparticle_Photon_mixing_General}
The macroscopic axion-Maxwell equations for a three-dimensional TMI are~\cite{2010NatPh...6..284L}
\begin{eqnarray}
		\nabla\cdot \vc{D}&=&\rho_f-\frac{\alphaEM}{\pi}\nabla(\dyAQ+\stAQ)\cdot\vc{B} \, ,\label{GaussE}\\
		\nabla\times\vc{H}-\partial_t\vc{D}&=&\vc{J}_f+\frac{\alphaEM}{\pi} \, (\vc{B}\partial_t(\dyAQ+\stAQ)-\vc{E}\times\nabla( \dyAQ+\stAQ)) \, ,\label{Ampere}\\
		\nabla\cdot\vc{B}&=&0 \, , \label{GaussB}\\
		\nabla\times\vc{E}+\partial_t\vc{B}&=&0 \, ,\label{Faraday}\\
		\partial_t^2\dyAQ-v_i^2\partial_i^2\dyAQ+\mAQ^2\dyAQ&=&\Lambda\vc{E}\cdot\vc{B}\label{KleinGordon} \, ,
\end{eqnarray}
where $\dyAQ$ is the pseudoscalar axion quasiparticle~(AQ) field, $\stAQ\in[0,\pi]$ a constant,  
$\fAQ^2$ the AQ decay constant, $v_i$~(with $i=x,y,z$) is the spin wave velocity, $\mAQ$ the spin wave mass, $\vc{E}$ is the electric field, $\vc{B}$ the magnetic flux density, $\vc{D}$ the displacement field, $\vc{H}$ the magnetic field strength, $\rho_f$ the free charge density, and $\vc{J}_f$ the free current density, which fulfill the continuity equation ${\nabla \cdot \vc{J}_f+\dot{\rho}_f = 0}$ as in usual electrodynamics. In what follows we often use the linear constitutive relations 
\begin{eqnarray}
\vc{D}=\epsilon\vc{E} \text{ and } \vc{H}=\mu^{-1}\vc{B},
\label{eq:constitutive_equations}
\end{eqnarray}
where $\epsilon$ and $\mu$ are the scalar permittivity and permeability, respectively. Note that it is important to include the $\stAQ$ term in the equations above: while $\stAQ$ is some constant in the TMI, it is always zero in vacuum. Applying the nabla operator can therefore give a delta function at the boundaries of the TMI, i.e.\ a boundary charge term.

Equations~\eqref{GaussE} and~\eqref{Ampere} can be written such that the terms including the dynamical AQ field $\dyAQ$ can be interpreted as additional contributions to polarization and magnetization, i.e.\
	\begin{eqnarray}
		\nabla\cdot \vc{D}_\Theta&=&\rho_f \, ,\label{GaussE_pol}\\
		\nabla\times\vc{H}_\Theta-\partial_t\vc{D}_\Theta&=&\vc{J}_f\, ,\label{Ampere_pol}\\
		\nabla\cdot\vc{B}&=&0\,,\label{GaussB_pol}\\
		\nabla\times\vc{E}+\partial_t\vc{B}&=&0\label{Faraday_pol}\, ,
	\end{eqnarray}
where we define
\begin{eqnarray}
\vc{D}_\Theta=\vc{D}+\frac{\alphaEM}{\pi} \, (\stAQ+\dyAQ) \, \vc{B}\, , \\
\vc{H}_\Theta = \vc{H} - \frac{\alphaEM}{\pi} \, (\stAQ+\dyAQ) \, \vc{E} \, .
\end{eqnarray}
To derive interface conditions for the electromagnetic fields, we consider two domains labeled~$1$ and~$2$. Both domains have different $\epsilon$, $\mu$, and $\stAQ$. Transforming Eqs.~\eqref{GaussE_pol}--\eqref{Faraday_pol} into their integral representation, and applying Gauss's~(Stokes') theorem to an infinitesimal volume~(surface) element, leads to the following interface conditions for the electromagnetic fields:
\begin{eqnarray}
\vc{n}\times (\vc{E}_2-\vc{E}_1)&=&0\,,\label{eq:InterfaceE}\\
\vc{n}\cdot (\vc{D}_{\Theta,2}-\vc{D}_{\Theta,1})&=&\sigma_S\,,\label{eq:InterfaceD}\\
\vc{n}\cdot (\vc{B}_{2}-\vc{B}_{1})&=&0\,,\label{eq:InterfaceB}\\
\vc{n}\times (\vc{H}_{\Theta,2}-\vc{H}_{\Theta,1})&=&\vc{J}_S\,,\label{eq:InterfaceH}
\end{eqnarray}
where $\sigma_S$ and $\vc{J}_S$ are free surface charge and current densities~(both assumed to be zero in what follows) and $\vc{n}$ is a unit vector pointing from domain~1 to domain~2. It is important to stress that Eqs.~\eqref{eq:InterfaceE}--\eqref{eq:InterfaceH} are interface conditions, not boundary conditions.

As described above, interface conditions follow from the differential equation in their integral form. In contrast, boundary conditions can be applied at the boundary domains for which a partial differential equation is solved, and do not follow from the integral representation of the differential equation. This is also the reason why the interface conditions are specified only for the electromagnetic fields, and not for the dynamical axion field $\dyAQ$. In this section we only consider the case of a TMI surrounded by a non-topological material/vacuum with $\stAQ=0$. We then only need to impose interface conditions for the electromagnetic fields; while they exist in both the TMI and the adjacent region, the dynamical AQ only exists in the TMI. The AQ is therefore only subject to boundary conditions. We revisit and deepen this discussion in Section~\ref{sec:transmission}, in the context of calculating reflection and transmission coefficients for a layer of TMI surrounded by vacuum.

\subsubsection{One dimensional model}\label{sec:Quasiparticle_Photon_mixing_OneDimensionalModel}
To develop a one-dimensional model, we assume that all fields only depend on the $z$-coordinate and time. Furthermore, all fields are taken to be transverse fields, i.e.\ $B_z=H_z=D_z=E_z=0$. Then, in a domain with constant $\stAQ$, Eqs.~\eqref{GaussE}--\eqref{KleinGordon} reduce to:
\begin{eqnarray}
	\partial_z\begin{pmatrix}-H_y\\H_x\end{pmatrix}-\partial_t\begin{pmatrix} D_x\\ D_y\end{pmatrix}-\vc{J}_f&=&\frac{\alphaEM}{\pi}\Bigg[\begin{pmatrix}B_x\\B_y\end{pmatrix}\partial_t\dyAQ+\begin{pmatrix}-E_y\\E_x\end{pmatrix}\partial_z\dyAQ\Bigg]\, ,\label{Ampere_1Df}\\
	\partial_z\begin{pmatrix}-E_y\\E_x\end{pmatrix}+\partial_t\begin{pmatrix}B_x\\B_y\end{pmatrix}&=&0\, ,\label{Faraday_1Df}\\
	\partial_t^2\dyAQ-v_z^2\partial_z^2\dyAQ+\mAQ^2\dyAQ&=&\Lambda(E_xB_x+E_yB_y)\label{KleinGordon_1Df}\, ,
\end{eqnarray}
where we assumed that no free static charges exist, i.e.\ $\rho_f=0$.\footnote{This does not mean that $\vc{J}_f$ vanishes. Since $\rho_f$ and $\vc{J}_f$ are connected via a continuity equation, $\vc{J}_f$ only has to fulfil $\nabla\cdot\vc{J}_f=0$ if $\rho_f=0$.} The interface conditions~\eqref{eq:InterfaceD} and~\eqref{eq:InterfaceB} are trivially fulfilled in the one-dimensional model since the $z$-components of all electromagnetic fields vanish, and $\vc{n}=\hat{\vc{e}}_z$.

\subsubsection{Linearization}\label{sec:Quasiparticle_Photon_mixing_Linearization}
The sources in Eqs.~\eqref{Ampere_1Df} and~\eqref{KleinGordon_1Df} are non-linear and, therefore, finding analytic solutions is in general not possible. However, we are interested in the special case of solving the equations in presence of a strong, static external $B$-field $\vc{B}_e=B_e\hat{\vc{e}}_y$. We may therefore separate the total $B$-field into a static and a dynamical part, i.e.\ $\vc{B}\rightarrow \vc{B}_e+ \vc{B}(\vc{x},t)$. Similarly, the free current $\vc{J}_f$ can be split into a part which sources $\vc{B}_e$, and an additional reaction current, i.e.\ $\vc{J}_f\rightarrow \vc{J}_{f0} + \vc{J}_f$. Physically, the reaction current describes losses of the electromagnetic fields in the materials. Note that $\vc{B}_e$ fulfils $\nabla\times\vc{H}_e=\vc{J}_{f0}$, and $\vc{J}_{f0}$ satisfies the continuity equation $\nabla\cdot\vc{J}_{f0}=0$. With these assumptions the resulting equations are: 
\begin{eqnarray}
	\partial_z\begin{pmatrix}-H_y\\H_x\end{pmatrix}-\partial_t\begin{pmatrix} D_x \, \\ D_y\end{pmatrix}-\sigma\begin{pmatrix}E_x\\E_y\end{pmatrix}&=&\frac{\alphaEM}{\pi}\Bigg[\begin{pmatrix}B_x\\B_{e}\end{pmatrix}\partial_t\dyAQ+\begin{pmatrix}-E_y\\E_x\end{pmatrix}\partial_z\dyAQ\Bigg],\label{Ampere_1Dl1}\\
		\partial_z\begin{pmatrix}-E_y\\E_x\end{pmatrix}+\partial_t\begin{pmatrix}B_x\\B_y\end{pmatrix}&=&0 \, ,\label{Faraday_1Dl1}\\
		\partial_t^2\dyAQ-v_z^2\partial_z^2\dyAQ+m^2\dyAQ&=&\Lambda(E_xB_x+E_yB_{e})\label{KleinGordon_1Dl1} \, ,
\end{eqnarray}
where we substitute the reaction current $\vc{J}_f$ with the loss term $\sigma\vc{E}$~(Ohm's law). When deriving Eqs.~\eqref{Ampere_1Dl1} and \eqref{KleinGordon_1Dl1}, we used that the external field $B_e$ is much larger than the $y$-component of the reaction $B$-field, $B_y$.
Note that it is straightforward to include an external source field in $\vc{E}$ and $\vc{B}$.

Let us now justify why the non-linear terms on the right-hand side in Eqs.~\eqref{Ampere_1Dl1} and~\eqref{KleinGordon_1Dl1} can be linearized. Consider the two distinct cases where a strong external laser field is parallel or orthogonal to the static external $B$-field: first, assume that the external laser field is parallel to $\vc{B}_e = B_e\hat{\vc{e}}_y$. Note that
\begin{eqnarray}
	B_e\frac{\partial_t\dyAQ}{\partial_z \dyAQ}\approx \SI{3e4}{\frac{\V}{\m}} \, \left(\frac{B_e}{\SI{1}{\tesla}}\right) \, ,
	\label{eq:waek_laser_condition}
\end{eqnarray}
where we approximated $\frac{\partial_t\dyAQ}{\partial_z \dyAQ}$ with a typical spin wave velocity, which is on the order of $v_s=10^{-4}$~\cite{Pickart}. 
Typical THz sources have a power around $P = \SI{e-5}{\watt}$, which leads to $E_y = \SI{27}{\frac{\V}{\m}}$ for a beam surface area of $\SI{10}{\mm^2}$. Equation~\eqref{eq:waek_laser_condition} is therefore fulfilled for sufficiently large external $B$-fields. With these considerations we see directly that $B_e\partial_t\dyAQ\gg \partial_z\dyAQ E_x$ since $E_x$ is even smaller than $E_y$. It follows that the non-linear term in the second component on the right-hand side in Eq.~\eqref{Ampere_1Dl1} can be neglected.

Next, we consider the two source~terms in the first equation in the reft-hand side of Eq.~\eqref{Ampere_1Dl1}. The term $E_y\partial_z\dyAQ$ dominates over the term $B_x\partial_t\dyAQ$ since $E_y$ contains the external laser source. However, the large source~term $B_e\partial_t\dyAQ$ in the term in Eq.~\eqref{Ampere_1Dl1} is larger than the dominating source in the first term: $B_e\partial_t\dyAQ \gg E_y\partial_z\dyAQ$, cf.\ Eq.~\eqref{eq:waek_laser_condition}. 
From Eq.~\eqref{Faraday_1Dl1} it is clear that $\partial_tB_y=-\partial_zE_x$ and therefore due to $H_y\sim B_y$ the source of the first component in~\eqref{Ampere_1Dl1} sources the $E_y$-component. Therefore we can ignore the non-linear sources in the first equation in~\eqref{Ampere_1Dl1} and focus only on the $E_y$-component, e.g.\ the large linear source in the second equation in~\eqref{Ampere_1Dl1}.
The non-linear term $E_xB_x$ in Eq.~\eqref{KleinGordon_1Dl1} can also be neglected since it is much smaller than the term $E_yB_{e}$, which includes two external fields.

Second, in the case that the external laser field is orthogonal to $\vc{B}_e=B_e\hat{\vc{e}}_y$, the dominating source of the Klein-Gordon equation, cf.\ Eq.~\eqref{KleinGordon_1Dl1} is the linear term $E_yB_e$. Note that the fields $B_x$ and $E_y$ can only be induced by polarization rotation and are both on the order of $\frac{\alphaEM}{\pi}$. However, since $\SI{3e8}{\frac{\V}{\m}} \, \left(\frac{B_e}{\SI{1}{\tesla}}\right) \gg E_x$, we can linearize the source term of the Klein-Gordon, cf.\ Eq.~\eqref{KleinGordon_1Dl1}, i.e. $E_x B_x \ll E_y B_e$.
The second component of Eq.~\eqref{Ampere_1Dl1} can be linearized because any available THz~lasers has an amplitude that is below the limit in Eq.~\eqref{eq:waek_laser_condition}. The first component of Eq.~\eqref{Ampere_1Dl1} can also be linearized, i.e.\ the source terms are neglected since both source terms include electromagnetic fields that are only generated via polarization rotation.

In summary, whether an external laser $E$-field is parallel or orthogonal to $\vc{B}_e$, the equations can be linearized, and they reduce to:
\begin{align}
\partial_z^2 E_x -n^2\partial_t^2 E_x-\mu\sigma\partial_t E_x &= 0 \, ,\label{eq:eqofmotionEx}\\
\partial_z^2E_y-n^2\partial_t^2 E_y-\mu\sigma\partial_t E_y &= \frac{\alphaEM}{\pi} \mu B_e\partial_t^2\dyAQ \, ,\label{eq:eqofmotionEy}\\
		\partial_t^2\dyAQ-v_z^2\partial_z^2\dyAQ+\mAQ^2\dyAQ &= \Lambda E_yB_{e} \, ,\label{eq:eqofmotiondelTheta}
\end{align}
where we explicitly use the linear constitutive relations, cf.\ Eq.~\eqref{eq:constitutive_equations}. Furthermore the refractive index is given by
\begin{eqnarray}
n^2=\epsilon\,\mu\label{eq:refractiveIndex} \, .
\end{eqnarray}
The material properties $\mu$, $\epsilon$, $\mAQ$, $\sigma$, $\stAQ$, $v_z$, and $\Lambda$ are constants in the equations of motion. Regions with different material properties are linked by using interface conditions for the fields. 

The corresponding interface conditions are given in Eqs.~\eqref{eq:InterfaceE} and~\eqref{eq:InterfaceH} with $\vc{n}=\hat{\vc{e}}_z$. Equation~\eqref{eq:InterfaceE} remains unchanged after linearization, while the definition of $\vc{H}_\Theta$ in Eq.~\eqref{eq:InterfaceH} changes due to the linearization to $\vc{H}_\Theta=\vc{H}+\frac{\alphaEM}{\pi} \stAQ \vc{E}$. 

\subsubsection{Losses}\label{sec:Quasiparticle_Photon_mixing_Losses}
Losses can appear in the linearized equations of motion~\eqref{eq:eqofmotionEx}--\eqref{eq:eqofmotiondelTheta} in case of a finite conductivity~$\sigma$. However, magnon losses, and losses that mix between magnons and photons, are not included. We now generalize Eqs.~\eqref{eq:eqofmotionEx}--\eqref{eq:eqofmotiondelTheta} to include all possible types of losses. The equations then read:
\begin{equation}
\mat{K}\partial_t^2 \vc{X} - \mat{\Gamma}\partial_t\vc{X} + \mat{M}\vc{X} = 0 \, , \label{eq:eom_with_losses}
\end{equation} 
where we define 
\begin{eqnarray}
\vc{X}=\begin{pmatrix}
E_x\\E_y\\\dyAQ
\end{pmatrix} \, ,~~\mat{K}&=&\begin{pmatrix}
1&0&0\\
0&1&\frac{\alphaEM}{\pi}\frac{ B_e}{\epsilon}\\
0&0&1
\end{pmatrix} \, ,~~\bm{\Gamma}=\begin{pmatrix}
\Gamma_\rho&0&0\\
0&\Gamma_\rho&\Gamma_{\times,1}\\
0&\Gamma_{\times,2}&\Gamma_m
\end{pmatrix},\nonumber\\\mat{M}&=&\begin{pmatrix}
\frac{k^2}{n^2}&0&0\\
0&\frac{k^2}{n^2}&0\\
0&-\Lambda B_e&v_z^2k^2+\mAQ^2
\end{pmatrix} \, , \label{eq:AQ_P_mixing_M_G_X_def}
\end{eqnarray}
and where $\Gamma_\rho=\sigma/\epsilon$ is the photon loss, $\Gamma_m$ is the equivalent loss for magnons, and $\Gamma_{\times,1/2}$ are mixed losses that can arise when photons and magnons interact. We retain these for the most general treatment, and set them to zero later. Note that not all $\Gamma$s have the same mass dimension since $[\Gamma_\rho]=[\Gamma_m] =1$, while $[\Gamma_{\times,1}]=3$ and $[\Gamma_{\times,2}]=-1$. The approach also gives the possibility to define different refractive indices $n$ and photon losses $\Gamma_\rho$ for the $E_x$ and $E_y$ components. However, these effects can only become important when polarization rotation effects are discussed in detail.  In the following, polarization rotation effect are computed, however they are not discussed at a level of detail, such that including different refractive indices for different polarizations would not change the results significantly.

The interface conditions~\eqref{eq:InterfaceE}--\eqref{eq:InterfaceH} remain the same in the presence of losses, because it is assumed that all losses are bulk losses.
\FloatBarrier
\subsection{Transmission and Reflection Coefficients}\label{sec:transmission}
The presence of an AQ leads to a gap in the dispersion relation, which does not include any propagating modes. Based on this, Li et al.~\cite{2010NatPh...6..284L} proposed a transmission measurement~(cf.\ Fig.~\ref{fig:ATR}) to determine the band gap in a TMI polariton spectrum, opened by the presence of the AQ (cf.\ Fig.~\ref{fig:Disp_relat_vz_zero}). We now compute the transmission and reflection coefficients, and we demonstrate how to experimentally determine the parameters of interest -- in particular the relevant terms of the loss matrix~$\mat{\Gamma}$.

\subsubsection{Solution of linearized equations}\label{sec:AQ_P_mixing_solutions}
Our strategy for solving the linearized equations is as follows: we solve the equations for each spatial domain of constant material properties. We then apply the appropriate interface conditions to match the solutions in the different domains.

\paragraph{Lossless case ($\boldsymbol{\Gamma = 0}$).} The dispersion relation for the $E_x$-component, see Eq.~\eqref{eq:eqofmotionEx}, is the usual photon dispersion relation: 
\begin{eqnarray}
	k^2 = n^2\omega^2 \equiv \kp^2 \, .
	\label{eq:PhotonDispersion_withoutLosses}
\end{eqnarray}
The $E_y$-component mixes with the AQ and, in the $v_z = 0$ case, we find a typical polariton dispersion~\cite{Mills_1974,2010NatPh...6..284L}:
\begin{equation}
	\omega^2_{\pm}=\frac{1}{2}\Big[\omLO^2+\frac{k^2}{n^2}\Big]\pm \frac{1}{2}\Big[\big(\omLO^2 - \frac{k^2}{n^2}\big)^2+4b^2\frac{k^2}{n^2}\Big]^{1/2} \, ,
	\label{eq:dispersion_om_vz0}
\end{equation}
where we have defined
\begin{eqnarray}
 b^2 \equiv\frac{\alphaEM}{\pi} \frac{\Lambda B_e^2}{\epsilon},\label{eq:AQ_photon_mixing_bDef}\\
 \omLO^2 \equiv b^2 + \mAQ^2 \, .
\end{eqnarray}
The case $v_z\neq 0$ is discussed later since $v_z$ is on the order of the spin wave velocity $10^{-4}$ and therefore the expected effect is small.

\begin{figure}
	\centering
	\includegraphics[width=0.49\textwidth]{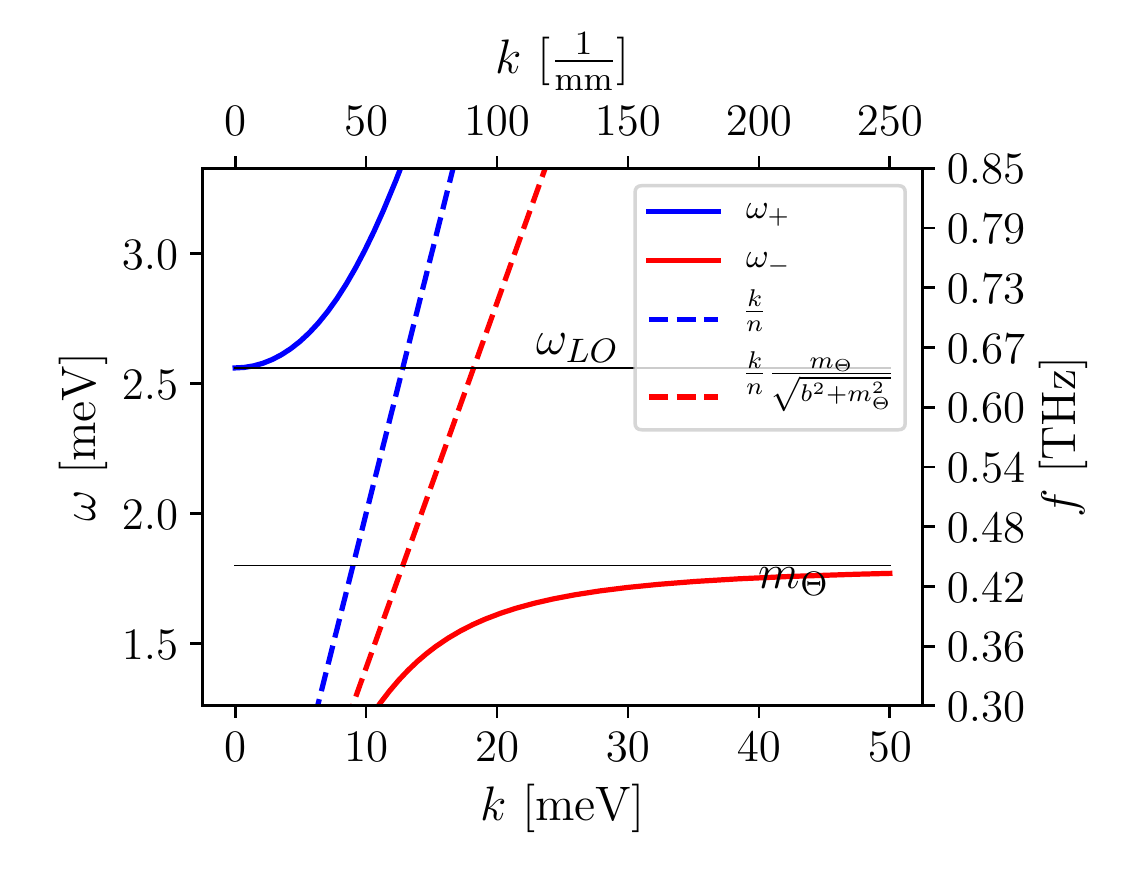}
	\includegraphics[width=0.49\textwidth]{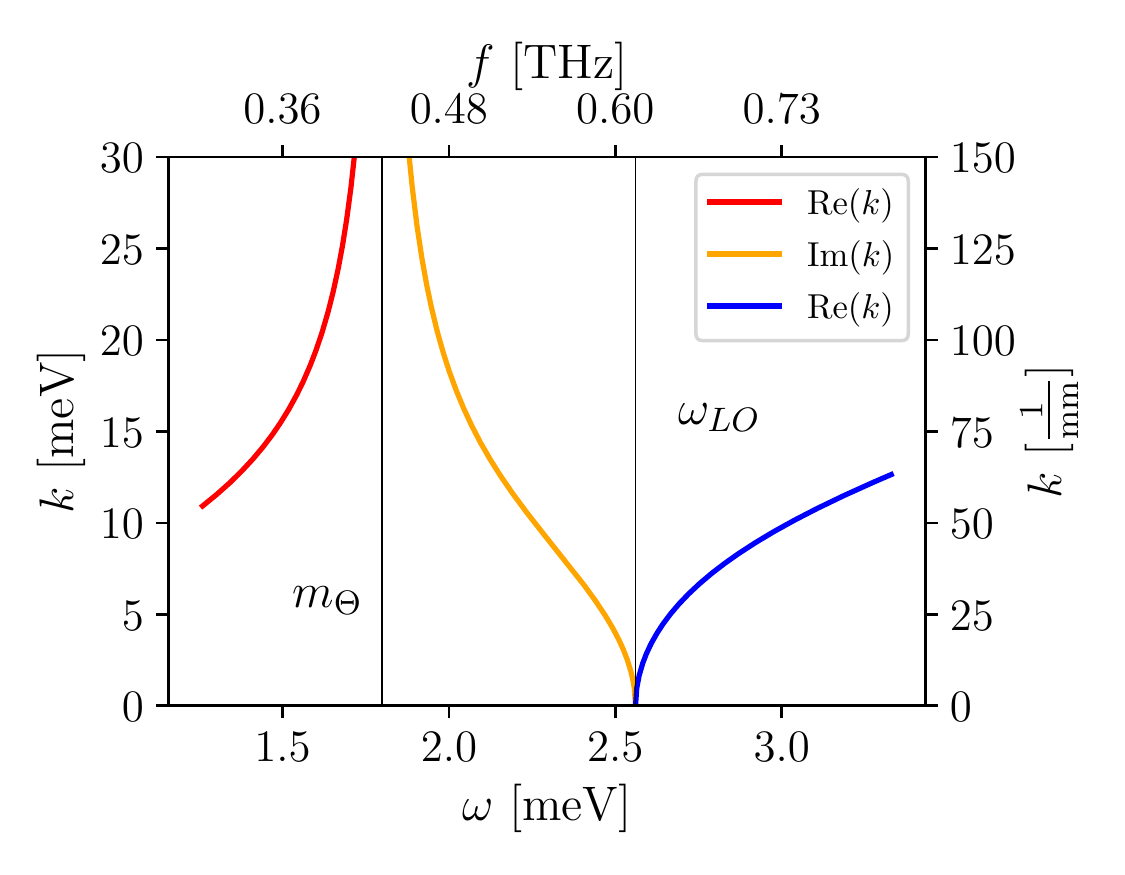}
	\caption{Polariton dispersion relation, arising from the mixing of AQs and photons for a spin wave velocity $v_z=0$. \defaultSetup. The left panel shows the $\omega_{\pm}$ mode, which has a bandgap between $\mAQ$ and $\omLO$ (horizontal lines). The right panel illustrates the inverse of the dispersion relation for~$k$. Inside the bandgap (vertical lines), $k$ is only imaginary, and hence no propagating modes exist.}
	\label{fig:Disp_relat_vz_zero}
\end{figure}
We show $\omega_{\pm}$ as a function of the wave number~$k$ in the left panel of Fig.~\ref{fig:Disp_relat_vz_zero}. The horizontal black lines indicate the gap between $\mAQ$ and $\omLO$, where total reflection is expected. The resulting frequencies for $\mAQ$ and $\omLO$ are in the THz regime what makes clear why THz sources are needed to probe the gap in the dispersion relation. $\omega_+$ converges for large $k$ to a photon dispersion (dashed blue line). $\omega_-$ has for small $k$ an almost photon-like dispersion $\omega_-=\frac{k}{n}\frac{m}{\sqrt{b^2+m^2}}$ (dashed red line).

Inverting Eq.~\eqref{eq:dispersion_om_vz0} gives:
\begin{equation}
	k^2 = n^2\omega^2\Big[1-\frac{b^2}{\omega^2-\mAQ^2}\Big] \equiv \kAQ^2\equiv n_\Theta^2\omega^2\, .
	\label{eq:dispersion_k_vz0}
\end{equation}
We show $k$ as a function of $\omega$ in the right panel of Fig.~\ref{fig:Disp_relat_vz_zero}. In the limit of $b\rightarrow 0$, Eq.~\eqref{eq:dispersion_k_vz0} becomes the usual photon dispersion relation. For $\omega^2$ we have two solutions, while the solution for $k^2$ can be described by a single function. Inside the bandgap, $k^2$ is negative, thus $k$ is purely imaginary, and no propagating mode is present. In the following section it is explicitly shown that this leads to total reflection and zero transmission.

The most general ansatz for the field evolution in a TMI medium are
\begin{eqnarray}
E_x(z)&=&\hat{E}^{+}_x e^{i \kp z}+\hat{E}^{-}_x e^{-i \kp z} \, , \label{eq:vz0_ExGeneralSol}\\
E_y(z)&=&\hat{E}^{+}_y e^{i\kAQ z}+\hat{E}^{-}_y e^{-i\kAQ z} \, ,\label{eq:vz0_EyGeneralSol}\\
\dyAQ(z)&=&\delta\hat{\Theta}^{+} e^{i\kAQ z}+\delta\hat{\Theta}^{-} e^{-i\kAQ z} \, ,\label{eq:vz0_DelThetaGeneralSol}
\end{eqnarray}
where we omitted the time dependence $e^{-i\omega t}$ in each line. After plugging the solutions into the equations of motion, cf.\ Eq.~\eqref{eq:eom_with_losses} the following relations are obtained:
\begin{equation}
	\delta\hat{\Theta}^{\pm} = \Theta_E\hat{E}^{\pm}_y\, , \quad \Theta_E=\frac{\Lambda B_e}{\mAQ^2 - \omega^2} \, , \label{eq:vz0_relation_Theta}\\
\end{equation}
or, equivalently,
\begin{equation}
	\hat{E}^{\pm}_y = E_\Theta\delta\hat{\Theta}^{\pm}\, , \quad E_\Theta=-\frac{\alphaEM}{\pi}\frac{\mu\omega^2 B_e}{\kp^2-\kAQ^2} \, .\label{eq:vz0_relation_Ey}
\end{equation}
In the following, the relations in Eq.~\eqref{eq:vz0_relation_Theta} are used to reduce the number of unknowns in the ansatz~\eqref{eq:vz0_DelThetaGeneralSol}:
\begin{eqnarray}
\dyAQ(z)&=&\Theta_E \hat{E}^{+}_y e^{i\kAQ z}+\Theta_E \hat{E}^{-}_y e^{-i\kAQ z}.\label{eq:vz0_DelThetaGeneralSol_reduced}
\end{eqnarray}
The remaining constants $\hat{E}_y^\pm$ can be determined by using the interface conditions~(explicitly shown in Section~\ref{sec:AnalogueAxionPhoton_MatrixFormalism}). The AQ field $\dyAQ$ is completely determined, cf.\ Eq.~\eqref{eq:vz0_DelThetaGeneralSol_reduced}, and no boundary conditions for $\dyAQ$ have to be applied when, for example, a layer of TMI surrounded by vacuum is considered. It will become clear in the following that this is a consequence of the $v_z = 0$ limit.

Note that the relations in Eq.~\eqref{eq:vz0_relation_Ey} could have also been used to reduce the constants in Eq.~\eqref{eq:vz0_EyGeneralSol}. However, a short calculation reveals that this would result in the same outcome, regardless whether the relations in Eq.~\eqref{eq:vz0_relation_Theta} or~\eqref{eq:vz0_relation_Ey} was used to reduce the constants.

\begin{figure}
	\centering
	\includegraphics[width=0.49\textwidth]{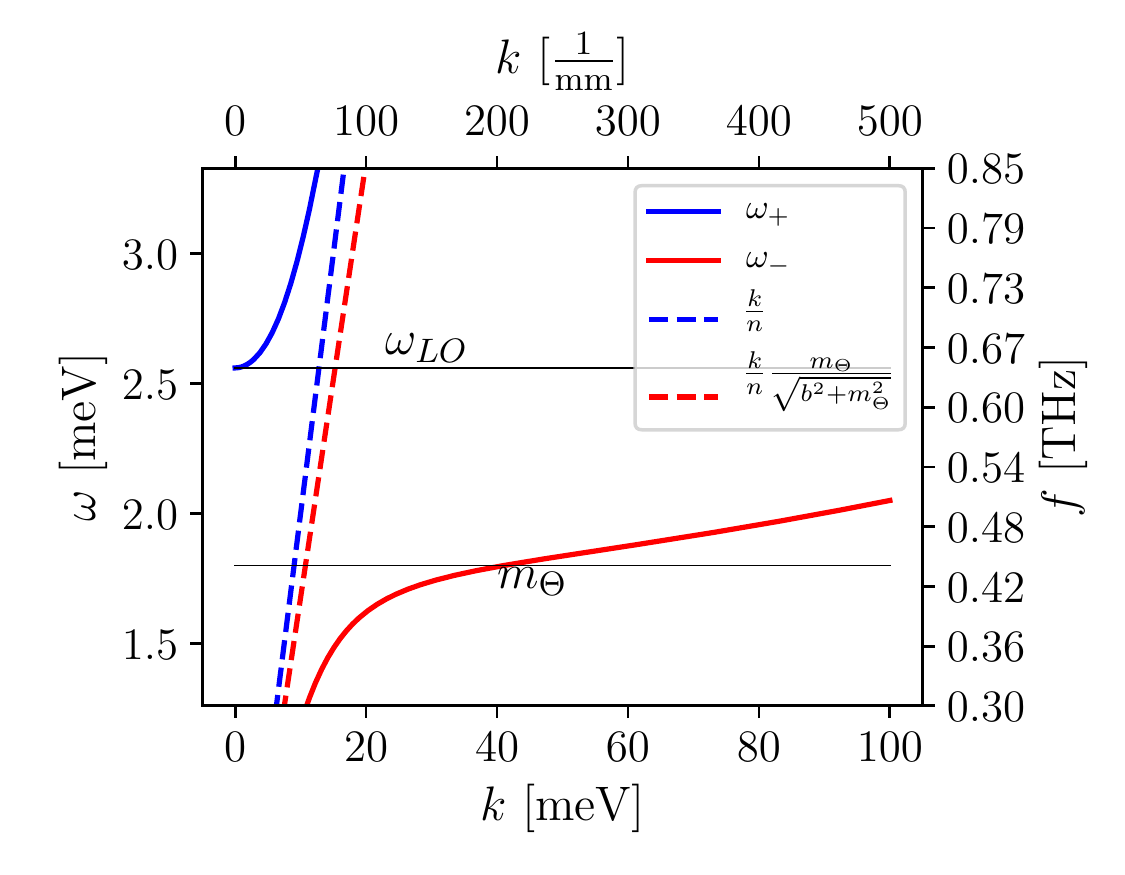}
	\includegraphics[width=0.49\textwidth]{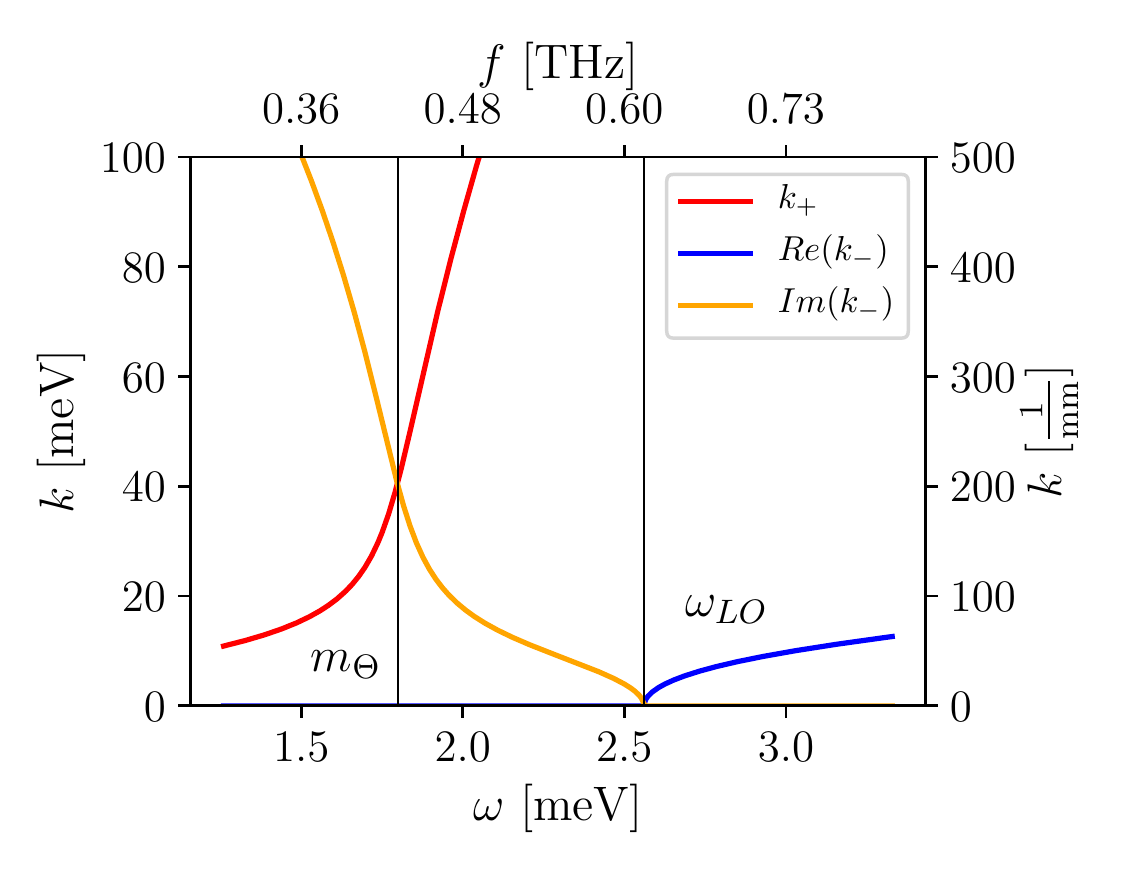}
	\caption{Dispersion relation for a non-zero spin wave velocity of~$v_z = 0.01$. This exaggerated value was chosen because for realistic value of $10^{-4}$ the effect of band crossing of the $+$~mode is not visible. \defaultSetup.}
	\label{fig:Disp_relat_vz_Nonzero}
\end{figure}

A finite spin wave velocity, $v_z\neq 0$, leads a slightly modified dispersion relation:
\begin{equation}
\omega^2_{\pm}=\frac{1}{2}\Big[\omLO^2+k^2(v_z^2+\frac{1}{n^2})\Big]\pm \frac{1}{2}\Big[\big(\omLO^2+k^2(v_z^2-\frac{1}{n^2})\big)^2+4b^2\frac{k^2}{n^2}\Big]^{1/2}.
\label{eq:vz_nonzero_omega}
\end{equation}
Equation~\eqref{eq:vz_nonzero_omega} is not a typical polariton dispersion relation, since the sign of~$v_z$ under the square root is positive, not negative. The dispersion relation for $\omega_\pm$ from Eq.~\ref{eq:vz_nonzero_omega} is shown in the left panel of Fig.~\ref{fig:Disp_relat_vz_Nonzero}, where we used an unrealistically large vale of $v_z = 0.01$ for illustrative purposes. Typical values for $v_z$ are on the order of~\num{e-4}. A~non-zero value of $v_z$ leads to a gap-crossing of the $\omega_-$ mode. However, due to the smallness of the spin wave velocity compared to the speed of light, the gap crossing happens at large values of the wave number~$k$.

Inverting Eq.~\eqref{eq:vz_nonzero_omega} yields two modes for $k^2$,
\begin{eqnarray}
k^2_{\pm} = \frac{1}{2v_z^2} \left[ \omega^2 - \mAQ^2  + n^2\omega^2v_z^2 \pm \Big(\big(\mAQ^2+\omega^2(n^2 v_z^2-1)\big)^2+4\omega^2n^2b^2v_z^2\Big)^{1/2}\right] \, ,
\label{eq:vz_nonzero_k}
\end{eqnarray}
whereas we only obtained one mode for $k^2$ in the $v_z=0$ case, cf.\ Eq.~\eqref{eq:dispersion_k_vz0}. The functional dependence of Eq.~\eqref{eq:vz_nonzero_k} is shown in the right panel of Fig.~\ref{fig:Disp_relat_vz_Nonzero}. The imaginary part of the $k_-$ mode, which for $v_z = 0$ was only present inside the gap, now keeps rising outside of the gap for frequencies $\omega < \mAQ$. The $k_+$ mode crosses the gap such that for $\omega > \omLO$ two propagating modes exist. However the wavelength of the $k_+$~mode is always much shorter than the wavelength of the $k_-$~mode. 

The most general ansatz in the case of non-vanishing spin wave velocity is:
\begin{eqnarray}
E_x(z)&=&\hat{E}^{+}_x e^{i \kp z}+\hat{E}^{-}_x e^{-i \kp z},\label{eq:AQ_P_mixing_vzNonZero_AnsatzEx}\\
E_y(z)&=&\hat{E}^{++}_y e^{ik_+z}+\hat{E}^{+-}_y e^{-ik_+z}+\hat{E}^{-+}_y e^{ik_-z}+\hat{E}^{--}_y e^{-ik_-z},\\
\dyAQ(z)&=&\delta\hat{\Theta}^{++} e^{ik_+z}+\delta\hat{\Theta}^{+-} e^{-ik_+z}+\delta\hat{\Theta}^{-+} e^{ik_-z}+\delta\hat{\Theta}^{--} e^{-ik_-z} \, .\label{eq:AQ_P_mixing_vzNonZero_AnsatzDelTheta}
\end{eqnarray}

Relations for the unknown constants in the ansatz~\eqref{eq:AQ_P_mixing_vzNonZero_AnsatzEx}--\eqref{eq:AQ_P_mixing_vzNonZero_AnsatzDelTheta} can be derived in complete analogy to the $v_z=0$ case. However, we would now need to specify boundary conditions for the dynamical axion in order to determine all constants. We do not perform the explicit calculation here since we expect the difference to the $v_z=0$ case to be minimal, thanks to the smallness of the spin wave velocity. To see this, consider the following argument:

Let an incoming electromagnetic wave in vacuum be described by $A_0e^{i \kp z}$. In the TMI material with $v_z\neq 0$, two modes are present. Around $\omega \sim \omLO$, the first mode $k_s$ has a wavelength that is much shorter than $\kp$, while the second mode $k_l$ has a much longer wavelength than $\kp$, i.e.\ $|k_s|\gg|\kp|\gg |k_l|$. This is exactly the situation that we face~(cf.\ Fig.~\ref{fig:Disp_relat_vz_Nonzero}), where $k_s=k_+$ and $k_l=k_-$.\footnote{Note that due to the fact that we plot $k_+$ only up to $\SI{100}{\meV}$, the much larger values of $k_+$ around $\omLO$ are not visible in the right panel of Fig.~\ref{fig:Disp_relat_vz_Nonzero}.} Neglecting reflections, the fraction of the amplitudes of the two modes in medium~1 are $\left|\frac{A_1^l}{A_1^s}\right|=\left|\frac{k_a-\kp}{\kp-k_l}\right|\approx\left|\frac{k_s}{\kp}\right|\gg 1$, where the index 1 refers to medium 1. Therefore the amplitude of long wavelength mode $A_1^l$ is much larger than the amplitude of the short wavelength mode $A_1^s$. Based on these arguments, the contribution of the $k_+$~mode can therefore be neglected -- even though it is in principle present. In what follows, we will consequently assume that $v_z=0$.

\paragraph{Case with losses ($\boldsymbol{\Gamma \neq 0}$).}
If material losses are included, the dispersion relations \eqref{eq:PhotonDispersion_withoutLosses} and \eqref{eq:dispersion_k_vz0} are modified. The dispersion relation of the $E_x$-component is
\begin{eqnarray}
	k^2 = n^2 \, \omega^2 \left(1+i\frac{\Gamma_\rho}{\omega}\right) =: \kp^2
	\label{eq:PhotonDispersion_withLosses}
\end{eqnarray}
and the dispersion relation for the mixed system of $E_y$ and $\dyAQ$ is
\begin{eqnarray}
\kAQ^2\equiv k^2&=&n^2\omega^2 \left(1+\frac{b^2}{-i \Gamma_m \omega+\mAQ^2-\omega^2}+i \frac{\Gamma_\rho}{\omega}\right)\nonumber\\&+&n^2 \omega \left(\frac{i B_e \left(\frac{\alphaEM}{\pi}\frac{\Gamma_{\times,2}  \omega^2}{\epsilon}+\Lambda \Gamma_{\times,1}\right)-\omega\Gamma_{\times,1} \Gamma_{\times,2}}{-i \Gamma_m \omega+\mAQ^2-\omega^2}\right).
\label{eq:dispersion_relation_vz0_losses}
\end{eqnarray}
The first part of the dispersion relation in Eq.~\eqref{eq:dispersion_relation_vz0_losses} only includes the diagonal losses $\Gamma_m$ and $\Gamma_\rho$, while the second part also includes mixed losses. We argued in Section~\ref{sec:damping} that mixed losses are smaller than the diagonal losses $\Gamma_\rho$ and $\Gamma_m$. We therefore neglect mixed losses in what follows.

Rewriting the dispersion relation~\eqref{eq:dispersion_relation_vz0_losses} without mixed losses gives:
\begin{equation}\boxed{
\kAQ^2\equiv k^2=n^2\omega^2 \left(1+\frac{(m^2-\omega^2)\omega b^2}{ \Gamma_m^2 \omega^2+(\mAQ^2-\omega^2)^2} + i \, \frac{\Gamma_m\omega b^2}{ \Gamma_m^2 \omega^2+(\mAQ^2-\omega^2)^2} + i \, \frac{\Gamma_\rho}{\omega}\right) \, .
\label{eq:dispersion_relation_vz0_losses_withoutMixed}
}
\end{equation}

\begin{figure}
	\centering
	\includegraphics[width=0.32\textwidth]{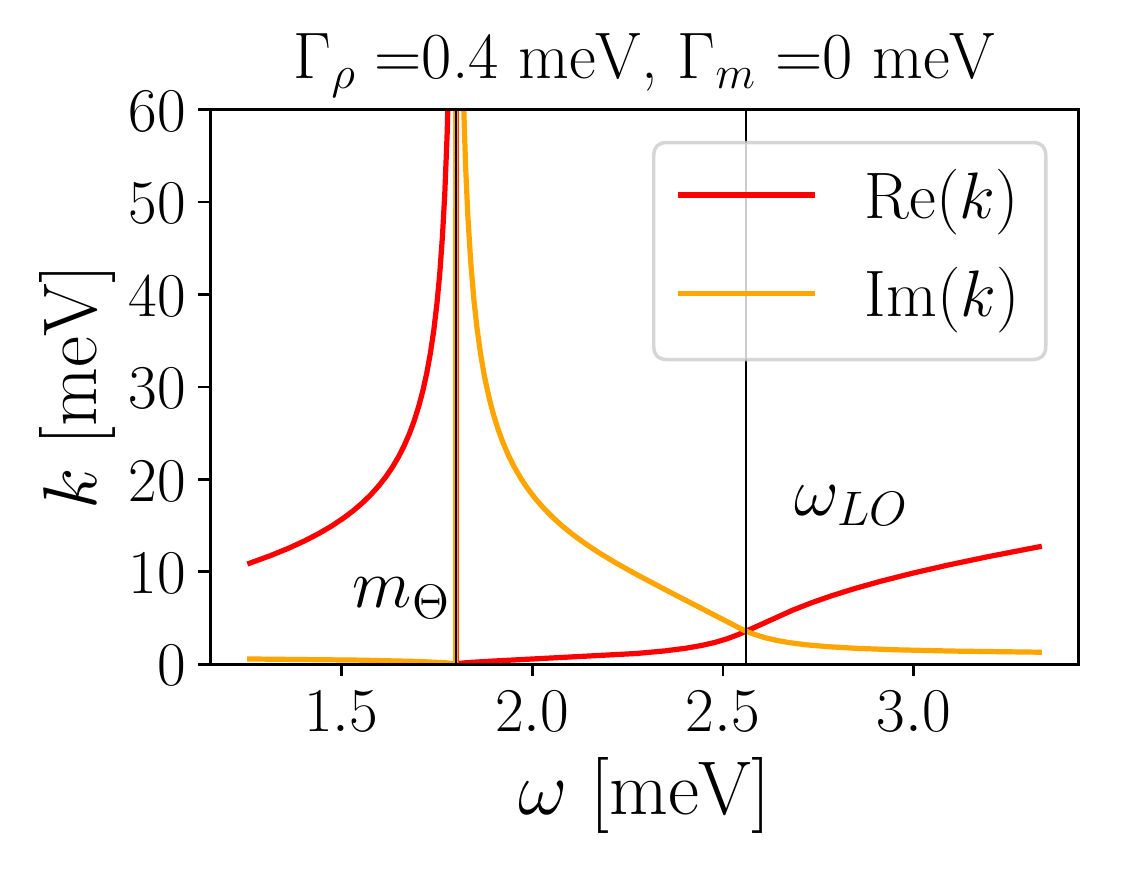}
	\includegraphics[width=0.32\textwidth]{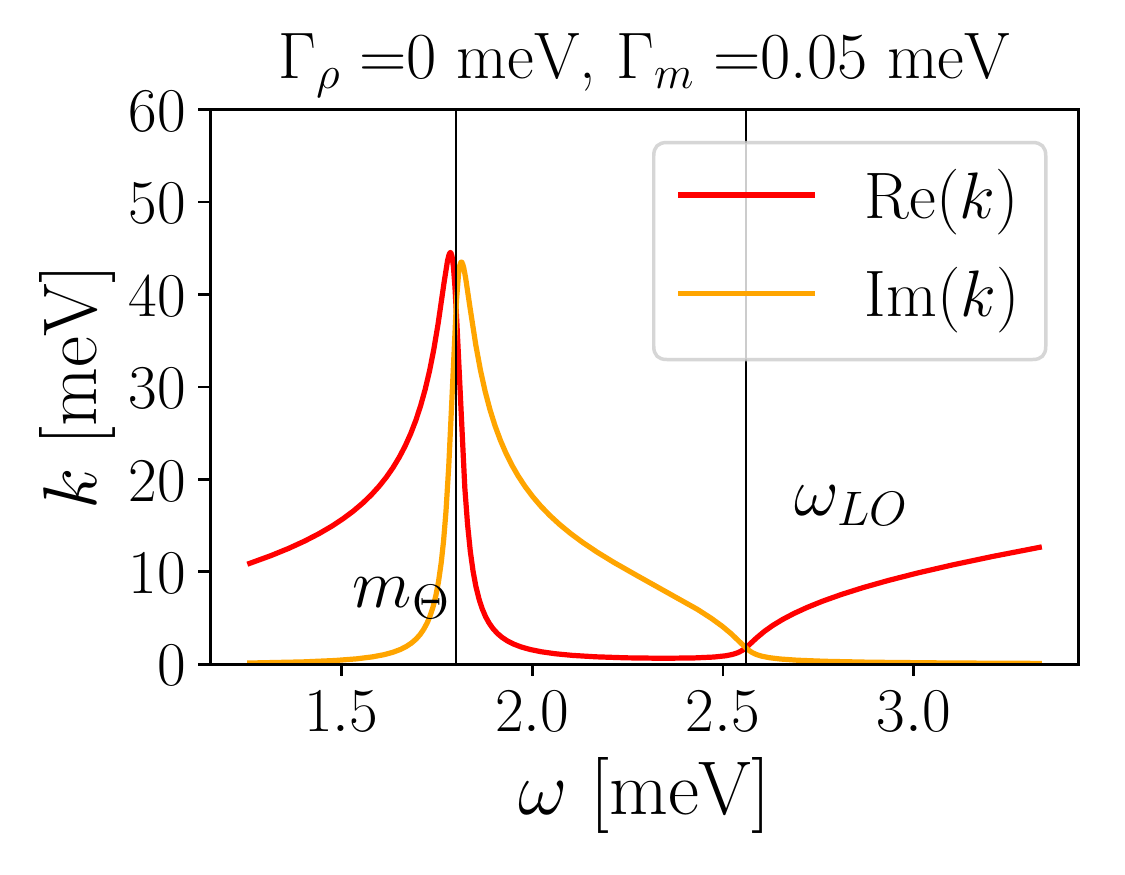}
	\includegraphics[width=0.32\textwidth]{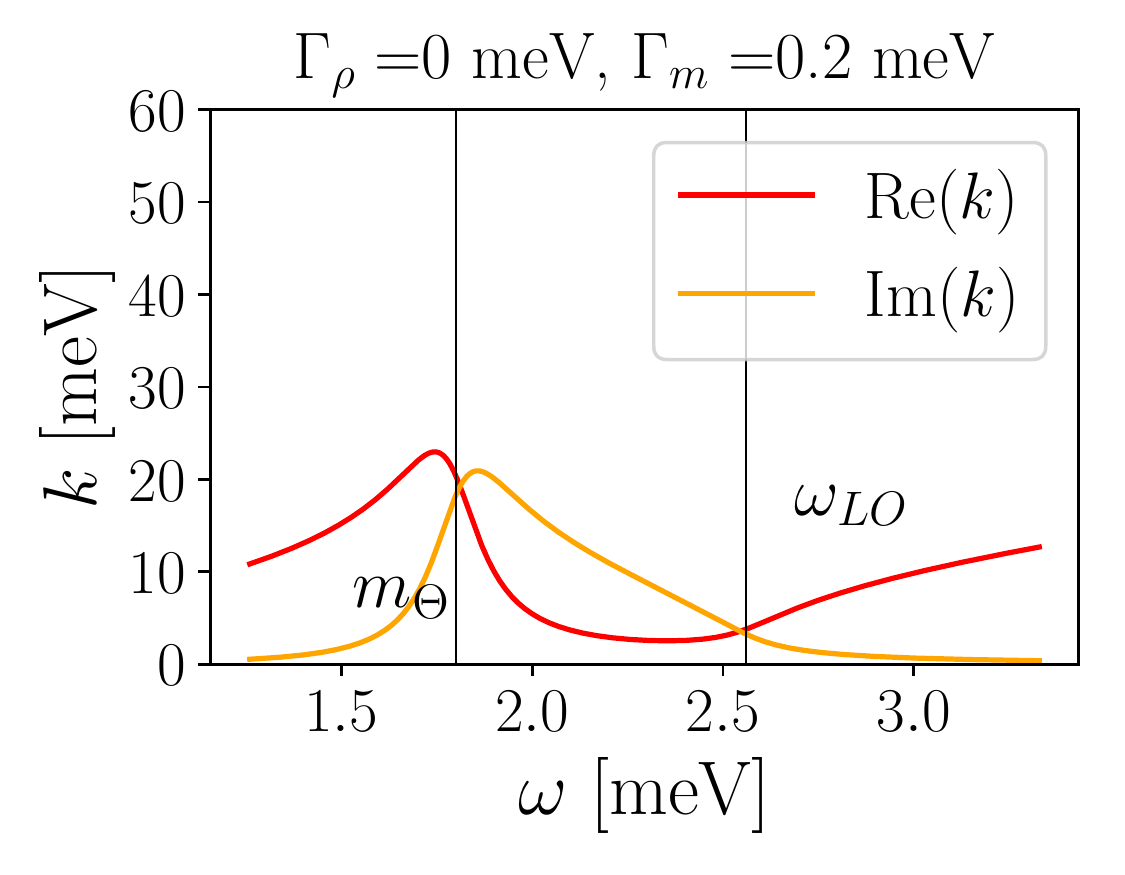}
	\caption{Dispersion relation of the axion-polariton for magnon and photon losses, $\Gamma_m$ and $\Gamma_\rho$. Mixed losses are neglected. \textit{Left:} $\Gamma_\rho = \SI{0.4}{\meV}$ and $\Gamma_m = 0$. Photon losses introduce an almost constant imaginary part to the dispersion relation if the chosen frequency interval is not too large. \textit{Middle:} $\Gamma_\rho = 0$ and $\Gamma_m=\SI{0.05}{\meV}$, \textit{right:} $\Gamma_\rho = 0$ and $\Gamma_m=\SI{0.2}{\meV}$. The larger the magnon loss, the larger the FWHM of the imaginary part in the dispersion relation. \defaultSetup.}
	\label{fig:Disp_relat_vz_zero_losses}
\end{figure}
Equation~\eqref{eq:dispersion_relation_vz0_losses_withoutMixed} shows that the $\Gamma_\rho$~contribution is unaffected by any other material properties, and it stays approximately constant when $\omega$ does not vary too much. We show an example for $\Gamma_m=0$ in the left panel of Fig.~\ref{fig:Disp_relat_vz_zero_losses}. While the peak of the resonance is not affected much by the losses, $\Gamma_\rho$ introduces an almost constant imaginary for all frequencies. In contrast, magnon losses $\Gamma_m$ are dominant around $\mAQ$. This can be seen from the third term in Eq.~\eqref{eq:dispersion_relation_vz0_losses_withoutMixed} which represents a Lorentzian curve that peaks around $\omega=\mAQ$ and has a full width at half maximum~(FWHM) of $\Gamma_m$. In the middle and right panels of Fig.~\ref{fig:Disp_relat_vz_zero_losses} we show examples for $\Gamma_\rho = 0$. The larger~$\Gamma_m$, the larger the FWHM of the imaginary part in the dispersion relation. In other words, frequencies away from the gap are damped more strongly when $\GammaM$ is large. Furthermore, the resonance becomes less pronounced for large $\GammaM$. As a consequence, it will be difficult to confirm the existence of the gap in the spectrum, and the presence of a dynamical AQ, when large losses are present. We investigate this more quantitatively in Section~\ref{sec:AQ_P_mixing_Ref_Trans}, where we calculate the reflection and transmission coefficients for a single TMI layer.

In the presence of losses the most general solution, cf.\ Eq.~\eqref{eq:vz0_ExGeneralSol}--\eqref{eq:vz0_DelThetaGeneralSol}, is still valid. However, the relations in the Eqs.~\eqref{eq:vz0_relation_Theta} and~\eqref{eq:vz0_relation_Ey} are modified:
\begin{equation}
\delta\hat{\Theta}^{\pm} = \Theta_E\hat{E}^{\pm} \, , \quad \Theta_E=\frac{\Lambda B_e+i\omega\Gamma_{\times,2}}{-\omega^2+\mAQ^2-i\omega\Gamma_m} \, ,\label{eq:vz0_losses_relation_Theta}
\end{equation}
or, equivalently,
\begin{equation}
\hat{E}_y^{\pm} = E_\Theta\delta\hat{\Theta}^{\pm} \, , \quad E_\Theta=\frac{\alphaEM}{\pi}\frac{\omega^2\mu B_e + i n^2\omega\Gamma_{\times,1}}{\kAQ^2-\kp^2} \, .\label{eq:vz0_losses_relation_Ey}
\end{equation}
It can be checked that Eqs.~\eqref{eq:vz0_losses_relation_Theta} and~\eqref{eq:vz0_losses_relation_Ey} reduce to Eqs.~\eqref{eq:vz0_relation_Theta} and~\eqref{eq:vz0_relation_Ey} in the limit of $\mat{\Gamma} \rightarrow 0$. In complete analogy to the case without losses, Eq.~\eqref{eq:vz0_losses_relation_Theta} determines the dynamical AQ field, cf.\ Eq.~\eqref{eq:vz0_DelThetaGeneralSol_reduced}.

\subsubsection{Matrix formalism for many interfaces}\label{sec:AnalogueAxionPhoton_MatrixFormalism}
\begin{figure}
	\centering
	\includegraphics[width=0.6\textwidth]{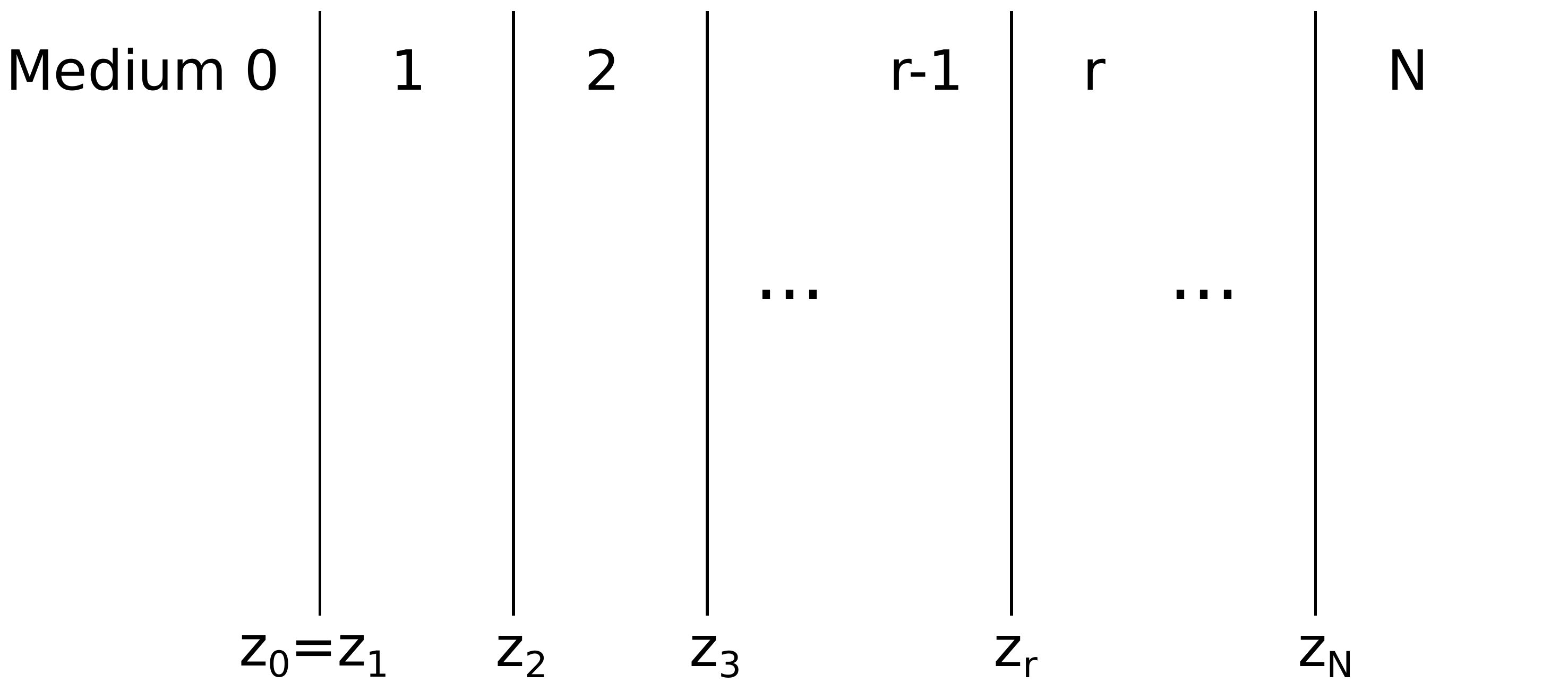}
	\caption{Multilayer system of different materials. Each medium is characterized by $\epsilon_r$, $\mu_r$, $\mat{\Gamma}_r$, and $\stAQ_r$. The external $B$-field has the same strength and polarization in each medium.}
	\label{fig:matrix_formalism}
\end{figure}
In the previous section, we discussed the solutions of the one-dimensional axion-Maxwell equations in a homogeneous~TMI. Here, we consider $N+1$~media, separated by $N$ interfaces, as shown in Fig.~\ref{fig:matrix_formalism}. Let the first interface be located at $z_0 = z_1$, and the last interface at $z_N$. We label each medium with an index~$r$, i.e.\ $r = 0,\ldots, N$. For example, the permittivity and permeability of medium~$r$ are thus denoted by $\mu_r$ and $\epsilon_r$, respectively. Recall that, in all media, we set $v_z = 0$ and define the constant external $B$-field to be $\vc{B}_e = B_e\,\hat{\vc{e}}_y$.

We now develop a matrix formalism to link the solutions in different materials to each other. This makes it possible to compute the scattering of incoming electromagnetic radiation from a multilayer system. The simplest application is the computation of the reflection and transmission coefficients for THz~radiation that hits a layer of TMI; we discuss this case at the end of this section.

The most general ansatz in medium~$r$ is given by:
\begin{align}
E_x^r &= \hat{E}_{x,r}^+e^{i\kp^r (z-z_r)}+\hat{E}_{x,r}^-e^{-i\kp^r (z-z_r)} \, , \nonumber\\
E_y^r &=\hat{E}_{y,r}^+e^{ik^r_\Theta (z-z_r)}+\hat{E}_{y,r}^-e^{-ik^i_\Theta (z-z_r)} \, , \nonumber\\
\dyAQ^r &= \Theta_E^r\hat{E}_{y,r}^+e^{ik^r_\Theta (z-z_r)}+\Theta_E^r\hat{E}_{y,r}^-e^{-ik^r_\Theta (z-z_r)} \, ,
\label{eq:AQ_P_mixing_MatrixFormalism_Ansatz}
\end{align}
where, compared to Eqs.~\eqref{eq:vz0_ExGeneralSol}--\eqref{eq:vz0_DelThetaGeneralSol}, we introduce different phase shifts~$z_r$ for each medium.
The expressions for $\kp$, $\kAQ$, and $E_\Theta$ were derived already in Eq.~\eqref{eq:PhotonDispersion_withoutLosses}, \eqref{eq:dispersion_k_vz0}, and~\eqref{eq:vz0_relation_Theta} for the case $\mat{\Gamma}=0$ and in Eq.~\eqref{eq:PhotonDispersion_withLosses}, \eqref{eq:dispersion_relation_vz0_losses}, and~\eqref{eq:vz0_losses_relation_Theta}, for the case $\mat{\Gamma}\neq 0$.
Applying the interface conditions~\eqref{eq:InterfaceE} and~\eqref{eq:InterfaceH} for the electromagnetic fields at $z_r$ yields the following system of equations:
\begin{eqnarray}
\vc{t}_r=\mat{M}_r^{-1} \, \mat{M}_{r-1} \, \mat{P}_{r-1} \vc{t}_{r-1} \, ,
\end{eqnarray}
with 
\begin{eqnarray}
\vc{t}_{r} = \begin{pmatrix} \hat{E}_{x,r}^+\\ \hat{E}_{x,r}^-\\ \hat{E}_{y,r}^+\\ \hat{E}_{y,r}^- \end{pmatrix} \, , \quad
\mat{M}_{r}=\begin{pmatrix}
1&1&0&0\\
0&0&1&1\\
\frac{\kp^{r}}{\omega\mu_{r}}&-\frac{\kp^{r}}{\omega\mu_{r}}&-\frac{\alphaEM}{\pi}\stAQ_{r}&-\frac{\alphaEM}{\pi}\stAQ_{r}\\
-\frac{\alphaEM}{\pi}\stAQ_{r}&-\frac{\alphaEM}{\pi}\stAQ_{r}&-\frac{k^{r}_\Theta}{\omega\mu_{r}}&\frac{k^{r}_\Theta}{\omega\mu_{r}}
\end{pmatrix} \, ,
\label{eq:MatrixFormalism_DefMatrix}
\end{eqnarray}
and
\begin{equation}
\mat{P}_{r} = \text{diag}\left(e^{i\Delta^p_{r}}, \, e^{-i\Delta^p_{r}}, \, e^{i\Delta_{r}^\Theta}, \, e^{-i\Delta_{r}^\Theta}\right) \, .
\end{equation}
The phases are defined as: $\Delta_{r}^\Theta\equiv k^{r}_\Theta(z_{r+1}-z_{r})$ and $\Delta_{r}^p\equiv k^{r}_p(z_{r+1}-z_{r})$.

Let us define the matrix $\mat{S}$ to relate the incoming field amplitude from medium~$0$ to the outgoing field amplitude in medium~$N$:
\begin{equation}
	\vc{t}_N=\mat{S}\,\vc{t}_0 \, .
\end{equation}
For instance, for a single interface, $\mat{S}$ is given by
\begin{equation}
\mat{S}=\mat{M}_1^{-1}\,\mat{M}_0\,\mat{P}_0
\label{eq:AQ_P_mixing_matrix_S_single_interface}
\end{equation}
and, for two interfaces, $\mat{S}$ is given by
\begin{equation}
\mat{S}=\mat{M}_2^{-1} \, \mat{M}_1 \, \mat{P}_1\mat{M}_1^{-1}\,\mat{M}_0\,\mat{P}_0 \, .
\label{eq:AQ_P_mixing_matrix_S_two_interface}
\end{equation}
Finally, for $N$ interfaces, we find $\mat{S}$ to be given by
\begin{equation}
\mat{S}=\mat{M}_N^{-1}\,\mat{M}_{N-1}\,\mat{P}_{N-1}\,\mat{M}_{N-1}^{-1}\,\mat{M}_{N-1}\mat{P}_{N-2}\,\mat{M}_{N-2}\cdots\mat{M}_2^{-1}\,\mat{M}_1\,\mat{P}_1\mat{M}_1^{-1}\,\mat{M}_0\,\mat{P}_0 \, .
\label{eq:AQ_P_mixing_matrix_S_N_interface}
\end{equation}

For electromagnetic radiation coming into the system from medium~$0$, $\hat{E}_{x0}^+$ and $\hat{E}_{y,0}^+$ are known and $\hat{E}_{x,N}^- = \hat{E}_{y,N}^- = 0$. The other unknown field values can be determined from the elements of $\mat{S}$, i.e.\ $S_{ij}$, via
\begin{eqnarray}
\begin{pmatrix}
\hat{E}^+_{x,N}\\
\hat{E}^-_{x,0}\\
\hat{E}^+_{y,N}\\
\hat{E}^-_{y,0}
\end{pmatrix}
=
\begin{pmatrix}
-1&S_{12}&0&S_{14}\\
0&S_{22}&0&S_{24}\\
0&S_{32}&-1&S_{34}\\
0&S_{42}&0&S_{44}
\end{pmatrix}^{-1}\cdot\begin{pmatrix}
-S_{11}\hat{E}^+_{x0}-S_{13} \hat{E}^+_{y0}\\
-S_{21}\hat{E}^+_{x0}-S_{23} \hat{E}^+_{y0}\\
-S_{31}\hat{E}^+_{x0}-S_{33} \hat{E}^+_{y0}\\
-S_{41}\hat{E}^+_{x0}-S_{43} \hat{E}^+_{y0}\\
\end{pmatrix} \, .
\label{eq:Matrix_Formalism_solutions}
\end{eqnarray}

\subsubsection{Layer of topological magnetic insulator}\label{sec:AQ_P_mixing_Ref_Trans}
Let us now apply the matrix formalism to a system with~$N=2$. However, note that the matrix approach developed here is able to describe more complicated systems, consisting of many layers. One particular example could be a layered system of different topological insulators with different material properties. The matrix formalism with $N>2$ could be useful in DM searches to increase the boost factor using additional layers of TMI or dielectric.

We now calculate the reflection and transmission coefficients for one TMI layer. In the language of the matrix formalism the system has $N=2$ boundaries, and hence three media. Media~$0$ and~$2$ are vacuum while medium~$1$ is a TMI, hosting a dynamical AQ. The THz~laser radiation is coming from medium $0$ and hits the layer of TMI. In what follows, we omit the subscripts~$r$ that label the materials because the only non-vacuum medium is the TMI, i.e.\ medium~$1$.

We assume that the laser polarization is oriented in the $y$-direction, parallel to the external $B$-field. In this case, we obtain -- to lowest order in $\frac{\alphaEM}{\pi}\stAQ$ -- the following reflection and transmission coefficients:
\begin{eqnarray}
T_y&=&\frac{2i  \tilde{k} }{\left(\tilde{k}^2+1\right)\sin\Delta+2i \tilde{k} \cos\Delta  }+\mathcal{O}\left(\left(\frac{\alphaEM}{\pi}\stAQ\right)^2\right),\label{eq:AQ_Photon_mixing_Ty}\\
R_y&=&-\frac{\left(\tilde{k}^2-1\right) \sin\Delta }{\left(\tilde{k}^2+1\right)\sin\Delta +2 i \tilde{k}\cos\Delta}+\mathcal{O}\left(\left(\frac{\alphaEM}{\pi}\stAQ\right)^2\right),\label{eq:AQ_Photon_mixing_Ry}
\end{eqnarray}
where $\tilde{k}=\frac{\kAQ}{\omega \mu}, \Delta\equiv d\kAQ$ and $d=z_2-z_1$ is the thickness of the layer. Note that, although $\tilde{k}$ depends on the expansion parameter we did not expand $\tilde{k}$ because otherwise the expansion for the transmission and reflection coefficients would not be valid around the resonance. The calculated transmission and reflection coefficients are valid for both the case with and without losses since we assume all losses to be bulk losses. $T_y$ and $R_y$ agree with the normal transmission and reflection coefficients of a dielectric disk~\cite{Millar:2016cjp} if the coupling $b$ of the AQ to the photon is set to zero, i.e.\ $\kAQ\rightarrow  n \omega$, corresponding to $\fAQ\rightarrow\infty$. 
 
 \begin{figure}
 	\centering
 	\includegraphics[width=0.49\textwidth]{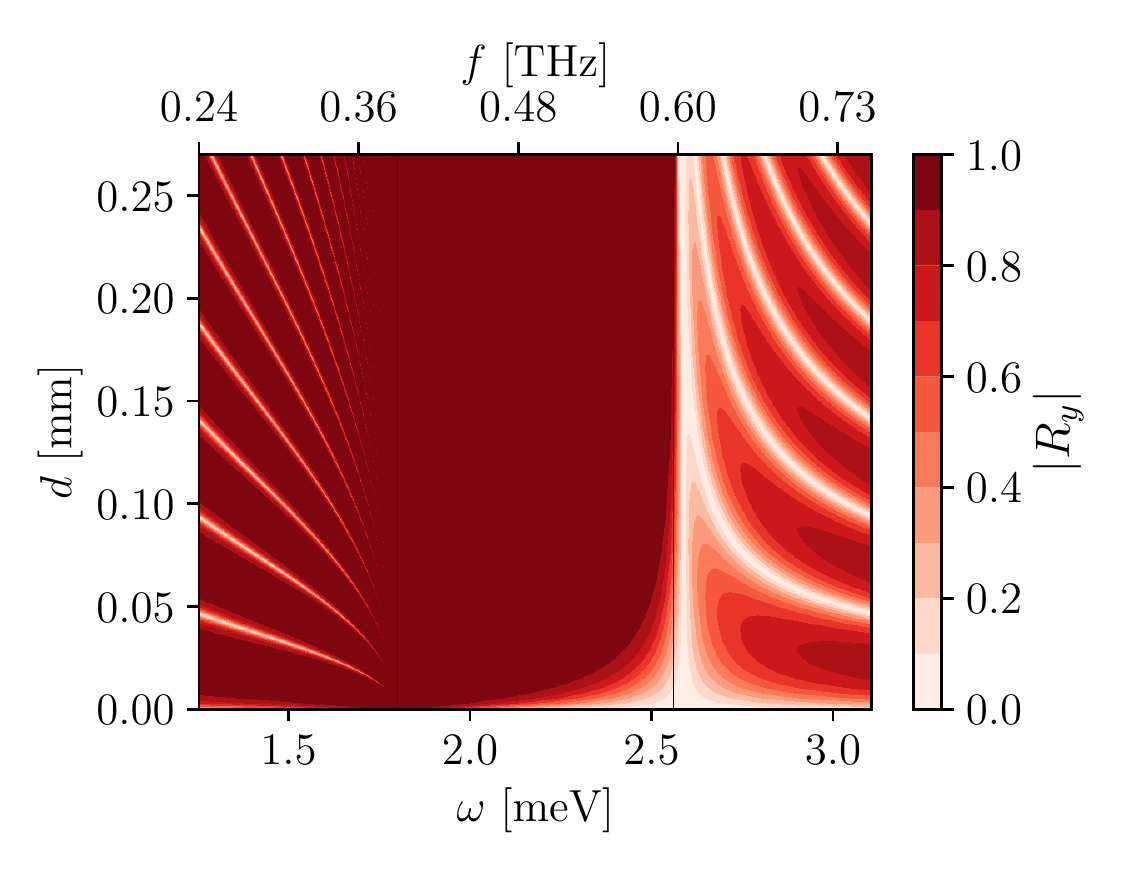}
 	\includegraphics[width=0.49\textwidth]{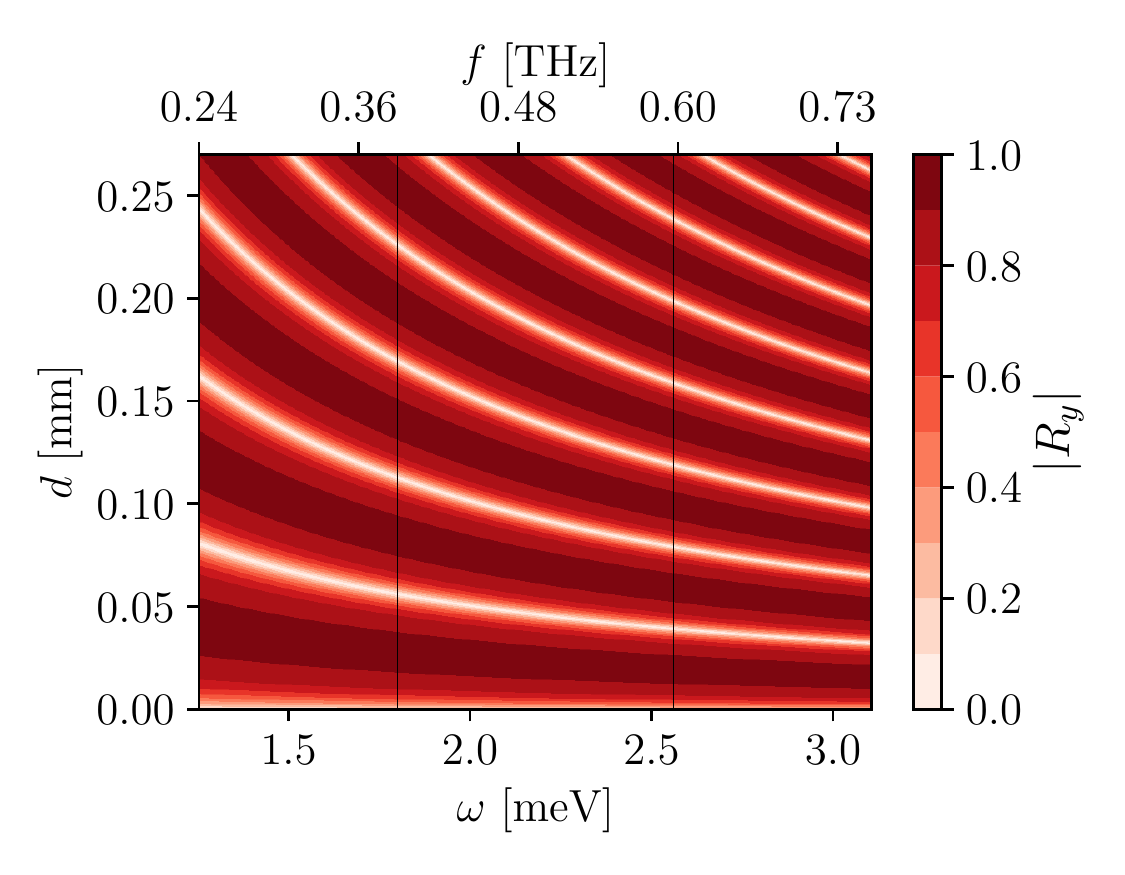}\\
 	\includegraphics[width=0.49\textwidth]{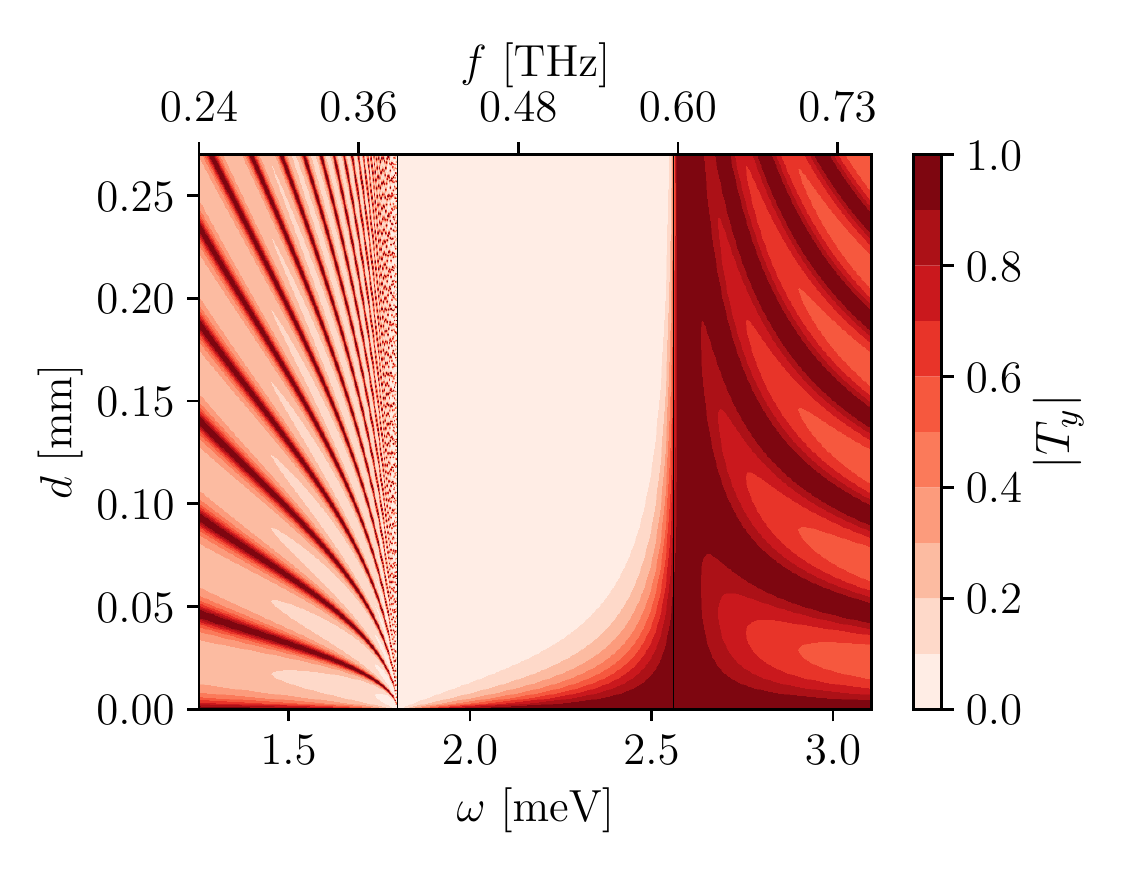}
 	\includegraphics[width=0.49\textwidth]{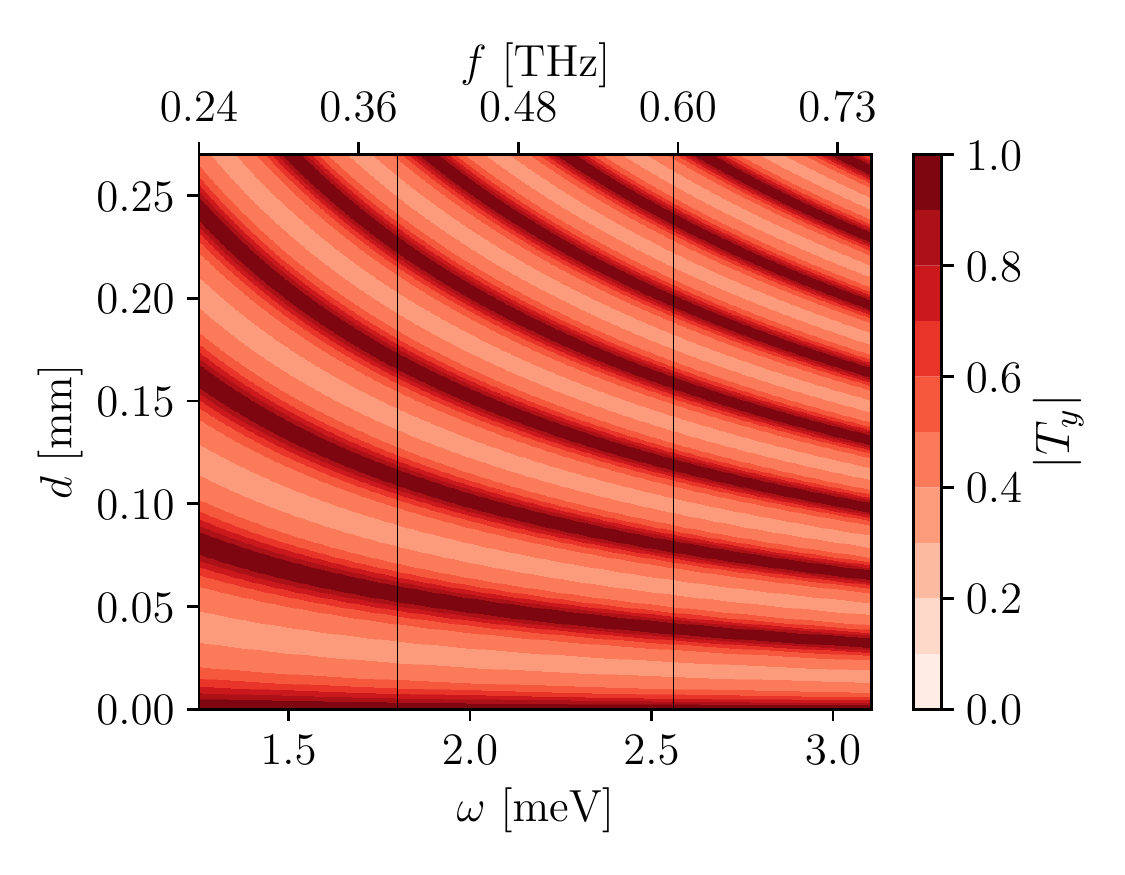}\\
 	\includegraphics[width=0.49\textwidth]{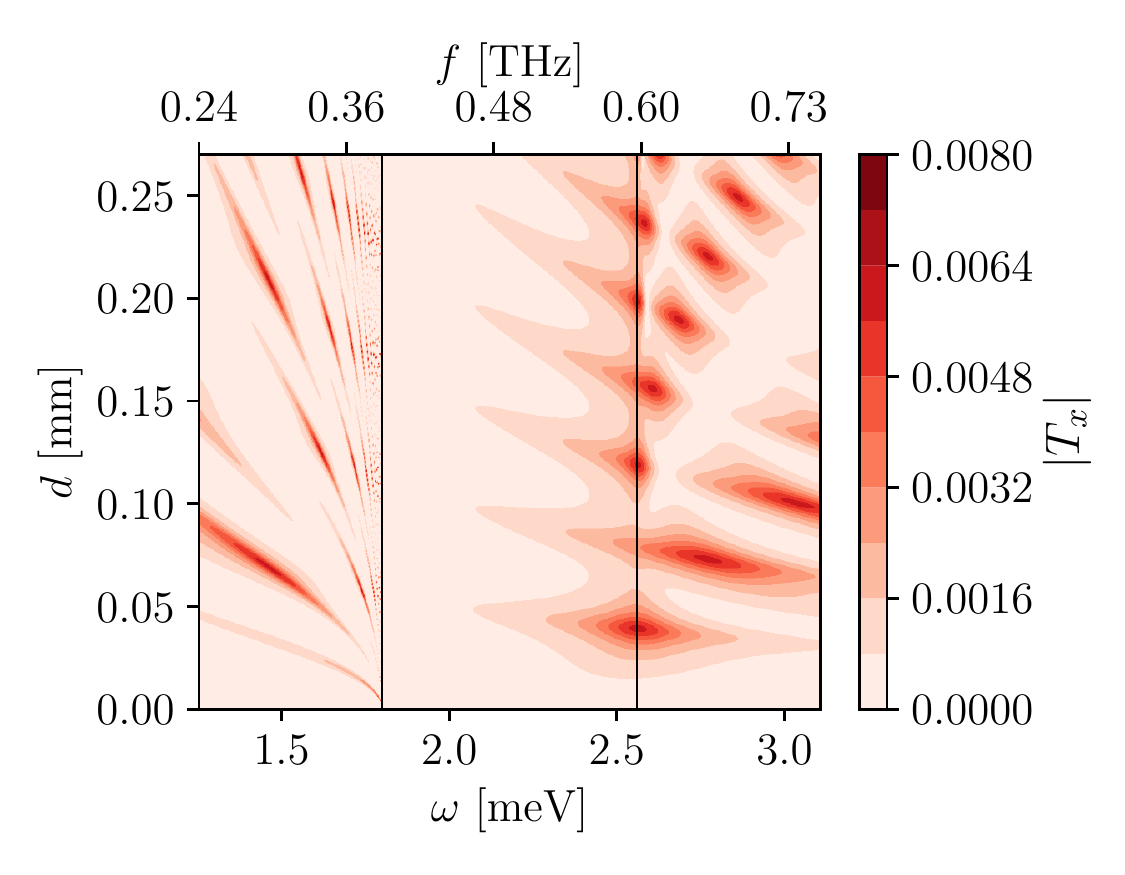}
 	\includegraphics[width=0.49\textwidth]{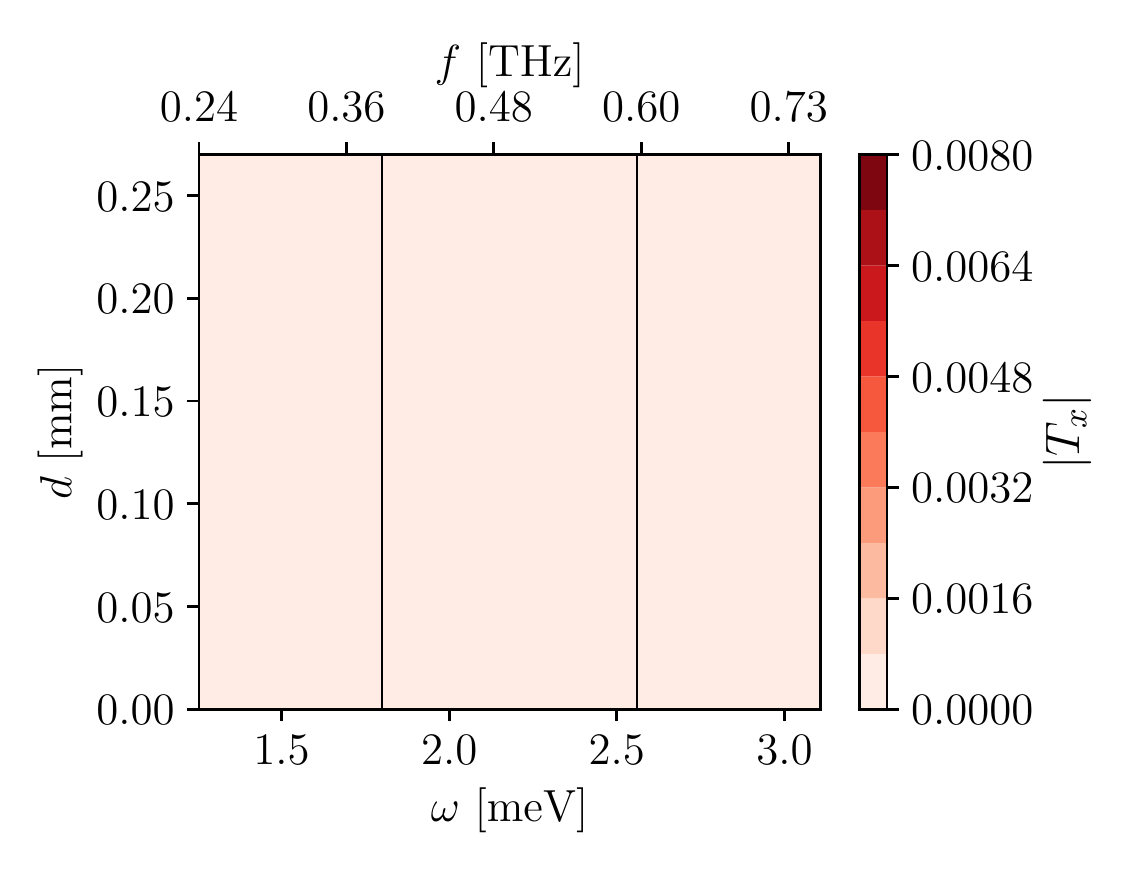}
 	\caption{Reflection and transmission coefficients for a laser that hits a material with~(\textit{left}) and without~(\textit{right}) a dynamical AQ. The laser polarization is in the $y$-direction, parallel to the external $B$-field. The materials have $\stAQ=0.8\,\pi$, $n=5$, and $\mu=1$. Typical material values for a Mn$_2$Bi$_2$Te$_5$ TMI with an external $B$-field $B_e$ of $\SI{2}{\tesla}$ are chosen, cf.\ Table~\ref{tab:derived_params} and Eq.~\eqref{eqn:b_meV}. $\mAQ$ and $\omLO$ are marked with the black vertical lines. The $R_y$ ($T_y$) coefficient is always close to one (zero) inside the gap if a dynamical AQ is present. The $T_x$ coefficient is only non-zero is a AQ is present. However the effect of an AQ in the $E_x$-component is much smaller than in the $E_y$-component since the $E_x$-component can only be induced via polarization rotation with non-zero $\stAQ$.}
 	\label{fig:TwoInterfaces_BParallelE_refTrans}
 \end{figure}
We show the full functions for the reflection and transmission coefficients without losses ($\mat{\Gamma}=0$) in Fig.~\ref{fig:TwoInterfaces_BParallelE_refTrans}. The coefficients are shown for different values of the laser frequency~$\omega$ and sample thickness~$d$. The left column assumes the presence of a dynamical AQ, while the figures in the right column show the case when no dynamical AQ is present. Note that in both cases $\stAQ =0.8\, \pi$ is assumed although the shown results for $T_y$ and $R_y$ do not depend on $\stAQ$ to lowest order, cf.\ equations~\eqref{eq:AQ_Photon_mixing_Ty} and~\eqref{eq:AQ_Photon_mixing_Ry}.

First, we discuss the figures in the top and middle row, which show the reflection and transmission coefficients for the $E_y$-components.
If a dynamical AQ is present, the dispersion relation $\kAQ$ becomes imaginary between $\mAQ$ and~$\omLO$. The gap between these two frequencies is marked with the two vertical lines.
For large thicknesses $d$, all frequencies in the gap are reflected, and none are transmitted.
This is a direct consequence of the purely imaginary $\kAQ$ in the gap. For small values of the thickness, the gap size is reduced. This happens around $\omLO$, i.e.\ the upper part of the gap, since the imaginary part gets reduced (the skin depth becomes larger) the more $\omLO$ is approached from smaller frequencies, cf.\ right panel of Fig.~\ref{fig:Disp_relat_vz_zero}.
When going away from the gap, the figures in the left and right columns agree more and more.
This is as expected since the dispersion relation $\kAQ$ differs only significantly from a normal photon dispersion around the gap. In the case of no dynamical AQ (left panel) we notice a clear non-zero reflection and transmission inside the gap. Comparing the figures on the left-\ and right-hand side, it is clear that the AQ causes an $\order(1)$~modification of the $T_y$ and $R_y$ coefficients compared to the spectrum when no dynamical AQ is present. 

Next, we discuss the bottom row of Fig.~\ref{fig:TwoInterfaces_BParallelE_refTrans}, which shows the transmission coefficient for the $E_x$-component. If no dynamical AQ field is present~(right panel) but we have a topological material with $\stAQ = 0.8\,\pi$, the transmission $T_x$ vanishes. This may be surprising at fist glance because there is mixing at the interface of ordinary TIs and, hence, also a polarization rotation. However, the transmission in the $x$-component vanishes since the polarization rotations at the two interfaces cancel each other. If in addition to the static $\stAQ=0.8\,\pi$ a dynamical AQ is present (left panel), we get a small non-zero transmission~$T_x$. The signal is much smaller than in the case of the $T_y$ coefficient. This is because the incoming laser is polarized in the $E_y$ component and a non-zero $E_x$-component can only be induced due to a nonzero $\stAQ$, i.e.\ mixing at the interfaces, which is proportional to the small parameter $\frac{\alphaEM}{\pi}\stAQ$. In conclusion, we should first look for the AQ by studying the $E_y$-component because the AQ~modification of this component is much larger than for the $E_x$-component.

However, once the AQ is found, one can also use the $E_x$-component to determine, for example, $\stAQ$ of the material by reflection and transmission measurements. Is is also possible to study the influence of non-linear effects with the $x$-components. In Eq.~\eqref{Ampere_1Dl1} it was shown that the laser sources the $x$-component in a non-linear fashion. This effect is neglected here because the equations are linearized. 

\begin{figure}
	{\centering
		\includegraphics[width=0.49\textwidth]{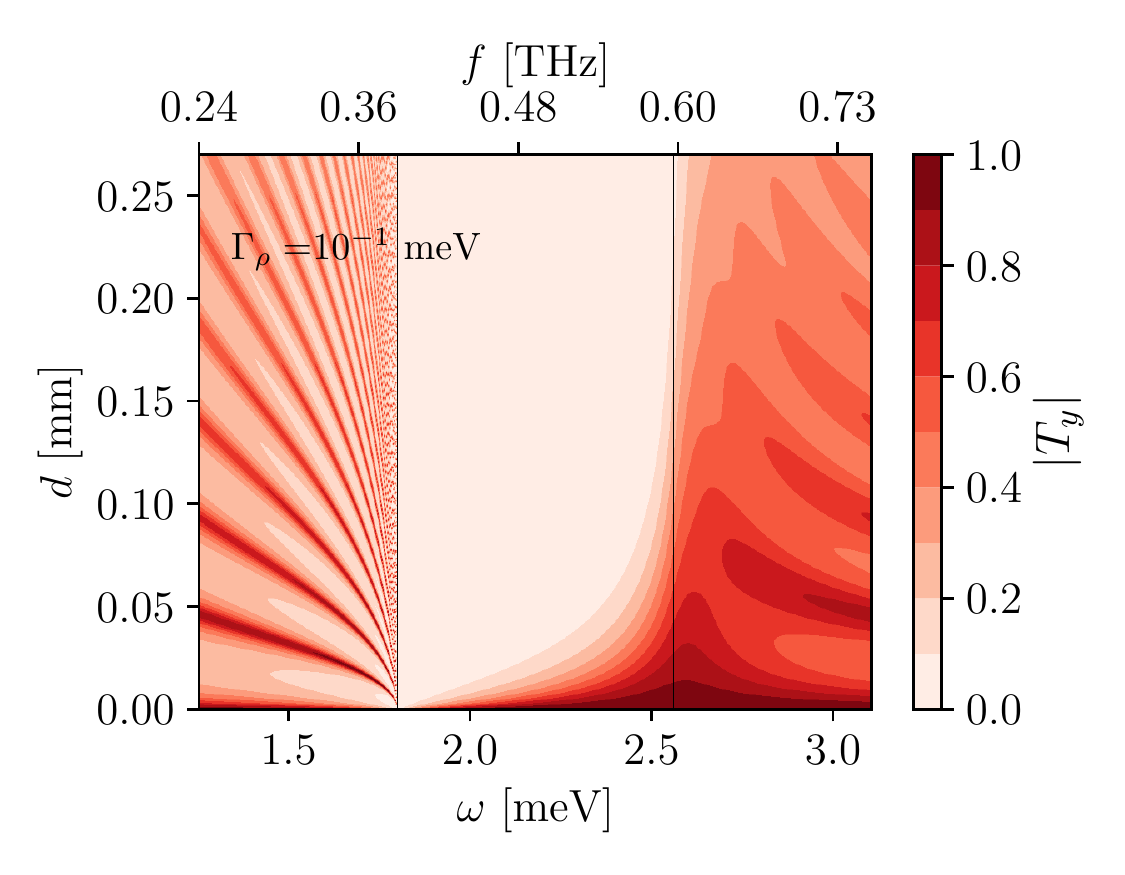}
		\includegraphics[width=0.49\textwidth]{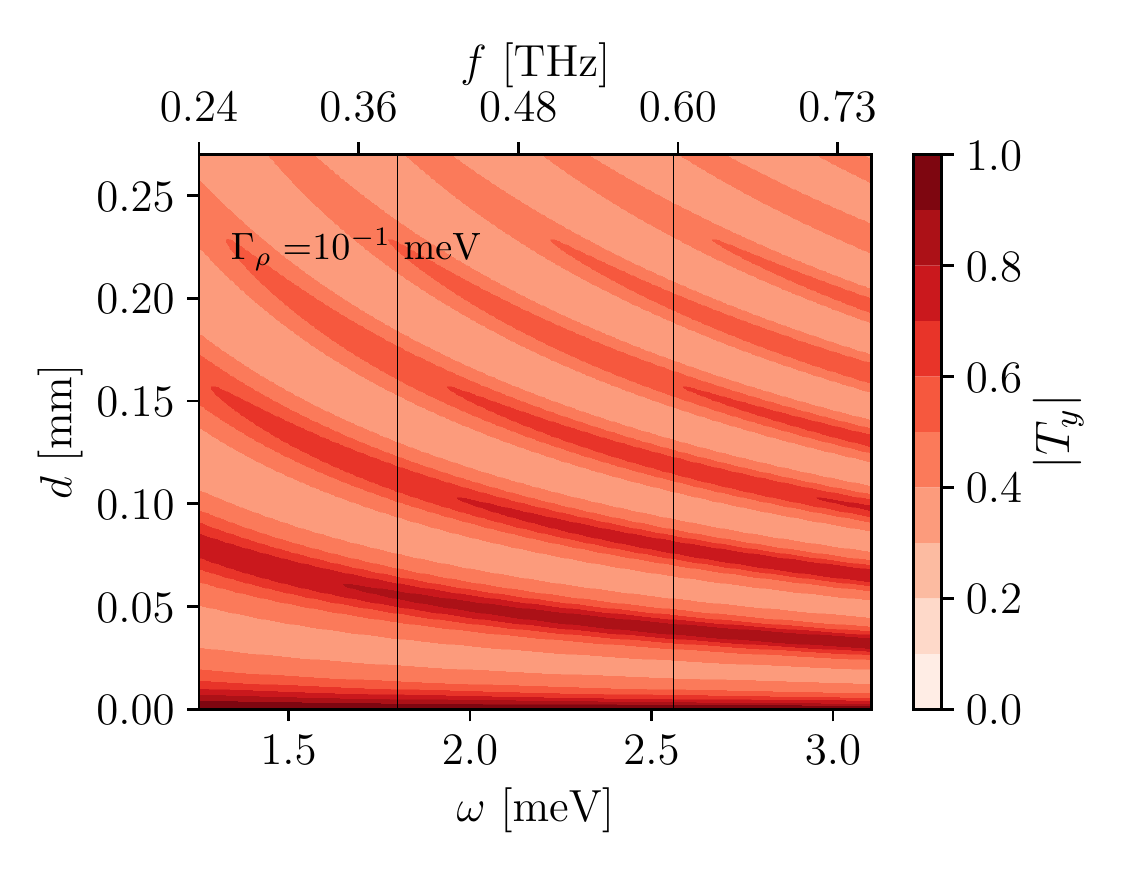}}\\
	\includegraphics[width=0.49\textwidth]{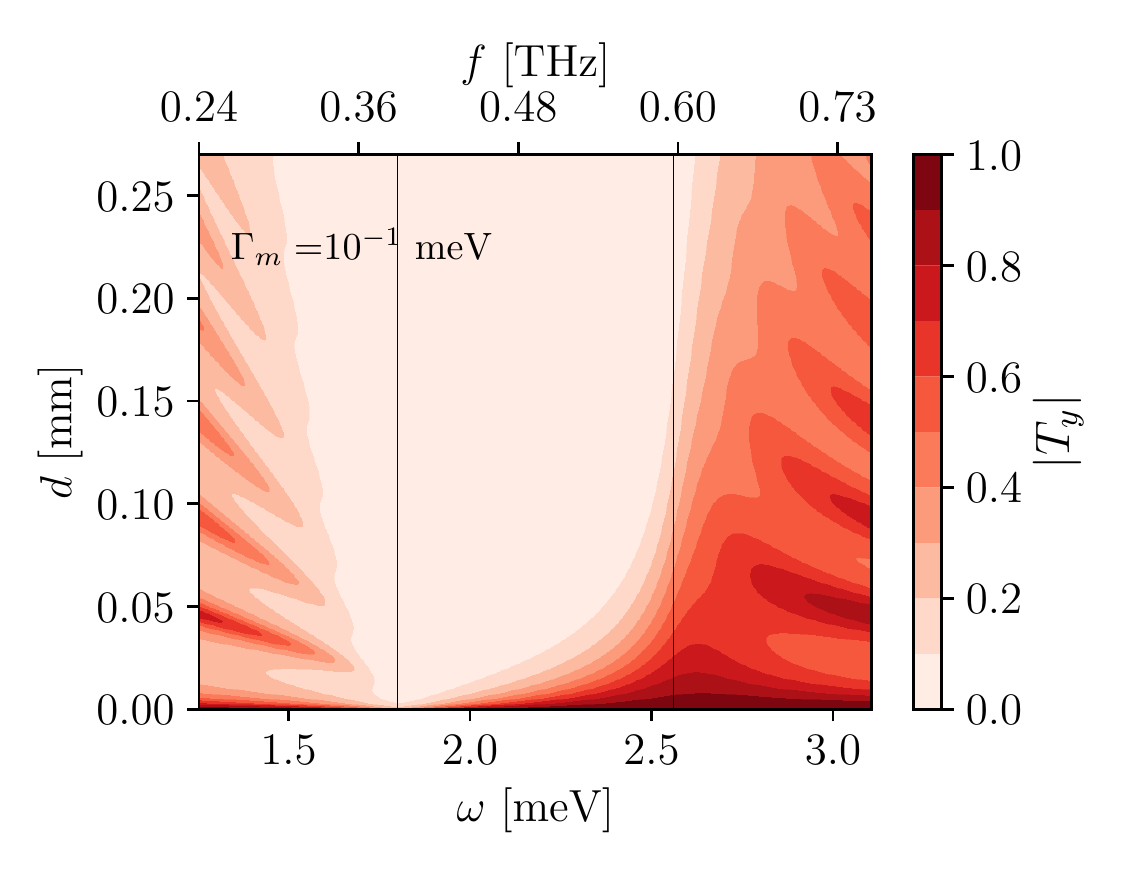} 
	\caption{Transmission coefficients for the $E_y$-component (parallel to the external $B$-field) for exaggerated photon and magnon losses, $\Gamma_\rho$ and $\Gamma_m$. We show the results for when a dynamical AQ field~(\textit{left}) and if no dynamical AQ is present; (\textit{right}). In both cases we have $\stAQ=0.8\,\pi$. \defaultSetup. $\mAQ$ and $\omLO$ are marked with the black vertical lines. }
	\label{fig:TwoInterfaces_BParallelE_Ty_Gr}
\end{figure}
Figure~\ref{fig:TwoInterfaces_BParallelE_Ty_Gr} shows the transmission coefficient $T_y$ for different losses. The figures are produced for our benchmark material Mn$_2$Bi$_2$Te$_5$ with $n=5,\mu=1$ and $\stAQ=0.8\,\pi$. In the top row we illustrate the influence of photon losses $\Gamma_\rho$ for a TMI with a dynamical AQ~(left panel) and for a normal TI~(right panel). 
The transmission at large layer thicknesses becomes smaller independently of the resonance. This is due to the fact that $\Gamma_\rho$ appears in the dispersion relation~\eqref{eq:dispersion_relation_vz0_losses_withoutMixed} as an additional term which is approximately constant in the small shown frequency interval. The skin depth is of the order of~$\Gamma_\rho^{-1}$. It is therefore advantageous to have thin material samples for distinguishing between the case of a DA~(left panel) and no DA~(right panel). However, should not be too thin. For very small thicknesses the frequencies inside the gap lead to a transmission coefficient $T_y$ that is not very small anymore. Note that the effect of losses for large $d$ becomes more pronounced for larger refractive indices $n$.

In the bottom row of Fig.~\ref{fig:TwoInterfaces_BParallelE_Ty_Gr} photon losses are zero and the effect of magnon losses $\Gamma_m$ is illustrated. We do not show the case without an AQ because without AQ there are no magnon losses and one should compare to the already existing Fig.~\ref{fig:TwoInterfaces_BParallelE_refTrans} (middle row, right). The larger the magnon losses, the more pronounced is the widening of the gap. This can be understood by looking at the dispersion relation in Eq.~\eqref{eq:dispersion_relation_vz0_losses_withoutMixed}. Magnon losses $\Gamma_m$ introduce a Lorentzian shaped imaginary part to the dispersion relation. The width of the Lorentzian is proportional to $\Gamma_m$. Due to the Lorentzian shape of the damping imaginary part in the dispersion relation also frequencies that are not directly in the gap -- but close to the gap -- can become highly damped. This effect becomes more pronounced the thicker the sample is. 

From the previous discussion it becomes clear that finding the AQ will depend very sensitively on the losses and thickness of the material. The losses that we show in Fig.~\ref{fig:TwoInterfaces_BParallelE_Ty_Gr} are exaggerated and in reality we expect them to be much smaller. Therefore from Fig.~\ref{fig:TwoInterfaces_BParallelE_Ty_Gr} we find that with a layer thickness on the order of $\SI{0.03}{\milli\metre}$ and $\SI{0.3}{\milli\metre}$ AQ can be most effectively be detected.

Once the AQ is detected the characterization of the parameters of the TMI is of huge importance. In Section~\ref{sec:AQ_DA_calculations} and~\ref{sec:DM-detection} it is shown how a TMI can be used as dark matter axion detector. To estimate the induced photon signal from a DA, AQ and photon mixing the parameters of the TMI have to be known precisely. 
In the following we demonstrate that fitting measurements to the presented results can determine the parameters of the TMI.

\begin{figure}
	\centering
	\includegraphics[width=\textwidth]{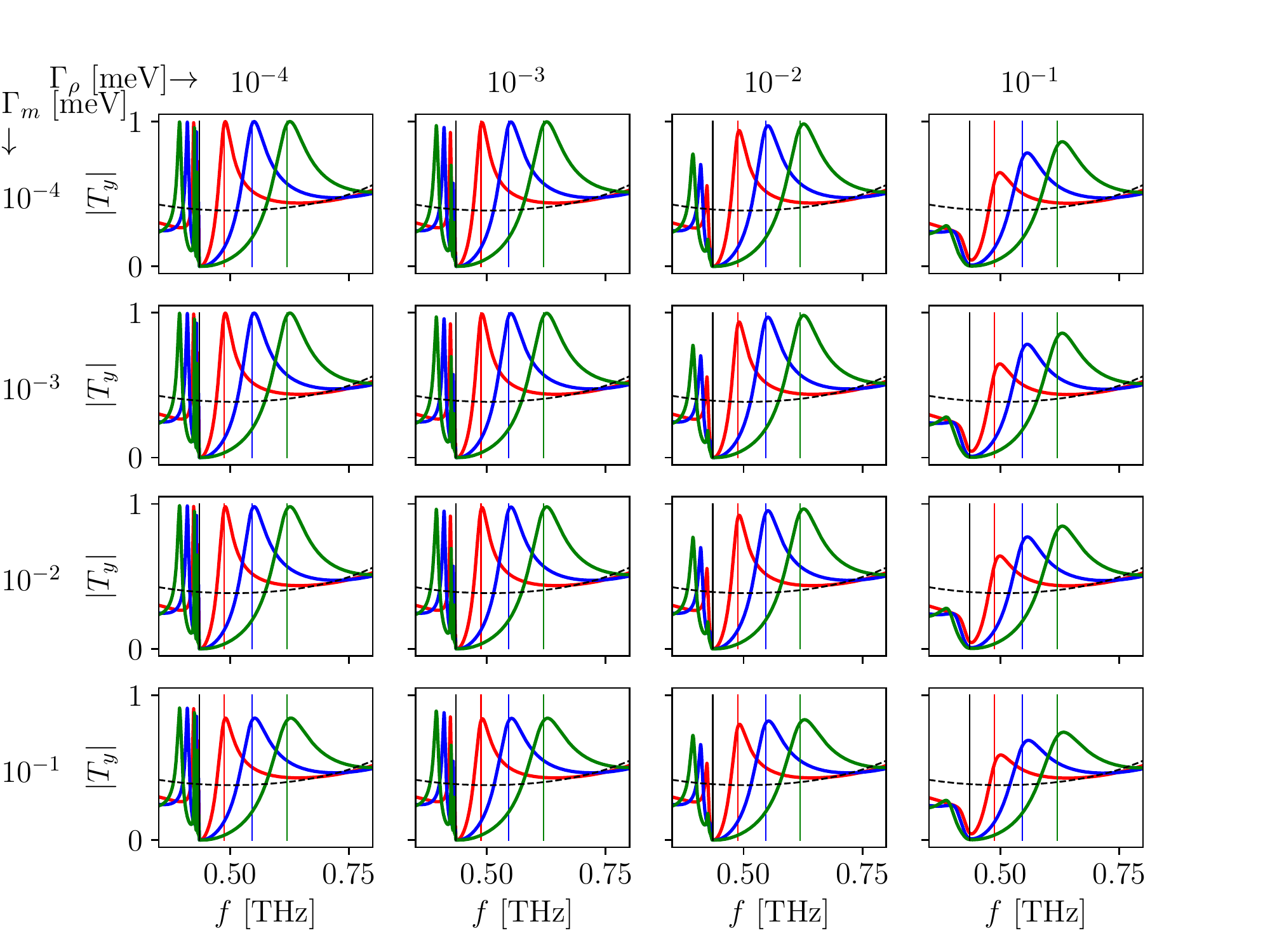} \\
	\includegraphics[width=0.8\textwidth]{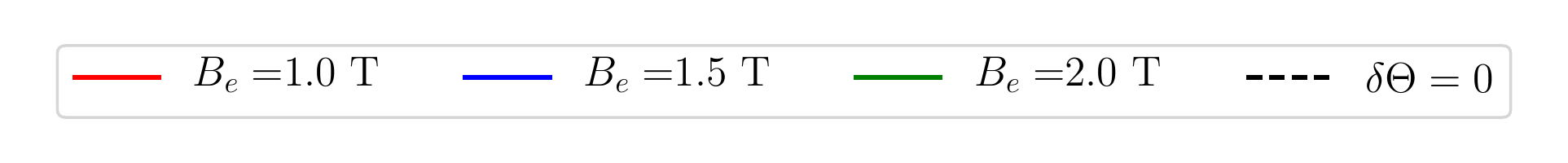} 
	\caption{Transmission coefficient $T_y$ for a layer of TMI with thickness $d=\SI{0.03}{mm}$ for different $B$-fields (colours) are shown. Panels vary the losses $\Gamma_\rho$~(from left to right) and $\Gamma_m$~(from top to bottom). The vertical black line indicates the value of $\mAQ = \SI{2}{\milli\electronvolt}$, while the other vertical lines indicate the value of $\omLO$ for different values of the external $B$-field. The larger the external $B$-field, the larger the gap between $\mAQ$ and $\omLO$. \defaultSetupTMI. The dashed black line shows the result when no dynamical AQ $\dyAQ$ is present, while the solid coloured lines are for the case with a dynamical~AQ. 
	}
	\label{fig:TwoInterfaces_DifferentLosses_and_mB}
\end{figure}
In Fig.~\ref{fig:TwoInterfaces_DifferentLosses_and_mB}, photon losses $\Gamma_\rho$ are varied between $\Gamma_\rho = \num{3e-4}\,{\rm meV}$ and $0.3\,{\rm meV}$, and magnon losses between $\Gamma_m = \num{3e-4}\,{\rm meV}$ and $0.3\,{\rm meV}$. The layer thickness is fixed to $d=\SI{0.03}{\milli\metre}$. In each figure, we show three different external $B$-field values. The values of $\mAQ$ and $\omLO$ are marked with a vertical black and coloured lines, respectively. The weaker the external $B$-field, the smaller the gap.

Figure~\ref{fig:TwoInterfaces_DifferentLosses_and_mB} makes again clear that the larger the losses the harder it is to distinguish the case where a dynamical AQ is present (solid coloured curves) from the case that not dynamical AQ (dashed black line) is present. For relatively small losses, the distinction between the curves is very clear. We therefore conclude that comparing these results to future measurements will make it possible to explicitly determine the material parameters, i.e. losses, refractive index, $\stAQ$, and the parameters that enter the AQ mass $\mAQ$ and the gap size parameter $b$.

We now investigate further the resonance around $\omLO$, cf.\ Fig.~\ref{fig:TwoInterfaces_DifferentLosses_and_mB}. When the losses in Fig.~\ref{fig:TwoInterfaces_DifferentLosses_and_mB} are small the resonance frequency $f_{\text{res}}=\frac{\omega_{\rm res}}{2\pi}$ corresponds to $\omLO$. However, with higher losses, the resonance frequency  $\omega_{\text{res}}$ moves to higher frequencies, i.e.\ $f_{\text{res}}>\frac{\omLO}{2\pi}$. With increasing losses, the resonance smears out until it vanishes completely. Figure~\eqref{fig:TwoInterfaces_DifferentLosses_and_mB} allows us to directly read off the amount of losses that would still be acceptable AQ detection~(for a sample of thickness $d = \SI{0.03}{mm}$). The resonance peaks to the right of $\omLO$ in Fig.~\ref{fig:TwoInterfaces_DifferentLosses_and_mB} are not symmetric. We therefore define the width of the resonance peak, $\Gamma_{\text{res}}$, as two times the frequency interval that ranges from the frequency at the transmission maximum down to the smaller frequency at half the transmission maximum. 

\begin{figure}
	\centering
	\includegraphics[width=0.5\textwidth]{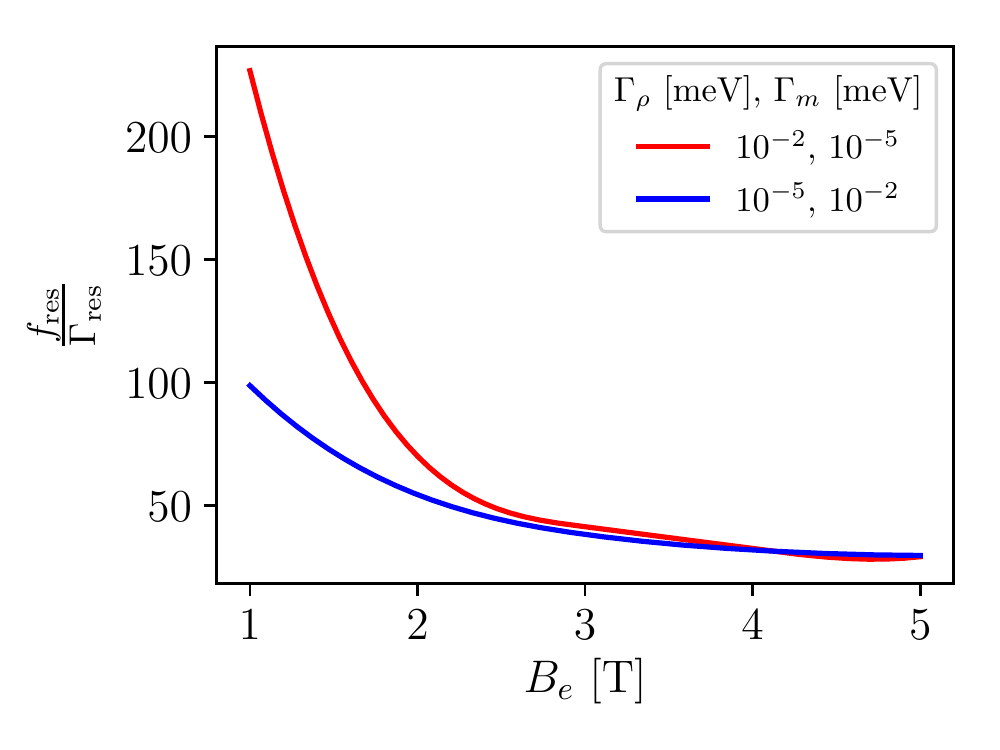} \\
	\caption{TMI Quality factor, $Q = f_{\text{res}}/\Gamma_\text{res}$, for different losses, with respect to the external $B$-fields. $f_\text{res}$ is the frequency of the maximal transmission peaks around $\omLO$, cf.\ Fig.~\ref{fig:TwoInterfaces_DifferentLosses_and_mB}. The TMI layer has a thickness of $d=\SI{0.03}{mm}$.}
	\label{fig:TwoInerfaces_RefTransmission_Q_factor}
\end{figure}

The ratio $\frac{f_{\text{res}}}{\Gamma_{\text{res}}}$ is called the $Q$-factor. It describes the quality of the resonance in the sense that large $Q$-factors give rise to a well-defined resonance, whereas low $Q$-factors show that the resonance is highly damped. In Fig.~\ref{fig:TwoInerfaces_RefTransmission_Q_factor}, the $Q$-factor is shown with respect to the applied external $B$-field for different losses. We consider  the case of dominant conductive losses (red) and dominant magnon losses (blue). The largest $Q$-factor is observed at small external $B$-field, and the low magnon losses lead to the largest $Q$. This is consistent with the intuition that at low $B$-field the polariton is largely magnon-like. For larger external $B$-fields the difference between the two cases becomes small, as both sources of loss contribute almost equally. 

\FloatBarrier

\section{Axion Dark Matter and Axion Quasiparticles}\label{sec:AQ_DA_calculations}

\subsection{Dark axion, axion quasiparticle and photon mixing}
Paper~I proposed using dynamical AQs in TMIs to detect DAs. This is possible since DAs can mix resonantly with axion polaritons. Compared to Paper~I, we work out a more detailed calculation for the emitted photon signal by taking into account the correct interface conditions and material losses. This, in turn, allows us to present a more rigorous calculation of the sensitivity reach for DA searches using TMIs. 

As a starting point, we use the three-dimensional equations of motion, Eqs.~\eqref{GaussE}--\eqref{KleinGordon}. We linearize these and derive a one-dimensional model in analogy to Section~\ref{sec:Quasiparticle_Photon_mixing_Linearization}. In what follows, the one-dimensional model is used to derive the photon signal generated by DAs passing through a magnetized TMI that hosts dynamical AQs.

\subsubsection{General formulation}
To describe the threefold mixing between AQs, DAs, and photons, we need to add the Klein-Gordon equation for DAs, which is sourced by the electromagnetic fields, to Eqs.~\eqref{GaussE}--\eqref{KleinGordon}. Additional source terms, arising due to the presence of DAs, have therefore to be added to Eqs.~\eqref{GaussE} and~\eqref{Ampere}. Doing so results in the following equations of motion:
\begin{align}
		\nabla\cdot \vc{D} &= \rho_f - \frac{\alphaEM}{\pi}\nabla(\delta\Theta+\stAQ)\cdot\vc{B} - \gDA \nabla a\cdot\vc{B} \, ,\label{GaussE_aq_da}\\
		\nabla\times\vc{H} - \partial_t\vc{D} &= \bm{J}_f + \frac{\alphaEM}{\pi} \left[\vc{B}\partial_t(\delta\Theta+\stAQ)-\vc{E}\times\nabla( \delta\Theta+\stAQ)\right] \nonumber\\
		&\mathrel{\phantom{=\bm{J}_f}} + \gDA \, (\vc{B}\partial_t a-\vc{E}\times\nabla a) \, , \label{Ampere_aq_da}\\
		\nabla\cdot\vc{B} &= 0 \, , \label{GaussB_aq_da}\\
		\nabla\times\vc{E}+\partial_t\vc{B} &= 0 \, , \label{Faraday_aq_da} \\
		\partial_t^2\delta\Theta-v_i^2\partial_i^2\delta\Theta+m_\Theta^2\delta\Theta &= \Lambda \, \vc{E}\cdot\vc{B} \, , \label{KleinGordon__aq_da}\\
		(\partial_t^2-\nabla^2+m_a^2) \, a &=  \gDA \, \vc{E}\cdot \vc{B} \, . \label{KleinGordon__aq_da2}
\end{align}
where $a$ is the pseudoscalar DA field, $g_{a\gamma}$ is the DA-photon coupling, and $\mDA$ is the DA mass, in addition to the other variables already defined in Eqs.~\eqref{GaussE}--\eqref{KleinGordon}.

In Section~\ref{sec:Quasiparticle_Photon_mixing_General} we already noted that one cannot obtain interface conditions from the Klein-Gordon equation for an interface between media with and without AQs. However, DAs are expected to permeate any medium due to the necessarily feeble interactions of dark matter, and their presence in the Galaxy. Therefore, for two media that both contain DAs, Eq.~\eqref{KleinGordon__aq_da2} can be used to derive an interface condition for the DA~field.

Consider an infinitesimal volume element between two media, say, between medium~$1$ and medium~$2$. We integrate over this infinitesimal volume element and apply the divergence theorem. It follows that the normal derivative of the DA~field between two interfaces has to be continuous,
\begin{equation}
	\vc{n}\cdot (\nabla a_1 - \nabla a_2) = 0 \, . \label{eq_DA_AP_P_mixing_Da}
\end{equation}
Furthermore, we require that the DA field be continuous over the interface:
\begin{equation}
	a_1 - a_2 = 0 \, . \label{eq_DA_AP_P_mixing_a}
\end{equation} 
We stress that the continuity of the axion field in Eq.~\eqref{eq_DA_AP_P_mixing_a} does not follow from the axion-Maxwell equations, but is a reasonable approximation. 
In other words, as DAs only interacts with matter through very small couplings, and we are interested in the conversion of axions to photons again by a very small coupling, any modification due to the axion interacting with the interface is at higher order, and thus negligible.

\subsubsection{Linearized one-dimensional model}\label{sec:DA:P_AQ_mixing_eom}
Let us again assume the presence of a strong and static external $B$-field, $\vc{B}_e$. Without loss of generality, let $\vc{B}_e$ be polarized in the $y$-direction, i.e.\ $\vc{B}_e = B_e \, \hat{\vc{e}}_y$. Then, similar steps as in Sections~\ref{sec:Quasiparticle_Photon_mixing_OneDimensionalModel} and~\ref{sec:Quasiparticle_Photon_mixing_Linearization} lead to the following linearized equations of motion:
\begin{eqnarray}
	\left(\partial_z^2-n^2\partial_t^2-\sigma\mu\partial_t\right)E_y&=&\mu B_e\partial_t^2\left(\frac{\alphaEM}{\pi}\delta\Theta+g_{a\gamma}a\right) \, ,\label{eq:eqOfMotion_ThreefoldMixing_E} \\
	\left(v_z^2\partial_z^2-\partial_t^2-\mAQ^2\right)\delta\Theta&=&-\Lambda B_e E_y,\label{eq:eqOfMotion_ThreefoldMixing_T} \\
	\left(\partial_z^2-\partial_t^2-\mDA^2\right)a&=&-g_{a\gamma}B_eE_y \, .\label{eq:eqOfMotion_ThreefoldMixing_a}
\end{eqnarray}

The photon signal in the $E_x$-component, induced by DAs, is always an order $\frac{\alphaEM}{\pi}\stAQ$ smaller than the $E_y$-component. This is due to the fact that only the $E_y$-component mixes with DAs and AQs. The $E_x$-component can only be generated due to the mixing at the interface, which is proportional to $\frac{\alphaEM}{\pi}\stAQ$. The main photon signal is therefore polarized parallel to the external $B$-field in experimental DA searches, i.e.\ in the $E_y$-component. Due to the suppression of the $E_x$-component, it will be even more challenging to detect a signal in the $E_x$-component. This justifies neglecting the $E_x$-component in what follows.

In addition to the linearization assumptions made in Section~\ref{sec:Quasiparticle_Photon_mixing_Linearization}, we further assume that the non-linear terms, which include the DA, can also be linearized. This assumption is justified because of the small coupling and non-relativistic nature of Galactic DAs, for which $\partial_ta/\partial_za \approx 10^{-3}$.

The interface conditions for the electromagnetic fields after linearization are obtained with the linearized fields $\vc{D}_{\Theta} = \vc{D} + \frac{\alphaEM}{\pi} \, (\stAQ+\delta\Theta) \, \vc{B}_e$ and $\vc{H}_{\Theta}=\vc{H}-\frac{\alphaEM}{\pi}\stAQ\vc{E}$. In the one-dimensional model, the conditions  $\bm{n}\cdot (\vc{D}_{\Theta 2}-\vc{D}_{\Theta 1})=0$ and $\bm{n}\cdot (\vc{B}_2-\vc{B}_1)=0$ are always fulfilled, since transverse waves are assumed to vanish, i.e.\ $B_z = 0$ and $E_z = 0$. The only non-trivial interface conditions are:
\begin{eqnarray}
	\bm{n}\times(\vc{H}_{\Theta,2}-\vc{H}_{\Theta,1})&=&0 \, ,\label{eq:DA_AQ_P_mixing_LinearizedInterface_H_1}\\
	\bm{n}\times(\vc{E}_2-\vc{E}_1)&=&0 \, ,\label{eq:DA_AQ_P_mixing_LinearizedInterface_E}
\end{eqnarray}
where $\vc{n} = \hat{\vc{e}}_z$ and it is assumed again that no free surface charges and currents are present.

Including bulk losses in the one-dimensional model does not change the interface conditions. The magnon losses $\Gamma_m$, photon losses $\Gamma_\rho$, and mixed losses $\Gamma_{\times,1}$ and $\Gamma_{\times,2}$ are included in complete analogy to Section~\ref{sec:Quasiparticle_Photon_mixing_Losses}. The resulting equations of motion are:
\begin{eqnarray}
\mat{K} \, \partial_t^2 \vc{X} - \mat{\Gamma} \, \partial_t \vc{X} + \mat{M} \, \vc{X} = 0 \, , \label{eq:dm_exp:eom}
\end{eqnarray} 
where we define
\begin{eqnarray}
\vc{X}=\begin{pmatrix}
E_y\\\delta\Theta\\a
\end{pmatrix}\, , \quad \mat{K}&=&\begin{pmatrix}
1&\frac{\alphaEM}{\pi}\frac{ B_e}{\epsilon}&\frac{\gDA B_e}{\epsilon}\\
0&1&0\\
0&0&1
\end{pmatrix} \, , \quad \mat{\Gamma}=\begin{pmatrix}
\Gamma_\rho&\Gamma_{\times,1}&0\\
\Gamma_{\times,2}&\Gamma_m&0\\
0&0&0
\end{pmatrix} \, , \nonumber\\
\mat{M}&=&\begin{pmatrix}
\frac{k^2}{n^2}&0&0\\
-\Lambda B_e&v_z^2k^2+\mAQ^2&0\\
-g_{a\gamma}B_e&0&k^2+\mDA^2
\end{pmatrix} \, .
\end{eqnarray}
No losses for the DA are included since a valid DM candidate must, by necessity, have an astronomically long lifetime (indeed, the QCD axion in the mass range of interest satisfies this constraint by many orders of magnitude).

\subsection{Dark matter signal calculation}
In this section, we solve the linearized equations of motion~\eqref{eq:dm_exp:eom}. We first consider the lossless case and then generalize the solutions to include losses. Material properties are always considered piecewise homogeneous, We introduce a matrix formalism to calculate emitted photon and axion power from an experimental setup with multiple TMI layers. We apply the matrix formalism to our benchmark setup, a single-TMI layer surrounded by vacuum. Using a multi-layer might be able to boost the signal, similar to multi-layer proposals for the MADMAX haloscope~\cite{Millar:2016cjp}, although the higher frequencies considered here would lead to significant mechanical challenges if any tuning was required. Note that we set $v_z = 0$ in our calculations, cf.\ Section~\ref{sec:AQ_calculations} for an explanation.

\subsubsection{Solution of the one-dimensional model}\label{sec:AQ_DA_P_mixing_SolOneDimModel}

\paragraph{Lossless case ($\boldsymbol{\Gamma=0}$).}
We first focus on the case without losses. The dispersion relation implied by Eq.~\eqref{eq:dm_exp:eom} is:
\begin{eqnarray}
	\left(k^2-\frac{k_a^2+\kAQ^2}{2}\right)^2&=&b_a^2k_p^2+\left(\frac{k_a^2-\kAQ^2}{2}\right)^2 \, , \label{eq:Mixing_AQ_DA_P_dispersion_withoutLoss}
\end{eqnarray}
where $b_a^2 = \frac{\gDA^2 B_e^2}{\epsilon}$ was defined in analogy to $b^2 = \frac{\alphaEM}{\pi}\frac{\Lambda B_e^2}{\epsilon}$. The dispersion relations, up to leading order in the DA-photon coupling, therefore are 
\begin{align}
k_+ &= \kAQ + \mathcal{O}(\gDA^2) \, ,\\
k_- &= k_a + \mathcal{O}(\gDA^2) \, .
\end{align}
The most general ansatz for the fields is:
\begin{align}
E &= \hat{E}^{++}e^{i\kAQ z}+\hat{E}^{+-}e^{-ik_+ z}+\hat{E}^{-+}e^{ik_- z}+\hat{E}^{--}e^{-ik_- z} \, , \nonumber\\
\delta\Theta &= \delta\hat{\Theta}^{++}e^{ik_+ z}+\delta\hat{\Theta}^{+-}e^{-ik_+ z}+\delta\hat{\Theta}^{-+}e^{ik_- z}+\delta\hat{\Theta}^{--}e^{-ik_- z} \, ,\nonumber\\
a &= s\hat{a}^{++}e^{ik_+ z}+\hat{a}^{+-}e^{-ik_+ z}+\hat{a}^{-+}e^{ik_- z}+\hat{a}^{--}e^{-ik_- z} \, . \label{eq:Ansatz_v_G0}
\end{align}
In the following, we focus on the DA zero-velocity limit, i.e.\ $k_a = 0$. This is an appropriate approximation for dark matter and the most general ansatz in Eq.~\eqref{eq:Ansatz_v_G0} reduces to:
\begin{align}
E &= \hat{E}^{++}e^{i\kAQ z}+\hat{E}^{+-}e^{-ik_+ z}+\hat{E}^{-},\nonumber\\
\delta\Theta &= \delta\hat{\Theta}^{++}e^{ik_+ z}+\delta\hat{\Theta}^{+-}e^{-ik_+ z}+\delta\hat{\Theta}^{-} \, ,\nonumber\\
a &= \hat{a}^{++}e^{ik_+ z}+\hat{a}^{+-}e^{-ik_+ z}+\hat{a}^{-} \, ,\label{eq:Ansatz_v0_G0}
\end{align}
where we omit the $y$~index of the $E$-field since we ignore the $E_x$-component. The case of finite axion velocity was explored in Ref.~\cite{SchtteEngel:450146}.

Plugging Eq.~\eqref{eq:Ansatz_v0_G0} into the equations of motion, Eqs.~\eqref{eq:eqOfMotion_ThreefoldMixing_E}--\eqref{eq:eqOfMotion_ThreefoldMixing_a}, we obtain relations between the constants in the general ansatz. After plugging these relations back into the ansatz~\eqref{eq:Ansatz_v0_G0}, we obtain:
\begin{eqnarray}
\begin{pmatrix}
E\\\delta\Theta\\a
\end{pmatrix}&=&
\hat{E}^{++}\begin{pmatrix}
1\\\Theta^+_E\\a^+_E
\end{pmatrix}e^{ik_+z}+
\hat{E}^{+-}\begin{pmatrix}
1\\\Theta^+_E\\a^+_E
\end{pmatrix}e^{-ik_+z}+
\hat{a}^{-}\begin{pmatrix}
E_a^-\\\Theta^-_a\\1
\end{pmatrix} \, ,
\label{eq:Ansatz_v0_G0_rel}
\end{eqnarray}
where the following variables were defined:
\begin{eqnarray}
\Theta_E&=&\frac{\Lambda B_e}{\mAQ^2-\omega^2} \, , \quad a_E=\frac{g_{a\gamma} B_e}{k^2} \, ,\label{eq:DA_AQ_P_mixing_ET_Ea_Def}\\
E_a&=&\frac{\omega^2\mu g_{a\gamma}B_e}{k^2-\kAQ^2}\, , \quad  \Theta_a=\Theta_E E_a \, .\label{eq:DA_AQ_P_mixing_aT_aE_Def}
\end{eqnarray}
From Eq.~\eqref{eq:Ansatz_v0_G0_rel} it becomes clear that the dynamical AQ is completely determined by fixing the variables $\hat{E}^{++}$, $\hat{E}^{+-}$, and $\hat{a}^{-}$. In the next section we show that these variables can be fully determined by using the interface conditions for the electromagnetic and DA fields. Therefore no boundary conditions for the AQ need to be applied.\footnote{If we were to consider a finite spin~wave velocity, we would obtain three modes $k_{1,2,3}^2$. In this case, the most general ansatz would have six unknowns per field, and we would have to specify boundary conditions for the AQ.}

\paragraph{The case with losses ($\mat{\Gamma}\bm{\neq 0}$).}
When losses are included, the full dispersion relation $k_{\pm}^2$ takes on a more complicated form. However, in the limit $\gDA \rightarrow 0$ we find that $k_-^2 \rightarrow k_a^2$ and $k_+^2 \rightarrow \kAQ^2$, where $\kAQ^2$ is given by Eq.~\eqref{eq:dispersion_relation_vz0_losses}. In what follows, $a_E$ and $E_a$ are needed also in the case of losses. $a_E$ in Eq.~\eqref{eq:DA_AQ_P_mixing_ET_Ea_Def} does not get modified in the case of losses, and $E_a$ has the same form as in Eq.~\eqref{eq:DA_AQ_P_mixing_aT_aE_Def}. However, we now require the full form of $\kAQ$ from Eq.~\eqref{eq:dispersion_relation_vz0_losses}.

\subsubsection{Matrix formalism}\label{sec:AQ_DA_P_mixing__matrixFormalism}
In the previous section, we described the solution of the linearized equations in a homogeneous medium. Here, we discuss solutions for the fields in a multilayer system that consists of $N+1$ media, cf.\ Fig.~\ref{fig:matrix_formalism}. We use the same labels for the media as in Section~\ref{sec:AnalogueAxionPhoton_MatrixFormalism}. There are $N$ interfaces, which we label by $r=0,\dots, N$. The first interface is at $z_0=z_1$ and the last interface is at $z_N$. The material properties in Eq.~\eqref{eq:Ansatz_v0_G0_rel} of each medium are labeled with the corresponding index~$r$ as a subscript. The constant $\stAQ$ does not influence the emitted photon signal at lowest order and is therefore neglected in the following. We further introduce a phase similar to the case of AQ-photon mixing is introduced in the ansatz, cf.\ Eq.~\eqref{eq:AQ_P_mixing_MatrixFormalism_Ansatz}. The external $B$-field $B_e$ is the same in all media and is polarized in the $y$-direction. Recall that we consider the DA zero-velocity limit with a zero spin wave velocity.

Applying the interface conditions for the electromagnetic fields, cf.\ Eq.~\eqref{eq:DA_AQ_P_mixing_LinearizedInterface_H_1},~\eqref{eq:DA_AQ_P_mixing_LinearizedInterface_E}, and~\eqref{eq:DA_AQ_P_mixing_LinearizedInterface_E} and for the DA, cf.\ Eq.~\eqref{eq_DA_AP_P_mixing_Da}, at $z_r$ between medium $r-1$ and $r$ we obtain the following system of equations:
\begin{eqnarray}
\vc{t}_r=\mat{M}_r^{-1} \, \mat{M}_{r-1} \, \mat{P}_{r-1} \,  \vc{t}_{r-1} \, ,
\end{eqnarray}
with 
\begin{eqnarray}
\mat{M}_{r}=\begin{pmatrix}
1&1&E_{a,r}^-\\
\frac{k^{r}_+}{\mu_r}&-\frac{k^{r}_+}{\mu_r}&0\\
a_{E,r}^+&a_{E,r}^+&1
\end{pmatrix}\, , \quad  \vc{t}_{r}=\begin{pmatrix}
\hat{E}_{r}^{++}\\\hat{E}_{r}^{+-}\\\hat{a}_{r}^{-}
\end{pmatrix}
\end{eqnarray}
and, defining $\Delta_{r}^+ \equiv k^{r}_+(z_{r+1}-z_{r})$ \, ,
\begin{equation}
	\mat{P}_{r} = \mathrm{diag}(\eul^{i\Delta_{r}^+}, \, \eul^{-i\Delta_{r}^+}, \, 1) \, .
\end{equation}

In complete analogy to Section~\ref{sec:AnalogueAxionPhoton_MatrixFormalism} the $S$-matrix, which relates the states in media $0$ and $N$ to each other, is defined via
\begin{eqnarray}
\vc{t}_N=\mat{S}\,\vc{t}_0 \, .
\end{eqnarray}

The expressions for one, two, \ldots, $N$ interfaces are the same as in Eqs.~\eqref{eq:AQ_P_mixing_matrix_S_single_interface}--\eqref{eq:AQ_P_mixing_matrix_S_N_interface}. The unknown fields can be calculated from the $S$-matrix as follows:
\begin{eqnarray}
\begin{pmatrix}
\hat{E}^{++}_{N}\\
\hat{E}^{+-}_{0}\\
\hat{a}^{-}_N
\end{pmatrix}
=
-\hat{a}^{-}_0\begin{pmatrix}
-1&S_{12}&0\\
0&S_{22}&0\\
0&S_{32}&-1
\end{pmatrix}^{-1}\cdot\begin{pmatrix}
S_{13}\\
S_{23}\\
S_{33}
\end{pmatrix} \, ,
\end{eqnarray}
where the amplitude of the DAs is known and, has to lowest order the same magnitude in each medium $|\hat{a}^-_0| = |\hat{a}^-_r|$ for all $r = 1,\ldots, N$.
The emitted $E$-field in medium~$N$ that propagates in the positive $z$-direction is called $\hat{E}^{++}_{N}$. The emitted $E$-field that propagates in the negative $z$-direction is called $\hat{E}^{+-}_{0}$. 

\subsubsection{Layer of topological insulator}\label{sec:one_layer_tmi}
Let us now consider the case of a single TMI layer~(hosting a dynamical AQ) surrounded by vacuum. Dark axions are present in the form of an background field that oscillates in time with a frequency that is determined by the DA mass, $m_a$. The DAs mix with the AQs and photons. In terms of the matrix formalism, there are two interfaces~($N=2$), with media~$0$ and~$2$ are vacuum, and medium~$1$ is a TMI of thickness $d$. The TMI has constant refractive index $n^2=\epsilon$,\footnote{A nontrivial permeability, $\mu \neq 1$, can be incorporated straight-forwardly into the matrix formalism, described in the previous sections. However, we set $\mu = 1$ for simplicity and because this is a good approximation for the TMI materials discussed in Section~\ref{sec:params}.} and losses $\mat{\Gamma}$. The external $B$-field is present in all media. The DAs have the same magnitude in each medium, which is determined by the axion dark matter density $\rho_a$: $|\hat{a}^-_0|^2 = |\hat{a}^-_r|^2=2\rho_a/m_a^2$.

The three-way mixing between DAs, AQs, and photons produces a photon at the boundary, which propagates away from the TMI layer. Note that, since we neglect the spin wave velocity, the system behaves essentially as a two-level system of massive photons and DAs. The emitted $E$-fields in media~$0$ and~$2$ are denoted by $\hat{E}_{0}^{+-}$ and $\hat{E}_{2}^{++}$, respectively. Recall that $\hat{E}_{0}^{+-}$ is the $E$-field amplitude that is emitted in negative $z$-direction in medium~$0$ and $\hat{E}_{2}^{++}$ is the emitted photon signal emitted in the positive $z$-direction in medium~$2$. We assume that the DA particles are effectively at rest. In this limit there is no preferred direction and the magnitudes of $\hat{E}_{2}^{++}$ and $\hat{E}_{0}^{+-}$ are the same. 

\paragraph{Lossless case ($\boldsymbol{\Gamma = 0}$).}
The full formula for $\hat{E}_{2}^{++}$ from the matrix formalism is impractical. We therefore quote the result first order in the DA-photon coupling, which, assuming that $\gDA$~is sufficiently small, should be a good approximation: 
\begin{equation}
\hat{E}^{++}_2 = \hat{a}^{-}_0 \, \frac{ \sin(\Delta/2) \left(\nAQ^2-1\right)}{\nAQ \left(\nAQ  \sin(\Delta/2)+i \cos(\Delta/2)\right)} \, \gDA \, B_e + \mathcal{O}\left((\gDA \, B_e)^2\right) \, ,
\end{equation}
where we define the phase depth $\Delta = d \kAQ=d\omega \nAQ$ (where $\kAQ$ is the lossless solution to the dispersion relation, eq.~\eqref{eq:dispersion_k_vz0}) and the effective refractive index is
\begin{equation}\boxed{
\nAQ^2 = n^2\left(1-\frac{b^2}{\omega^2-\mAQ^2}\right) \, .
}
\end{equation}
Furthermore, we used in the language of the matrix formalism, such that $a_{E,0}^+ = \frac{\gDA \, B_e}{\omega^2} = a_{E,2}^+$, $E_{a,0}^-=-\gDA \, B_e=E_{a,2}$, and $a_{E,1}^+ = \frac{\gDA \, B_e}{\nAQ^2\omega^2}$, $E_{a,1}^- = -\frac{\gDA\,B_e}{\nAQ^2}$. From now on, terms of the order $\mathcal{O}\left((\gDA\,B_e)^2\right)$ are omitted to simplify the expressions.
We also normalize the field amplitude $\hat{E}_2^{++}$ to the DA-induced field in vacuum, $E_0 = \gDA \, B_e \, a^{-}_0$,
\begin{equation}
\frac{\hat{E}^{++}_2}{E_0}=-\frac{ \sin(\Delta/2) \left(1-\nAQ^2\right)}{\nAQ \left(\nAQ  \sin(\Delta/2)+i \cos(\Delta/2)\right)} \, .
\label{eq:DA_AQ_P_mixing_LayerBoost}
\end{equation}
Note that Eq.~\eqref{eq:DA_AQ_P_mixing_LayerBoost} has the same form as in the case of fields emitted from as a dielectric disk~\cite{Millar:2016cjp}, with the  effective refractive index $\nAQ$, which is equivalent to introducing a photon mass. 

From analysing Eq.~\eqref{eq:DA_AQ_P_mixing_LayerBoost}, it becomes clear that a resonance occurs if the condition
\begin{eqnarray}
\Delta = \Delta_j=\nAQ(\omega_j) \, \omega_j d= (2j+1) \, \pi \, , \quad j\in\mathbb{N}_0 \, ,
\label{eq:resonance_condition}
\end{eqnarray}
is fulfilled. Here, $\omega_j$ are the resonance frequencies, which are located at
\begin{eqnarray}
	\omega_j^2=\frac{\omLO^2}{2}+\sqrt{\frac{\omLO^4}{4}+\Delta_j^2\frac{b^2}{n^2d^2}}=\omLO^2+\delta\omega_j^2+\mathcal{O}\left(\frac{4\Delta_j^2b^2}{n^2d^2\omLO^4}\right) \, ,
\label{eq_AQ_DA_P_mixing_res_frequencies}
\end{eqnarray}
where we have defined
\begin{eqnarray}
\delta\omega_j^2\equiv\frac{\Delta_j^2b^2}{n^2d^2\omLO^2} \, .
\end{eqnarray}
From Eq.~\eqref{eq_AQ_DA_P_mixing_res_frequencies}, it is evident that -- in the lossless limit -- the resonance frequencies are always larger than $\omLO$, i.e.\ $\omLO < \omega_0 < \omega_1 < \ldots$. As the thickness of the TMI increases, $d\rightarrow\infty$, the resonant frequencies converge to the limiting value, i.e.\ $\omega_j \rightarrow \omLO$. Furthermore, Eq.~\eqref{eq_AQ_DA_P_mixing_res_frequencies} implies that the resonance frequency $\omega_j$ can be \emph{tuned} via the external $B$-field, since $b\propto B$. In Section~\ref{sec:ScanningRange} we investigate the frequency and, equivalently, DA mass range that can be scanned with our benchmark materials and realistic external $B$-fields.

To understand why the frequencies defined via Eq.~\eqref{eq:resonance_condition} are indeed resonance frequencies, consider the following: Equation~\eqref{eq:resonance_condition} implies that $\cos\Delta_j = 0$ and, hence, the emitted field in Eq.~\eqref{eq:DA_AQ_P_mixing_LayerBoost} is $\hat{E}_2^{++}\sim 1/\nAQ^2 -1 \sim 1/\nAQ^2$ for small~$\nAQ$. In fact, the smaller~$\nAQ$ the more pronounced the resonance and, from Fig.~\ref{fig:Disp_relat_vz_zero}, we can see that this is the case when $\omega_j \sim \omLO$. Consequently, resonances that are further away from~$\omLO$~(i.e.\ $j>0$) have less pronounced peak values, such that the maximal value of $\hat{E}_2^{++}$ is always obtained for~$j=0$.
Furthermore, Eq.~\eqref{eq_AQ_DA_P_mixing_res_frequencies} reveals that resonances are more pronounced for larger sample thickness~$d$ of a given TMI.
We now investigate the resonances in more detail and provide analytical expressions for their widths and maximum values. 

Around the resonances we have $|\nAQ|\ll 1$, and Eq.~\eqref{eq:DA_AQ_P_mixing_LayerBoost} can be approximated as:
\begin{eqnarray}
	\frac{\hat{E}^{++}_2}{E_0} = -\frac{1}{\nAQ^2 + i \, \nAQ\cot\left(\frac{\Delta}{2}\right)} \, .
\label{eq:DA_AQ_P_mixing_LayerBoost_res}
\end{eqnarray}
Expanding $\nAQ^2$ around $\omega_j^2$ yields, to lowest order,
\begin{eqnarray}
	\nAQ^2 = n^2\frac{\delta\omega_j^2}{b^2} \, ,
\end{eqnarray}
where we require that $b^2 > \delta\omega_j^2$. The expansion of $\cot(\Delta/2)$ leads, to lowest order, to
\begin{eqnarray}
\cot\left(\frac{\Delta}{2}\right)=-\frac{1}{2}\frac{\omega_j d n^2}{2\nAQ(\omega_j)b^2}\left(\omega^2-\omega_j^2\right) \, .
\end{eqnarray}
The emitted fields in Eq.~\eqref{eq:DA_AQ_P_mixing_LayerBoost_res} can then be approximated about the resonances as follows:
\begin{eqnarray}
	\frac{\hat{E}^{++}_2}{E_0}=-\frac{iA_j}{i\gamma_j\omega_j+(\omega^2-\omega_j^2)} \, ,
\label{eq:DA_AQ_P_mixing_LayerBoost_res_simp}
\end{eqnarray}
with
\begin{eqnarray}
\gamma_j&=&\frac{4}{d}\frac{\omega_j^2-\omLO^2}{\omega_j^2}\approx\frac{4\Delta_j^2b^2}{n^2d^3\omLO^4} \, ,\label{eq:TwoInterfaces_FWHM} \\
A_j&=&\frac{4b^2}{n^2\omega_jd}\approx \frac{4b^2}{n^2\omLO d} \, ,
\end{eqnarray}
where we used that in a resonant case $\omega_j$ is close to $\omLO$. 

The power output on resonance is:
\begin{eqnarray}
P = \frac{|E_0|^2}{2}\beta^2 A  \, ,
\label{eqn:boost_power}
\end{eqnarray}
where $A$ is the surface area of the TMI layer and where we used that the Poynting vector in $z$-direction has magnitude $\frac{1}{2}|\hat{E}^{++}_2|^2$. The power boost factor~$\beta^2$ is defined as
\begin{eqnarray}
\beta^2=\left|\frac{\hat{E}_2^{++}}{E_0}\right|^2 \, . \label{eq:beta_definition}
\end{eqnarray}
Following Ref.~\cite{Millar:2016cjp}, we refer to the unsquared~$\beta$ as the boost factor. 

The full width at half the maximum value~(FWHM) of $\beta^2$ about the resonance $\omega_j$ is given by $\gamma_j$. The highest value at the resonance, the peak amplitude, $\omega_j$ is given by
\begin{eqnarray}
	\beta^2(\omega_j)=\frac{A_j^2}{\gamma_j^2\omega_j^2}\approx\left(\frac{d\omLO}{\Delta_j}\right)^4\approx\frac{1}{\nAQ(\omega_j)^4}.
\label{eq:TwoInterfaces_Maximum_at_resonance}
\end{eqnarray}
\begin{figure}
\centering 
    \includegraphics[width=0.49\textwidth]{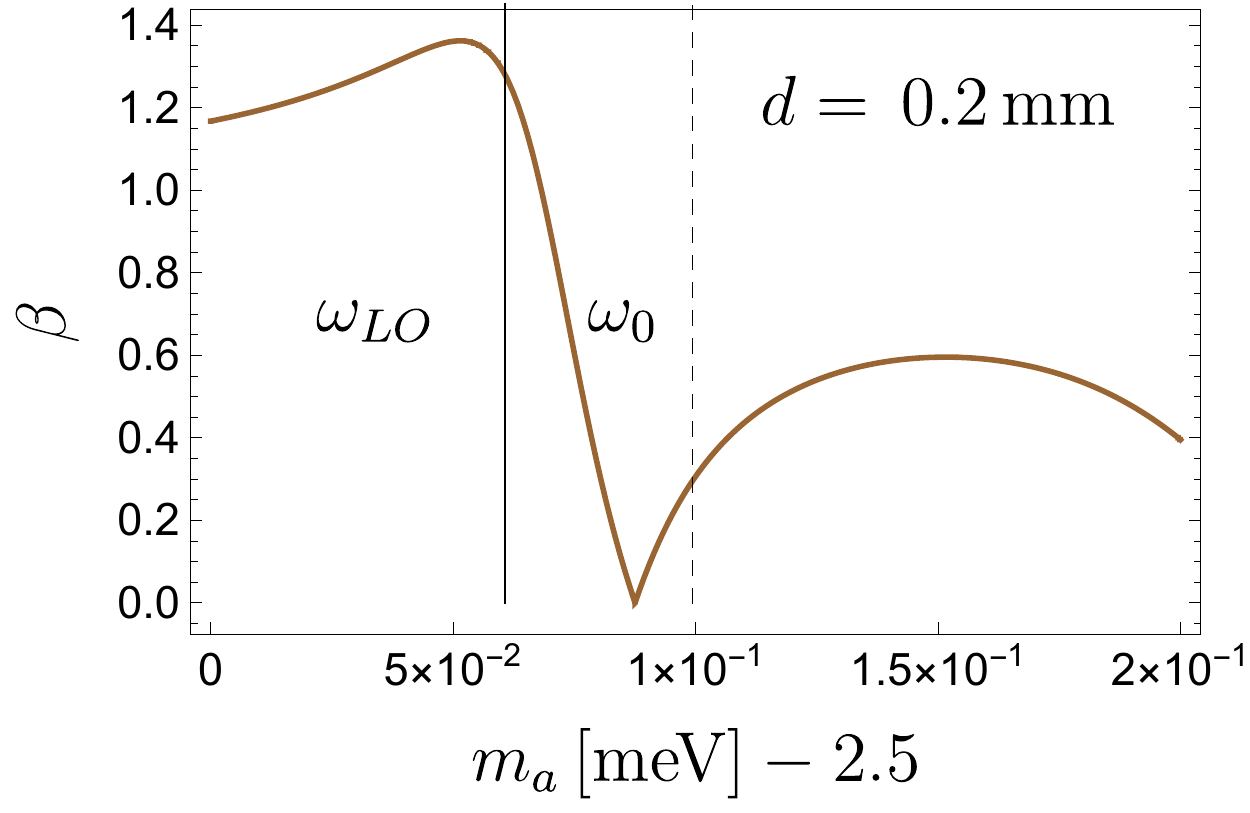}
    \includegraphics[width=0.49\textwidth]{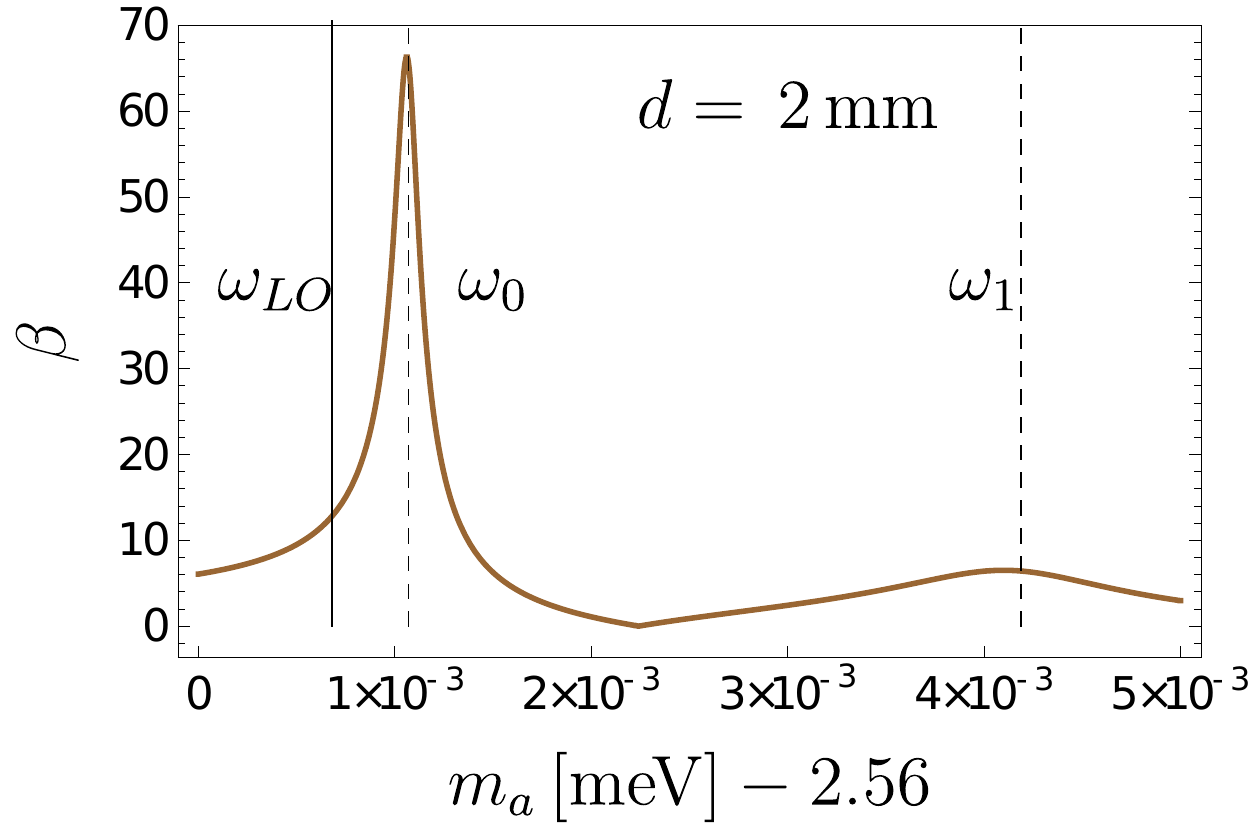}\\
    \includegraphics[width=0.49\textwidth]{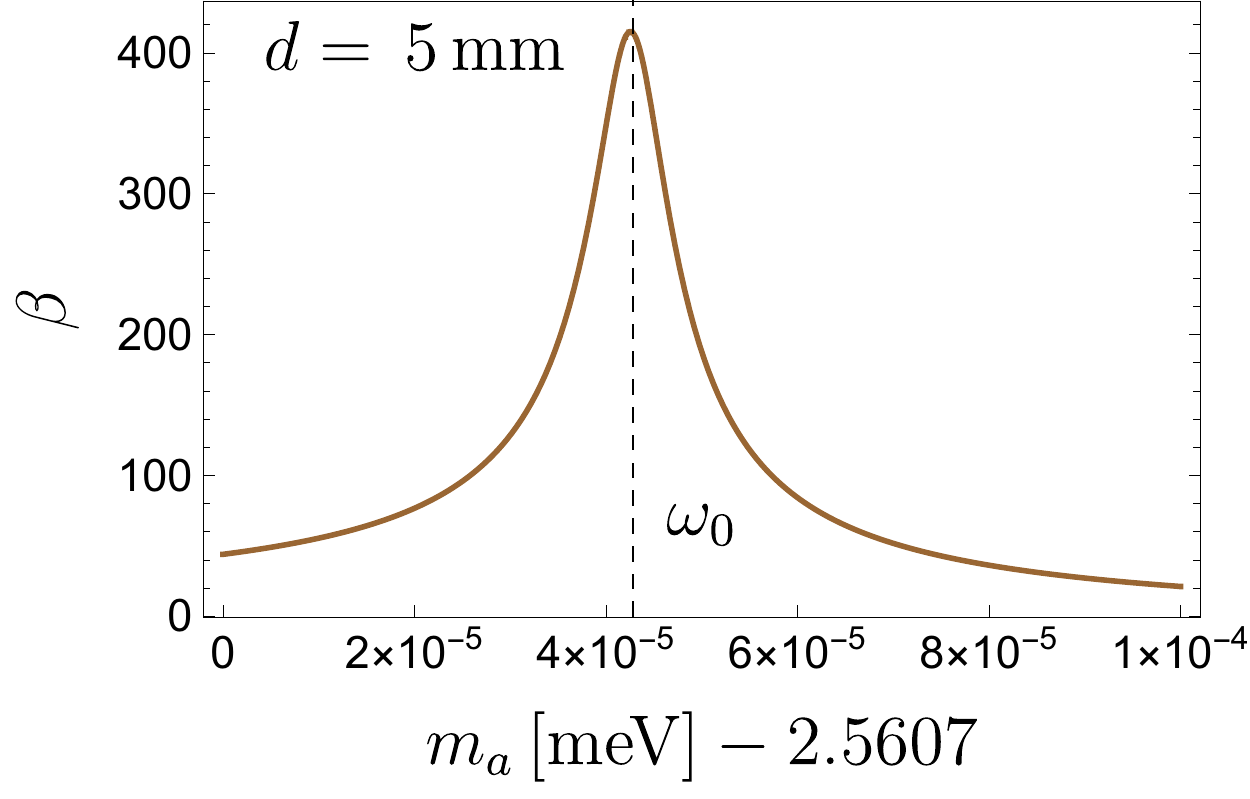}
    \includegraphics[width=0.49\textwidth]{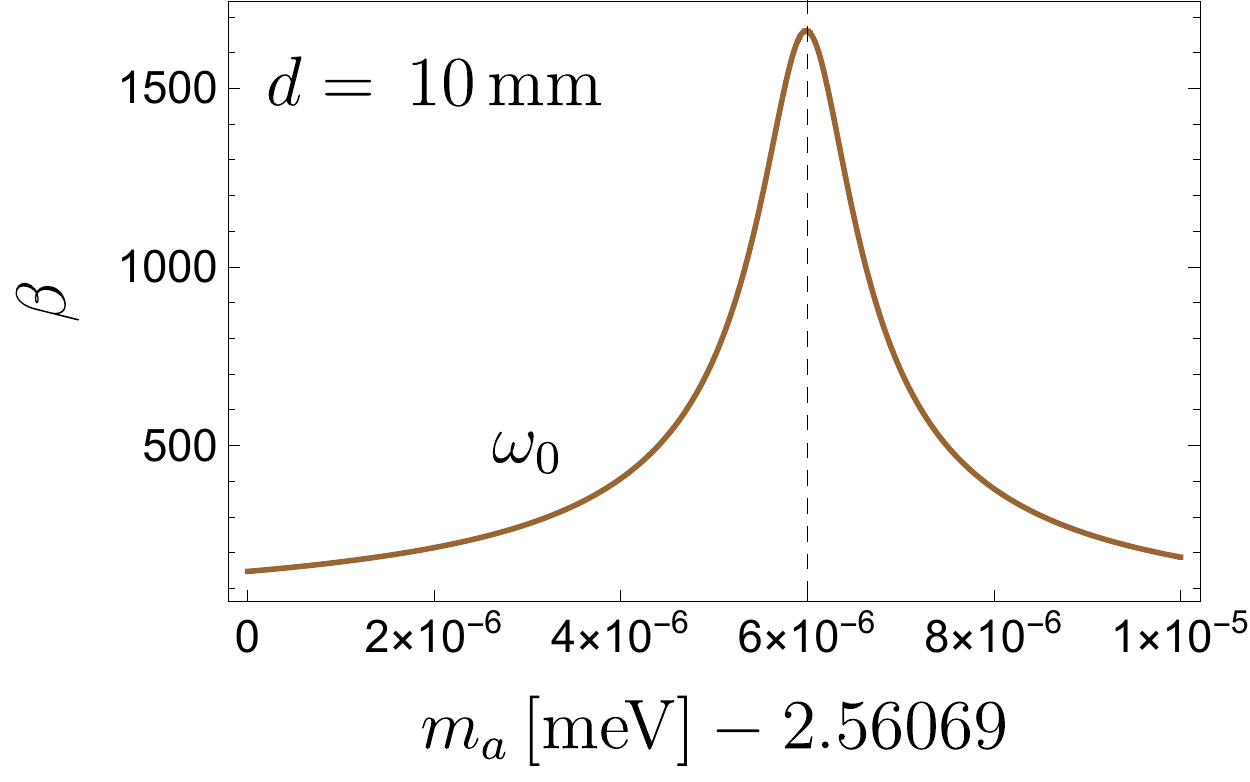}\\ 
 	\caption{\textbf{Effect of material thickness, $d$, on the boost factor.} In the lossless limit for one layer $\beta$ is given by Eq.~\eqref{eq:DA_AQ_P_mixing_LayerBoost}. We assume zero DA velocity ($\vdm=0$, valid when the resonance is wide compared to the DA linewidth). Typical material values for \MnBiTe{2}{2}{5} TMI with an external $B$-field $B_e$ of $\SI{2}{\tesla}$ are chosen, cf.\ Table~\ref{tab:derived_params} and Eq.~\eqref{eqn:b_meV}. Vertical lines mark the frequencies $\omLO$ and the resonance frequencies $\omega_j$. The resonance boost increases, and bandwidth decreases, as the thickness $d$ increases. }
	\label{fig:DA_AQ_P_mixing_Signal_validation}
\end{figure}

With Eq.~\eqref{eq:TwoInterfaces_Maximum_at_resonance} it now becomes explicitly clear that the higher modes have a lower maximum resonance value, since $\Delta_{j}<\Delta_{j+1}$. Also large layer thicknesses $d$ increases the maximal emitted $E$-field on resonance. Therefore to achieve a certain amount of signal boost from one layer of TMI a relatively large layer thickness $d>1/\omega$ is needed. Equation~\eqref{eq:TwoInterfaces_FWHM} tells us the necessary information about the width of the resonance. First note, that going to larger modes $j$ or larger $b$ will increase the FWHM $\gamma_j$. Relatively thick TMI layers, i.e.\ large $d$, yield a very narrow resonance. Therefore a good balance for $d$ has to be found because $d$ should be relatively large to reach a high resonance value. The refractive index $n$ does not affect the maximum value of the resonance, cf.\ Eq.~\eqref{eq:TwoInterfaces_Maximum_at_resonance}. However large $n$ makes the FWHM $\gamma_j$ very small. We therefore conclude that it is advantageous to have low $n$ materials fro broadband response.

In Fig.~\ref{fig:DA_AQ_P_mixing_Signal_validation} the boost amplitude $\beta$ is shown for our benchmark material \MnBiTe{2}{2}{5} for four different layer thicknesses. For the thinnest case, $d=\SI{0.2}{\milli\metre}$, no clear enhancement of the boost factor is reached, since for this relatively small thickness we obtain a resonance frequency $\omega_0$ (dashed vertical line) that is too far away from $\omLO$ (solid vertical line) such that $\nAQ(\omega_0)$ is not much smaller than unity. For $d=\SI{2}{\milli\metre}$ we have a resonance at $\omega_0$. A larger width, but a lower maximum value, is realized at the second resonance peak $\omega_1$. According to Eq.~\eqref{eq:TwoInterfaces_FWHM}, the width of the resonance around $\omega_1$ should be broader by a factor of about $(\Delta_1/\Delta_0)^2 = 9$ than the resonance width around $\omega_0$. 

For thicker samples still, $d = \SI{5}{\milli\metre}$, $d=\SI{10}{\milli\metre}$ the resonant boost at $\omega_0$ increases further. In particular from Eq.~\eqref{eq:TwoInterfaces_Maximum_at_resonance} we find that the peak heights scale as $(d_1/d_2)^2$. Furthermore, the width of the peak around $\omega_0$ shrinks as $d$ increases. From Eq.~\eqref{eq:TwoInterfaces_FWHM} we can directly read off that the width shrinks by a factor $(d_2/d_1)^3$. 
For $d=\SI{10}{\milli\metre}$ the linewidth of $\beta$ is on the order of the DA linewidth, $10^{-6}m_a$. Without taking into account material losses the linewidth of the power boost factor is larger that the axion linewidth if
\be
\gamma_j>10^{-6}\omega_j\approx 10^{-6}\omLO.
\label{eq:Grammaj_eq_AqxionLinewidth}
\ee
Equation~\eqref{eq:Grammaj_eq_AqxionLinewidth} tells us the requirements for the material parameters such that the power boost factor bandwidth is larger than the axion linewidth. 

\begin{figure}
\centering 
    \includegraphics[width=0.49\textwidth]{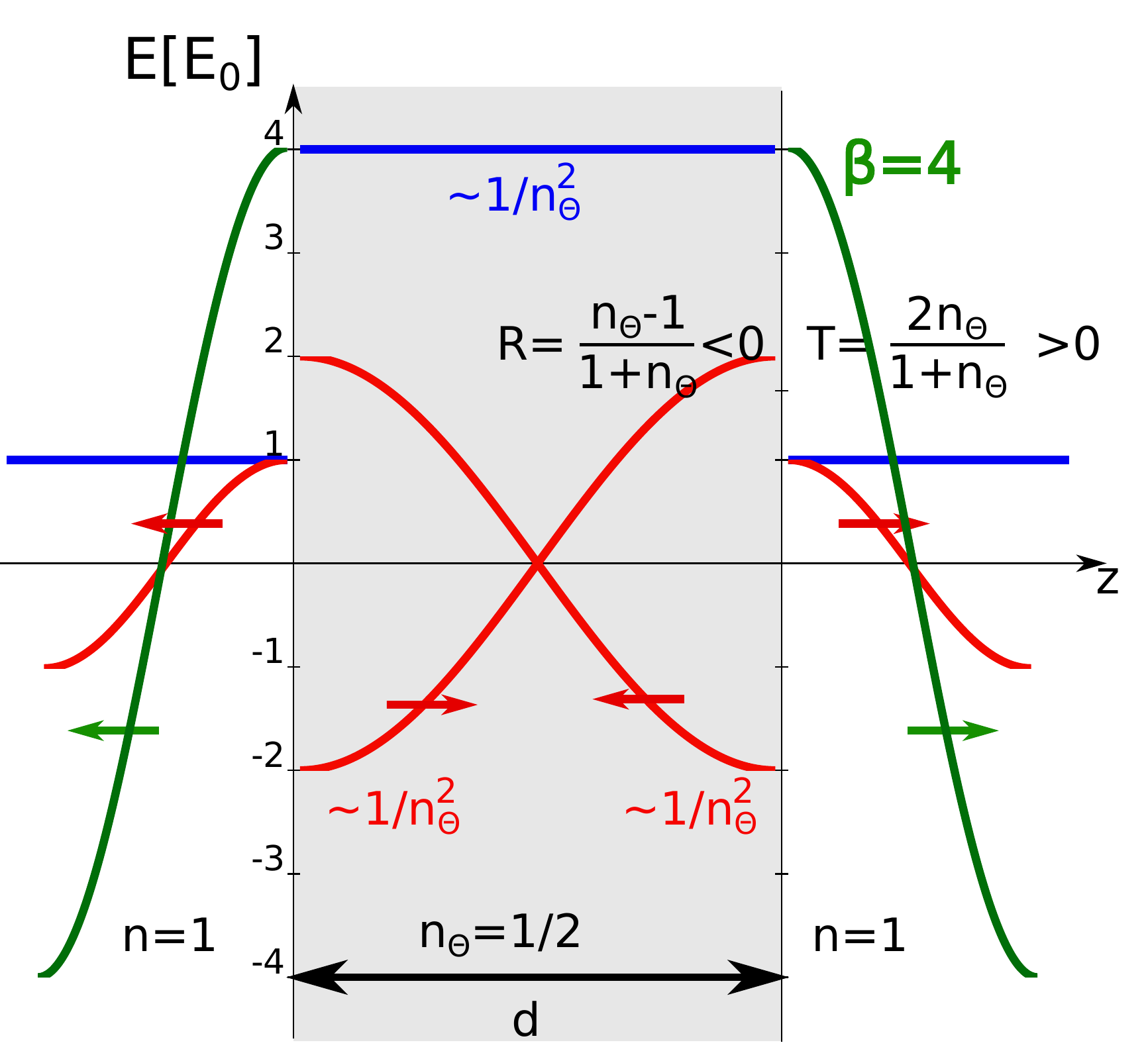}
 	\caption{Physical understanding of the resonant enhancement of the emitted electromagnetic fields from the mixing between DAs, AQs and photons. The TMI layer (gray) of thickness $d$ has an effective refractive index $\nAQ=\frac{1}{2}<1$ and is surrounded by vacuum. The axion induced field (blue) is enhances inside the TMI and since the total electric field has to be continuous over the two interfaces propagating modes (red) are emitted off both interfaces. The $E$-fields are given in units of the axion-induced field in vacuum, $E_0$. Due to the smallness of $\nAQ$ around the resonance the transmission coefficient for the fields which propagate inside the TMI is small, while the reflection coefficient is large. Therefore effectively the TMI works as a cavity. After bouncing many times between the two interfaces the effective emitted field is proportinal to $\beta\gg 1$. In the specific example of $\nAQ$ the emitted field is four times larger than the axion induced field in vacuum ($\beta=4$). A more detailed description can be found in the text.}
  	\label{fig:DA_AQ_P_interpretation}
\end{figure}

Before we discuss the influence of material losses we want to give a clearer physical picture of the observed resonances. In Fig.~\ref{fig:DA_AQ_P_interpretation} we consider three domains. The middle domain is a TMI layer with thickness $d$ and with an effective refractive index $n_\Theta=\frac{1}{2}$. The two outer domains are vacuum with $n=1$. The axion induced field, which is shown in blue, is one in vaccum and enhanced inside the TMI, cf.\ Eq.~\eqref{eq:DA_AQ_P_mixing_ET_Ea_Def}. The enhancement of the axion induced field in the TMI layer is proportional to $\frac{1}{n_\Theta^2}$. Consider now the interface between the TMI and the left vacuum. To fulfil the continuity requirement of the total electric field, propagating modes (red) are emitted to both sides. One can check that this is indeed the case by adding the red and blue amplitudes at the interface. The emitted amplitude in vacuum is one, while the propagating fields inside the TMI are enhanced, since they are proportional to $\frac{1}{n_\Theta^2}$.
The outlined scenario happens at both interfaces of the TMI. Let us first consider the emitted radiation that propagates inside the TMI from left to right. The radiation hits the right interface. The transmission and reflection coefficients determine the fraction of the radiation, which is transmitted to the outside or reflected. For plane waves we have $T=\frac{2 \nAQ}{1+\nAQ}$ and $R=\frac{n_\Theta-1}{1+n_\Theta}$.
The important point is that the transmitted radiation is added in phase ($T>0$) to the radiation which is emitted from the right interface to the outside. The part of radiation which is reflected at the right interface receives a phase shift since $R<0$. Therefore the reflected radiation is coherently added to the radiation which is emitted from the right surface to the left in the TMI. A similar scenario happens at the left interface.
Now since $n_\Theta\ll 1$ the transmission coefficient is small and the reflection coefficient is large. Therefore the radiation bounces many times between both interfaces. After each bounce a small fraction of the radiation is transmitted to the outside. This is exactly how a cavity works and due to the fact that the transmitted fields to the outsides are all added coherently the total emitted field is enhanced by the boost factor $\beta=1/\nAQ^2$. The total emitted field is shown in green in Fig.~\ref{fig:DA_AQ_P_mixing_Signal_validation}.
With this physical picture in mind we can also understand why larger thicknesses $d$ lead to larger $\beta$'s on resonance. To fulfill the resonance condition $\nAQ(\omega_j)\omega_j d=\pi$ we need a smaller $\nAQ(\omega_j)$ the larger we make $d$. However making $\nAQ(\omega_j)$ smaller leads to a larger axion induced field and therefore also to a larger total emitted field.

\paragraph{Case with losses ($\boldsymbol{\Gamma \neq 0}$).}
In the case of losses we can use that $g_{a\gamma}$ is a relatively small coupling and therefore $k_+\rightarrow \kAQ$, where $\kAQ$ now includes losses, cf.\ Eq.~\eqref{eq:dispersion_relation_vz0_losses}. $k_-\rightarrow k_a=0$ in the axion zero velocity limit. Furthermore we have shown in Section~\ref{sec:AQ_DA_P_mixing_SolOneDimModel} that the relations for $a_E$ and $E_a$ still hold when we include the losses into the effective refractive index. 
In conclusion we can use Eq.~\eqref{eq:DA_AQ_P_mixing_LayerBoost} also if losses are present. The only thing that we have to do is to use the effective refractive index which includes losses:
\begin{eqnarray}\boxed{
\nAQ^2=n^2\left(1+\frac{b^2}{m^2_\Theta-\omega^2-i\omega\Gamma_m}+i\frac{\Gamma_\rho}{\omega}\right).
}
\label{eq:AQ_DA_P_mixing_effective_n_losses}
\end{eqnarray}
We subsequently neglect mixed losses.

\begin{figure}
\centering 
 \includegraphics[width=0.49\textwidth]{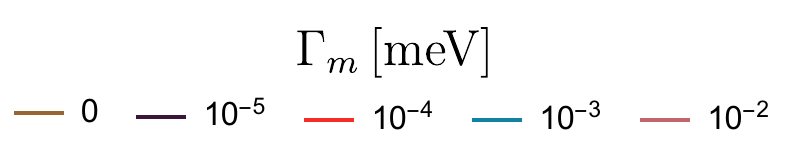}~~~~~~
 \includegraphics[width=0.49\textwidth]{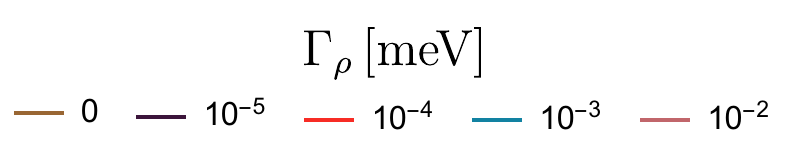}\\
    \includegraphics[width=0.49\textwidth]{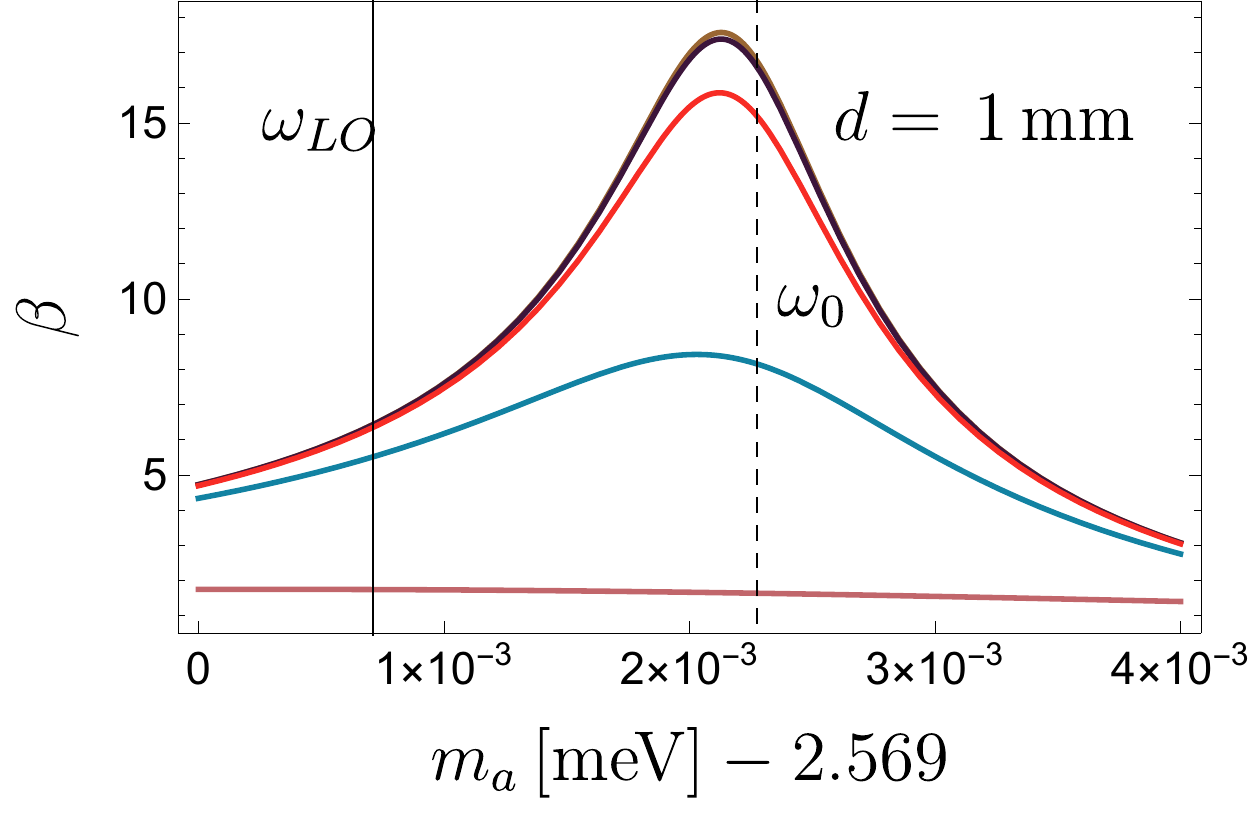}
    \includegraphics[width=0.49\textwidth]{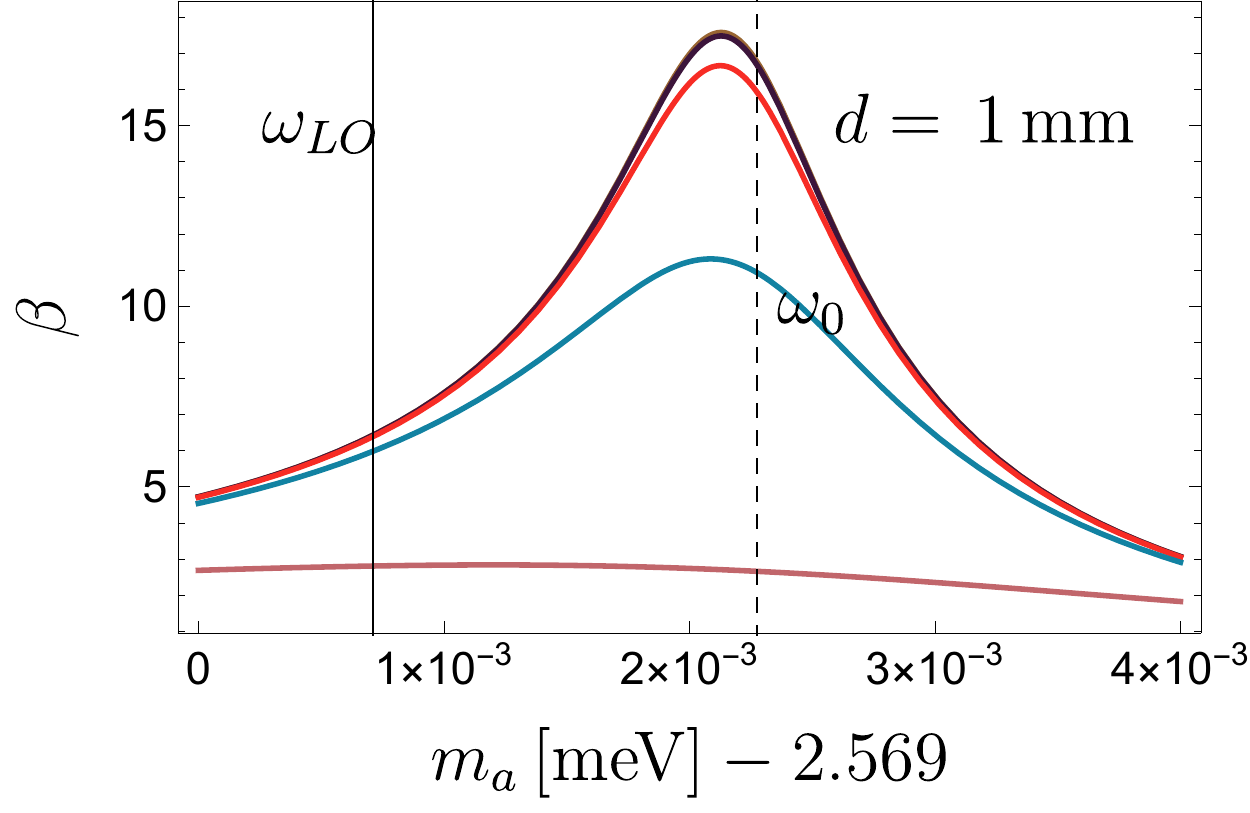}\\
  
 	\caption{\textbf{Effect of losses on the boost factor.} TMI layer of thickness $d=\SI{1}{\milli\metre}$. Other parameters are as Fig.~\ref{fig:DA_AQ_P_mixing_Signal_validation}. \emph{Left}: Varying magnon losses $\Gamma_m$, with $\Gamma_\rho=0$. \emph{Right}: Varying photon losses $\Gamma_\rho$ with $\Gamma_m=0$.}
  	\label{fig:DA_AQ_P_mixing_Signal_Losses_main}
\end{figure}

We begin by expanding the boost factor around the resonance frequencies $\omega_j$, which remain unmodified by the losses, cf.\ Eq.~\eqref{eq:resonance_condition}. Then an expansion can be done in complete analogy to the lossless case. For $|1-\nAQ^2(\omega_j)|\approx 1$ the emitted field takes the same form as in the lossless case, cf.\ Eq.~\eqref{eq:DA_AQ_P_mixing_LayerBoost_res}, where $\nAQ$ is given now by Eq.~\eqref{eq:AQ_DA_P_mixing_effective_n_losses}. We now expand the two terms, $\nAQ^2$ and $\cot\left(\frac{\Delta}{2}\right)$, that appear in the denominator in Eq.~\eqref{eq:DA_AQ_P_mixing_LayerBoost_res}. We find
\begin{eqnarray}
\nAQ^2(\omega_j)=n^2\left(\frac{\delta\omega_j^2}{b^2}+i\tilde{\Gamma}_j^2\right)
\label{eq:AQ_DA_P_mixing_losses_nAQ_expainsion}
\end{eqnarray}
with $\tilde{\Gamma}_j^2\equiv\frac{\Gamma_\rho}{\omega_j}+\frac{\omega_j\Gamma_m}{b^2}$. In deriving Eq.~\eqref{eq:AQ_DA_P_mixing_losses_nAQ_expainsion} we have assumed that $b^2>\delta\omega_j^2$ and $b^2>\omega_j\Gamma_m$. If such a condition is not fulfilled, the material is likely too lossy to be useful in DA detection. Next we consider the nearly lossless limit, i.e. $\frac{\delta\omega_j^2}{b^2}>\tilde{\Gamma}_j^2$. In this case the relevant expressions for us are $\nAQ(\omega_j)=n\frac{\delta\omega_j}{b}\left(1+i\frac{1}{2}\frac{\tilde{\Gamma}_j^2}{\delta\omega_j^2}b^2\right)$ and $\nAQ^2(\omega_j)=n^2\frac{\delta\omega_j^2}{b^2}$, where we have written down only the important leading order terms.
Next we expand the $\cot\left(\frac{\Delta}{2}\right)$ term:
\begin{eqnarray}
\cot\left(\frac{\Delta}{2}\right)=\cot\left(\frac{\Delta }{2}\right)_{\omega^2=\omega_j^2}+\left[\frac{\partial}{\partial \omega^2}\cot\left(\frac{\Delta}{2}\right)\right]_{\omega^2=\omega_j^2}\left(\omega^2-\omega_j^2\right)+\cdots
\label{eq:DA_AQ_P_mixing_losses_cotExpainsion}
\end{eqnarray}
where we approximate $\omega\approx\omega_j$ for the linear dependencies and the dots represent higher order terms, which we do not have to consider for a reasonable expansion. The expansion in Eq.~\eqref{eq:DA_AQ_P_mixing_losses_cotExpainsion} can be simplified in the small thickness limit, ${\rm Im}\left[\Delta(\omega_j)\right]<1$.
Equation~\eqref{eq:DA_AQ_P_mixing_losses_cotExpainsion} then simplifies to:
\begin{eqnarray}
\cot\left(\frac{\Delta}{2}\right)=-i\frac{{\rm Im}\left[\Delta(\omega_j)\right]}{2}-\frac{1}{4}\frac{\Delta_j}{\delta\omega_j^2}\left(\omega^2-\omega_j^2\right)
\end{eqnarray}
Putting everything together we obtain -- as in the lossless case, cf.\ Eq.~\eqref{eq:DA_AQ_P_mixing_LayerBoost_res_simp} -- a Lorentzian shaped functional dependence around the resonance frequencies. The width of the curve receives an additional term in the presence of losses:
\begin{equation}
	\gamma_j=\frac{4b^2\Delta_j^2}{n^2\omLO^4d^3}+\left(\Gamma_m+\frac{b^2}{\omLO^2}\Gamma_\rho\right)\,. \label{eq:gammaloss}
\end{equation}
and the amplitude $A_j$ remains unchanged with respect to the lossless case.

In Fig.~\ref{fig:DA_AQ_P_mixing_Signal_Losses_main} we show the boost factor $\beta$ around the first resonance $\omega_0$ for a layer thicknesses of $d=\SI{1}{\milli\metre}$ for different values of the loss parameters $\Gamma_\rho$ and $\Gamma_m$.
We observe that each loss parameter has a similar quantitative effect on redicing the boost factor peak, with magnon losses being only slightly more important. This is due to the fact that $\Gamma_m$ directly enters the resonance, cf.\ Eq.~\eqref{eq:AQ_DA_P_mixing_effective_n_losses}, while photon losses $\Gamma_\rho$ only enter via an additional term that is added to the other terms of the dispersion relation.

\begin{figure}
\centering 
 \includegraphics[width=0.49\textwidth]{{figures/DA_AP_P_mixing/LineLegendM}.pdf}~~~~~~
 \includegraphics[width=0.49\textwidth]{{figures/DA_AP_P_mixing/LineLegendR}.pdf}\\
    \includegraphics[width=0.49\textwidth]{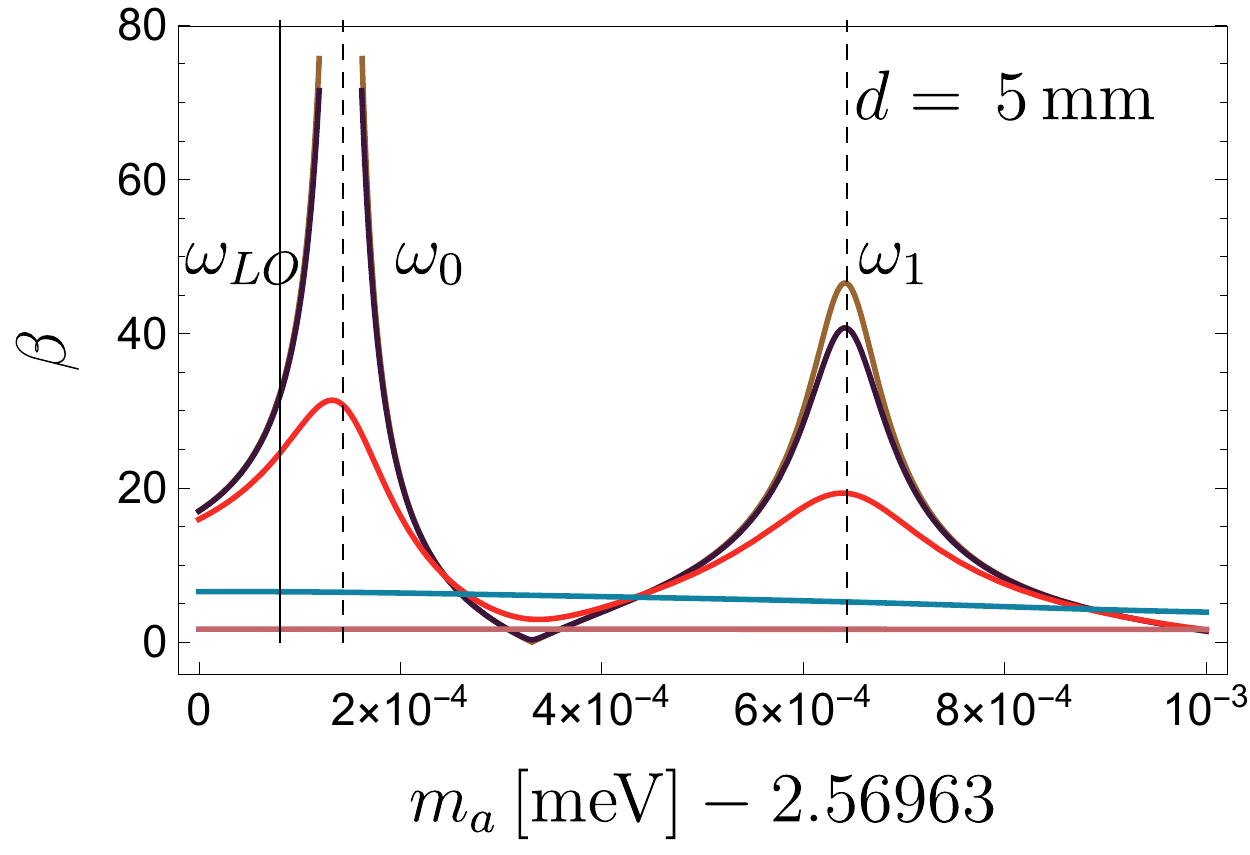}
    \includegraphics[width=0.49\textwidth]{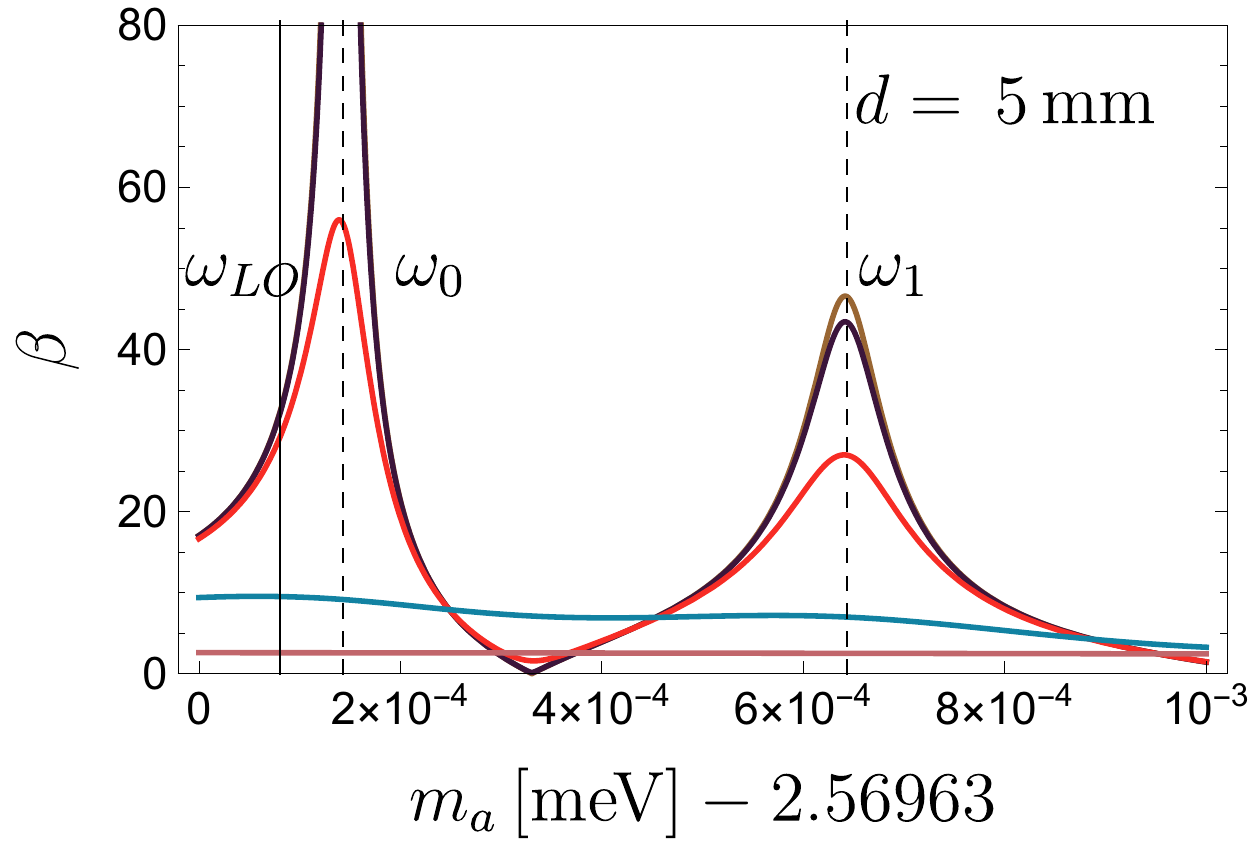}
  
 	\caption{\textbf{Effect of losses on the higher resonance peaks.} TMI layer of thickness $d=\SI{5}{\milli\metre}$. Other parameters are as Fig.~\ref{fig:DA_AQ_P_mixing_Signal_Losses_main}. The value of $\beta$ at $\omega_0$ without losses is around $\beta\approx 500$, cf.\ Fig.~\ref{fig:DA_AQ_P_mixing_Signal_validation} (bottom left). Therefore the relative reduction due to losses at the resonant frequencies is larger at $\omega_0$ than at $\omega_1$.}
  	\label{fig:DA_AQ_P_mixing_Signal_Losses_higher_res}
\end{figure}

Next let us discuss the effect of losses on the higher resonance frequencies. In Fig.~\ref{fig:DA_AQ_P_mixing_Signal_Losses_higher_res} we show the first two resonance peaks at $\omega_0$ and $\omega_1$ for $d=\SI{5}{\milli\metre}$.
In the lossless case we have $\beta(\omega_0)\approx 500$, cf.\ Fig.~\ref{fig:DA_AQ_P_mixing_Signal_validation}. The reduction of the $\omega_0$ resonance is therefore more severe than the resonance at $\omega_1$. This is simply the case because the system is more resonant at $\omega_0$ and losses lead to a larger reduction. We conclude that losses may lead to a scanning strategy in the end that uses a higher resonance mode $j>0$. However, the final scanning strategy can only be given when the losses are determined experimentally.

We can see from Eq.~\eqref{eq:gammaloss} that the point where losses dominate is a function of the refractive index, thickness of the material and intrinsic losses. While, in the lossless case, increasing the thickness of the layer increases the resonance, we can see that this is limited by the losses, which give a width independent of $d$. Looking at the height of the resonance in the loss dominated limit is quite revealing
\begin{equation}
\beta(\omega_j)=\frac{A_j}{\gamma_j\omega_j}=\frac{4b^2}{n^2\omLO^2 d}\frac{1}{\frac{4b^2\Delta_j^2}{n^2\omLO^4d^3}+\frac{b^2}{\omLO}\left(\frac{\omLO\Gamma_m}{b^2}+\frac{\Gamma_\rho}{\omLO }\right)}\overset{\text{loss dom.}}{\approx} \frac{4}{dn^2\left(\frac{\omLO^2}{b^2}\Gamma_m+\Gamma_\rho\right)}\,.
\label{eq:DA_AQ_P_betaomegaj_losses}
\end{equation}
Unlike the lossless case, increasing $d$ now hinders the resonance height, if not its width. Similarly, while $n$ does not effect $\gamma_j$, again the height is significantly reduced on resonance, further discouraging high $n$ materials. Once the loss term of a material is known, the optimal thickness can be found by requiring that losses do not dominate.

\begin{figure}
\centering
\includegraphics[width=0.7\textwidth]{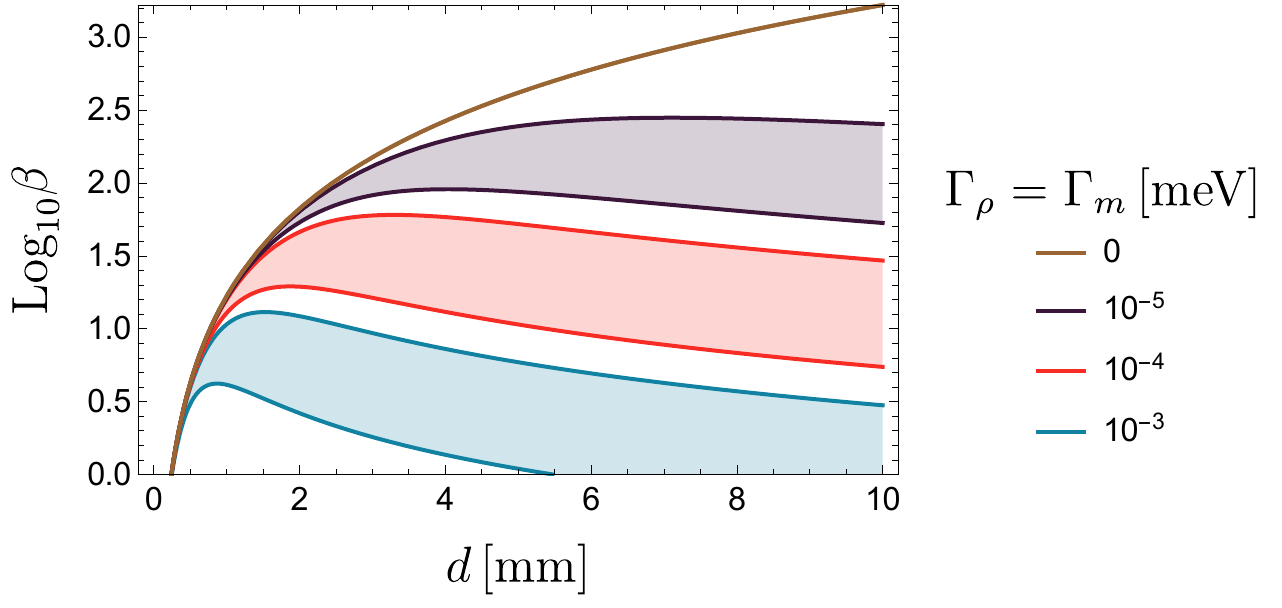} 
 	\caption{Maximal boost factor $\beta$ on resonance at $\omega_0$ with respect to the TMI layer of thickness $d$ for different levels of loss. Material parameters correspond to \MnBiTe{2}{2}{5} with $B=\SI{2}{T}$. The bands show variation of the refractive index from $n=3$ (upper curve of the bands) to $n=7$ (lower curve of each band). }
  	\label{fig:betaloss} 
\end{figure}

To get an idea of the scale of the maximum losses that still allow for useful DA detection, we plot the maximum of $\beta$ at the first resonance $\omega_0$ as a function of $d$ in Fig.~\ref{fig:betaloss}. 
The different colours indicate different losses. We also vary the refractive index around our best guess value, $n=5$: the upper band for each colour corresponds to $n=3$, while the lower to $n=7$. Lower values of $n$ lead to greater boost factor maxima.

From Fig.~\ref{fig:betaloss} we can read off that for given material parameters there is an optimal thickness, which maximizes $\beta$ on resonance. Analytically one can show
\begin{eqnarray}\boxed{
d_{\rm opt}=\frac{2}{\omLO}\left(\frac{ \Delta_j}{n}\right)^{\frac{2}{3}}\left(\frac{1}{\frac{ \Gamma_\rho}{\omLO} +\frac{\Gamma_m \omLO}{b^2}}\right)^{\frac{1}{3}}.
}
\label{eq:DA_AQ_P_optimalThickness}
\end{eqnarray}
Note that the otimal thickness $d_{\rm opt}$ gives a thickness that is consistent with the small $d$ limit that we have used in the expansion only if $2\left(n^2\tilde{\Gamma}_j^2\Delta_j\right)^{\frac{1}{3}}<1$. This inequality is fulfilled for the cases that we are interested in.
In the zero loss limit the optimal thickness $d_{\rm opt}$ diverges . However it is important to stress that in this limit $\gamma_j\to 0$ and our optimal thickness has to be understood as the thickness that maximizes $\beta$ on resonance. Which thickness will be the optimal one with respect to a scanning strategy and sensitivity reach will be discussed in the next section. If the losses are finite they enter with the third root. Also note that scanning different frequencies changes $\omLO$ and therefore in principle for each scanning frequency a different optimal layer thickness is needed. It is not possible to change the layer thickness for each scanning frequency, and therefore the true optimal thickness will depend on the details of the frequencies to be scanned and the scan strategy.

Note that in Fig.~\ref{fig:betaloss} we are approximating the boost factor as a resonance. However, as $\beta\to1$ the boost factor is no longer well described by a Lorentzian. For the purposes of an experiment, all advantage over say a dish antenna is then lost. To estimate the highest allowable losses, we can note that the strongest resonance occurs when $\delta_j=Pi$ and $d=d_{\rm opt}$. By requiring that $\beta\gg 1$ we then find the requirement
\begin{equation}
\frac{\Gamma_\rho}{\omega_{\rm LO}}+\frac{\Gamma_m\omega_{\rm LO}}{ b^2}\ll \frac{1}{2n^2}\,.
\end{equation}
Thus the highest allowable losses are actually set by the refractive index of the material, at least in order to ensure a resonance occurs.

\section{Dark Matter Discovery Potential}\label{sec:DM-detection}

In this section, we review the suitability of TMIs hosting AQs for DA detection. We systematically investigate the discovery reach of the proposed single-TMI-layer benchmark experiment, and the necessary requirements for THz detectors. Astrophysical limits on the axion mass and coupling, and motivation for axion DM in the milliectronvolt range, is reviewed in Appendix~\ref{sec:axion_dm}. 

\subsection{Scanning range}\label{sec:ScanningRange}
Before considering THz detection technology and the reach of the proposed experiment in terms of the DA coupling, $\gDA$, we first determine the range of DA masses that can in principle be accessed using TMIs. Recall from Section~\ref{sec:one_layer_tmi} that the resonance frequencies of the experiment, $\omega_j$, can be tuned by changing the external magnetic field, $\Bext$. To estimate the resulting range, we look at the first resonance $\omega_0$ since $|\beta (\omega_0)| > |\beta (\omega_j)|$ for all $j > 0$. Doing so, we find that
\begin{equation}
	\omega_0(\Bext) \approx \omLO(\Bext)=\sqrt{\mAQ^2 + b^2(\Bext)} \, ,
\label{eq:Scanning_range}
\end{equation}
where $\mAQ$ is the AQ~mass and $b$ is given by Eq.~\eqref{eqn:b_meV}. In the limit of $\Bext \rightarrow 0$ we simply have $\omega_0 \rightarrow \mAQ$, while for very strong $B$-fields of $\Bext = \SI{10}{\tesla}$ and our benchmark parameters in Table~\ref{tab:toorad_params} (our best approximation to \MnBiTe{2}{2}{5}), we find that
\begin{equation}
	\SI{1.8}{\meV} = \mAQ < \omega_0 < \SI{8.2}{\meV} \, . \label{eq:numerical_scanning_range}
\end{equation}

As of now, the DA mass is unknown, so it is desirable to cover a wide range of axion masses with a given TMI crystal. Since typical magnetic fields in the lab are restricted to the order of a few tesla, we cannot arbitrarily increase $\Bext$ and hence need to maximise the relative response of the AQ to $\Bext$, viz.
\begin{equation}
	\frac{1}{\mAQ}\frac{\dd b}{\dd \Bext} = \frac{\alphaEM}{\pi\sqrt{2}} \frac{1}{\sqrt{\epsilon}\fAQ\mAQ} \approx \frac{0.46}{\si{\tesla}} \, \left(\frac{25}{\epsilon} \right)^{1/2} \left(\frac{\SI{70}{\eV}}{\fAQ}\right) \left(\frac{\SI{1.8}{\meV}}{\mAQ}\right) \, . \label{eq:tunability}
\end{equation}
This means that smaller $\fAQ$, $\mAQ$, or $\epsilon$ are beneficial for a TMI in order to cover a larger range of frequencies for a given maximum possible value of the applied $B$-field.

A large relative AQ response in Eq.~\eqref{eq:tunability} is only beneficial if the applied $B$-field value can be controlled to sufficiently high accuracy over the course of the measurement. This is because fluctuations in $\Bext$ will translate into fluctuations in $\omega_0$, which might result in the resonance around $\omega_0$ fluctuating in and out of the bandwidth of the detector. The magnet design for TOORAD will thus require relatively precise control of the $B$-field, and could be a limitation in cost, field strength, or total volume.


\subsection{Detectors for THz Radiation}\label{sec:detectors}
Searching for dark DAs is challenging because the resultant photon signal is very weak and can be hidden in wide range of frequencies, since the DA mass in unknown. To improve our chance of success, we need to understand the intrinsic and extrinsic background noise of our photon detection system, coupling efficiency of photon detectors to our proposed experimental setup, and scalability in collecting more photons from the material that hosts the AQ. In earlier sections, we have discussed the generation of electromagnetic radiation in the THz~(millimeter wave) regime using AQs for DA detection. We will focus here to the available technology to detect these photons with energies from 0.01 to a few THz. 

Detectors that have high sensitivity for the search of DAs in our frequency range of interest include amplifiers, heterodyne mixers, bolometers, and single-photon detectors~(SPDs). We shall consider experiments performing at temperatures much lower than the frequency, i.e.\ $T \ll \omega$ to avoid the thermal photons from the blackbody radiation and to focus on the fundamental limit of photon detection~\cite{Caves:1982hd, Mather:1984wt, Tucker:1985bd, Clerk:2010dh}. 
\begin{table}
	\caption{Comparison of detector technologies for searching dark matter using quasiparticle axions. See the main text for explanations of the symbols.\label{tab:DetectorComp} }
	\centering
	\begin{tabular}{llll} 
		\toprule
		Detector type & \multicolumn{2}{l}{Fundamental noise limit} & Metric \\
		\midrule
		\makecell[l]{Amplifiers \&\\ \hspace{0.5em} heterodyne mixers} & quantum noise & $\hbar\omega$~\cite{Caves:1982hd, Tucker:1985bd, Clerk:2010dh} & $T_{Q}$ \\
		Bolometers & thermal fluctuations & $\sqrt{4G_
		\text{th} k_\text{B} T_0}$~\cite{Mather:1984wt, 2019_Lee} & $\mathrm{NEP}$ \\
		Calorimetric SPDs & \makecell[l]{energy resolution\\ for finite bandwidth} & $\sqrt{C_\text{th}k_\text{B}T_0^2}$~\cite{MCCAMMON:2005br, Walsh:2017kk} & dark count rate \\
		\bottomrule
	\end{tabular}
\end{table}
Since amplifiers and heterodyne mixers, e.g.\ superconductor\hyp{}insulator\hyp{}superconductor~(SIS) and hot-electron bolometric mixers, are sensitive to the voltage or the electric field of the signal, we can put them into one category, while bolometers and SPDs go into another. We present a comparison of all detector types discussed in what follows in Table~\ref{tab:DetectorComp}.

State-of-the-art amplifiers and mixers can detect a very weak signal from as little as a few photons by parametric amplification or non-linear mixing processes. As the signal-to-noise ratio is given by the ratio of the number of photons in the signal to that in the amplifier noise, the amplifier noise can be quantified naturally in units of photon quanta, i.e.\ $\hbar\omega$, or amplifier noise temperature, i.e.\ $T_{Q} = \hbar\omega/k_\text{B}$. For the linear, phase-preserving amplification, the minimum amplifier noise is one quanta~\cite{Caves:1982hd,Tucker:1985bd, Clerk:2010dh}. Half of this comes from the quantum fluctuation from the parametric pumping port used in modulation for the amplification gain, whereas another half from the quantum noise in the signal port. At lower microwave frequencies, quantum noise-limited amplifiers have been achieved based on parametric effects~\cite{Yurke:1989ev, Brubaker:2017ee, Aumentado:2020gf} and, at higher frequencies, in SIS detectors~\cite{Tucker:1985bd, McGrath:1998jj}, hot electron bolometric mixers~\cite{Zhang:2010gn, LaraAvila:2019gs}, and plasmonic mixers~\cite{Wang:2019gz}. However, the insertion loss and insufficient first amplification gain may degrade the overall performance, resulting in a higher system noise temperature $T_{\rm sys} \geq T_{Q}$. For a total measurement time, $t_{\rm meas}$, and measurement bandwidth, BW, the average noise is given by the Dicke radiometer formula,
\begin{equation}
	\text{Noise} = \frac{T_{\rm sys}}{\sqrt{ {\rm BW}t_{\rm meas}}}
\end{equation}

Instead of amplifying the voltage, bolometers are high-sensitivity, square-law detectors that measure the power of microwave and far-infrared radiation. They operate by first absorbing the incident radiation and subsequently inferring the radiation power from the temperature rise due to the increase of its internal energy. The bolometer sensitivity is quantified by noise equivalent power~(NEP), measured in units of \si{\WpsqHz}, i.e. the power fluctuations of the bolometer in absence of any incident power during a 1-second averaging window. Previous experiments project NEP values as low as \SI{e-21}{\WpsqHz}~\cite{Wei:2008jw}. The sensitivity of this technique is not limited by quantum fluctuations, but rather by the fundamental thermal fluctuations~\cite{Mather:1984wt, 2019_Lee}. This fundamental-fluctuation-limited NEP is given by $\sqrt{4G_\text{th}k_\text{B}T_0}$ with $G_\text{th}$ being the thermal conductance of the bolometric material to the thermal bath, and $T_0$ is the bolometer (bath) temperature. Therefore, bolometers for DA detection will require to operate at the lowest achievable temperatures with the least thermal conductance to its surrounding.

In addition to power detection by bolometer, single-photon detectors (SPD) is another viable option to capture the photons generated from DAs. Efficient DA searches will require the SPD to have simultaneously a high quantum efficiency to register every precious photon, and a low dark count rate to minimize the false positive signal. Naturally, these two requirements are competing against each other because a higher quantum efficiency also means the detector can be triggered by noises to produce a count in the absence of photons. Fortunately at cryogenic temperatures, we can employ superconductors to detect photons efficiently and accurately. When the photon energy is larger than the superconducting gap energy, $\Delta_{\text{S}}$, the incident photons can break Cooper pairs and produce a sizable number, $\eta_{\text{d}}\hbar\omega/\Delta_{\text{S}}$, of quasiparticles, with $\eta_{\text{d}} < 1$ being the energy downconversion efficiency \cite{Day:2003tl}. These quasiparticles can then transduce into a readout signal of resistance, temperature, kinetic inductance, or excess current in SPDs such as superconducting nanowire SPD \cite{Goltsman:2001eaa, Hochberg:2019dj}, transition edge sensor \cite{Irwin:2005book}, microbolometer \cite{Wei:2008jw,doi:10.1063/1.4739839}, Josephson junction SPD \cite{Walsh:2020ty}, kinetic inductance detector \cite{Day:2003tl}, and superconducting tunneling junction detector \cite{Peacock:1996io}. SPDs based on this mechanism have been highly successful especially in the near infrared domain when the relatively high photon energy can produce a considerable amount of quasiparticles. Single-photon detection in THz regime is a lot more challenging. Microbolometers based on the superconducting nanowire have demonstrated experimentally energy-resolved, single-photon detection down to \SI{38}{\THz}~\cite{doi:10.1063/1.4739839} and projection give energy resolutions as low as \SI{0.12}{\THz}~\cite{Wei:2008jw}. Recently, quantum capacitance detector \cite{Echternach:2018iw} has demonstrated the detection of \SI{1.5}{\THz} by sensitively sensing the change of quantum capacitance from the quasiparticle through a resonator.

Since the photon detection mechanism depending on Cooper pair breaking will inevitably become more and more challenging at lower millimeter wave frequencies, we can also exploit the giant thermal response in graphene~\cite{Vora:2012cs, Fong:2012ut, McKitterick:2013ue, Walsh:2017kk} to absorb the incident photon first before measuring the temperature rise. The concept of calorimetric SPDs has been developed for x-ray detection and superconducting SPDs~\cite{Moseley:1984wh, MCCAMMON:2005br}. In contrast to bolometers, single photon detection not strictly limited by thermal fluctuations if a large enough detection bandwidth and high sensitivity temperature transducer is available. Intuitively, this is because when the photon impinges into the detector, SPDs produce a sharp rising signal that can be observed with a wide-bandwidth detector. Quantitatively, this is due to the improvement of signal-to-noise ratio through a matched filter that is tailored to the shape of the expected signal from a single photon. Nevertheless, the energy resolution, $\Delta\epsilon = \sqrt{C_\text{th}k_\text{B}T_0^2}$ with $C_\text{th}$ being the thermal heat capacity of the calorimeter, is still a good benchmark for calorimeter SPDs. When the NEP is white-noise limited, we can use~\cite{Moseley:1984wh} 
\begin{equation}
	\Delta\epsilon = {\rm NEP}\sqrt{\frac{C_\text{th}}{G_\text{th}}}
\end{equation}
to compare the sensitivity with bolometers. For DA searches in wide frequency bandwidth, graphene-based single photon detection also has an advantage in wide bandwidth photon coupling by impedance matching the input to the photon absorber with an antenna. Spiral, log-periodic, and bow-tie antennas have been implemented for graphene detectors~\cite{Vicarelli:2012de, Bandurin:2018cx}. As graphene-based bolometers have been demonstrated recently in the microwave regime~\cite{2019_Lee, Kokkoniemi:2020dk} with energy resolution projections to a few \SI{10}{\GHz}, it can potentially complement SPDs by operating at millimeter wave frequencies.

In addition to superconductor-based and calorimeter SPD, superconducting qubits and quantum dots can also detect single photons \cite{Dixit:2020tv, Komiyama:2000eo}. These nano fabricated devices have discrete energy states and can serve effectively as artificial atoms. When incident photons promote the qubit or quantum dot to an excited state, they can be detected by measuring the state of the artificial atoms. Detection of single photons has been demonstrated using superconducting qubits at microwave frequencies \cite{Johnson:2010tn, Kono:2018ij} and using quantum dots as low as \SI{1.5}{\THz} \cite{Komiyama:2000eo} with photon coupling through superconducting resonators and dipole antenna, respectively. 

Table~\ref{tab:DetectorComp} compares the fundamental limit of detectors that will be useful for dark matter detectors; since the quantum noise rises linearly with frequency, SPDs will have an advantage over amplifiers for the search of higher axion mass~\cite{Lamoreaux:2013dv}. A dark count rate $\lambda_\text{d} \sim \SI{1}{\milli\Hz}$ has been demostrated experimentally~\cite{Komiyama:2000eo, Komiyama:2011wma} for a quantum dot detector. Note, however, that the realised experimental efficiency for that detector was only $\eta = 0.01$~\cite{Komiyama:2000eo,Komiyama:2011wma}. Overall, it is desirable to obtain a detector with the optimal combination of low dark count rate and high efficiency, as this will ultimately determine the sensitivity of TOORAD. Detectors that feature a better efficiency typically have a worse dark count rate than the detector from Ref.~\cite{Komiyama:2000eo} considered above. We will therefore define a pessimistic~(optimistic) scenario by setting $\lambda_\text{d} \sim \SI{1}{\milli\Hz}$ with a detection efficiency of $\eta = 0.01$~($\eta = 1$).

Last but not least, we shall consider how to put the photon detector together with the material that hosts the AQs. The efficiency of the dark matter search relies on this system integration. The goal is to maximize the photon coupling as the axion quasiparticle material scales up. Therefore we will need to design an antenna that can collect the photons emitted from the AQs to the detector with the least inert loss. This will be an important factor to select a potential detector technology to develop. Ultimately, to detect a small signal from DA, the detector metric should be the total experimental averaging time for an experiment to reach a statistical significance and will depend on both efficiency and sensitivity. To improve our chance of detecting dark matter, we need more research on detector technologies, which are also be useful in other applications including radio~astronomy, spectroscopy, and medical imaging~\cite{Echternach:2018iw, Mittleman:2017dq, Lewis:2019dl}.

\subsection{Experimental Sensitivity and Forecasts}\label{sec:sensitivity_reach}
As discussed above, the signal from the DA-polariton-photon conversion may be detected using an SPD, which is superior compared to heterodyne power detection in \si{\THz}. In this section we quantify the sensitivity and discovery reach for a photon-counting experiment.

The detection of individual photon events is governed by Poisson statistics i.e.\ the likelihood of detecting $\mathds{N}$~photons given model parameters~$\vc{x}$~(the set of DA and material properties) is given by
\begin{equation}
	p(\mathds{N}|\vc{x}) = \frac{\left(\eta \, n_\text{s} + n_\text{d} \right)^\mathds{N}}{\mathds{N}!} \; \mathrm{e}^{-\eta \, n_\text{s} - n_\text{d}} \, , \label{eq:poisson_stats}
\end{equation}
where $n_\text{s} = \lambda_\text{s} \tau$ and $n_\text{d} = \lambda_\text{d} \tau$ are the number of expected signal and dark count events, respectively, as calculated from their respective rates, $\lambda_\text{s}$ and $\lambda_\text{d}$, and total observation time~$\tau$. The parameter $\eta$ describes the total detector efficiency i.e.\ takes into account the intrinsic efficiency of the detector as well as any other imperfections in the experimental setup. Note that Eq.~\eqref{eq:poisson_stats} assumes that there are no external backgrounds present. While we do not use a likelihood approach based on Eq.~\eqref{eq:poisson_stats} directly for our estimates, it should be noted that the form above is a better approach than the asymptotic, approximate equations used in what follows. For the case of a single-bin Poisson distribution without any nuisance parameters -- i.e.\ assuming that the material and detector properties are perfectly known -- we performed a Monte Carlo simulation to check the validity of the asymptotic formulae that we employ. We found them to be conservative and, hence, suitable for the purpose of estimating TOORAD's sensitivity. For an actual analysis of experimental data, however, a likelihood-based approach should be used.

\subsubsection{Sensitivity}

In order to compute the sensitivity, we assume that no significant signal over background is found. The significance is $S=2(\sqrt{n_s+n_d}-\sqrt{n_d})$, where $n_s$ is the number of signal events and $n_d$ the number of dark count events~\cite{doi:10.1142/S0217732398003442,Bityukov:2000tt,PhysRevD.82.115018}. Then the exclusion limit at 95\%~C.L. for photon counting based on Poisson statistics (Eq.~\eqref{eq:poisson_stats}) is obtained from $S<2$, i.e.\ $\lambda_\text{s} < \frac{1}{\tau} + 2\sqrt{\frac{\lambda_\text{d}}{\tau}}$, where $\lambda_\text{d}$ is the dark count rate, $\lambda_\text{s}$ the signal rate and $\tau$ the measurement time. For a discovery one would require $S>5$. In Section~\ref{sec:detectors} we argued that $\lambda_\text{d} = \SI{1}{\mHz}$ is reasonable. In the following we estimate the sensitivity in two scenarios. The case $\frac{1}{\tau}<2\sqrt{\frac{\lambda_\text{d}}{\tau}}$ can be achieved for sufficiently long measurement times and is called the background dominated scenario, i.e.\ $\tau>\frac{1}{4 \lambda_\text{d}}=\SI{250}{\second}$. If the measurement time is short $\tau<\frac{1}{4 \lambda_\text{d}}=\SI{250}{\second}$ then it is not background dominated. 

First, we investigate the case that the measurement is not background dominated. The number of signal photons per measurement time is $\lambda_\text{s}= \eta\frac{|E_0|^2}{2 \omega} A \, \beta^2 \,$ where $A$ is the surface area of the TMI, $\eta$ the photon counting efficiency and $\omega=\omega_j\approx \omLO$ is the resonance frequency where the power boost factor peaks. 
 The power boost factor is the emitted electromagnetic field normalized to the axion induced field $E_0$, which is determined by the local axion dark matter density $\rho_a$, the axion photon coupling $\gDA$ and the strength of the external $B$-field: $E_0= \gDA \Bext a_0^{-} \simeq \gDA \,  \frac{\sqrt{2\rho_a}}{\mDA} \, \Bext \, $.
Putting everything together we obtain the sensitivity estimate: 
\begin{eqnarray}
	\gDA &>& \num{4.4}\times \SI{e-11}{\GeV^{-1}} \left(\frac{0.01}{\eta}\right)^{\frac{1}{2}} \left(\frac{\SI{2}{\tesla}}{\Bext}\right) \,  \left(\frac{100}{\beta}\right) \,  \left(\frac{(\SI{0.2}{\m})^2}{A}\right)^{\frac{1}{2}} \, \left(\frac{\SI{4}{\minute}}{\tau}\right)^{\frac{1}{2}}\times \nonumber\\ &\times&\left(\frac{\SI{0.3}{\frac{\GeV}{\centi\metre^3}}}{\rho_a}\right)^{\frac{1}{2}} \, \left(\frac{\mDA}{\SI{2.83}{\meV}}\right)^{\frac{3}{2}}\, , \quad \text{(negligible backgrounds, $\tau<\lambda_d^{-1}$)}
		\label{eq:sensitivity_NotbackgroundDominated}
\end{eqnarray}
where an axion mass $m_a=\SI{2.83}{\meV}$ corresponds to the scanned axion mass with a $\SI{2}{\tesla}$ external $B$-field under the assumption of the benchmark material ($n=5,f_\Theta=64\,{\rm eV}$ and $\mAQ=\SI{2}{\meV}$).
The reference area in Eq.~\eqref{eq:sensitivity_NotbackgroundDominated}, $A=(\SI{0.2}{\m})$. $\SI{0.2}{\m}$, is around half of the square de Broglie wavelength for an axion with velocity $v=10^{-3}c$ and mass $\SI{2.83}{\milli\electronvolt}$. Single crystals of \MnBiTe{}{2}{4} grown in Ref.~\cite{PhysRevX.9.041038} are on the order of cm$^2$. Reaching large surface area will thus require tiling and machining many crystals together. Tiling is known to introduce significant complications for dielectric haloscopes like MADMAX~\cite{Knirck_2019,Beurthey:2020yuq}. Further, as the axion gives an opening angle of $v\sim 10^{-3}$ the collecting area of the THz detector must be large. We anticipate that this problem can be overcome with the correct antenna.

When there are finite losses, we can use the peak value from Eq.~\eqref{eq:DA_AQ_P_betaomegaj_losses} to eliminate $\beta$ and obtain:
\begin{eqnarray}
	\gDA &>& \num{4.95}\times \SI{e-11}{\GeV^{-1}} \,\left(\frac{0.01}{\eta}\right)^{\frac{1}{2}} \left(\frac{\SI{2}{\tesla}}{\Bext}\right) \,    \left(\frac{(\SI{0.2}{\m})^2}{A}\right)^{\frac{1}{2}} \, \left(\frac{\SI{4}{\minute}}{\tau}\right)^{\frac{1}{2}}\,\left(\frac{\SI{0.3}{\frac{\GeV}{\centi\metre^3}}}{\rho_a}\right)^{\frac{1}{2}} \,\times \nonumber\\ &\times& \left(\frac{\SI{2.83}{\meV}}{\mDA}\right)^{\frac{1}{2}}\,\left(\frac{\Delta_j}{\pi}\right)^2\left(\frac{\SI{2}{\milli\metre}}{d}\right)^2\,\times\, \Sigma.
	\label{eq:sensitivity_NotbackgroundDominated_explicit}
\end{eqnarray}
where we have defined in the dimensionless quantity:
\begin{eqnarray}
\Sigma\equiv1+2\left(\frac{d}{d_\text{opt}}\right)^3.
\label{eq:Sigma}
\end{eqnarray}
We did not plug in any specific value for $\Sigma$ in the sensitivity estimate because when the thickness is chosen to be close or equal to the optimal thickness $\Sigma$ is of the order one. The losses, AQ decay constant, $\fAQ$, and refractive index, n, all appear implicitly via the determination of $d_{\rm opt}$, the optimal material thickness.

Compare the sensitivity Eq.~\eqref{eq:sensitivity_NotbackgroundDominated_explicit} to that obtained with heterodyne detection. In this case we use the Dicke radiometer equation with noise temperature $T$. The signal over noise ratio is given by SNR $=\frac{P_s}{T_{\rm sys}}\sqrt{\frac{\tau}{\Delta \nu_a}}$, where $\Delta\nu_a=10^{-6}\, m_a$ is the DA linewidth, and $P_s$. If the physical system temperature is low enough, cf.\ Section~\ref{sec:detectors}, $T_{\rm sys}$ is limited by the standard quantum limit (SQL) $T_{\rm sys}=\omega=m_a$. 
The resulting sensitivity is:
\begin{eqnarray}
g_{a\gamma}&>&\num{1.1}\times \SI{e-9}{\GeV^{-1}}\left(\frac{\rm SNR}{2}\right)^{\frac{1}{2}}\,\left(\frac{m_a}{\SI{2.83}{\milli\electronvolt}}\right)^{\frac{7}{4}} \, \left(\frac{\SI{2}{\tesla}}{\Bext}\right)\, \left(\frac{\SI{0.3}{\frac{\GeV}{\centi\metre^3}}}{\rho_a}\right)^{\frac{1}{2}}\times\nonumber\\ &\times&\left(\frac{100}{\beta}\right) \, \left(\frac{(\SI{0.2}{\m})^2}{A}\right)^{\frac{1}{2}} \, \left(\frac{\SI{4}{\minute}}{\tau}\right)^{\frac{1}{4}} \quad\text{(heterodyne SQL).}
\end{eqnarray}
The sensitivity is worse than the SPD, cf.\ Eq.~\eqref{eq:sensitivity_NotbackgroundDominated}, by approximately an order of magnitude. This is as expected since for high frequencies the SQL pushes $T$ to large values. The SQL can, however, be overcome by ``squeezing''~\cite{Backes:2020ajv}.

Next we focus on the case that the measurement is background dominated ($\lambda_\text{s}<2\sqrt{\frac{\lambda_\text{d}}{\tau}}$). For our benchmark dark count rate of $\lambda_d$ this gives $\tau>250\,{\rm s}$. Long measurement times on a fixed frequency could be adopted in a ``hint'' scenario where the axion mass is thought to be known by some other means (for example, an astrophysical hint, or highly accurate relic density prediction), and a resonant DM search is required to verify the hint. To consider this scenario, we take the measurement time on each frequency to be $\tau=3\,$yr, i.e. an entire experimental campaign. The sensitivity in this case is: 
\begin{eqnarray}
	\gDA &>& \num{1.63}\times \SI{e-12}{\GeV^{-1}} \left(\frac{0.01}{\eta}\right)^{\frac{1}{2}} \left(\frac{\SI{2}{\tesla}}{\Bext}\right) \,  \left(\frac{100}{\beta}\right) \,  \left(\frac{(\SI{0.2}{\m})^2}{A}\right)^{\frac{1}{2}} \, \left(\frac{\lambda_\text{d}}{\SI{e-3}{\hertz}}\right)^{\frac{1}{4}}\, \left(\frac{3\,\text{yr}}{\tau}\right)^{\frac{1}{4}}\times \nonumber\\ &\times&\left(\frac{\SI{0.3}{\frac{\GeV}{\centi\metre^3}}}{\rho_a}\right)^{\frac{1}{2}} \, \left(\frac{\mDA}{\SI{2.83}{\meV}}\right)^{\frac{3}{2}} \quad \text{(background dominated)}.
	\label{eq:sensitivity_backgroundDominated}
\end{eqnarray}
Using now again the maximum peak value from Eq.~\eqref{eq:TwoInterfaces_Maximum_at_resonance} to eliminate $\beta$ in the previous equation we obtain the sensitivity estimate
\begin{eqnarray}
	\gDA &>& \num{1.94}\times \SI{e-12}{\GeV^{-1}}\, \left(\frac{0.01}{\eta}\right)^{\frac{1}{2}} \left(\frac{\SI{2}{\tesla}}{\Bext}\right) \,    \left(\frac{(\SI{0.2}{\m})^2}{A}\right)^{\frac{1}{2}} \, \left(\frac{\lambda_d}{\SI{e-3}{\hertz}}\right)^{\frac{1}{4}}\, \left(\frac{3\,\text{yr}}{\tau}\right)^{\frac{1}{4}}\,\times \nonumber\\ &\times&\left(\frac{\SI{0.3}{\frac{\GeV}{\centi\metre^3}}}{\rho_a}\right)^{\frac{1}{2}} \, \left(\frac{\SI{2.83}{\meV}}{\mDA}\right)^{\frac{1}{2}}\,\,\left(\frac{\Delta_j}{\pi}\right)^2\left(\frac{\SI{2}{\milli\metre}}{d}\right)^2\times \Sigma \,.
	\label{eq:sensitivity_backgroundDominated_explicit}
\end{eqnarray}

To complete our discussion we also estimate the sensitivity for bolometric detectors whose performance is specified by the NEP. The minimal detectable signal power which such a detector can detect is $P_s>{\rm NEP}/\sqrt{\tau}$. Evaluating this leads to the sensitivity:
\begin{eqnarray}
\gDA&>&\num{9.7}\times \SI{e-13}{\GeV^{-1}}\,\left(\frac{\rm NEP}{10^{-21}{\rm W}/\sqrt{\rm Hz}}\right)^{\frac{1}{2}} \, \left(\frac{(\SI{0.2}{\m})^2}{A}\right)^{\frac{1}{2}} \, \left(\frac{3\,\text{yr}}{\tau}\right)^{\frac{1}{4}}\,\times\nonumber\\ 
&\times& \left(\frac{\mDA}{\SI{2.83}{\meV}}\right)\,\left(\frac{\SI{0.3}{\frac{\GeV}{\centi\metre^3}}}{\rho_a}\right)^{\frac{1}{2}} \left(\frac{100}{\beta}\right)
\label{eq:sensitivity_NEP}
\end{eqnarray}
The sensitivity estimate in Eq.~\eqref{eq:sensitivity_NEP} has a similar order of magnitude as the SPD sensitivity, cf.\ Eq.~\eqref{eq:sensitivity_backgroundDominated} but a slightly different scaling with the axion mass. 

\begin{figure}\centering
\includegraphics[width=0.55\textwidth]{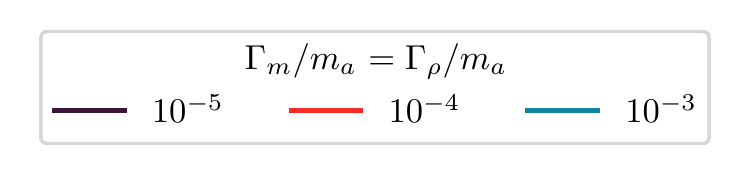} \\
    \includegraphics[width=0.99\textwidth]{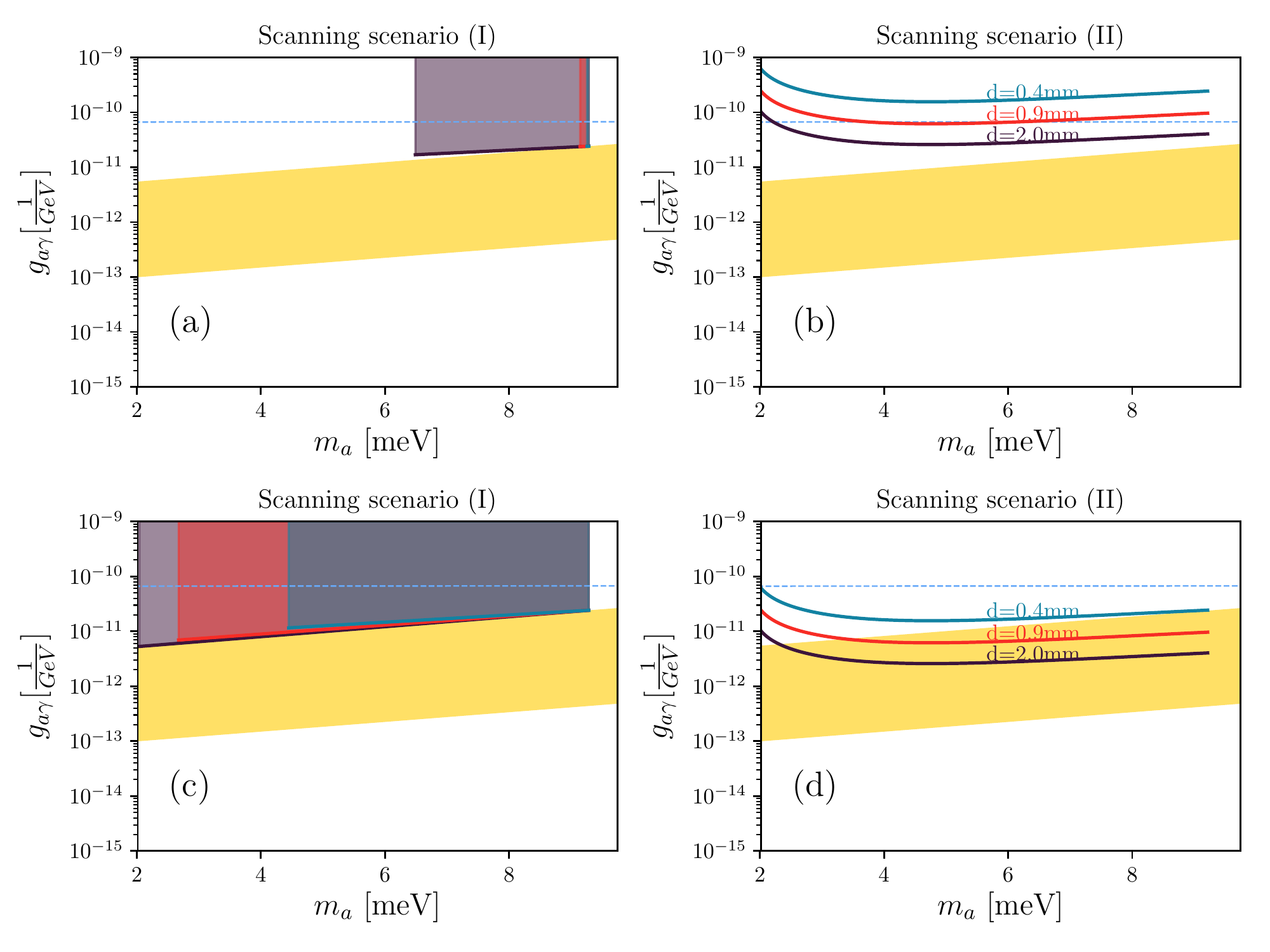}
		\caption{Sensitivity for ``Material 2'' baseline parameters (see text for details) for various loss values $\Gamma$. Top row: $\eta=0.01$. Bottom row: $\eta=1$. We fix the dark count rate $\lambda_d=10^{-3}\,$Hz. The yellow band shows QCD axion models, and the dashed blue line the CAST exclusion on $\gDA$. The scanning scenarios are defined in the text.}
		\label{fig:Scanning_Sensitivity}
\end{figure}

\begin{figure}\centering
\includegraphics[width=0.55\textwidth]{figures/TwoInterfaces_Scanning_and_Hint_Legend.pdf} \\
    \includegraphics[width=0.55\textwidth]{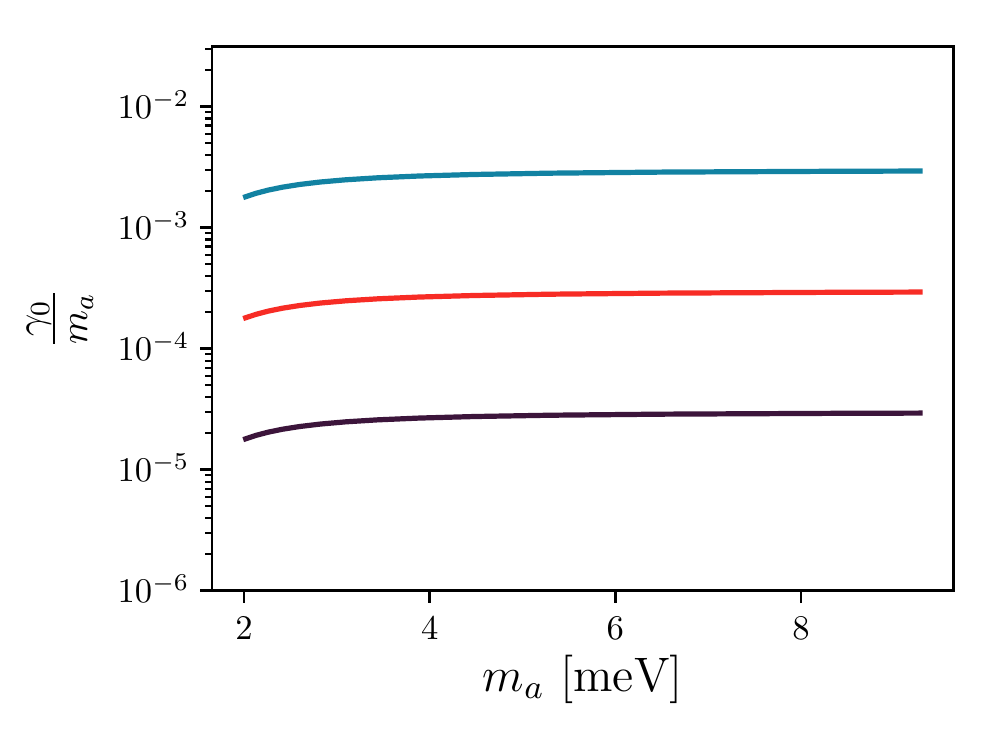}
		\caption{Linewidth of the boost parameter for ``Material 2'' baseline parameters (see text for details) for various loss values $\Gamma$. }
		\label{fig:linewidth}
\end{figure}

\subsubsection{Scanning Strategies}

We now compute forecasts for the baseline parameters of ``Material 2'' (best approximation to \MnBiTe{2}{2}{5}, with refractive index $n=5$ and $\mu=1$), and consider three possibilities for the losses,\,\footnote{Note that we now take into account the frequency scaling of the losses which we have found in section~\ref{sec:damping}. In all previous results of this paper we did not take into account this scaling since it was not necessary in order to understand the physical picture. However, here we want to estimate a realistic sensitivity of a DA search and therefore we take the frequency scaling of the losses into account.} $\Gamma_m /\omega=\Gamma_\rho /\omega=10^{-5}, 10^{-4},10^{-3}$. 

 Assuming a fixed ratio $\Gamma/\omega$ is consistent with our model for the impurity based losses, and assumes $\epsilon_2$ is approximately constant in the relevant range. For fixed ratio $\Gamma/\omega$, there are larger losses at higher frequencies. We first assumed SPD efficiency of $\eta=0.01$ and dark count rate $\lambda_d=10^{-3}\,$Hz, which has been demonstrated. We also show a more optimistic sensitivity estimate with $\eta=1$ (dotted line) for $\Gamma_\rho/\omega=\Gamma_m/\omega=10^{-4}$. The surface  area of the TMI layer is fixed to be $A=(0.2\,{\rm m})^2$, where $0.2\,{\rm m}$ is on the order of half of the de Broglie wavelength. Furthermore we use the main resonance $j=0$ for the sensitivity estimate.

We consider two different scanning scenarios, with $B$-field values from $1\,$T to $10\,$T: 
\begin{itemize}
\item Scanning I. We begin at the highest frequency with the largest $B$-field where the base power is largest and the QCD band is at the largest $\gDA$. We scan to the top of the QCD band. We then move by the width $\gamma_0$ on to the next frequency at lower $B$, and repeat for a total scan time of 3 years.\,\footnote{Note that we assume the peak power is achieved over the width $\gamma_0$. While this is less conservative than assuming the minimal value (i.e., half the maximum), note that power also exists outside of $\gamma_0$, which would still be integrated over during the scan. As this is an estimate, rather than a detailed exclusion limit of an experiment, the final limit would likely fall somewhere between the two.} Fig.~(\ref{fig:Scanning_Sensitivity},a). We compute the optimal thickness with the largest axion mass within the scanned region.
\item Scanning II. We scan for a fixed time set equal on all frequencies and repeat for a total scan time of 3 years. In each step we move in frequency by the width $\gamma_0$, Fig.~\ref{fig:linewidth} (Eq.~\ref{eq:gammaloss}). We compute the optimal thickness with the axion mass that is in the middle of the scanned interval.
\end{itemize}
In each case the limit is found for signal to noise equal to two, 95\% C.L. exclusion.

In the Scanning II case we assume that each individual scan takes the same amount of time $\tau$. Then with the bandwidth from Eq.~\eqref{eq:gammaloss} (Fig.~\ref{fig:linewidth}) we can calculate the total number of scans. From this we then calculate the scan time for each individual scan such that the total scanning time for each case is $t_\text{scan}=3\,$years. Depending on the individual scan time $\tau$ we calculate the sensitivity in the right limit, cf.\ Eq.~\eqref{eq:sensitivity_NotbackgroundDominated_explicit} and~\eqref{eq:sensitivity_backgroundDominated_explicit}.

In the Scanning I scenario for $\eta=0.01$, we find that a wide range of the QCD band can only be covered in the case with extremely small losses, $\Gamma/\omega=10^{-5}$. With this scanning strategy, $\eta=1$ detection efficiency allows a wide range of the top of QCD band to be scanned for all loss parameters. In the Scanning II scenario the QCD band cannot be reached with $\eta=0.01$. However, with $\eta=1$ we find that a reasonable portion of the upper part of the QCD band can be scanned with $\Gamma/\omega=10^{-4}$. With very low losses $\Gamma/\omega=10^{-5}$ and $\eta=1$ the Scanning II scenario reaches almost KSVZ sensitivity across a wide range of masses. We also considered the intermediate case $\eta=0.1$, which allows some sensitivity to the QCD axion band with $\Gamma/\omega=10^{-4}$. We conclude that a successful QCD-sensitive experiment requires high efficiency SPDs.

\subsection{Parameter Study}

\begin{table}
	\caption{Parameter reference values and ranges. Our benchmark material is ``Material 2'', based on \MnBiTe{2}{2}{5}.\label{tab:toorad_params} }
	\centering
	\begin{tabular}{llll}
		\toprule
		\multicolumn{2}{l}{Parameter name \& symbol} & Range & Benchmark \\
		\midrule
		\multicolumn{4}{l}{\textit{TMI parameters}}\\
		\addlinespace[2pt]
		Decay constant & $\fAQ$ & [50, 200]\,\si{\eV} & \SI{70}{\eV} \\
		AQ mass & $\mAQ$ & $\sim\order(\si{\meV})$ & \SI{1.8}{\meV} \\
		Permittivity & $\epsilon$ &[9,49] & 25  \\
		Magnetic permeability & $\mu$ & $\sim\order(1)$ & 1\\
		Magnon losses & $\GammaM$ & [\num{e-5}, \num{e-3}]\,\si{\meV} &  \\
		Specific conductance & $\GammaRho$ & [\num{e-5}, \num{e-3}]\,\si{\meV} &  \\
		Area of crystal face & $A$ &$(0.2\,{\rm m})^2$ &  \\
		Thickness & $d$ & $d_{\text{opt}}$, cf.\ Eq.~\eqref{eq:DA_AQ_P_optimalThickness}\, &  \\
		\midrule
		\multicolumn{4}{l}{\textit{Experimental parameters}}\\
		\addlinespace[2pt]
		External $B$-field & $\Bext$ & [1, 10]\,\si{\tesla} & \SI{2}{\tesla} \\
		Detection effciency & $\eta$ & [0.01, 1] & 0.01\\
		Dark count rate & $\lambda_\text{d}$ & $\gtrsim \SI{1}{\milli\hertz}$ &  \SI{1}{\milli\hertz} \\
		\bottomrule
	\end{tabular}
\end{table}

We now wish to investigate how the sensitivity and scan range depend on the yet unknown material parameters of the TMIs. In this section we consider only the scanning II scenario. In Table~\ref{tab:toorad_params} we list the unknown parameters, and reasonable ranges they might take in different materials within our rough approximations to the theoretical uncertainties. The ranges for the parameters have been motivated in Section~\ref{sec:params}. 

\begin{figure}
\centering
\includegraphics[width=0.55\textwidth]{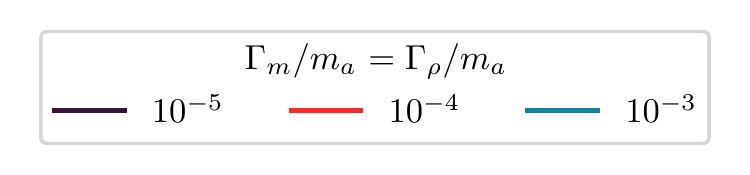} \\
\includegraphics[width=1.1\textwidth]{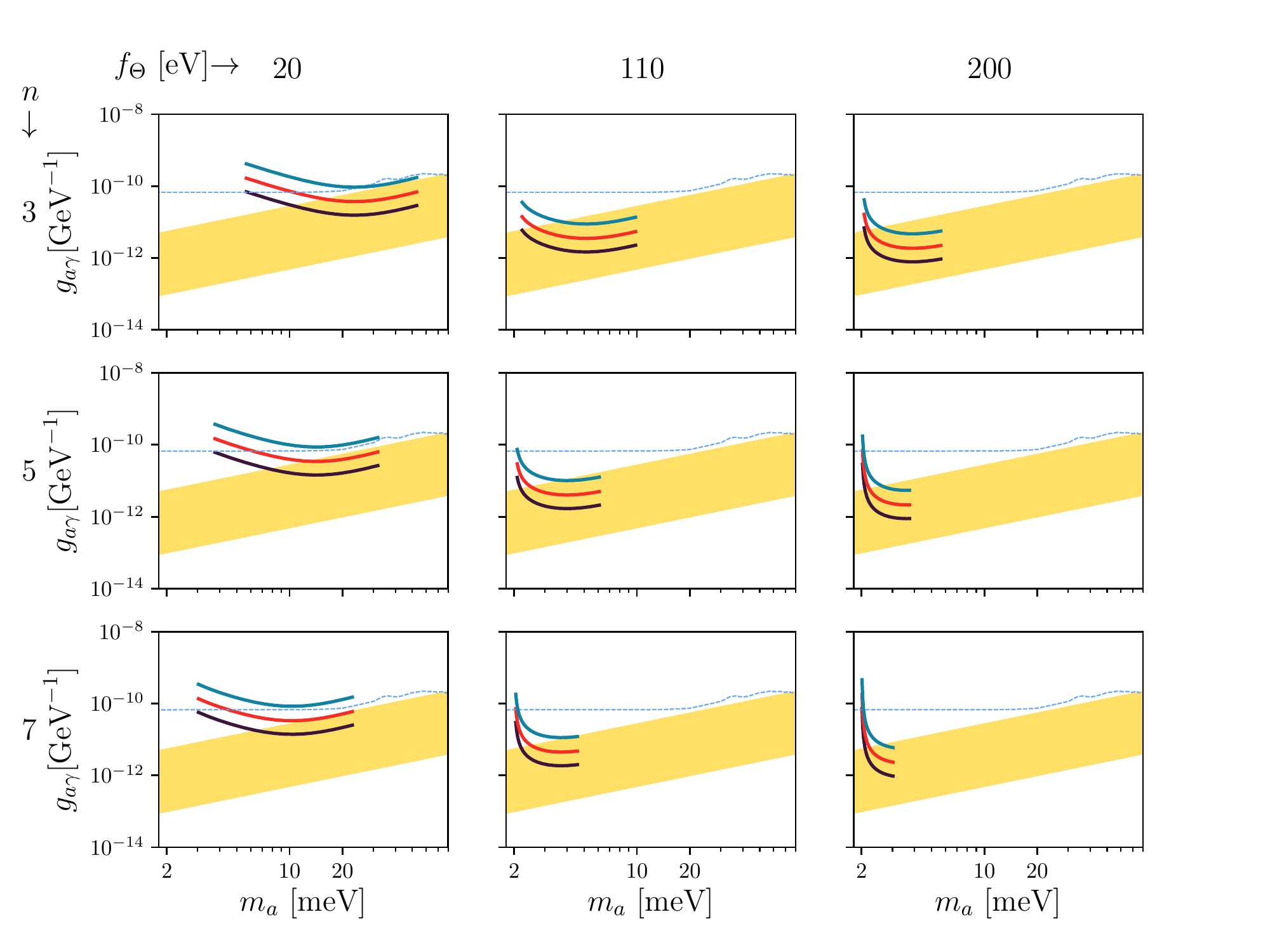} 
 	\caption{Sensitivity estimate for the DA-photon coupling $g_{a\gamma}$ varying the external $B$-field from $\SI{1}{\tesla}$ to $\SI{10}{\tesla}$. The surface area is fixed to $A=(0.2\,{\rm m})^2$. The thickness $d$ is set to the optimal thickness, cf.\ Eq.~\eqref{eq:DA_AQ_P_optimalThickness}. 
    We assume each frequency is scanned for the same amount of time, and the total scanning time is $t_\text{scan}=3\,$years. For the detector $n_b=10^{-3}\,$Hz and efficiency $\eta=1$. The yellow band represents the QCD band with $C_{a\gamma}=12.75\cdots 0.25$, cf.\ eq.~\eqref{eq:def_gag} for the definition of $C_{a\gamma}$. The dashed blue line shows the CAST limit.}
  	\label{fig:Parameterstudy} 
\end{figure}

In Fig.~\ref{fig:Parameterstudy} we study the effect of varying the AQ decay constant $f_\Theta$ and the refractive index $n$ on the scan range and sensitivity (we do not vary the AQ mass, since this has the trivial effect of changing the lower limit of the scan range). The sensitivity and other parameters are fixed as described in the previous subsection. 
Let us first discuss the scanning range. The smaller $n$ and $f_\Theta$ the larger is the axion mass range that can be probed. This is because the upper range of the scanned axion mass is determined by Eq.~\eqref{eq:Scanning_range}. 

To understand the effect of $n$ and $\fAQ$ on the sensitivity, it is enlightening to study the behaviour of the sensitivity estimates in the limit that the external $B$-field is very large, i.e. $m_a\approx\omLO\approx b\sim \frac{B_e}{n f_\Theta}$. Both sensitivity estimates in the background dominated, cf.\ Eq.~\eqref{eq:sensitivity_backgroundDominated}, and in the non-background dominated limit, cf.\ Eq.~\eqref{eq:sensitivity_NotbackgroundDominated_explicit}, are proportional to $g_{a\gamma}\sim\frac{1}{B_e}\frac{1}{d_\text{opt}^2}$, where we have assumed that $\Sigma$ does not vary too much.\footnote{Remember that $\Sigma$ would be exactly $3$, if we would choose for each axion mass that is scanned the exact optimal thickness. However, in a scanning scenario this will for practical reasons not be possible and we choose $d$ to be the optimal thickness that corresponds to the axion mass that is in the center of all axion masses that are scanned. As a consequence $\Sigma$ can also be slightly larger than $3$ in the whole axion mass that is being scanned.} Plugging in the optimal thickness we obtain the scaling behaviour: 
\begin{eqnarray}
g_{a\gamma}\sim \left(\frac{1}{B_e}\right)^{\frac{1}{6}}\, \left(\frac{1}{\fAQ}\right)^{\frac{5}{6}}\,\sqrt{n}.
\label{eq:sensitivity_scaling}
\end{eqnarray}
The strongest scaling is induced by the AQ decay constant $f_\Theta$. This view is also confirmed by the plots in Fig.~\ref{fig:Parameterstudy}. However increasing $f_\Theta$ also leads to a smaller scanning interval such that the reached $C_{a\gamma}$ in the QCD band is almost constant.
The refractive index $n$ enters in the sensitivity only weakly with a square root dependence. However for fixed $f_\Theta$ it is visible from the plots in Fig.~\ref{fig:Parameterstudy} that decreasing $n$ gives a slightly better limit on the DA-photon coupling. Furthermore, the scaling in eq.~\eqref{eq:sensitivity_scaling} only applies so long as the approximation $m_a\approx\omLO\approx b\sim \frac{B_e}{n \fAQ}$ holds. At large $\fAQ$ this approximation breaks down for suitable values of $B_e$ (either the experimental maximum, or spin flop field, whichever is lower).

\begin{figure}
	\centering
	\includegraphics[width=5.9in]{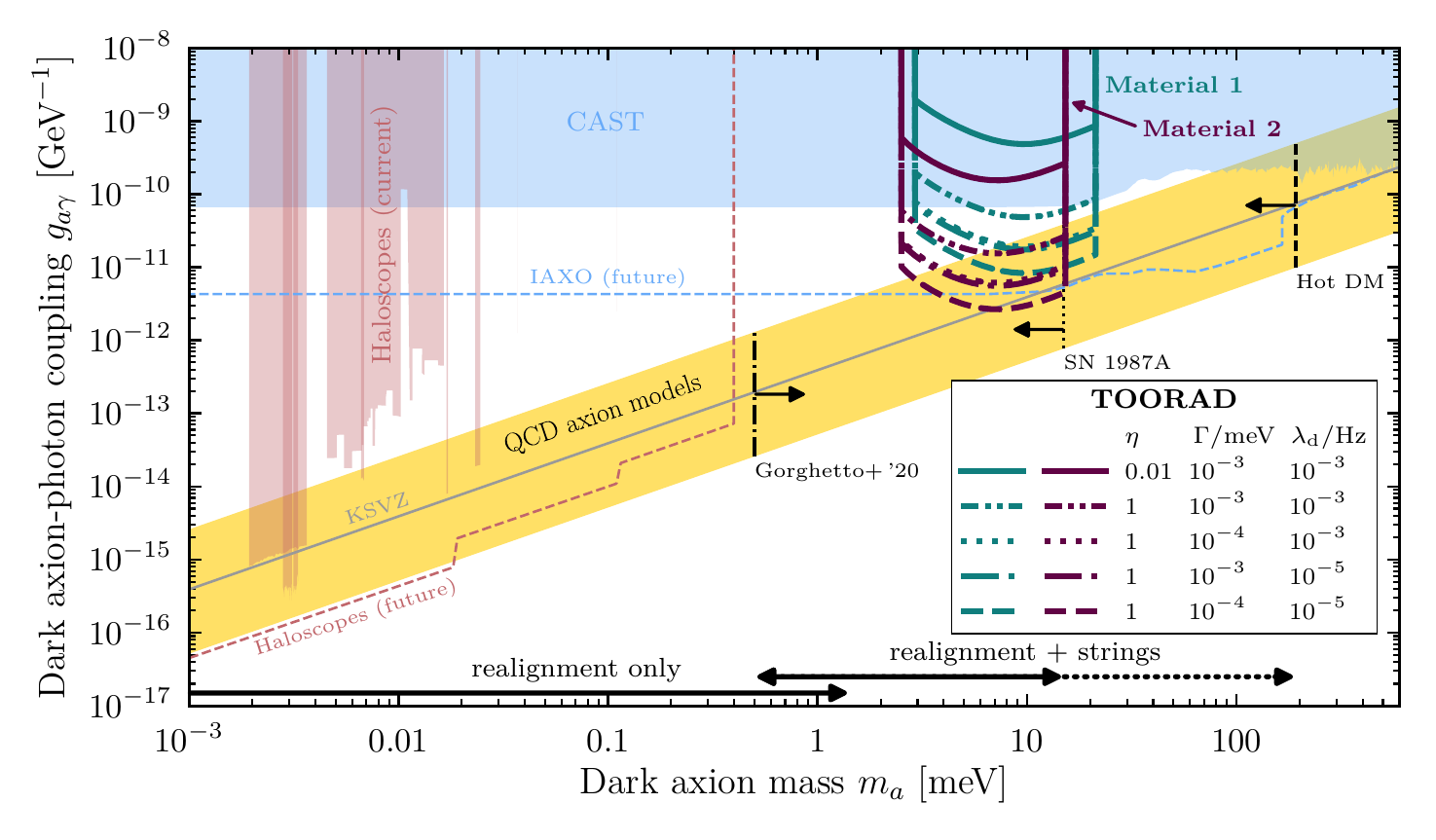}
	\caption{The projected TOORAD sensitivity for Material 1 [\BiFeSe-inspired] and Material 2 (\MnBiTe{2}{2}{5}-inspired) for different losses ($\Gamma_m=\Gamma=\Gamma_\rho$) and detector sensitivities. See Table~\ref{tab:toorad_params} for all other benchmark parameter values. We show limits and forecasts~\cite{2020_Zenodo_OHare} for CAST~\cite{hep-ex/0702006,1705.02290}, IAXO~\cite{1904.09155}, and various haloscopes~\cite{DePanfilis:1987dk,0910.5914,1706.00209,1803.03690,Du:2018uak,1903.06547,Lawson:2019brd,Braine:2019fqb,2001.05102,Beurthey:2020yuq,Backes:2020ajv}~(for $\rho_\text{loc} = \SI{0.3}{\GeV/\centi\m^3}$) as well as the bounds from hot dark matter constraints~\cite{2011.14704}, energy loss arguments in SN1987A~\cite{1906.11844}. The preferred regions cold dark matter~\cite{Aghanim:2018eyx} in the realignment scenario, and with the latest cosmic string decay calculations~\cite{2007.04990} are also indicated as horizontal arrows. The QCD~axion band encompasses all ``preferred'' KSVZ-type axion models as defined in ref.~\cite{1705.05370}, in addition to the original KSVZ and DFSZ models.\label{fig:summary_plot}}
\end{figure}

With these effects in mind, we revisit the candidate AQ material \BiFeSe\, (``Material 1''), considered in Paper~I. We estimate that this material has slightly smaller $\fAQ$, and will thus have a slightly worse sensitivity to $\gDA$ than the alternative Material 2, although it will have a narrower possible scan range. To be more optimistic with Material 1, we adopt $n=3$ for presentation (although this has a very small effect). Our results are collected in fig.~\ref{fig:summary_plot}.\footnote{Appendix~\ref{sec:axion_dm} gives more details about the QCD axion model assumptions indicated in this figure.}

\FloatBarrier

\section{Discussion and Conclusions}\label{sec:conclusions}

\subsection{Summary of Results}

The present work has developed the theory of axion quasiparticles in topological magnetic insulators, and how such materials can be used to detect axion dark matter. 

\textbf{Model of Axion Quasiparticles:} We first presented in some detail the symmetry criteria for the existence of axion quasiparticles, and the Dirac model for their realisation in topological magnetic insulators. While already known in the literature (e.g. refs.~\cite{2010NatPh...6..284L,Sekine:2014xva,Zhang_2020,Sekine2020}), these have not been shown in detail in relation to axion DM, and provide important background to the subsequent results. We laid out carefully the symmetry criteria necessary for a material to posses an AQ. Our exploration of the model sheds light on the nature of the AQ as a longitudinal magnon, i.e. a spatially and temporally varying AF spin fluctuation. It is non-linearly related to the transverse magnons of ordinary AFMR.

In order to estimate the parameters $\fAQ$ and $\mAQ$ of the model, we used the result of the \emph{ab initio} calculation given in ref.~\cite{2010NatPh...6..284L} for \BiFeSe on a cubic lattice. We rescaled the results to use updated values of the material parameters of \BiFeSe, and \MnBiTe{2}{2}{5}, for which there is not a result available in the literature. More accurate \emph{ab initio} calculations of the parameters for both \BiFeSe\, and \MnBiTe{2}{2}{5} are highly desirable. We considered multiple possible sources of loss in these materials, and attempted to estimate the contributions to the polariton linewidth. This often involved extrapolation of results obtained at different frequencies and only measured in related materials. Direct spectroscopic measurement of all these parameters is thus necessary. 

\textbf{Axion Quasiparticle Detection:} We computed explicitly the transmission function of AQ materials. This transmission function displays a magnetic field-dependent gap, and a series of resonances, which depend on the size of the loss terms. By measuring the frequency of the upper and lower ends of this gap, and the linewidths of the resonances, one could determine the parameters of the model directly. Furthermore, the gap in the polariton spectrum, and the scaling of the gap size with field strength, demonstrate directly the existence of the AQ and its coupling to the electromagnetic field via a Chern-Simons interaction. Thus, THz transmission spectroscopy can be used to discover the AQ.

The considered material candidates that can host an AQ are all antiferromagnets. Antiferromagnets exhibit an antiferromagnetic resonance (AFMR) with typical resonance frequencies in the THz regime. This raises the question how one can distinguish the AFMR from the axion-polariton resonance in the transmission spectrum. It is well known how the AFMR frequency scales with a non-zero external $B$-field~\cite{PhysRev.85.329,Keffer1953,Mills_1974}. This scaling is distinct from that of the axion-polariton resonance, which consists of a fixed resonance at $\mAQ$, and a second one near $\omLO=\sqrt{\mAQ^2+b_0^2(B/B_0)^2}$ (where $b_0=b(B_0)$ and $B_0$ is a reference scale). We expect transmission spectra of the AF axion insulator \MnBiTe{}{2}{4} to show the single AFMR, while the AQ material \MnBiTe{2}{2}{5} will show both the axion polariton resonances and AFMR. Comparing results for both materials and the $B$-field dependence will help isolate the effect of the AQ. 

\textbf{Axion Dark Matter Detection:} We developed the computation of the power output of an AQ material in the presence of axion DM. The system bears many similarities to dielectric and plasma haloscopes, and is characterised by a boost amplitude, $\beta(\omega)$. The boost amplitude increases with thicker sample sizes, and the height and width of the boost are affected by magnon and photon losses. The power is amplified by $\beta^2$ compared to a magnetized mirror, and for realistic models of the loss $10^2\lesssim \beta^2\lesssim 10^3$ with a bandwidth of order $10^{-4}$ to $10^{-3}$. 

Figure~\ref{fig:summary_plot} shows our best estimates for the discovery potential of TOORAD compared to other constraints on axion dark matter, and proposals for future experiments. The present best estimate shows that TOORAD, using a material similar to \MnBiTe{2}{2}{5} could scan an $\mathcal{O}(1)$ range in the upper half of the QCD axion model band if the SPD efficiency is very good, $\eta\approx 1$. In the best case scenario  with low dark count rate detectors the KSVZ band can be reached.

The primary difference between the two material candidates considered lies in the estimated value of $\fAQ$, with slightly higher values being favourable in the scan depth, but having a slightly narrower total range. If the spin flop transition of the material is lower than the maximum 10 T field assumed, then the scans would begin at lower frequencies, and span a slightly smaller range of masses. 

\subsection{Discussion}

\textbf{Comparison to other axion detection proposals:} We have considered detecting the dark matter axion via the axion-photon coupling, $\gDA$, combined with the mixing between the photon and the AQ. It is interesting to note that if the dark matter axion also possess a coupling to electrons, $g_{ae}$, then this can excite AFMR in the TMI via the ``axion wind'' derivative interaction~\cite{Kakhidze:1990in} (this interaction has been successfully constrained with nuclear magnetic resonance~\cite{Garconeaax4539} and ferromagnetic resonance~\cite{2018EPJC...78..703C,Crescini:2020cvl}). The AFMR axion wind interaction opens the possibility that AQ materials could measure both couplings, $\gDA$ and $g_{ae}$, with the same material by tuning to different resonant modes. This could be used to perform model discrimination between the KSVZ model, with loop suppressed electron coupling, and the DFSZ model, with leading order electron coupling. This would be an interesting line of future research. 

Similarly to dielectric and plasma haloscopes, TOORAD aims to avoid the Compton wavelength limits imposed in traditional cylindrical cavities. Most experiments try to avoid this limit through breaking translation invariance on roughly half Compton wavelength scales. Examples include dielectric haloscopes~\cite{TheMADMAXWorkingGroup:2016hpc} like MADMAX~\cite{Brun:2019lyf} and LAMPOST~\cite{Baryakhtar:2018doz}, multicavity arrays \cite{Goryachev:2017wpw,Jeong:2018ljc} such as RADES~\cite{Melcon:2018dba,Melcon:2020xvj} and hybrid approaches using dielectric loaded resonators~\cite{Morris:1984nu,Quiskamp:2020yrx} such as Orpheus~\cite{Carosi:2020akt}. In contrast, TOORAD aims to give the photon an effective mass (in the low spin wave momentum limit). In this sense, the most similar analogue in axion experimental design is a plasma haloscope~\cite{Lawson:2019brd}, which directly gives the photon a mass in the form of a plasma frequency.

The THz regime represents a unique challenge for axion detection, as it represents an intermediate regime between scales and technologies. Dielectric haloscopes have been proposed at lower~\cite{Brun:2019lyf} and higher~\cite{Baryakhtar:2018doz} frequencies. THz represents a middle ground between the use of discrete, movable disks and ${\cal O}(1000)$ layer deposited thin films implying unique engineering challenges to cover the available parameter space. 

Dish antennas~\cite{2013JCAP...04..016H} are the simplest structure to target THz, due to their broadband nature, however they lack resonant enhancement that could allow a more targeted search at higher signal to noise. Currently the only proposed dish antenna in this range is BRASS~\cite{BRASS}. 

A more recent idea in the meV range is to use the axion's coupling to phonon polaritons or magnons~\cite{Mitridate:2020kly}, however the resonance frequency in this proposal is not easily tuned, which makes scanning axion masses difficult. To cover a range of axion masses, different materials of high quality would need to be measured. Further, the single quanta measurement of such particles remains challenging~\cite{Mitridate:2020kly}.

As the field of THz axion detection is still very young, and each approach has different material or engineering challenges, it is important to have a wide range of ideas in order have a chance to look in this well motivated, but very difficult, parameter space. 

\textbf{Materials Science:} In terms of material research we have revealed there is a stark contrast between conventional strong dynamical axion response in solids and dynamical axion quasiparticle response suitable for DM detection discussed here.
\begin{itemize}
\item The axion quasiparticles for DM detection favour longitudinal spin waves with linear coupling to photons. In contrast, the heterogeneous dynamical axion field present in the chiral magnetic effect or antiferromagnetic resonance of the standard transversal spin modes does not provide within minimal models for such a coupling~\cite{Zhang_2020}. 
\item While conventional large axion response can be achieved close to the magnetic phase transition~\cite{Zhang_2020}, a DM search favours lower temperatures, ensuring sharper resonance linewidth free of thermal and scattering disorder. 
\item The static quantised axion insulators are protected by axion odd symmetries such as spatial inversion (parity). Our dynamical axion quasiparticles favour $\mathcal{PT}$ symmetric systems: $\mathcal{PT}$ allows for Dirac quasiparticles enhancing the (dynamical) nonquantized AQ response by allowing tunability close to the topological phase transition.
\end{itemize}
Antiferromagnetism is favourable in many ways for axion DM detection. Reasons for this include its compatibility with tunable axionic Dirac quasiparticles~\cite{Smejkal2016}, availability of semiconducting band-structure with potentially large band-gaps, high critical temperatures, and large spin-flop fields. Furthermore, multi-sublattice systems can provide for a combination of separated heavy atomic elements with strong spin-orbit interaction and lighter magnetic elements.

\textbf{Materials wishlist:} We close with stating the desirable properties of an AQ material for axion DM detection.
\begin{itemize}
\item Longitudinal spin wave mass, $\mAQ$, in the meV range. The goal is to detect the QCD axion in this mass range. With much smaller $\mAQ$ there are already existing technologies, while for much larger values the QCD axion is already excluded.
\item Decay constant, $\fAQ$, in the 10 to 100 eV range.\footnote{Recall that in the Dirac model $\fAQ^2=2 M_0^2 J$ where $M_0$ is the bandgap and $J$ is the spin wave stiffness.} For $\fAQ$ much larger than 100 eV the AQ is not strongly coupled enough to the $\Theta$ term for efficient mixing. Another way to express this requirement is that the polariton gap for fields of order 1 T should be of order $\mAQ$.
\item Low refractive index ($n\lesssim 5$) and high resistivity ($\rho>10^3$ meV$^{-1}$) in THz, preferably measured from the axion-polariton spectrum resonance.
\item Low impurity density: impurity separation scale of microns or larger.
\item High spin flop field. This should definitely exceed 1 T for sufficiently large power output. Larger spin flop fields permit a wider scan range.
\item High N\'{e}el temperature. The experiment can be operated in a dilution refrigerator with $T\ll 4$ K. However, the further this is below the N\'{e}el temperature, the better, since we expect magnon losses to decrease for $T\ll T_N$.
\item Ability to manufacture samples with thickness in excess of 1 mm. Ultimately one must also machine multiple samples together into a large surface area disk.
\end{itemize}
We have shown that, with plausible assumptions, \MnBiTe{2}{2}{5} and \BiFeSe both satisfy many of these requirements, although we expect the AQ phase of \MnBiTe{2}{2}{5} to be more stable, since it does not require magnetic doping. If it can be proven that any material satisfies the above requirements, then, in combination with existing detector and magnet technology, such a material can be used to make an effective search for axion dark matter in the theoretically well-motivated mass range near 1 meV.

\acknowledgments
We thank Bobby Acharya, Caterina Braggio, {Nicol\`o} Crescini, Matthew Lawson, Erik Lentz, Chang Liu, Eduardo Neto, Naomi Nimubona, Alireza Qaiumzadeh, Andreas Ringwald, and David Tong for useful discussions. DJEM, SH, and MA are supported by the Alexander von Humboldt Foundation and the German Federal Ministry of Education and Research. JSE is supported through Germany's ExcellenceStrategy - EXC 2121 ``Quantum Universe" - 390833306. KCF was supported in part by Army Research Office under Cooperative Agreement Number W911NF-17-1-0574. FC-D is supported by STFC grant ST/P001246/1, Stephen Hawking Fellowship EP/T01668X/1. EH is supported by STFC grant ST/T000988/1. AM is supported by the
European Research Council under Grant No. 742104 and  by the Swedish Research Council (VR) under Dnr 2019-02337 ``Detecting Axion Dark Matter In The Sky And In The Lab (AxionDM)". AS is supported by the Special Postdoctoral Researcher Program of RIKEN. LS acknowledges the EU FET Open RIA Grant No. 766566, the Elasto-Q-Mat (DFG SFB TRR 288), Czech Science Foundation Grant No. 19-28375X, and Sino-German DFG project DISTOMAT. This research was supported by the Munich Institute for Astro- and Particle Physics (MIAPP) which is funded by the Deutsche Forschungsgemeinschaft (DFG, German Research Foundation) under Germany´s Excellence Strategy – EXC-2094 – 390783311.

\FloatBarrier
\appendix

\section{Antiferromagnetic Resonance and Magnons for Particle Physicists}\label{appendix:AFMR}
\subsection{Effective Field Theory of AFMR}\label{appendix:eft_afmr}

We follow Refs.~\cite{Hofmann:1998pp,Burgess:1998ku}, and present the effective field theory of antiferromagnetic resonance (EFT of AFMR), which we believe is illuminating, especially from a particle physics perspective.

The EFT of AFMR considers the dynamics of the AF magnetization $\vc{n}$ considered as a field in the continuum limit of the Heisenberg model of the magnetic lattice, which is equivalent to the Hubbard model in the half-filling limit, as discussed in section \ref{sec:params}. The magnetic lattice consists of $A$ sites and $B$ sites, with spins $\vc{S}_A$ and $\vc{S}_B$ at each site, and $\vc{n}=(\vc{S}_A-\vc{S}_B)/2$. The symmetry group $G=SO(3)$ is related to the internal rotations of $\vc{n}$ (not spatial rotations). This symmetry is broken by the groundstate AF order, $\langle \vc{n}\rangle = (\langle \vc{S}_A\rangle-\langle \vc{S}_B\rangle)/2$ (which can be normalised to unity) and is invariant under the group $H=SO(2)$ of rotations about the axis. Magnetic order implies that the groundstate breaks time translation invariance, $\mathcal{T}$, which flips the spin orientations. However, the groundstate preserves an effective time translation invariance $\widetilde{\mathcal{T}}=\mathcal{T}\mathcal{S}$, where $\mathcal{S}$ swaps the $A$ lattice sites for the $B$ lattice sites. This leads, as we shall see, to a ``relativistic'' dispersion relation for AF spin waves. Spin-orbit effects (finite electron mass corrections) lead to explicit breaking of $SO(3)$, which can be considered as a perturbation, and leads to a preferred ``easy axis'' related to a direction in the crystal lattice.

The Lagrangian for fluctuations in $\vc{n}$ must be invariant under the coset space $G/H$, which has the symmetry group of rotations on the surface of the two-sphere, $S^2$, and imposes the restriction $\vc{n}\cdot\vc{n}=1$. This restriction can be imposed as a constraint and expanded for small perturbations in Cartesian coordinates for $\vc{n}$, which is sufficient to derive the normal modes and dispersion relation. More generally, the constraint can be imposed by the correct choice of coordinates and metric, in this case the $SO(3)$ invariant metric on $S^2$, and leads to the full non-linear model in polar coordinates. We begin with the first case, since we can align the coordinates with the spacetime directions and arrive at well known results quickly, while the second case is illuminating since it preserves the symmetries manifestly, and leads to insights into the nature of the longitudinal mode. 

\subsubsection{AFMR in Cartesian Coordinates}

The Lagrangian at leading order in derivatives is:
\be
\mathcal{L} = \frac{F_1^2}{2}\dot{\vc{n}}\cdot\dot{\vc{n}}-\frac{F_2^2}{2}\nabla\vc{n}\cdot\nabla\vc{n} \, ,
\label{eqn:eft_L0}
\ee
where $F_1^2$ is the spin wave stiffness, and $F_2^2=v^2 F_1^2$ with $v$ the spin wave speed. The external fields are the applied field, $\vc{H}_0$, the probe photon with fields $\vc{E}_\gamma$, $\vc{H}_\gamma$ and wavevector $\vc{k}_\gamma$, and the anisotropy field, $\vc{H}_A$, which defines the easy-axis in the material. In the simplest AFMR geometry we consider the applied field to be parallel to the $z$-axis, which is also parallel to the anisotropy field. We further consider the probe photon (RF-field) moving along the positive $z$-axis, polarised in the $y$-direction. The fields are thus:
\bea
\vc{k}_\gamma&=(0,0,k)\, ,\notag \\
\vc{H}_\gamma&=(H_\gamma,0,0)\, , \notag\\
\vc{E}_\gamma&=(0,E_\gamma,0)\, , \notag\\
\vc{H}_0&=(0,0,H_0)\, , \notag\\
\vc{H}_A&=(0,0,H_A)\, .
\eea
For ordinary AFMR, the photon electric field is decoupled from the system. 

The applied field $H_0$ and the photon magnetic field are coupled into the Lagrangian Eq.~\eqref{eqn:eft_L0} by replacing the derivatives with $SO(3)\cong SU(2)$ covariant derivatives:
\be
\partial_\mu n_a\rightarrow D_\mu n_a = \partial_\mu n_a+ \epsilon_{abc}f_{\mu b}n_c\, ,
\ee
where $n_a$ are the directions in the $SO(3)$ group space, $\mu=0,1,2,3$ as subscript is the spacetime index (which should not be confused with the Bohr magneton $\mu_B$), $\epsilon_{abc}$ is the antisymmetric symbol in three dimensions with $\epsilon_{123}=1$ (i.e. the structure constants of $SU(2)$), and $f_{\mu b}$ is the applied field. For an applied magnetic field we have $\mu_B H_i=f_{0i}$ which allows us to relate the group space index $a$ to the spacetime axis $i=1,2,3$. At lowest order in the applied fields, this result can be understood by appealing to the interaction Lagrangian:
\bea
\mathcal{L}_{\rm em} = -\mu_B \vc{s}\cdot \vc{H}\, , \quad \vc{s} = F_1^2 ( \dot{\vc{n}}\times \vc{n} )\, , \notag \\
\Rightarrow \mathcal{L}_{\rm em} = \mu_B F_1^2 \epsilon_{ijk} \dot{n}_i H_j n_k\, ,
\label{eqn:external_field}
\eea
where the spin density $\vc{s}$ follows from the leading order term in the derivative expansion of the Noether current due to the $SO(3)$ invariance.

The anisotropy field is included in the Lagrangian via a perturbation of the form $\Delta\mathcal{L}=\mathcal{O}^a n_a$ and for our field geometry is given by:
\be
\Delta\mathcal{L}=\mu_B \Sigma_s H_A n_3\,
\ee
where $\Sigma_s=S/V_{\rm u.c.}$ is the ``staggered magnetization'', $(S_A-S_B)/2$, in the unit cell.

In order to derive the dispersion relation (the propagator), we only require the quadratic Lagrangian. Anticipating the well-known Keffer-Kittel result for the AFMR polarisations~\cite{PhysRev.85.329} we use $n_1$ and $n_2$ as coordinates, and Taylor expand for small $n_3$ using the constraint, i.e. $n_3=(1-n_1^2-n_2^2)^{1/2}$. Momentum conservation demands that $\vc{k}=\vc{k}_\gamma$, and with the given geometry this simplifies the problem to effectively one-dimensional along the $z$(3)-axis. After some basic algebra, the quadratic Lagrangian is found to be:
\bea
\mathcal{L}&=&\frac{F_1^2}{2}\left[\dot{n}_1^2+\dot{n}_2^2\right]-\frac{F_2^2}{2}\left[(\partial_zn_1)^2+(\partial_zn_2)^2\right] \\
&& -F_1^2\mu_B H_\gamma [\dot{n}_2+\mu_B H_0 n_1]+F_1^2\mu_B H_0 [\dot{n}_2n_1-\dot{n}_1n_2+\mu_B H_0 (n_1^2+n_2^2)] -\frac{\mu_B \Sigma_sH_A}{2}(n_1^2+n_2^2)\, .\nonumber
\eea
The first line is the kinetic term, and the second line includes the effects of the external fields. The photon field has been considered a perturbation, and thus couples linearly to the fields $n_i$ in the Lagrangian. The photon field thus acts as an oscillating source term in the equations of motion. On the other hand $H_0$ and $H_A$ couple to quadratic combinations of $n_i$, and affect the dispersion relation.

The equations of motion are:
\bea
\ddot{n}_1-2\mu_B H_0\dot{n}_2+\left(v^2k^2+\frac{\mu_B\Sigma_sH_A}{F_1^2}-\mu_B^2H_0^2\right)n_1&=\mu_B^2H_\gamma H_0 \, , \\
\ddot{n}_2-2\mu_B H_0\dot{n}_1+\left(v^2k^2+\frac{\mu_B\Sigma_sH_A}{F_1^2}-\mu_B^2H_0^2\right)n_2&=\mu_B\dot{H}_\gamma \, .
\eea
To derive the dispersion relation, we consider the homogeneous equation with the right hand side set equal to zero, and move to frequency space by Fourier transforming $t\rightarrow \omega$. The system is diagonalised by the complex fields $n_\pm = n_1\pm i n_2$ leading to the system:
\be
\omega_\pm^2\mp 2\mu_B H_0 \omega_\pm -\left(v^2k^2+\frac{\mu_B\Sigma_sH_A}{F_1^2}-\mu_B^2H_0^2\right) = 0\, ,
\ee
which is solved by
\bea
\omega_+&=\mu_B H_0 \pm \sqrt{v^2k^2+\frac{\mu_B\Sigma_s H_A}{F_1^2}}\, , \nonumber\\
\omega_-&=-\mu_B H_0 \pm \sqrt{v^2k^2+\frac{\mu_B\Sigma_s H_A}{F_1^2}}\, .
\label{eqn:afmr_dispersion}
\eea

\begin{figure}
\center
\includegraphics[width=0.9\textwidth]{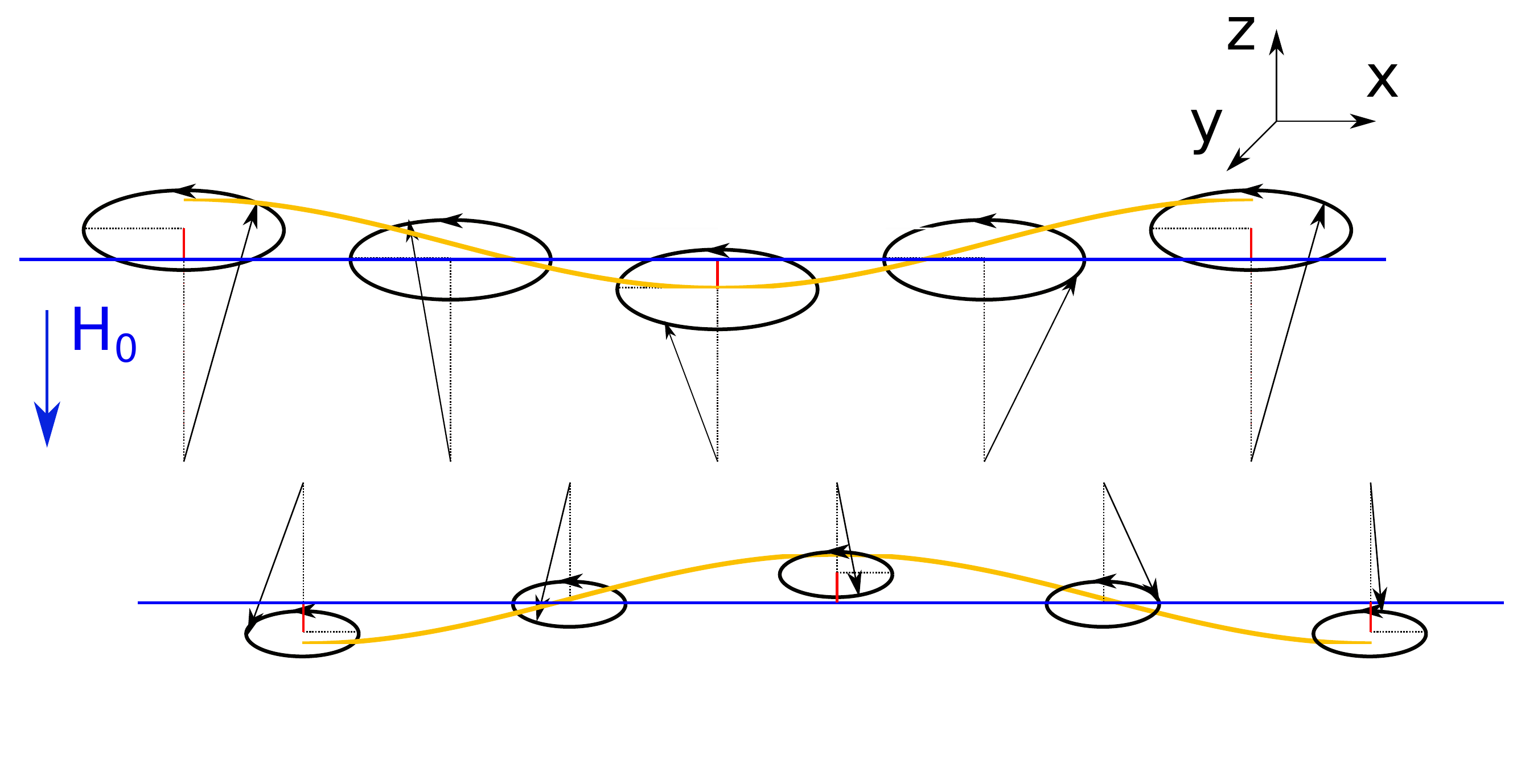}
\caption{AFMR spin wave, with $|\vc{n}|^2=1$, indicating the higher order change in $n_3$ associated with the spin precession. }
\label{fig:af_cartoon}
\end{figure}

The dispersion relation, Eq.~\eqref{eqn:afmr_dispersion} for the fields $n_\pm = n_1\pm i n_2$ displays all the well-known properties of AFMR. The two modes $n_\pm=n_1\pm in_2$ correspond to clockwise and anticlockwise precession of the N\'{e}el vector~\cite{PhysRev.85.329}. The resulting spin wave is depicted in Fig.~\ref{fig:af_cartoon}. The constraint $|\vc{n}|^2=1$ leads to an oscillation of $n_3$ accompanying the precession. As shown in Appendix~\ref{appendix:LLG}, $n_3$ in this case oscillates with a frequency twice that of the AFMR. However, if $|\vc{n}|^2=1$, then $n_3$ in not an independent polarisation and its fluctuation does not change the length of the N\'{e}el vector.

In the absence of $H_A$, the dispersion relation is linear in $k$. The application of $H_A$ induces a ``mass term'', i.e. a term inducing a gap and leading order quadratic piece in the dispersion relation near $k=0$:
\be
m_s^2=\frac{\mu_B\Sigma_s H_A}{F_1^2}
\ee
Rearranging, we find 
\be
m_s^2F_1^2 = \mu_B\Sigma_s H_A\, ,
\label{eqn:magnon_GelllMann}
\ee 
which has the form $m_s^2F_1^2 = \text{(spontaneous)}\times\text{(explicit)}$ symmetry breaking, and is the AF analogue of the Gell-Mann-Oakes-Renner relation~\cite{PhysRev.175.2195} for pions~\cite{Hofmann:1998pp} (and also the QCD axion). Furthermore, since $F_1^2\propto \Sigma_s$ this fits with the microscopic interpretation of $F_1^2$ as arising from the staggered magnetization angular momentum per unit cell mentioned above.

The applied field $H_0$, rather than leading to a mass term, instead induces a linear shift in the frequency, the ``Kittel shift'', which arises from an effective (anti-)damping term and ``negative mass squared'' in the equations of motion for $n_\pm$.

The exchange field, $H_E$, is not incorporated directly in our treatment of EFT. However, as noted in Ref.~\cite{Hofmann:1998pp}, we should fix the EFT parameters with reference to a microscopic theory. The microscopic theory (e.g. Ref.~\cite{Wagner}) gives the spin wave mass from the energy gap:
\be
m_s^2 = \mu_B^2H_A(2H_E+H_A)\, ,
\label{eqn:mass_beyond_eft}
\ee
where $H_E$ is the exchange (or Weiss) field. The second term of Eq.~\eqref{eqn:mass_beyond_eft} is not present in the EFT, which is linear in $H_A$. Indeed, EFT is valid in the limit $H_A/H_E\ll 1$, and breaks down for large anisotropy fields~\cite{Hofmann:1998pp}. Comparing Eq.~\eqref{eqn:magnon_GelllMann} with the first term of Eq.~\eqref{eqn:mass_beyond_eft}  we identify $H_E=\Sigma_s/2\mu_B F_1^2$ leading to:
\be
F_1^2 = \frac{\Sigma_s}{\mu_B H_E} = \frac{S}{\mu_B H_E V_{\rm u.c.}}\, . 
\ee

The EFT of AFMR is based on the mean field Heisenberg model. The Heisenberg model is the strong coupling limit of the Hubbard model (the fundamental model on which our theory of the AQ is based), with different perturbative degrees of freedom. In the Heisenberg model with nearest neighbour interactions the Hamiltonian is:
\begin{equation}
H = J_H \sum_{<i j>} {\bf \sigma}_i \cdot {\bf \sigma}_j - D \sum_i (\sigma_i^z)^2
\label{eqn:heisenberg}
\end{equation}
where the first sum is over the spins ${\bf \sigma}_i$ and ${\bf \sigma}_j$ on adjacent lattice sites. The anisotropy field is given by $H_A = \frac{2 S D}{g \mu_B}$. The anisotropy field arises due to the spin orbit coupling which explicitly breaks the $SO(3)$ symmetry of the Heisenberg model due to finite electron mass corrections. The Heisenberg EFT is valid for weak spin orbit coupling $H_A\ll H_E$.

\subsubsection{AFMR in Polar Coordinates}

In the following we explicitly follow the treatment of Ref.~\cite{Burgess:1998ku}, and use the field geometry: 
\bea
\vc{k}_\gamma&=(k,0,0)\, , \notag\\
\vc{H}_\gamma&=(0,0,H_\gamma)\, , \notag\\
\vc{E}_\gamma&=(0,E_\gamma,0)\, , \notag\\
\vc{H}_0&=(H_0,0,0)\, , \notag\\
\vc{H}_A&=(H_A,0,0)\, . \label{eqn:geometry_afmr_polar}
\eea
In terms of polar coordinates, we have
\be
n_1 = \sin\theta\cos\phi \, , \quad n_2 =\sin\theta\sin\phi \, , \quad n_3 = \cos\theta\, ,
\label{eqn:polar_coordinates}
\ee
AF order breaks the $SO(3)$ internal symmetry of the spins down to the coset space $SO(3)/SO(2)$ which has the geometry of $S^2$. The dynamics of the Goldstone modes can be expressed using the polar coordinates. The easy axis has coordinates $\theta_0$, $\phi_0$, and we normalise the order parameter to unity. The Goldstone mode Lagrangian at lowest order in derivatives is:
\be
\mathcal{L} = \frac{F_1^2}{2}\gamma_{ab}\dot{\vartheta}^a\dot{\vartheta}^b - \frac{F_2^2}{2}\gamma_{ab}\nabla\vartheta^a\cdot\nabla\vartheta^b\, ,
\ee 
where $a,b=\theta,\phi$ and the metric $\gamma_{ab}$ is the round metric on the sphere, $\gamma_{ab}=\text{diag}[1,\sin^2\theta]$. The dynamics is easiest to express choosing $n_1$ to be the easy axis, $\theta_0=\pi/2$, $\phi_0=0$. We then find trivially that, at leading order in fluctuations:
\be
n_1 = 1-\frac{\delta\theta^2}{2}-\frac{\delta\phi^2}{2} \, , \quad n_2 =\delta\phi \, , \quad n_3 = -\delta\theta\, .
\ee
The longitudinal fluctuation, i.e. the change in $\vc{n}$ projected along the direction of travel of the spin wave, $n_1$ in this case, is quadratic in the Goldstone modes, while the transverse fluctuations are linear. Note, however, that in these coordinates we always have explicitly two polarizations and no change in the length of the N\'{e}el vector. The anisotropy field perturbs the Lagrangian as above, $\Delta \mathcal{L}_A=\mu_B\Sigma_s H_A n_1$, and induces a mass term for $\delta\theta$ and $\delta\phi$. The interaction with applied fields follows exactly as in the Cartesian case using the relations in Eq.~\eqref{eqn:polar_coordinates}.

\subsubsection{Longitudinal Spin Waves in the Heisenberg Model}

The AQ is related to longitudinal fluctuations of the N\'{e}el vector (i.e. those in the direction of the anisotropy field), but this is \emph{not} equivalent to a third \emph{longitudinal polarisation} that changes the length of the vector. Such a ``true'' longitudinal mode is the mode that breaks $SO(3)$ giving rise to AF order, i.e. the Higgs-like radial mode (see also refs.~\cite{Bar:2003ip,Imboden:2003jpa}). When writing down the model, we need to be careful that it respects all the symmetries. The field $\vec{\phi}=(\langle\vc{S}_A\rangle-\langle\vc{S}_B\rangle)/2$ is the total (staggered) magnetization, the N\'{e}el vector, and we can write it as:
\be
\vec{\phi} = \rho(x)(n_1(x),n_2(x),n_3(x)) = \rho(x)\vc{n}\, .
\ee
The field $\rho$ is the longitudinal polarization, while $\vc{n}$ is the AFMR field introduced above with $|\vc{n}|^2=1$. There is a maximum magentization given by the spin density, and a minimum value pointing in the opposite direction. In our conventions, $\phi$ is dimensionless and normalized to a maximum of unity, thus $|\phi|^2\leq 1$~\cite{Haldane1983}. 

The constraint $|\phi|^2\leq 1$ can be enforced naturally by considering the EFT given by the $SO(3)$ invariant metric on $S^3$ with unit radius. We use the field coordinates:
\be
\vec{\varphi} = (\alpha,\theta,\phi)\, ,
\ee
where $\theta$, $\phi$ are the AFMR variables in polar coordinates introduced above, and $\alpha$ is a third polar angle. The metric is
\be
ds^2 = g_{AB}d\varphi^Ad\varphi^B= d\alpha^2 + \sin^2\alpha d\Omega_2\, ,
\ee
where $A,B=\alpha,\theta,\phi$, and $d\Omega_2=\gamma_{ab}d\vartheta^ad\vartheta^b$ is the round metric on $S^2$, and $\vec{\vartheta}=(\theta,\phi)$ as above. We see that the polar angle $\alpha$ gives the radius of the $S^2$ submanifold of $S^3$, as desired and with the correct normalisation, $\rho=\sin\alpha$. We can interpret $\alpha$ as the angle between the spins in a ``bending mode''.

The Lagrangian in the absence of external fields is:
\be
\mathcal{L} = \frac{F_1^2}{2}g_{AB}\dot{\varphi}^A\dot{\varphi}^B-\frac{F_2^2}{2}g_{AB}\nabla\varphi^A\nabla\varphi^B\, .
\ee
Specifying the anisotropy field allows us to identify the polar axis as $n_1$ as above. The anisotropy field introduces explicit symmetry breaking and a potential for $\alpha$, $V(\alpha)\propto -n_1\propto -\sin\alpha$, which is minimized at $\alpha=\pi/2$. To consider the fluctuations, we write $\alpha=\pi/2-\sigma$ and $\sigma$ is the angular field giving rise to the fluctuation in $\rho$, i.e. the third magnon polarization. It has quadratic Lagrangian:
\be
\mathcal{L} = \frac{F_1^2}{2}\dot{\sigma}^2-\frac{F_2^2}{2}\nabla\sigma^2-\mu_B\Sigma_s H_A\frac{\sigma^2}{2}, .
\ee
The  field $\sigma$ couples to the other AFMR fields via the metric $g_{AB}$:
\be
\mathcal{L} = \cos^2\sigma\left[ \frac{F_1^2}{2}\gamma_{ab}\dot{\vartheta}^a\dot{\vartheta}^b-\frac{F_2^2}{2}g_{ab}\nabla\vartheta^a\nabla\vartheta^b\right]\, .
\ee
Expanding $\cos^2\sigma = 1-\sigma^2$ for the quadratic Lagrangian we see that at leading order we obtain the angular AFMR theory from above, and $\sigma$ is decoupled. Similarly, $\sigma$ is decoupled from the external fields in the quadratic Lagrangian, since the spin density, $\vc{s}=F_1^2 \dot{\phi}\times \phi$, only contains $\sigma$ at cubic order. Thus, in this $S^3$ EFT of  of the Heisenberg model, the $\sigma$ degree of freedom corresponding to changes in the length of the N\'{e}el vector is stabilised by the anisotropy field, and is neither excited by external fields nor mixes with the transverse AFMR polarisations. Could this mode be the AQ? We take the general expression for $\dyAQ$ in eq.~\eqref{eqn:AQ_from_n} and expand $n_A$ in the angular fields. Once again, $\dyAQ$ is quadratic in all the variables of this model, including $\sigma$. We have not been able to obtain a quadratic kinetic term for $\dyAQ$ from an $SO(3)$ invariant EFT including only the N\'{e}el order parameter. 

The preceding discussion suggests a possible solution to the problem of the EFT of the AQ. We notice that $S^3$ is in fact the spin group $Sp(1)=Spin(3)=SU(2)$. Furthermore $SU(2)\cong SO(3)/\mathbb{Z}_2$, and for the AQ we are concerned with models that break the discrete symmetries $P$ and $T$. This suggests using a \emph{complex} field $\phi$ in the fundamental 2-dimensional representation of a \emph{chiral} $SU(2)$ to represent the AF order parameter, which now has four real degrees of freedom. Thus, after SSB this would give \emph{three} goldstone modes: two ``charged'' goldstones, giving the transverse magnons, and one ``neutral'' goldstone, which we assume will be the longitudinal magnon. Each goldstone corresponds to a $U(1)$ subgroup of $SU(2)$. The neutral goldstone is a pseudoscalar, and thus this $U(1)$ group is itself chiral, i.e. a Peccei-Quinn symmetry. The Dirac fermions in the band structure should be charged under this symmetry, such that they acquire chiral rotations (``$m_5$'' mass) governed by the longitudinal mode. Just like the axion and the neutral pion, this new goldstone mode can now couple to $\vc{E}\cdot\vc{B}$ via the chiral anomaly. We have not, unfortunately, been able to work out this theory completely.

\subsection{The Landau-Lifshitz Equations}\label{appendix:LLG} 
In this appendix, following Refs.~\cite{Sekine2016a,Sekine2017}, we describe an antiferromagnetic resonance (AFMR) state using the Landau-Lifshitz equation.
We consider the action of the N\'{e}el field described by the non-linear sigma model \cite{Haldane1983},
\begin{align}
S_{\mathrm{AF}}=g^2 J\int dtd^3r \left[(\partial_\mu \bm{n})\cdot(\partial^\mu \bm{n})-\Delta_0^2\bm{n}^2\right].\label{Action-NLSigma}
\end{align}
In order to implement a little more realistic condition in Eq.~(\ref{Action-NLSigma}), we take into account a small net magnetization $\bm{m}$ satisfying the constraint $\bm{n}\cdot\bm{m}=0$ with $|\bm{n}|=1$ and $|\bm{m}|\ll 1$.
Furthermore, we assume the case of AF insulators with easy-axis anisotropy.
Then a modification of Eq. (\ref{Action-NLSigma}) gives the free energy of such AF insulators as \cite{LL-book,Hals2011}
\begin{align}
F_{\mathrm{AF}}=\int d^3r \left[\frac{a}{2}\bm{m}^2+\frac{A}{2}\sum_{i=x,y,z}(\partial_i\bm{n})^2-\frac{K}{2}n_z^2-\bm{H}\cdot\bm{m}\right],
\label{free-energy-AF}
\end{align}
where $a$ and $A$ are the homogeneous and inhomogeneous exchange constants, respectively, and $K$ is the easy-axis anisotropy along the $z$ direction.
The fourth term is the Zeeman coupling with $\bm{H}=g\mu_B\bm{B}$ being an external magnetic field.

In the case in which a dc magnetic field $\bm{H}_0$ and an ac magnetic field (i.e., RF field) $\bm{h}(t)$ are applied to the AF insulator, the total magnetic field in Eq.~(\ref{free-energy-AF}) is
\begin{align}
\bm{H}=\bm{H}_0+\bm{h}(t),
\end{align}
where $\bm{H}_0=g\mu_B B\bm{e}_z$ with $B$ being much weaker than both the AF exchange coupling and easy-axial anisotropy and $\bm{h}(t)=\bm{h}_{\mathrm{RF}}e^{-i\omega_{0}t}$.
Here, $\bm{e}_z$ is the unit vector parallel to the easy axis of the AF order.
Now we study the dynamics of $\bm{m}$ and $\bm{n}$ phenomenologically, i.e., based on the Landau-Lifshitz-Gilbert (LLG) equation \cite{Hals2011,Sekine2016a,Sekine2017}.
From the free energy of the system $F_{\rm AF}$, the effective fields for $\bm{n}$ and $\bm{m}$ are given by
\begin{align}
\bm{f}_n=-\frac{\delta F_{\rm AF}}{\delta\bm{n}}&=A\bm{n}\times(\nabla^2\bm{n}\times\bm{n})+Kn_z\bm{e}_z-(\bm{n}\cdot\bm{H})\bm{m},\nonumber\\
\bm{f}_m=-\frac{\delta F_{\rm AF}}{\delta\bm{m}}&=-a\bm{m}+\bm{n}\times(\bm{H}\times\bm{n}),
\label{effective-fields}
\end{align}
The LLG equation is given by
\begin{align}
\dot{\bm{n}}&=(\gamma\bm{f}_m-G_1\dot{\bm{m}})\times\bm{n},\nonumber\\
\dot{\bm{m}}&=(\gamma\bm{f}_n-G_2\dot{\bm{n}})\times\bm{n}+(\gamma\bm{f}_m-G_1\dot{\bm{m}})\times\bm{m},
\label{LLG-eq}
\end{align}
where $\gamma=1/\hbar$ and $G_1$ and $G_2$ are dimensionless Gilbert damping constants.
For the purpose of deriving the AFMR state, we may neglect the Gilbert damping constants.
Then, the LLG Eq.~(\ref{LLG-eq}) is simplified as
\begin{subequations}
\begin{align}
\dot{\bm{n}}&=\gamma(-a\bm{m}+\bm{H})\times\bm{n},\label{LLG-eq2a}\\
\dot{\bm{m}}&=\gamma Kn_z\bm{e}_z\times\bm{n}+\gamma\bm{H}\times\bm{m},
\label{LLG-eq2b}
\end{align}
\end{subequations}
where we have assumed that $\bm{n}$ is spatially uniform, and we have used $|\bm{n}|^2=1$ and an identity for matrices $\bm{A}\times(\bm{B}\times\bm{C})=(\bm{A}\cdot\bm{C})\bm{B}-(\bm{A}\cdot\bm{B})\bm{C}$.
After some straightforward matrix algebra, we arrive at the following equation for the N\'{e}el field:
\begin{align}
\bm{n}\times\ddot{\bm{n}}+\omega_K\omega_a n_z\bm{e}_z\times\bm{n}-\gamma^2(\bm{n}\cdot\bm{H})\bm{H}\times\bm{n}+2\gamma(\bm{n}\cdot\bm{H})\dot{\bm{n}}+\gamma(\bm{n}\cdot\dot{\bm{H}})\bm{n}=\gamma\dot{\bm{H}}.
\label{Eq-for-n}
\end{align}

To obtain the AFMR state, where all the spins are fluctuating uniformly, we assume the dynamics of the N\'{e}el vector and the total magnetization around the easy axis as 
\begin{align}
\bm{n}(t)=\bm{e}_z+\delta\bm{n}(t)\ \ \ \ \mathrm{and}\ \ \ \ \bm{m}(t)=\delta\bm{m}(t), 
\end{align}
denoting that $\delta\bm{n}(t)$ and $\delta\bm{m}(t)$ are the small fluctuation components with $|\delta\bm{n}|, |\delta\bm{m}|\ll1$.
Substituting this form into Eq.~(\ref{Eq-for-n}), and then linearizing and Fourier transforming $\delta\bm{n}(t)=\int \delta\tilde{\bm{n}}(\omega)e^{-i\omega t}d\omega/(2\pi)$, Eq. (\ref{Eq-for-n}) reduces to \cite{Sekine2016a,Sekine2017}
\begin{align}
2i\omega_H\omega\delta\tilde{\bm{n}}/\omega_a
+\left[ \left(\omega^2+\omega_{H}^2\right)/\omega_{a}-\omega_{K}\right]
\bm{e}_z\times\delta\tilde{\bm{n}}
=\bm{\mathcal{D}}\delta(\omega_{0}-\omega),
\label{LLG-linearized}
\end{align}
where $\omega_{H}=\gamma g\mu_B B$, $\omega_{a}=\gamma a$, $\omega_{K}=\gamma K$, and $\omega_{0}$ is the frequency of the RF field [$\bm{h}(t)=\bm{h}_{\mathrm{RF}}e^{-i\omega_{0}t}$].
In Eq.~(\ref{LLG-linearized}), $\bm{\mathcal{D}}=-i\gamma\omega_0(h_{\mathrm{RF}}^x\bm{e}_x+h_{\mathrm{RF}}^y\bm{e}_y)$ is understood as the ``driving force'' vector causing the AFMR.
Equation~(\ref{LLG-linearized}) is rewritten in the matrix form
\begin{align}
\begin{bmatrix}
2i\omega\omega_{H}
& -\left(  \omega^2-\omega_{a}\omega_{K}+\omega_{H}^2\right) \\
\omega^2-\omega_{a}\omega_{K}+\omega_{H}^2
& 2i\omega\omega_{H}
\end{bmatrix}
\begin{bmatrix}
\delta \tilde{n}_{x}(\omega) \\ \delta \tilde{n}_{y}(\omega)
\end{bmatrix}
=\omega_{a}\delta(\omega_{0}-\omega)
\begin{bmatrix}
\mathcal{D}_x \\ \mathcal{D}_y
\end{bmatrix}.
\end{align}
Multiplying the inverse matrix from the left hand side, we obtain
\begin{align}
\begin{bmatrix}
\delta \tilde{n}_{x}(\omega) \\ \delta \tilde{n}_{y}(\omega)
\end{bmatrix}
=\ 
\begin{bmatrix}
\chi_1(\omega)
&  \chi_2(\omega) \\
-\chi_2(\omega)
& \chi_1(\omega)
\end{bmatrix}
\begin{bmatrix}
\mathcal{D}_x \\ \mathcal{D}_y
\end{bmatrix},
\label{transverse-AFMR-matrix}
\end{align}
where the susceptibility is defined as
\begin{align}
\begin{bmatrix}
\chi_1(\omega)
&  \chi_2(\omega) \\
-\chi_2(\omega)
& \chi_1(\omega)
\end{bmatrix}
=&\ 
\frac{
\omega_{a}\delta(\omega_{0}-\omega)
}{(\omega^2-\omega_+^2)(\omega^2-\omega_-^2)}
\begin{bmatrix}
2i\omega\omega_{H}
&  \omega^2-\omega_{a}\omega_{K}+\omega_{H}^2 \\
-\left( \omega^2-\omega_{a}\omega_{K}+\omega_{H}^2\right)
& 2i\omega\omega_{H}
\end{bmatrix}.
\label{susceptibility}
\end{align}
Here,
\begin{align}
\omega_{\pm}=\omega_{H}\pm\sqrt{\omega_{a}\omega_{K}}
\end{align}
are the resonance frequencies.
Note that these frequencies do not depend on the parameters of the driving force $\bm{\mathcal{D}}$.

Along with Eq.~(\ref{LLG-linearized}), the following equation is obtained from $2\gamma(\bm{n}\cdot\bm{H})\dot{\bm{n}}+\gamma(\bm{n}\cdot\dot{\bm{H}})\bm{n}$ in Eq.~(\ref{Eq-for-n}), which describes the ``longitudinal'' AFMR state:
\begin{align}
2\omega_H\delta \dot{n}_z\bm{e}_z=i\gamma\omega_0e^{-i\omega_0 t}(h_{\mathrm{RF}}^x\delta n_x+h_{\mathrm{RF}}^y\delta n_y)\bm{e}_z.
\end{align}
Fourier transforming this equation and substituting the solution for $\delta n_x$ and $\delta n_y$ [Eq.~(\ref{transverse-AFMR-matrix})] into it, we have
\begin{align}
\delta \tilde{n}_z(\omega)&\propto h_{\mathrm{RF}}^x\delta \tilde{n}_x(\omega-\omega_0)+h_{\mathrm{RF}}^y\delta \tilde{n}_y(\omega-\omega_0)
\propto \frac{
\delta(2\omega_{0}-\omega)}{[(\omega-\omega_0)^2-\omega_+^2][(\omega-\omega_0)^2-\omega_-^2]},
\label{Longitudinal-mode}
\end{align}
which indicates that the resonance frequencies of the longitudinal AFMR are $\omega=2\omega_0=2\omega_\pm$.
Eq.~(\ref{Longitudinal-mode}) reveals that the longitudinal mode is quadratic in the RF field, i.e., a second-order response to the RF field, while the transverse mode [Eq.~(\ref{transverse-AFMR-matrix})] is a linear response to the RF field.

\section{Axion Dark Matter and the Millielectronvolt Range}\label{sec:axion_dm}

Since their initial proposal as a solution for the Strong~$\mathcal{C}\mathcal{P}$ problem more than 40\,years ago~\cite{pecceiquinn1977,pecceiquinn1977_2,weinberg1978,wilczek1978}, (QCD)~axions have seen phases of growing interest due to a number of breakthroughs. The first was the realisation that axions are excellent dark matter candidates~\cite{1983PhLB..120..127P,1983PhLB..120..133A, 1983PhLB..120..137D,1986PhRvD..33..889T}, and that there are several ways to search for them experimentally~\cite{1983PhRvL..51.1415S,1985PhRvD..32.2988S,1989PhRvD..39.2089V,doi:10.1146/annurev.nucl.012809.104433}. Recently, there has been a huge growth of new ideas for axion searches~(see Ref.~\cite{1801.08127} for a review), which includes the present proposal (``Paper~I'') using topological insulators~\cite{2019PhRvL.123l1601M}.

The QCD~axion was originally proposed as the pseudo-Goldstone boson of a spontaneously broken global $U(1)$ symmetry, which couples to chiral fermions charged under the strong nuclear force, $SU(3)_c$ gauge symmetry (i.e.\ quarks). Such a global symmetry is known as a Peccei-Quinn~(PQ) symmetry, $U(1)_\text{PQ}$. More generally, QCD~axions can be regarded as pseudo-Goldstone bosons coupled to the QCD anomaly term, schematically $G\tilde{G}$, where $G$ is the gluon field strength tensor, and $\tilde{G}$ its dual.

The PQ symmetry breaking scale, $v_\text{PQ}$, is not predicted by the theory, although it is expected to be below the reduced Planck scale, $\Mpl = \SI{2.4e18}{\GeV}$~\cite{2007JHEP...06..060A}. The symmetry breaking scale sets the axion mass, which arises due to the axion's coupling to the QCD topological charge via $SU(3)_c$ instantons, and which reaches its zero-temperature value at temperatures lower than the QCD crossover temperature of around $\SI{157}{\MeV}$~\cite{1812.08235,2002.02821}. The axion mass at such temperatures is given by~\cite{1812.01008}
\be
	\mDA = \frac{\sqrt{\chi_0}}{\fDA} = \SI{5.69(5)}{\meV} \left( \frac{\SI{e9}{\GeV}}{\fDA} \right) \, ,
\ee
where $\chi_0$ is the zero-temperature QCD topological susceptibility, $\fDA = v_\text{PQ}/\mathcal{N}$, and $\mathcal{N}$ is the $SU(3)_c$ anomaly of $U(1)_\text{PQ}$. The value of $\chi_0$ can be calculated from chiral perturbation theory~\cite{weinberg1978,diCortona:2015ldu,1812.01008} while, at higher temperatures, it can be calculated using lattice quantum field theory~(see e.g.\
Ref.~\cite{1606.07494}), or using instanton methods~\cite{1981RvMP...53...43G,Wantz:2009it}.

The QCD~axion couples to the EM Chern-Simons term via two means. Firstly, by its model-independent mixing with pions, and secondly via the (model-dependent) electromagnetic anomaly~($\mathcal{E}$) of fermions charged under $U(1)_\text{PQ}$. The coupling is~\cite[e.g.][]{diCortona:2015ldu} 
\begin{equation}
	\Delta \mathcal{L} = \gDA \, a \, \vc{E}\cdot\vc{B}\, ,
\label{eqn:dark_axion_chern_simons}
\end{equation}
where $a \equiv \fDA \, \theta$ is the canonically normalised axion field and $\gDA$ is the axion-photon coupling, which is given by
\be
	\gDA = \frac{\alphaEM}{2\pi \fDA} \, C_{a\gamma} = \frac{\alphaEM}{\pi \fDA} \left[ \frac{\mathcal{E}}{\mathcal{N}} - 1.92(4) \right] \, ,
	\label{eq:def_gag}
\ee
where $\mathcal{N}$ is the $SU(3)_c$ anomaly of the PQ symmetry, and is equal to unity in the KSVZ model, while for the DFSZ model $\mathcal{N}=6$. The value of $\mathcal{E}/\mathcal{N}$ depends on the PQ charges and gauge group representations of fermions. We define the QCD model band according to the ``preferred'' models of ref.~\cite{1705.05370}, which corresponds to $5/3<\mathcal{E}/\mathcal{N}<44/3$. Experiments constrain $|\gDA|$, and so this band and encompasses the original KSVZ ($\mathcal{E}/\mathcal{N}=0$) and DFSZ ($\mathcal{E}/\mathcal{N}=8/3$) models. For a generic ``axion-like particle'', the coupling $\gDA$ is taken as a free parameter independent of $\mDA$. 

The QCD~axion mass is bounded from above and below by astrophysical constraints. The existence of BHs with masses of order ten solar masses with high spins, stable over astrophysical timescales, would be impossible if the QCD~axion existed and $\mDA\lesssim\SI{e-12}{\eV}$~\cite{2015PhRvD..91h4011A,Stott:2018opm,Stott:2020gjj}. In such a case, the axion Compton wavelength is resonant with the size of the BH ergoregion, causing axions to be abundantly created from vacuum fluctuations, and rapidly draining the spin of the BH. On the other end of the mass scale, the QCD~axion with $\mDA\gtrsim  \SI{0.02}{\eV}$ is excluded by observations of neutrinos coinciding with the galactic supernova SN1987A~\cite{2008LNP...741...51R,Chang:2018rso}. The QCD~axion couples to nuclei in the supernova, and axions are emitted by nuclear bremsstrahlung, cooling the supernova more rapidly and shortening the neutrino burst if the axion-nucleon coupling (proportional to $\mDA$) is too large. Since there is no statistically rigorous bound associated with SN1987A, we also mention that a looser upper limit on $\mDA$ can be derived from constraints on the relativistic energy density in the early Universe (parameterised as a hot DM component). Hot QCD axions are produced by their interaction with pions. The amount of hot axions produced is in conflict with the cosmic microwave background anisotropies as measured by the \emph{Planck} satellite~\cite{2016A&A...594A..13P,Aghanim:2018eyx} if $\mDA\gtrsim \SI{0.3}{\eV}$ (see e.g. refs.~\cite{2011.14704,DiLuzio:2020wdo}).

In the mass range of interest for TOORAD, the axion-photon coupling is mostly constrained by axion helioscopes~\cite{1705.02290} and cooling of Horizontal~Branch~(HB) stars through the ratio of HB and Red Giant Branch stars~\cite{1406.6053,1512.08108}, which both lead to limits of the order $\gDA \lesssim \SI{e-10}{\GeV^{-1}}$. This is far above the coupling for typical QCD axion models.

Assuming that the QCD~axion indeed composes the observed DM with cosmic density parameter $\Omega_d h^2=0.12$~\cite{Aghanim:2018eyx}, it is possible to analyse the value of $\mDA$ further (for a review, see Ref.~\cite{my_review}). If the maximum temperature of the Universe exceeds the PQ phase transition temperature, $T_\text{PQ}\sim v_\text{PQ}$, or if the Hubble scale during inflation $H_I > 2\pi v_\text{PQ}$, then the PQ symmetry is unbroken at the end of inflation. When it subsequently breaks, the Kibble mechanism leads to a network of topological strings that persists as the Universe expands and, when the axion mass becomes cosmologically relevant $\mDA\sim H$ (where $H$ is the Hubble parameter) domain walls form~\cite{1976JPhA....9.1387K,1980PhR....67..183K,1981PhRvD..24.2082V}. Such defects emit DM axions, and eventually decay when $H\ll \mDA$. In this case, if $\mathcal{N}=1$ (which imnolies the domain wall network is unstable, then the DM abundance is in principle calculable and depends only on $\mDA$. However the dynamics of the strings and domain walls are complex and the axion abundance can only accurately be determined by numerical simulations. This computation cannot be performed at the physical scale separation (between the string thickness and the Hubble length, which in turn sets the string tension), and so the results must be extrapolated. In the case $\mathcal{N}>1$, the DM abundance depends on an additonal biasing parameter required such that the domain wall network decays~\cite{Hiramatsu:2012sc}.

When $\mathcal{N}=1$, recent simulations~\cite{2007.04990} have placed a lower bound on the QCD~axion mass $\mDA \gtrsim \SI{0.5}{\meV}$ by considering axions emitted by the string network prior to the axion mass becoming relevant, a well controlled extrapolation to the physical scale separation, and a computation of non-linear effects after the network decays.\footnote{Other extrapolations to the physical scale separation lead to lower bounds on the mass~\cite{Klaer:2017ond,Buschmann:2019icd}.} This work also estimates the bound $\mDA \gtrsim \SI{3.5}{\meV}$ when $\mathcal{N}>1$, in agreement with the general expectation that the axion mass should be larger in this scenario~\cite{Hiramatsu:2012sc,Caputo:2019wsd,Ringwald:2015dsf}. One issue for axion direct detection in the post inflation scenario is the existence of axion ``miniclusters''~\cite{Hogan:1988mp,Kolb:1993zz,Kolb:1994fi}. Recent simulations of structure formation in this scenario suggest that a large fraction of the DM (50\% or more at the solar radius in the Milky Way) is bound in dense, low mass objects~\cite{2017JETP..125..434D,Vaquero:2018tib,Eggemeier:2019khm,Kavanagh:2020gcy}. These objects have a low collision cross section with the Earth, and reduce the effective value of the local DM density for a direct detection experiment.

In the alternative scenario for axion production, the PQ symmetry is broken in the very early Universe during the hypothetical period of inflation~\cite{1981PhRvD..23..347G,1982PhLB..108..389L,1982PhRvL..48.1220A}, and axions are subsequently produced when the initial vacuum state of the axion decays, in a process called ``realignment''. This scenario has more free parameters than just $\mDA$, and it is not possible to predict the axion mass based on the observed DM abundance. This scenario is incompatible with a large energy scale of inflation, and would be ruled out if primordial gravitational waves were observed~\cite{hertzberg2008}. If the initial vacuum value of $\theta$ is assumed to be of order 1 the axion mass in the pre-inflationary scenario is bounded to $\mDA \gtrsim \SI{0.7}{\micro\eV}$~\cite{1810.07192}.  However, anthropic pressure due to the need for DM to form galaxies can allow for much lower or higher values of the mass in this scenario~\cite{2004hep.ph....8167W,2006PhRvD..73b3505T}. Limits on the mass in this case are only imposed with additional assumptions on the energy scale of inflation, which limit the allowed level of tuning on the free parameter.

\section{Comparison to earlier results} 

\begin{figure}
\centering
\includegraphics[width=5in]{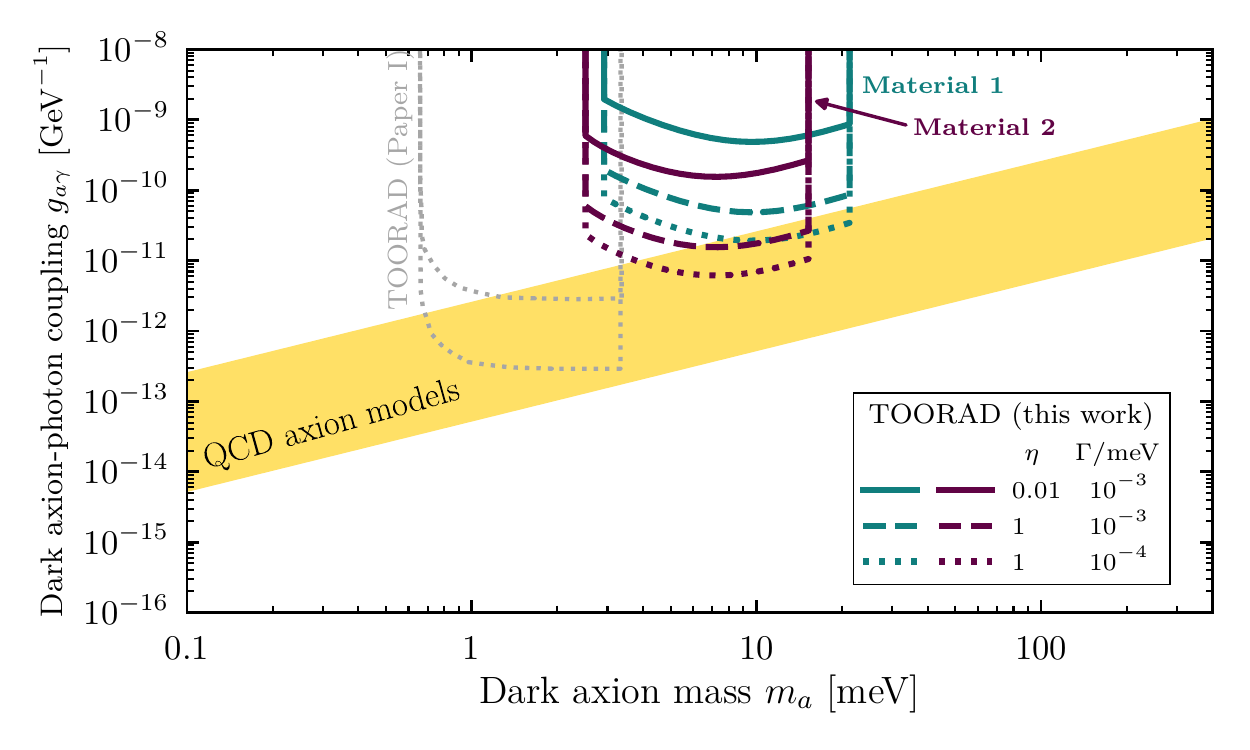}
\caption{Comparison between the forecasts made in Paper~I (gray) and those in the present work (coloured lines).}
\label{fig:paper_i_comparison}
\end{figure}

The forecasts shown in fig.~\ref{fig:summary_plot} differ in many respects from those in Paper~I, and we explain briefly why, see fig.~\ref{fig:paper_i_comparison}, which shows the same projection alongside those of Paper~I. For the detector, Paper~I assumed a coupling factor equivalent to efficiency $\eta=0.01$, and the same dark count rate as in the present work. The difference in the scan depth arises because the power output in Paper~I was computed in analogy to a resonant cavity, and this formula is incorrect for a medium with a polariton resonance. Allowing for a few translations, however, the results can be compared. All in all, however, we stress that the power formula in Paper~I was far too simplistic, and so should not be used. 

Our present corrected calculations have shown that the power does not scale with the material volume as assumed in Paper~I: rather it depends separately on the surface area 
and the material thickness. Losses lead to a maximum useful thickness, an effect which was not accounted for in Paper~I. For the models presently considered, this leads to a total useful material volume around 40 cm\,$^3$. Thus for comparison, we show the Paper~I estimates for ``Stage I'', which used a total volume of material 1 cm$^3$, and `Stage II'', which used a total volume of material 100 cm$^3$, equivalent to $d\approx 1\,\text{mm}$ in the present case. The ``Stage III'' volume considered in Paper~I is inaccessible due to the finite skin depth induced by realistic $\mathcal{O}(\mu{\rm eV})$ losses. 

Paper~I included losses only by a rough estimate for the bulk quality factor, which was taken as $Q=10^5$, roughly $\Gamma = \SI{e-5}{\meV}$ (this assumed power law decreases in $\Gamma_m$ at low $T$ as discussed in the present work, but neglected the impurity and conductance contributions). On the other hand, Paper~I assumed that the power was reduced by a polariton mode mixing factor, $f_+$, which is absent in the present treatment (mode mixing mostly affects the width of the resonance). Together, these amount to an assumed $\beta^2\approx10^4$ for some baseline parameters. Comparing to fig.~\ref{fig:betaloss}, using the correct computations from the present paper, such a large $\beta$ could only be achieved with losses $\Gamma = \SI{e-5}{\meV}$. These many considerations explain the different depth of the constraints in terms of $\gDA$ in the present work compared to Paper~I.

The difference in the scanned mass range in the present work compared to Paper~I is more useful and necessary to explain. It arises from the adopted values of $\mAQ$ and $\fAQ$. In the present work we take our preferred material as \MnBiTe{2}{2}{5}, while Paper~I uses \BiFeSe. Even so, there are differences to the \BiFeSe parameter estimates in Paper~I compared to the present work. In Paper~I we erroneously assumed that $\mAQ$ was equal to the AFMR frequency of the transverse magnon polarisations (with a reduction due to the doping required in \BiFeSe\,) leading to $\mAQ=\SI{0.6}{\meV}$, and thus a lower minimum value of $\mDA$ in the Paper~I treatment. Paper~I also incorrectly included the Kittel shift to $\mAQ$. As discussed, the longitudinal magnon is not simply related to the transverse modes and Paper~I should have used the value of $\mAQ$ computed by Ref.~\cite{2010NatPh...6..284L}, as we do in the present work. However, as we have noted, Ref.~\cite{2010NatPh...6..284L} used a square lattice approximation to compute $\mAQ$. We do not know the error induced by this assumption on the lower limit of the scannable mass range also in the present work. 

Furthermore, Paper~I assumed $\fAQ$ for \BiFeSe directly from Ref.~\cite{2010NatPh...6..284L}. In the present work we corrected this value using more up to date estimates of the bulk band gap of \BiFeSe including the effects of magnetic doping, leading to the ``Material 1'' parameter estimates with lower $\fAQ$. The lower value of $\fAQ$ in the present work allows for a wider range of masses to be scanned for the same range of $B$-field strengths.

\renewcommand{\baselinestretch}{0.25}\normalfont
\bibliographystyle{JHEP}
\bibliography{bibliography}

\end{document}